\documentclass[11pt,a4paper]{article}
\usepackage{journal_shortcuts}

\newcommand{\astroh}{{\it ASTRO-H}\,}
\def \nustar{{{\it NuSTAR}}~}
\def \xmm{{{\it XMM-Newton}}~}

\def \chandra{{{\it Chandra}}~}

\def \suzaku{{{\it Suzaku}}~}

\usepackage{ascmac}
\usepackage{amsmath}
\usepackage{amssymb}
\usepackage{bm}
\usepackage[footnotesize,bf]{caption}
\usepackage{fullpage}
\usepackage{color}
\usepackage{float}
\usepackage{graphicx}
\usepackage[]{natbib}
\usepackage{subfigure}
\usepackage{txfonts}
\usepackage{threeparttable}
\usepackage{wrapfig}
\usepackage[]{multicol}
\usepackage{wrapfig}
\usepackage{authblk}

\usepackage{floatflt}

\title{\large Complete list of the ASTRO-H Science Working Group}
\date{\vspace{-0.5cm}}
\newcommand{\MakeWhitePaperTitle}{
	\begin{center}
		\begin{figure}
			\vspace{1cm}
			\begin{center}
				\includegraphics[width=0.2\hsize]{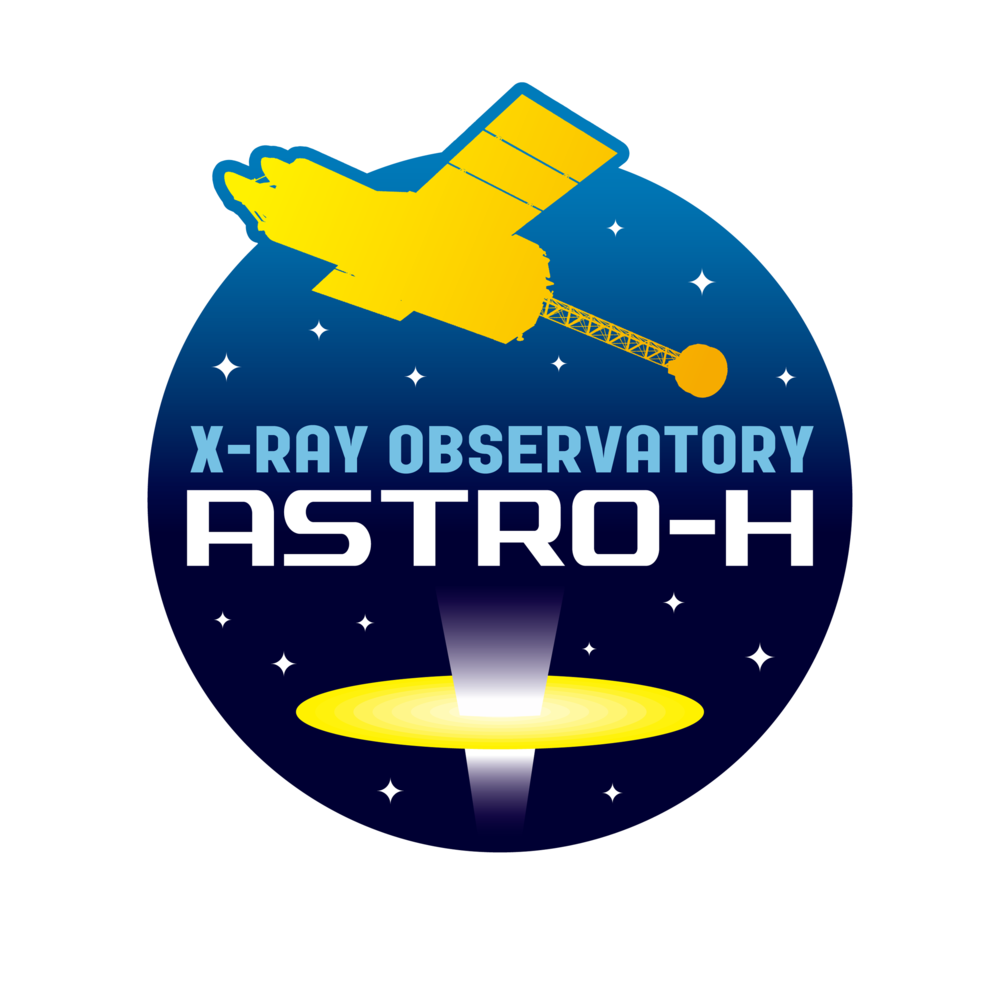}
			\end{center}
		\end{figure}
		\vspace{1cm}
		{\LARGE
		ASTRO-H Space X-ray Observatory\\
		White Paper\\
		}
		\vspace{5mm}
		{\large
		\WhitePaperTitle\\
		}
		\vspace{1cm}
		{
		\WhitePaperAuthors\\
		on behalf of the ASTRO-H Science Working Group
		}
	\end{center}
}

\usepackage{authblk}
\author[a]{Tadayuki~Takahashi}
\author[a]{Kazuhisa~Mitsuda}
\author[b]{Richard~Kelley}
\author[c]{Felix~Aharonian}
\author[d]{Hiroki~Akamatsu}
\author[e]{Fumie~Akimoto}
\author[f]{Steve~Allen}
\author[g]{Naohisa~Anabuki}
\author[b]{Lorella~Angelini}
\author[h]{Keith~Arnaud}
% ...
\affil[a]{Institute of Space and Astronautical Science (ISAS), Japan Aerospace Exploration Agency (JAXA), Kanagawa 252-5210, Japan}
\affil[b]{NASA/Goddard Space Flight Center, MD 20771, USA}
\affil[c]{Astronomy and Astrophysics Section, Dublin Institute for Advanced Studies, Dublin 2, Ireland}
\affil[d]{SRON Netherlands Institute for Space Research, Utrecht, The Netherlands}
\affil[e]{Department of Physics, Nagoya University, Aichi 338-8570, Japan}
\affil[f]{Kavli Institute for Particle Astrophysics and Cosmology, Stanford University, CA 94305, USA}
\affil[g]{Department of Earth and Space Science, Osaka University, Osaka 560-0043, Japan}
\affil[h]{Department of Astronomy, University of Maryland, MD 20742, USA}
% ...

\def\kms{km$\;$s$^{-1}$}
\def\lesssim{\mathrel{\hbox{\rlap{\hbox{\lower4pt\hbox{$\sim$}}}\hbox{$<$}}}}
\def\gtrsim{\mathrel{\hbox{\rlap{\hbox{\lower4pt\hbox{$\sim$}}}\hbox{$>$}}}}
\def\lax{\lesssim}
\def\gax{\gtrsim}
\def\grtsim{\mathrel{\hbox{\rlap{\hbox{\lower4pt\hbox{$\sim$}}}\hbox{$>$}}}}
\def\lesssim{\mathrel{\hbox{\rlap{\hbox{\lower4pt\hbox{$\sim$}}}\hbox{$<$}}}}

\begin{document}

\newcommand{\WhitePaperTitle}{Clusters of Galaxies and Related Science}
\newcommand{\WhitePaperAuthors}{
	T.~Kitayama~(Toho~University), M.~Bautz~(MIT),
	M.~Markevitch~(NASA/GSFC), K.~Matsushita~(Tokyo~University~of~Science),
	S.~Allen~(Stanford~University), M.~Kawaharada~(JAXA),
	B.~McNamara~(University~of~Waterloo), N.~Ota~(Nara~Women's~University), 
	 H.~Akamatsu~(SRON),  
	J.~de~Plaa~(SRON), M.~Galeazzi~(University of~Miami), 
	G.~Madejski~(Stanford~University), 
	R.~Main~(University~of~Waterloo),
	E.~Miller~(MIT), K.~Nakazawa~(University~of~Tokyo), 
	H. Russel (University of Waterloo\footnote{Also at Institute of Astronomy, University of Cambridge}), 
	K.~Sato~(Tokyo~University~of~Science), 
	N.~Sekiya~(JAXA/University~of~Tokyo), A.~Simionescu~(JAXA), T.~Tamura~(JAXA), 
	Y.~Uchida~(JAXA/University~of~Tokyo), E.~Ursino~(University of~Miami), 
	N.~Werner~(Stanford~University), I.~Zhuravleva~(Stanford~University), and J.~ZuHone (NASA/GSFC)
}

\MakeWhitePaperTitle

\begin{abstract}
The next generation X-ray observatory {\it ASTRO-H} will open up a new
dimension in the study of galaxy clusters by achieving for the first
time the spectral resolution required to measure velocities of the
intracluster plasma, and extending at the same time the spectral
coverage to energies well beyond 10 keV. This white paper provides an
overview of the capabilities of {\it ASTRO-H} for exploring gas motions in
galaxy clusters including their cosmological implications, the physics
of AGN feedback, dynamics of cluster mergers as well as associated
high-energy processes, chemical enrichment of the intracluster medium,
and the nature of missing baryons and unidentified dark matter.
\end{abstract}

\maketitle
\clearpage

\tableofcontents
\clearpage

\section{Executive Summary: Opening a New Dimension in Galaxy Cluster Research}

\noindent The {\it ASTRO-H} satellite, the first X-ray observatory with
a focal plane calorimeter, will open up a new dimension in the research
of galaxy clusters.  The high energy resolution of the non-dispersive
Soft X-ray Spectrometer (SXS) will make it possible to detect dozens of
emission lines from highly ionized ions and measure, for the first time,
their line profiles with sufficient accuracy to study gas motions.  The
Hard X-ray Imager (HXI) will extend the simultaneous
spectral coverage to energies well above 10 keV, which is critical for
studying both thermal and nonthermal gas in clusters.

The present white paper aims to demonstrate the above capabilities
explicitly on representative galaxy clusters to aid and encourage the
broader astrophysical community in developing \astroh science. It is a
compilation of documents showing detailed feasibility studies done by
the \astroh Science Working Group in an effort to select targets for the
Performance Verification phase of the mission.  As such, each section is
intended to provide a self-contained description of its subject and to
be accessible independently of the other sections. In this section, we
summarize a range of science topics covered in the paper and guide the
readers to the subsequent sections for more details of new results
expected from early observations of {\it ASTRO-H}.\footnote{Coordinators
of this section: M.~Markevitch, T.~Kitayama, S.~W.~Allen, K.~Matsushita}

\subsection*{From cluster dynamics to plasma microphysics}

Clusters of galaxies form via the infall and accretion of surrounding
matter, including smaller clusters and galaxy groups.  In the course of
a merger, a significant fraction of the energy of the infalling gas is
thermalized via shocks.  How efficiently this happens depends on the
microphysics of the intracluster medium (ICM) dominated by plasma with
temperatures $10^7-10^8$K.  The ICM is permeated by weak, tangled
magnetic fields, which makes its basic properties, such as its thermal
conductivity and viscosity, difficult to model. To complicate the
picture, many merging clusters host diffuse synchrotron radio halos,
produced by ultra-relativistic electrons coexisting with the thermal
ICM. These relativistic particles are believed to be re-accelerated by
ICM turbulence. Turbulence also reorders and amplifies the ICM magnetic
fields.

A critical, missing piece of this picture
is the direct observation of gas motions, both streaming and
turbulent. Turbulence should be sensitive to the ICM viscosity. By
measuring ICM bulk velocities and turbulence, and
comparing these data with detailed magnetohydrodynamic
(MHD) simulations, we can hope to improve our understanding of the most
important physical aspects of the ICM.

The brightest, nearest clusters, which provide the strongest signal and
the finest spatial information, are natural early targets for these
measurements.  Observations toward the Perseus and Coma clusters, for
example, will provide a first-ever detection of intracluster turbulence
and constrain its power spectrum in a relatively relaxed cluster and in
one that has experienced a recent merger, respectively
(Sec. \ref{sec-perseus} and \ref{sec-coma}). Observations of nearby
clusters harboring cold fronts, such as A3667, will
provide further information on the plasma physics
operating in intermediate stages of clusters growth.

These three clusters also exhibit (non-thermal) radio emission of
various morphologies.  While associated non-thermal X-ray emission has
so far been elusive, the deep HXI observations made simultaneously with
the SXS studies will provide new limits and will also
map the hottest thermal components recently discovered
in some clusters (Sec.  \ref{sec-highe}).

\subsection*{Direct contributions to cosmology}


Clusters are the largest virialized structures in the Universe, and are
thus sensitive probes for cosmology including the nature of dark
energy. The ability of clusters to constrain cosmology depends
critically on the accuracy with which cluster masses can be
determined. One of the most important techniques for measuring cluster
masses uses X-ray observations and relies on the assumption that the ICM
is in hydrostatic equilibrium. Any break of this key assumption, for
example due to bulk or turbulent motions in the ICM, can thus lead to a
bias in the cosmological constraints from X-ray clusters.

SXS observations of bright, nearby ($z<0.1$) clusters will place
 strong limits on deviations from hydrostatic equilibrium.  Pressure
 contributions from gas motions will be measured to an accuracy of a few
 percent in apparently relaxed clusters such as A2029
 (Sec. \ref{sec-nonth}). While a substantial investment of the \astroh\
 observing time is required,
measurements of turbulent and bulk motions up to radii $\sim r_{\rm
 2500}$ \footnote{$r_{\rm 2500}$ is the radius within which the average
 cluster matter density is 2500 times the critical density of the
 Universe; about 20\% of the virial radius. It is often chosen for
 cosmological tests because the X-ray emission is easily measurable and
 the hydrostatic equilibrium assumption is expected to hold there.}  for
 a sample of nearby relaxed clusters will contribute significantly to
 improved constraints on dark energy and other cosmological parameters.


High-resolution X-ray spectroscopy is also a unique tool for detecting
the Warm-Hot Intergalactic Medium (WHIM), which is predicted to contain
most of the cosmic baryons at low redshifts and which has eluded
detection so far.  Current searches are limited to absorption lines in
the spectra of distant quasars. SXS offers a chance to detect the denser
regions of WHIM in emission.  
Deep, well-selected pointings hold the potential to improve our
understanding of the WHIM (Sec. \ref{sec-whim}).

Finally, the nature of dark matter is among the most important
  unresolved questions in all of science.
Certain plausible warm dark matter candidates, such as a hypothetical
sterile neutrino in the $\sim$ keV mass range, are predicted to decay,
emitting a line in the soft X-ray energy band.  If present, SXS may
detect such a line from the dark matter concentrations being observed
for other purposes, e.g., in galaxy clusters and the Galactic Center
(Sec. \ref{sec-dm}).

\subsection*{The physics of AGN feedback}

Feedback from active galactic nuclei (AGN) plays an important role in
galaxy formation, preventing runaway cooling in cluster cores and
causing the largest, elliptical galaxies to appear `red and dead'.  AGN
inflate `bubbles' filled with relativistic plasma, displacing the
ambient X-ray 
gas and driving weak shocks, and dragging the coolest, lowest entropy
material (the fuel for future star formation) up and out of galaxy and
cluster centers.  We outline a strategy for studying feedback in the
bright Virgo and Perseus clusters, where ongoing feedback is observed in
their central galaxies, M87 and NGC~1275, respectively
(Sec. \ref{sec-perseus} and \ref{sec-virgo}).
For M87 in particular, high-resolution SXS spectra will reveal the
dynamics and microphysics of AGN feedback, allowing us to map the
line-of-sight velocity component of the gas uplifted by the AGN. The
data should also allow us to measure turbulence induced by the rise of
the radio bubbles and by AGN-driven shock fronts. In addition, SXS
observations will robustly measure how AGN-driven gas motions spread
metals produced in central galaxies out into the surrounding ICM.

\subsection*{The chemical composition of the ICM}

The ICM provides a reservoir storing essentially all of the metals ever
ejected from cluster galaxies. Measurements of metal abundances in the
ICM therefore provide a unique insight into the history of star
formation and the evolution of galaxies. Galaxies in clusters are mostly
early-type and have a different chemical evolution history than the
Milky Way. Most metals synthesized in cluster galaxies escape into the
ICM, but the deep gravitational potential wells of clusters keep them
locked within the virial radii.

The ICM is optically thin and in, or very close to, collisional
ionization equilibrium, with heavy elements highly ionized. Therefore,
spectral modeling of the ICM is relatively simple. With previous X-ray
satellites, emission lines of highly abundant elements, such as O, Si, S
and Fe in the ICM have been detected. \astroh\ will be able to detect
weaker lines from rare elements like Al, Na, Mn, and Cr from the
brightest cluster cores. 
The abundances of various elements provide sensitive tests of
the nature of the stellar population of the cluster galaxies and their
IMF, as well as information about the physics of SN Ia explosions.



For bright clusters, the above study can be done using the same
data sets as other science goals (Sec. \ref{sec-chem}). For example, due
to its relatively low ICM temperature, the Virgo cluster will produce
prominent lines below $E\sim 3$ keV, making it an ideal target for
precise measurements of the abundances of elements from N to Al. The
Perseus cluster is a target of choice for accurate abundance
measurements of the iron-peak elements, Mn, Cr, Fe and Ni.  Measurements
of the radial abundance profiles will also help us better understand the
enrichment history of the ICM.

\subsection*{Organization of the paper}

As indicated above, coordinated observations of a carefully selected
sample of targets are crucial for maximizing the scientific outcome of
the mission. Subsequent sections of this paper are thereby ordered as
follows. We first focus on two representative objects, the Perseus
cluster and the Virgo cluster in Sections \ref{sec-perseus} and
\ref{sec-virgo}, respectively; the former is the {\it brightest}
extragalactic extended X-ray source, which yields the highest photon
statistics essential for high-resolution spectroscopy, whereas the
latter is the {\it nearest} galaxy cluster which enables us to zoom into
the closest environments of the central AGN and associated feedback
processes. Thereafter, we discuss each of key science topics more
systematically for a sample of clusters. Section \ref{sec-nonth}
presents a search for turbulence at
larger radii in several relaxed clusters including A2029 and its
implications on the mass measurements. Sections \ref{sec-coma} and
\ref{sec-highe} explore the capabilities of SXS and HXI, respectively,
for studying dynamics, turbulent power
spectrum, and high energy processes
in merging clusters such as Coma and A3667. The significance and
prospects of elemental abundance measurements in a variety of clusters
are summarized in Section \ref{sec-chem}. Finally, Sections
\ref{sec-whim} and \ref{sec-dm} 
describe possible spectroscopic searches for missing
baryons and dark matter, respectively.

\bigskip 

\newcommand{\ion}[2]{#1\,{\sc{#2}}}
\def\disp {\displaystyle}

\section{Nature of Gas Motions in the X-ray Brightest Galaxy Cluster}
\label{sec-perseus}

\subsection*{Overview}

The Perseus Cluster, being the X-ray brightest cluster in the sky,
serves as a prime target for high-resolution spectroscopic studies by
\astroh.
Previous observations of this system have led to a series of landmark
discoveries and we expect \astroh observations to do the same.
\astroh will reveal in unprecedented detail the nature of gas
motions, and the thermodynamic and chemical structure of the ICM, which
are critically important for understanding galaxy formation and for
cosmological studies using clusters.  A mosaic of observations 
sampling the cluster from the core to $r_{2500}$ will
provide the first constraints on the typical spatial scales and
velocities of turbulent gas motions. The same data will simultaneously
probe resonance scattering in the core, providing additional constraints
on the anisotropy of gas motions.  The results for Perseus will provide
an essential benchmark for studies of more distant, relaxed clusters
used for evolutionary studies and cosmological tests, and address
directly the energy dissipation in cluster cores (crucial for keeping
the balance between heating and cooling), the chemical composition of
the ICM, and the bias in mass measurements due to turbulence and bulk
motions.\footnote{Coordinators of this section: N. Werner, S. W. Allen,
T. Tamura}

\subsection{Background and Previous Studies}

\subsubsection{Gas motions in cool core clusters}

To answer some of the most important questions regarding the formation
and evolution of galaxies and clusters of galaxies requires detailed
knowledge of the dynamics and physics of the hot, X-ray emitting
ICM. Critical questions include: how do AGN and merger induced bulk
motions, turbulent eddies, shocks and sound waves dissipate energy into
the surrounding medium?  How efficiently does mixing between the
different gas phases occur? Answering these questions requires
observational constraints on gas motions, viscosity, and thermal
conduction.

Constraints on gas motions from the broadening of emission lines have so
far only been obtained for the cooling cores of galaxy clusters with
strongly centrally peaked surface brightness distributions, using the
reflection grating spectrometers on board \xmm \citep[the best
constraints on turbulent velocities are of the order of few hundred
km~s$^{-1}$;][]{sanders2010,sanders2011turb,sanders2012}. Because the
\xmm gratings operate as slitless spectrometers, spectral lines are
broadened by 
the width of the spatial distribution of the emitting ion
in the source and the results therefore have large systematic
uncertainties.  Observations of line broadening with \astroh SXS will
not suffer from these uncertainties and will therefore
tightly constrain small scale ($\sim 25$ kpc) gas motions to a level more than an order of magnitude more
precise. Possible sources of small scale turbulence include AGN activity
(turbulence induced in the wakes of rising bubbles) and the shear
induced by gas sloshing. Observations of line shifts, on the other hand,
will allow us to measure coherent bulk motions of the gas in the core of
the cluster.

\subsubsection{The Perseus Cluster}

The Perseus Cluster
is a massive, relatively relaxed system at a distance of
only 68 Mpc ($z= 0.0179$).  \astroh observations of this target will
provide benchmark measurements of turbulent and ordered motions in the
hot X-ray emitting 
ICM, and of its thermal and chemical structure.  \astroh observations of
the Perseus Cluster will inform studies of more distant systems,
including those used for cosmological work.

Extended X-ray emission associated with the Perseus Cluster was
discovered by the Uhuru satellite \citep{gursky1971,forman1972} and this
source has been studied extensively by all the subsequent X-ray
missions. The cluster X-ray emission was found to be strongly peaked at
the position of the brightest cluster galaxy NGC~1275
\citep{fabian1981}, which harbors a powerful X-ray and radio bright
active galactic nucleus (AGN). ROSAT images of the region around
NGC~1275 revealed two `holes' in the X-ray emission coincident with the
radio lobes \citep{bohringer1993,mcnamara1996,churazov2000}. These
observations were critical in revealing how the radio bright jets
emanating from the central AGN inflate bubbles in the surrounding ICM,
displacing the thermal gas and heating the surrounding ICM. This AGN
feedback mechanism is now one of the most vigorously discussed topics in
astrophysics, with important implications for the physics of galaxy
formation, galaxy clusters, and growth of supermassive black holes.
More recent studies of the Perseus Cluster with {\it Chanrda}, 
{\it XMM-Newton}, and \suzaku further revolutionized our understanding of AGN
feedback and cluster physics. Observations of the bright cluster core
with \chandra revealed a series of X-ray faint cavities inflated by the
central AGN, as well as weak shocks and ripples (likely sound waves)
being driven by expanding cavities. These shocks and sound waves can in
principle heat the ICM isotropically. These data also showed a
continuous uplift of cooler multi-phase gas in the wakes of the rising
bubbles and ridges of high metallicity in the ICM
\citep{fabian2000,fabian2003,fabian2003b,sanders2004,sanders2007,fabian2011}.

Observations with \xmm provided intriguing suggestions for a lack of
resonance scattering in the Perseus Cluster core. If confirmed, this
would indicate differential gas motions along our line-of-sight with a
range of velocities of at least half of the sound speed
\citep{churazov2004}. If the inferred differential gas motions have the
character of small scale turbulence then its dissipation could provide
enough energy to compensate for the radiative cooling of the ICM.  
From the Fe-K line emission in spectra obtained by {\it Suzaku},
\cite{Tamura2014} placed constraints on bulk and turbulent motions, 
also finding hints of small bulk velocities ($<300$ km s$^{-1}$) at
$2'-4'$ west of the cluster center.

Extensive \suzaku observations of the Perseus Cluster, along 8 different
azimuths out to beyond its virial radius, offer the most detailed X-ray
observation of the ICM at large radii in any cluster to date, providing
important information about cluster growth and the virialization of the
ICM \citep{simionescu2011,simionescu2012,urban2013}. The observations
indicate that the sloshing/swirling ICM motions (preferentially along
the east-west axis of the cluster) extend out to very large radii and
suggest that the ICM in the outskirts may be clumpy.

Observations of the Perseus Cluster also provided significant
advancements in the studies of the chemical enrichment of the
ICM. Recent \suzaku observations e.g.  provided, for the first time,
significant detections of X-ray lines from the rare elements Cr and Mn
in addition to other metals previously seen in galaxy clusters
\citep{tamura2009}, as well as evidence that the ICM is enriched
homogeneously all the way out to the virial radius of the cluster
\citep{werner2013b}.

\subsection{Prospects and Strategy}
\label{sec-perseus-strat}

\begin{figure}[t]
\begin{center}
\begin{minipage}{0.8\textwidth}
\vspace{-5cm}
\includegraphics[width=12cm,clip=t,angle=0.]{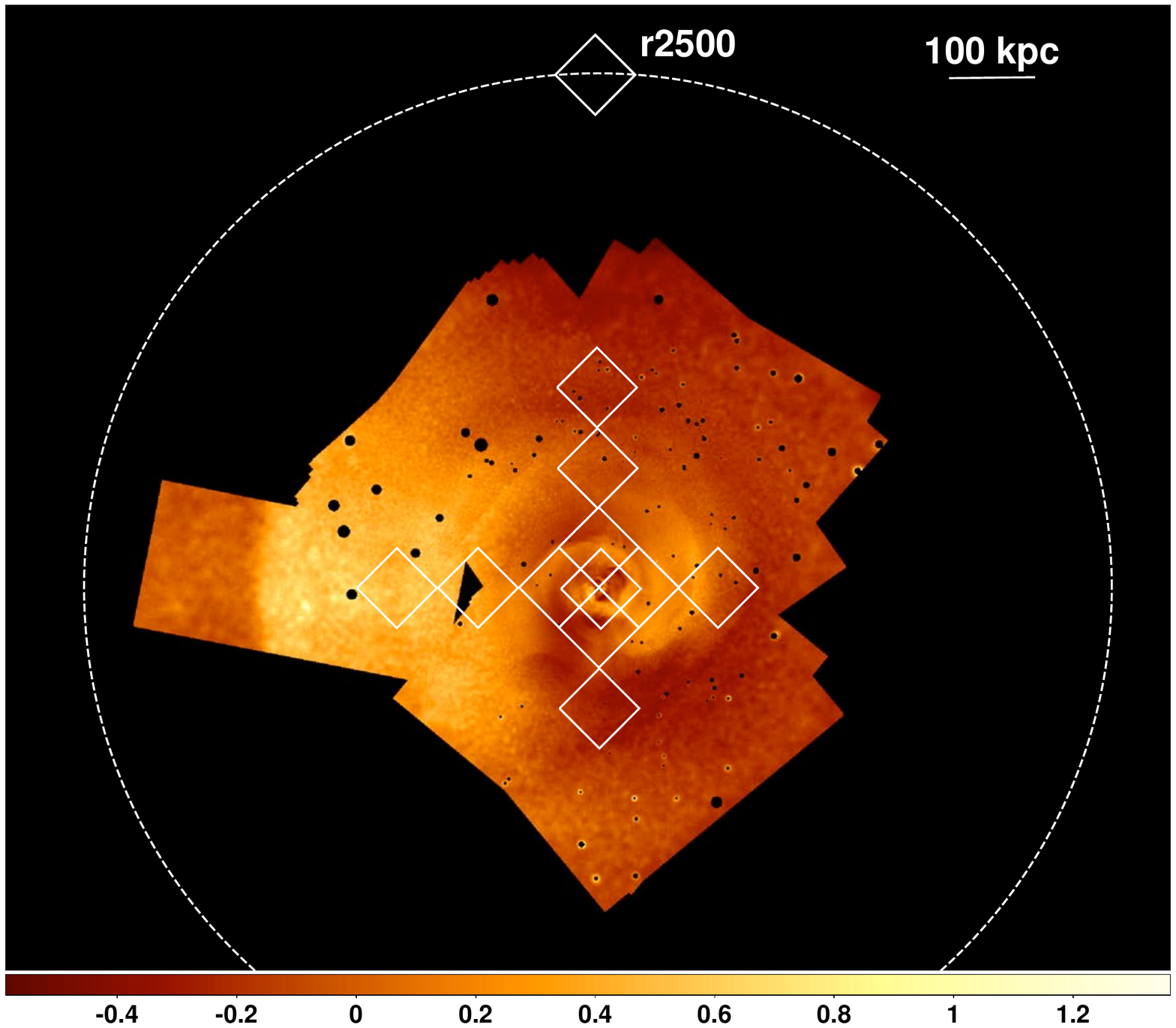}
\end{minipage}
\end{center}
\vspace{-3.5cm}
\caption{The pointing positions (boxes) over-plotted on the residual \chandra image of the cluster, from which a
radially symmetric model has been subtracted in order to emphasize the
substructure in the cluster core \citep{fabian2011}. The color
shows fractional variation in the X-ray surface brightness. We consider a deep pointing at
the cluster core, four pointings centered at $r=50$~kpc covering the
central $130\times130$~kpc$^2$ (these pointings will partially overlap with
the central pointing providing valuable information on systematics),
four pointings at $r=140$~kpc at approximately 90 degree azimuthal
intervals, two pointings at $r=240$~kpc to the north and east of the core,
and a pointing at $r_{2500}$ ($r=600$~kpc) along the northern
direction. The size of the boxes matches the $3'\times 3'$
field-of-view of SXS. For the five pointings spanning the
central $130\times130$~kpc$^2$ and the pointings at $r=140$~kpc, we will divide the field of view of the \astroh SXS into four approximately
independent regions, allowing us to study the projected physical
properties of the ICM on a spatial scale of 1.5~arcmin. We will study the $1\times1$ arcmin region centered on the AGN
independently.   }  \label{pointings2}
\end{figure}

\begin{figure*}[t]
\begin{center}
\begin{minipage}{0.9\textwidth}
\vspace{-4.cm}
\includegraphics[width=12cm]{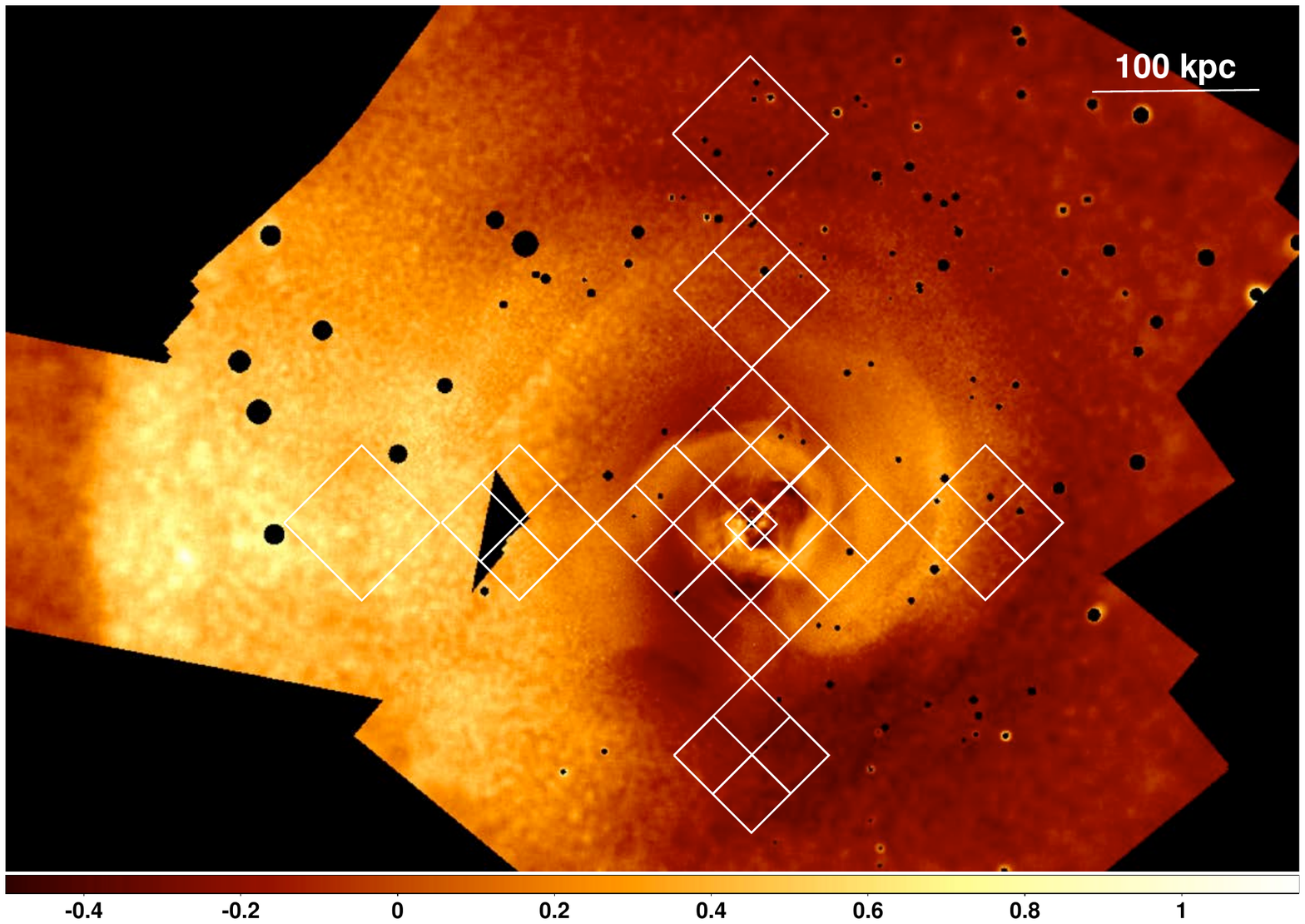}
\end{minipage}
\end{center}
\vspace{-4.cm}
\caption{Same as Figure~\ref{pointings2}, but zoomed in on the pointing positions near the cluster core.} \label{pointings1}
\end{figure*}

The high X-ray brightness, proximity, and the relatively relaxed morphology make the Perseus Cluster an ideal target to study the dynamics and physics of the hot ICM. 
We examine the feasibility of a set of twelve pointings
spanning a range of radii and azimuths. The surface brightness
distribution of the Perseus Cluster is asymmetric due to mild ICM
sloshing taking place primarily along the east-west major axis of the
cluster \citep{simionescu2012}, although the cluster is overall in a
relatively relaxed configuration. This gas sloshing produces a prominent
surface brightness discontinuity at $r\sim100$~kpc to the west of the
cluster center, and a strongly enhanced surface brightness extending all
the way to $r\sim650$~kpc to the east of the core. We
consider an initial deep pointing at the cluster core, four
pointings centered at $r=50$~kpc covering the central $130\times130$~kpc
(these pointings will partially overlap with the central pointing
providing valuable information on systematics), four pointings at
$r=140$~kpc at approximately 90 degree azimuthal intervals, two
pointings at $r=240$~kpc to the north and east of the core, and a
pointing at $r_{2500}$ ($r=600$~kpc) along the northern direction, which
appears to be relatively undisturbed (see Figure~\ref{pointings2}). 

\subsubsection{The nature of gas motions}

The high resolution \astroh SXS spectra of the 
Perseus Cluster will reveal in unprecedented detail the nature of gas
motions in the ICM. 
Pointed observations of the nucleus and selected locations
out to 
$r_{2500}$ will provide 
constraints on the typical spatial
scales and velocities of turbulent gas motions. The same data will
probe resonance scattering in the cluster core, providing additional
constraints on the anisotropy of gas motions. Performing this study in
the nearest, brightest, massive, relaxed system will provide a
benchmark for \astroh studies of more distant relaxed clusters used to
probe cosmology, and address directly the energy dissipation in
cluster cores and the bias on cluster mass measurements due to
turbulence (see Sec. \ref{sec-nonth} for details on the mass
measurement bias).


\subsubsection*{The cluster core}

The observations of the cluster core considered here
involve nine separate pointings: the first (100~ks) targeted on the
nucleus of NGC~1275 in the cluster core, four pointings (50 ks each) covering the central $130\times130$~kpc$^2$, and four pointings (50~ks
each) at approximately 90 degree azimuthal intervals,
centered at a radius of $r=140$~kpc from the cluster
center (see Figure~\ref{pointings1}). For the five pointings spanning the
central $130\times130$~kpc$^2$ and the pointings at $r=140$~kpc, we will divide the field of view of the \astroh SXS into four approximately
independent regions, allowing us to study the projected physical
properties of the ICM on a spatial scale of 1.5~arcmin. We will study the $1\times1$ arcmin region centered on the AGN
independently. 
Goals for these observations include: detecting line broadening due to small scale turbulence;
measuring ordered gas motions due to sloshing; obtaining independent
constraints on differential gas motions from measurements of resonance
scattering; measuring the temperature and chemical structure of the
cluster core.

\bigskip
\noindent{\it Line shifts and broadening}\\*[3pt]
The poinitng strategy mentioned above will enable a
search for coherent gas motions due to sloshing that might be
`rotational' in nature - due to the swirling of the gas in the
gravitational potential of the cluster. Such rotational motion can be
clearly seen in the central region of the cluster
(Figure~\ref{pointings1}). The western pointing of this set at $r=140$~kpc
probes a cold front previously identified in \xmm and \chandra data
\citep{churazov2003,fabian2011} and will allow
a search for a velocity shear underneath and across the surface
brightness discontinuity. Such velocity shears are predicted by
numerical simulations of gas sloshing and are expected to produce
turbulence and (re)-accelerate relativistic electrons responsible for
the observation of radio mini-halos \citep{ZuHone2013}. 
The bulk and turbulent velocities will be measured on
each side of the cold front and compare the measurements with those
obtained on the opposite side of the cluster. We will also search for
turbulent and coherent gas motions in the wakes of the buoyantly rising
bubbles along the north-south axis (the exact positions of the pointings
in the cluster core may be adjusted to better cover interesting features
seen in the high resolution \chandra images).

Interpretation of the \astroh data  will benefit from
the existing {\it Chandra} and {\it XMM-Newton} observations, which
provide complementary information about the history of AGN
outbursts producing turbulence. This will 
provide a more complete picture of how turbulence is driven and
eventually dissipated.

\bigskip
\noindent{\textit{Calibration between density and velocity 
fluctuations
}}\\*[3pt]
Deep {\it Chandra} and {\it XMM-Newton} observations of the Perseus
Cluster 
provide a probe of surface brightness and density
fluctuations over a range of length scales.  Detailed analysis of 
{\it Chandra} images shows that at radii 94--190 kpc, which are covered by
four \astroh\ pointings, the amplitude of density fluctuations
$\delta\rho/\rho$ is $\sim 7.5$--25\% on scales $\sim 14$--90~kpc
\citep{zhuravleva2014b}. Recently, it has been shown that the
amplitudes of density and velocity fluctuations are expected to be
approximately linearly related across a broad range of scales; 
analytical arguments predict a proportionality
coefficient between density and velocity fluctuations of $\eta\sim 1$,
while cosmological simulations of relaxed clusters show $\eta=1\pm 0.3$
\citep{zhuravleva2014}.  
Taking $\eta=1$, the characteristic amplitude of the one-component
velocity 
is predicted to be $\sim90$--250 km~s$^{-1}$ on
scales $\sim14$--90~kpc \citep{zhuravleva2014b}. Knowing the
characteristic velocities on different scales, we can predict the
expected line broadening \citep[see equation E9 in][]{zhuravleva2012}.

\astroh\ observations will 
further allow us to directly measure the velocity
broadening of emission lines and compare it to the values predicted
based on the observed density fluctuations. In particular, the
comparison of the predicted line broadening in a given radial range to
the measured average value, determined over several pointings,
should be useful for calibrating the statistical
relation between the density and velocity fluctuations. Moreover, the
comparison of the predicted line broadening in the individual pointings
with the direct measurements will
provide a clue to whether the perturbations driving the
turbulence are local (on scales comparable to the field of view of
\astroh), or the turbulence is driven on larger scales.

\bigskip
\noindent{\it Resonance scattering}\\*[3pt]
Although the hot ICM is typically optically thin, at the energies of the strongest resonant transitions, the hot plasma in the dense central regions of galaxies and galaxy clusters can become optically thick \citep{gilfanov1987}. The transition probabilities of strong resonance lines are large and, if the column density of the ion along a line-of-sight is sufficiently high, photons with the energies of these resonance lines will be absorbed and, within a short time interval, re-emitted in a different direction. Resonance scattering thus suppresses the line intensity in the cluster core and raises the line intensity at intermediate radii. 
\citet{gilfanov1987} pointed out that since the optical depth in the core of a resonance line depends on the characteristic velocity of small-scale motion, its measurement gives important information about the turbulent velocities in the hot plasma. Measuring the level of resonance scattering in clusters is thus a good way to determine the characteristic velocity of small-scale turbulence independently of line broadening. 

\citet{churazov2004} used this technique to obtain constraints on
differential velocities along our line-of-sight in the Perseus
Cluster. They analyzed \xmm EPIC data and compared the relative fluxes
of the 1s2p and 1s3p He-like Fe lines in the core and in an annulus
surrounding the core of the Perseus Cluster. The expected optical depth
of the 1s2p \ion{Fe}{XXV} resonance line at 6.7 keV is much larger than
that of the 1s3p line at 7.9 keV, therefore the ratio provides
information about the level of resonance scattering. \citet{xu2002},
\citet{werner2009}, and \citet{deplaa2012} used the \ion{Fe}{XVII} lines
at 15\AA\ and 17\AA\ to measure the level of resonance scattering in
giant ellipticals and groups of galaxies, placing interesting
constraints on turbulence in those systems. High-resolution \astroh SXS
spectra will allow to search for and study resonance scattering using
all of the lines in the available spectral band simultaneously.

Measurements of the suppression of spectral lines due to resonance
scattering and their velocity broadening provide two independent,
powerful probes of gas motions. These two diagnostics as a function of
radius have a different dependence on the directionality of gas motions
and therefore they allow us to reveal and place constraints on
anisotropy of gas motions in bright cluster cores
\citep[][]{zhuravleva2011,rebusco2008}. Measurements of the shapes of
strong spectral lines \citep{zhuravleva2011,shang2012,shang2012b} could
also be utilized as a third, independent probe of anisotropy.


\subsubsection*{Measurements on intermediate scales}

\begin{figure}
\begin{center}
\begin{minipage}{0.7\textwidth}
\includegraphics[width=12cm,clip=t,angle=0.]{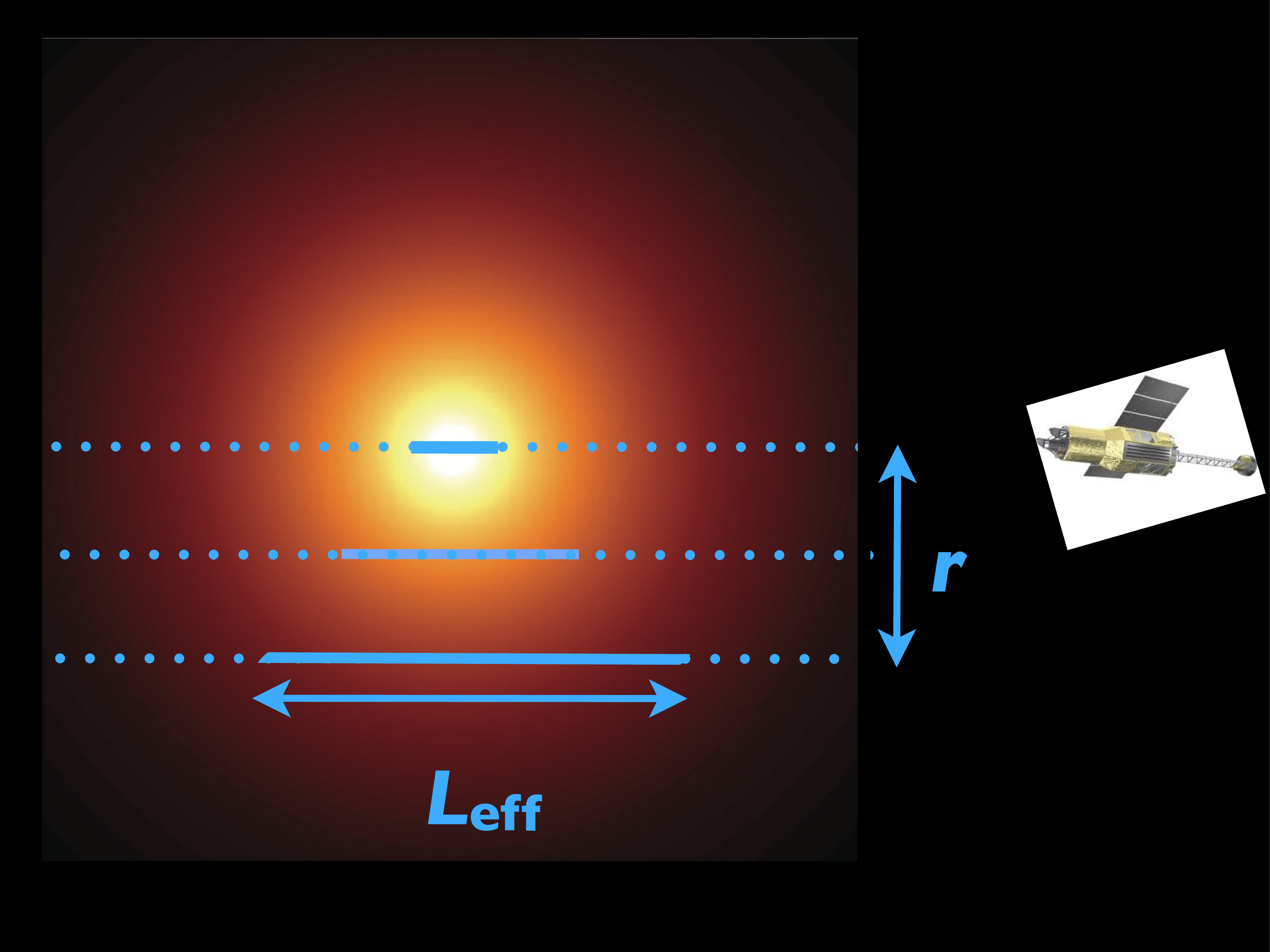}
\end{minipage}
\end{center}
\caption{An illustration of how in a cluster with a centrally peaked surface brightness distribution, the projected radius $r$ at which we observe determines the effective length, $L_{\rm eff}$, from which the largest fraction of line flux (and measured line width) arises. }
\label{Leff}
\end{figure}

As a potential strategy, we consider 
two pointings at $r=240$~kpc along the minor (northern) and major (eastern) axes, and a pointing at $r_{2500}$ ($r=600$~kpc) along the relatively undisturbed northern direction (Figure~\ref{pointings2}). 
Given the rapidly rising surface brightness profile in the core of the Perseus Cluster, the projected radius $r$ at which we observe determines the effective length, $L_{\rm eff}$, from which the largest fraction of line flux (and measured line width) arises \citep[see Figure~\ref{Leff};][]{zhuravleva2012}. This effective length increases as a function of radius. By measuring the width of spectral lines as a function of $r$, we therefore probe velocities on different spatial scales ($L_{\rm eff}$). The increase of the effective length, $L_{\rm eff}$, along the line-of-sight with growing projected distance $r$ implies that larger and larger eddies contribute to the observed line broadening. 

For gas motions on spatial scales $L < L_{\rm eff}$ we expect to see
significant line-of-sight velocity dispersions $\sigma_{\rm v}$,
resulting in line broadening, but no centroid shifts. The measured width
of spectral lines as a function of projected radius $\sigma_{\rm v}(r)$
therefore probes small scale motions.  On the other hand, for gas
motions on spatial scales $L > L_{\rm eff}$, we expect to see
significant centroid shifts. These measurements, in principle, 
provide the root-mean-square of the projected velocities $V_{\rm RMS}$
\citep[see][]{zhuravleva2012}. The observations on intermediate spatial
scales will thus tell us whether the turbulent motions are predominantly
small scale or large scale.

Since the projected velocity field mainly depends on large scale
motions, while the line broadening is more sensitive to small scale
motions, the ratio of the two is a good diagnostic of the shape of the
power spectrum of the 3D velocity field. In particular: (i) the radial variations of the line broadening $\sigma_{\rm v}$ are
closely related to the structure function of the velocity field; (ii)
the ratio of the spatial variations of the mean velocity $V_{\rm RMS}$ to the radial variations of the line broadening can be used as a proxy of the velocity field injection scales; (iii) the map of measured line shifts due to large-scale motions can be converted straightforwardly into the power spectrum of the velocity field
\citep{zhuravleva2012}. 

Even though it is unlikely that the turbulence follows
a single power spectrum throughout the inner volume of the cluster,
these observations will provide important constraints on the amplitude
of gas velocities as a function of spatial scale and distance from the
cluster center. The stratification of the cluster atmosphere, indicated
by the radial gradient in the entropy distribution, may lead to an
anisotropy in the turbulent velocity field, causing the largest eddies
to be predominantly tangential (on scales smaller than a few tens of
kpc, the turbulence is expected to be isotropic). Because potential
tangential motions will be probed relatively well by the offset
pointings (due to our line-of-sight being parallel to the tangential
direction), the large scale eddies should still be detected as a radial
increase of the velocity broadening $\sigma_{\rm V}$ or shifts in the
line centroid. A flattening of $\sigma_{\rm V}$ as a function of radius
should constrain the largest scale at which turbulence is injected into
the ICM inside of the given radius. The actual quality of observational
constraints will depend strongly on the intrinsic properties of the
velocity field (see Zhuravleva et al. 2012). The results will be
interpreted in combination with tailored numerical simulations and the
wealth of data obtained with the other X-ray missions.  


\bigskip
\noindent{\it{The measurement at $r_{\rm 2500}$}}\\*[3pt]
The pointing at $r_{2500}$ will provide a new benchmark measurement for cosmological studies with galaxy clusters. Measurements of the gas mass fraction, $f_{\rm gas}$ (the ratio of X-ray emitting gas mass to total mass) at intermediate radii in massive clusters provide a powerful tool for cosmology, enabling robust constraints to be placed on cosmological parameters including those describing dark energy \citep{allen2004,allen2008,mantz2014}. The Perseus Cluster is the nearest massive cluster suitable for such work. 

Recent work using {\it Chandra} X-ray observations \citep{mantz2014} has
 measured the intrinsic scatter in $f_{\rm gas}$ for a complete,
 rigorously selected sample of the most massive ($kT>5$~keV),
 morphologically-relaxed galaxy clusters known. The measurements are
 made in an optimized spherical shell spanning radii 0.8--1.2$r_{2500}$,
 finding a system-to-system scatter of $7.4\pm2.3$\%
 \citep{mantz2014}. This result places a firm upper limit on the
 cluster-to-cluster variation in turbulent pressure support at
 $r_{2500}$ that can be present in massive, relaxed clusters. In a
 cosmological context, a key task for \astroh\ is to
 determine the fraction of this scatter that is due to gas motions.

The pointing at $r_{2500}$ in the Perseus Cluster will allow us to
measure the line-of-sight velocity dispersion to a precision of $\sim
50~$km~s$^{-1}$. Such a precision is well matched to probing the
expected cluster-to-cluster scatter, which for a characteristic 1D
velocity of isotropic turbulence of $v_{\rm turb}=300$~km~s$^{-1}$
should be approximately $\pm 100$~km~s$^{-1}$ (68\% confidence
limit). We envisage that the $r_{2500}$ measurement for Perseus will be
among the first and (in the sense of mitigating
systematics associated with the \astroh\ PSF) most robust such
measurements for an eventual ensemble of such data gathered for bright,
nearby clusters over the first few years of the mission. Together these
data will determine the average turbulent pressure support in clusters,
providing a key reference point for cosmological studies.

The measurement at $r_{2500}$ will complement the full set of
observations for the northern arm (minor axis) providing reference
measurements of gas velocities, thermodynamics and element abundances
with a level of robustness that will be impossible for more distant
systems. 
As discussed in detail in Section
\ref{sec-nonth}, $r_{2500}$ is the largest radius out to which
interesting measurements of these properties can be made with \astroh\
in reasonable exposure times for local clusters.

\subsubsection{Temperature structure and cooling}

In addition to the electron temperature determined from the shape of the
continuum, high resolution X-ray spectra obtained with the \astroh SXS
will 
yield the ionization temperature based on line ratios of He- and
H-like ions. For the first time, the quality of the data will, in
principle, be sufficient to directly measure the ion temperature of the
ICM based on thermal line broadening.

Deep \chandra observations of the Perseus Cluster center
\citep{sanders2007} revealed multi-phase ICM spanning a range of
temperatures between $kT=0.5-8$~keV (see Figure~\ref{multitemp}). The
spectral resolution of \astroh SXS will allow us to study the
`multi-phasedness' of the ICM (the emission measure distribution as a
function of temperature) and constrain the cooling rate as a function of
location.

This effort will be helped by existing deep \suzaku and
\xmm data, which can be used to estimate the 3D deprojected
thermodynamic properties of the ICM, and by \chandra data that
provide the projected temperature distribution with a spatial
resolution of a few arc-seconds. Joint multi-mission analysis of the
data will be essential and will also help in constraining mixing and
conduction between the different gas phases.

\subsection{Targets and Feasibility}

The observing strategy described above involves a combination of
pointings at various radii and azimuths that will provide a
powerful tool to study the gas motions in the hot ICM, and determine
its temperature structure and chemical composition.  The feasibility of
the observations can be assessed using existing, 
deep \chandra, \suzaku, and \xmm observations, 
from which we determine the input parameters for the assumed spectral models. 
For the energy response of the SXS we use the response files 
`ah\_sxs\_5ev\_basefilt\_20100712.rmf' and `ah\_sxs\_7ev\_basefilt\_20090216.rmf', 
which include the current best estimates with 5~eV and 7~eV constant energy 
resolutions (FWHM), respectively, using baseline filters. For the effective area, 
we use the file `sxt-s\_100208\_ts02um\_of\_flatsky-vig1kev\_intallpxl.arf'. 
The effective area model assumes a sky with flat surface brightness 
and takes into account vignetting effects calculated at 1~keV. 
The flat sky ARFs available for the SXS are normalized so that the fitting results are 
provided per square degree, i.e. the true effective area has been multiplied by the field of view 
of the SXS in units of square degrees. To recover the true effective area we have thus 
multiplied the existing ARF by 1~deg$^{2} / 3\times3~{\rm sq~arcmin}=400$. For the detector 
and cosmic X-ray background, we use the parameters provided in the files `sxs-bck.fits' and 
`sxs-cxb.fits'\footnote{\astroh response files for spectral simulation can be 
downloaded from http://astro-h.isas.jaxa.jp/researchers/sim/response.html}. 
For the central pointing and the four pointings covering the central $130\times130$~kpc, 
we use exposure times of 100~ks and
$4\times50$~ks, respectively. For the four pointings at the radius $r=140$~kpc, 
we use an exposure time of $4\times50$~ks. 
These observations should gather a sufficient number of X-ray photons to
divide the field of view of the SXS into 4 approximately independent
region and perform spectral analysis. For the pointings centered at
radii of $r=240$~kpc, we adopt exposure times of 80~ks and 50~ks to the
X-ray fainter north and brighter east, respectively. For the pointing at
$r=600$~kpc ($r_{2500}$) we adopt an exposure time of 300~ks.

For our simulations, we use the SPEX spectral fitting package
\citep{kaastra1996}. We simulate absorbed thermal plasma models with
model parameters determined from fitting {\it Suzaku} spectra extracted
at the proposed pointing locations. To account for the AGN in NGC~1275,
for the central pointing, we also include a power-law like emission
component.  We assume isotropic turbulence with a characteristic
velocity of 300~km~s$^{-1}$.  Furthermore, we assume that the ion
temperatures (and thus the thermal broadening of the lines) and their
uncertainties are equal to the electron temperatures.  We find that by
fitting spectra extracted from the full field-of-view of the \astroh SXS
in the range of 0.3--10~keV, we will be able to measure model parameters
with the statistical precisions shown in Table~\ref{simtable} and
Table~\ref{simtable7ev} for energy resolution of 5~eV and 7~eV,
respectively. The precision of measurements in each of four regions will be
approximately a factor of 2 lower. The quoted chemical abundances are
with respect to the proto-Solar values of \citet{lodders2003}. We note
that the likelihood space is complex, with more than one local
minimum. Therefore when working with real data from the satellite, Monte
Carlo simulations will be needed to assess the statistical uncertainties
robustly.

\begin{table}[h]
\caption{Results of fits to simulated spectra assuming 5~eV resolution
for \astroh SXS. 
We assume spectra extracted from the full field-of view
of the \astroh SXS (dividing the field of view into four regions, the
precision of our measurements will be a factor of $\sim2$ lower) fitted
in the 0.3--10 keV energy range. $Y$ denotes the spectral
normalization and $kT_{\rm e}$ the temperature of the thermal plasma
model, $v_{\rm turb}$ and $v_{\rm bulk}$ are the turbulent and bulk
velocities, respectively, the elemental abundances are given relative
to the proto-Solar values by \citet{lodders2003}, $n_{\rm POW}$ is the
normalization and $\Gamma$ is the index  of the power-law like emission
component of the central AGN. For the best fit bulk and turbulent
 velocities we also present the results obtained by fitting the 6.7~keV Fe-K lines alone. The number of continuum subtracted counts in the Fe-K lines was calculated assuming an effective area of 300 cm$^{2}$ at the observed line energy (6.47--6.62~keV). Statistical error bars are given at the $\Delta$C=1 level. }\label{simtable}
\begin{center}
\begin{tabular}{lccccc}
\hline
Pointing position &  Center  & $r=50$~kpc & $r=140$~kpc  & $r=240$~kpc  & $r=600$~kpc \\
Exposure time (ks) &100& 50 & 50& 80 & 300 \\
\hline 
$Y$ ($10^{65}$~cm$^{-3}$)		& $72.80\pm0.09$ 	&	$64.00\pm0.01$	& $7.87\pm0.04$ 	& $2.63\pm0.02$ 	&$ 0.349\pm0.003$\\
$kT_{\rm e}$~(keV)				& $3.43\pm0.01$	& 	$4.04\pm0.01$ 	& $5.07\pm0.05$      & $5.54\pm0.09$ 	& $5.44\pm0.12$ \\
$v_{\rm bulk}$ (km~s$^{-1}$)		& $\pm3.7$ 		&	$\pm5.4$ 			& $\pm19$ 		& $\pm32$ 		& $\pm54$ \\
$v_{\rm bulk}$ (km~s$^{-1}$) by Fe-K & $\pm6.2$ 	&	$\pm7.9$			& $\pm23$ 		& $\pm37$ 		& $\pm66$ \\
$v_{\rm turb}$ (km~s$^{-1}$)		& $300\pm3.9$		&	$300\pm5.3$		& $300\pm15$		& $300\pm30$ 		& $300^{+53}_{-45}$\\
$v_{\rm turb}$ (km~s$^{-1}$) by Fe-K & $300\pm5.4$&	$300\pm6.8$		& $300\pm22$		& $300\pm35$ 		& $300^{+61}_{-51}$\\
numb. of phot. in Fe-K			& 14400			&	5800			& 600			&	250			&	100		\\
O 							& $0.65\pm0.02$	& $0.65\pm0.02$	& $0.5\pm0.06$	& $0.4\pm0.10$	& $0.3\pm0.14$\\
Ne 							& $0.65\pm0.02$	& $0.65\pm0.03$	& $0.5\pm0.07$	& $0.4\pm0.12$	& $0.3\pm0.16$ \\
Mg 							& $0.65\pm0.01$	& $0.65\pm0.02$	& $0.5\pm0.06$	& $0.4\pm0.09$	& $0.3\pm0.13$ \\
Si 							& $0.65\pm0.01$	& $0.65\pm0.01$	& $0.5\pm0.04$	& $0.4\pm0.06$	& $0.3\pm0.08$ \\
S 							& $0.65\pm0.01$	& $0.65\pm0.02$	& $0.5\pm0.05$	& $0.4\pm0.09$	& $0.3\pm0.11$ \\
Ar 							& $0.65\pm0.03$	& $0.65\pm0.04$	& $0.5\pm0.12$	& $0.4\pm0.20$	& $0.3\pm0.27$ \\
Ca 							& $0.65\pm0.03$	& $0.65\pm0.05$	& $0.5\pm0.13$	& $0.4\pm0.23$	& $0.3\pm0.30$ \\
Fe 							& $0.65\pm0.006$	& $0.65\pm0.007$	& $0.5\pm0.015$ 	& $0.4\pm0.02$	& $0.3\pm0.03$ \\
Ni 							& $0.65\pm0.03$	& $0.65\pm0.05$	& $0.5\pm0.14$	& $0.4\pm0.23$	& $0.3\pm0.30$ \\
$n_{\rm POW} 10^7$			& $6.0\pm0.02$ & - & - & - & - \\
$\Gamma$					& $1.72\pm0.008$ & - & - & - & - \\
\hline
\end{tabular}
\end{center}
\end{table}

\begin{table}[h]
\caption{Results for the same simulations as in Table~\ref{simtable}, assuming 7 eV resolution for \astroh SXS.}\label{simtable7ev}
\begin{center}
\begin{tabular}{lccccc}
\hline
Pointing position &  Center  & $r=50$~kpc & $r=140$~kpc  & $r=240$~kpc  & $r=600$~kpc \\
Exposure time (ks) &100& 50 & 50& 80 & 300 \\
\hline 
$Y$ ($10^{65}$~cm$^{-3}$)		& $72.80\pm0.09$ 	& $64.00\pm0.01$	& $7.87\pm0.04$ 	& $2.63\pm0.02$ 	&$ 0.349\pm0.003$\\
$kT_{\rm e}$~(keV)				& $3.43\pm0.01$	& $4.04\pm0.01$	& $5.07\pm0.05$      & $5.54\pm0.09$ 	& $5.44\pm0.12$ \\
$v_{\rm bulk}$ (km~s$^{-1}$)		& $\pm4.4$ 		& $\pm6.3$		& $\pm22$ 		& $\pm32$ 		& $\pm57$ \\
$v_{\rm bulk}$ (km~s$^{-1}$) by Fe-K& $\pm6.6$ 	& $\pm8.5$		& $\pm31$ 		& $\pm40$ 		& $\pm65$ \\
$v_{\rm turb}$ (km~s$^{-1}$)		& $300\pm4.8$		& $300\pm6.3$		& $300\pm21$		& $300\pm33$ 		& $300^{+59}_{-50}$\\
$v_{\rm turb}$ (km~s$^{-1}$) by Fe-K& $300\pm5.9$& $300\pm7.5$	& $300\pm23$		& $300\pm37$ 		& $300^{+66}_{-56}$\\
numb. of phot. in Fe-K			& 14400			&	5800			& 600			&	250			&	100		\\
O 							& $0.65\pm0.02$	& $0.65\pm0.02$	& $0.5\pm0.07$	& $0.4\pm0.10$	& $0.3\pm0.14$\\
Ne 							& $0.65\pm0.02$	& $0.65\pm0.03$	& $0.5\pm0.09$	& $0.4\pm0.13$	& $0.3\pm0.17$ \\
Mg 							& $0.65\pm0.01$	& $0.65\pm0.02$	& $0.5\pm0.07$	& $0.4\pm0.10$	& $0.3\pm0.13$ \\
Si 							& $0.65\pm0.01$	& $0.65\pm0.01$	& $0.5\pm0.05$	& $0.4\pm0.07$	& $0.3\pm0.08$ \\
S 							& $0.65\pm0.01$	& $0.65\pm0.02$	& $0.5\pm0.06$	& $0.4\pm0.09$	& $0.3\pm0.11$ \\
Ar 							& $0.65\pm0.03$	& $0.65\pm0.04$	& $0.5\pm0.14$	& $0.4\pm0.21$	& $0.3\pm0.27$ \\
Ca 							& $0.65\pm0.03$	& $0.65\pm0.05$	& $0.5\pm0.16$	& $0.4\pm0.23$	& $0.3\pm0.31$ \\
Fe 							& $0.65\pm0.008$	& $0.65\pm0.007$	& $0.5\pm0.02$ 	& $0.4\pm0.02$	& $0.3\pm0.03$ \\
Ni 							& $0.65\pm0.04$	& $0.65\pm0.05$	& $0.5\pm0.17$	& $0.4\pm0.24$	& $0.3\pm0.31$ \\
$n_{\rm POW} 10^7$			& $6.0\pm0.02$  & - & - & - & - \\
$\Gamma$					& $1.72\pm0.007$ & - & - & - & - \\
\hline
\end{tabular}
\end{center}
\end{table}

\subsubsection{Dynamics of the intra-cluster medium}

\begin{figure}\begin{center}
\begin{minipage}{0.47\textwidth}
\includegraphics[width=9cm]{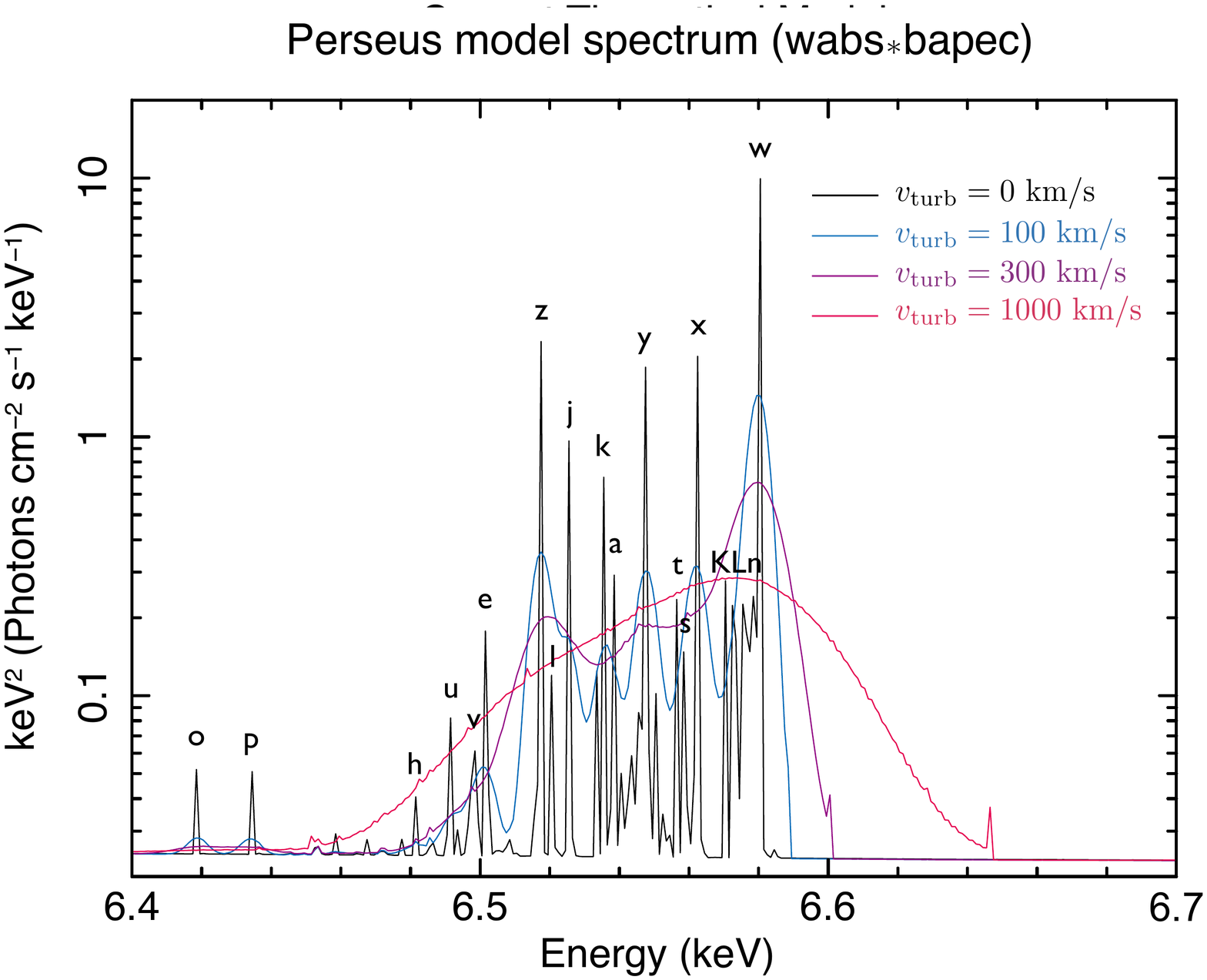}
\end{minipage}
\begin{minipage}{0.47\textwidth}
\includegraphics[width=9cm]{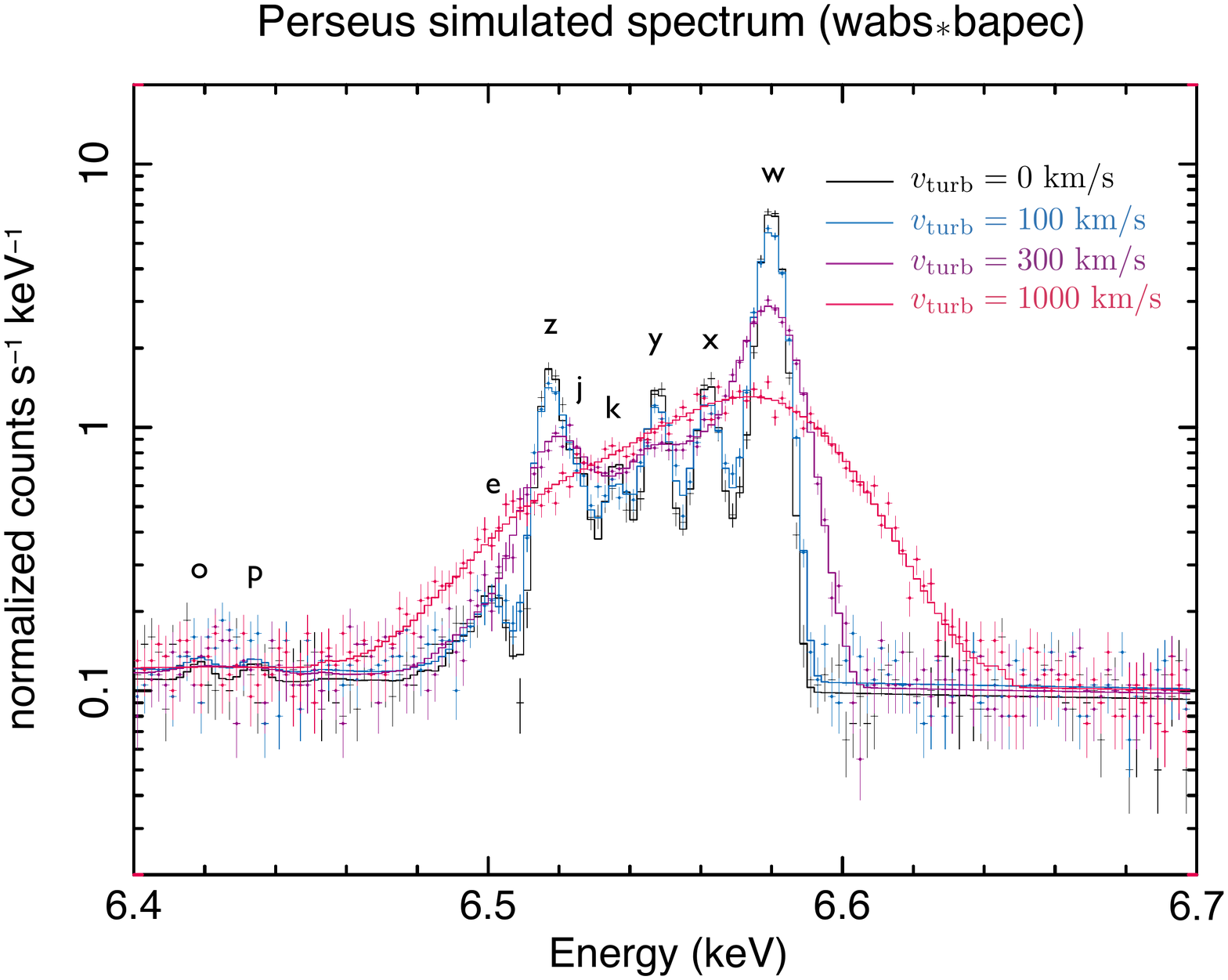}
\end{minipage}
\caption{{\it Left}: Single temperature plasma model with parameters
	       characteristic for the core of the Perseus Cluster at the
	       redshifted He-like Fe-K line energies for different
	       turbulent velocities. {\it Right}: Corresponding
	       simulated \astroh SXS spectra for a 100 ks
	       observation. Representative lines are marked by letters
	       as given in \cite{Gabriel1972}; e.g., the resonance line
	       (w), the intercombination lines (x and y), and the
	       forbidden line (z).}
\label{fea:core:fe-band}
 \end{center}
\end{figure}

\begin{figure}\begin{center}
\includegraphics[width=0.45\textwidth]{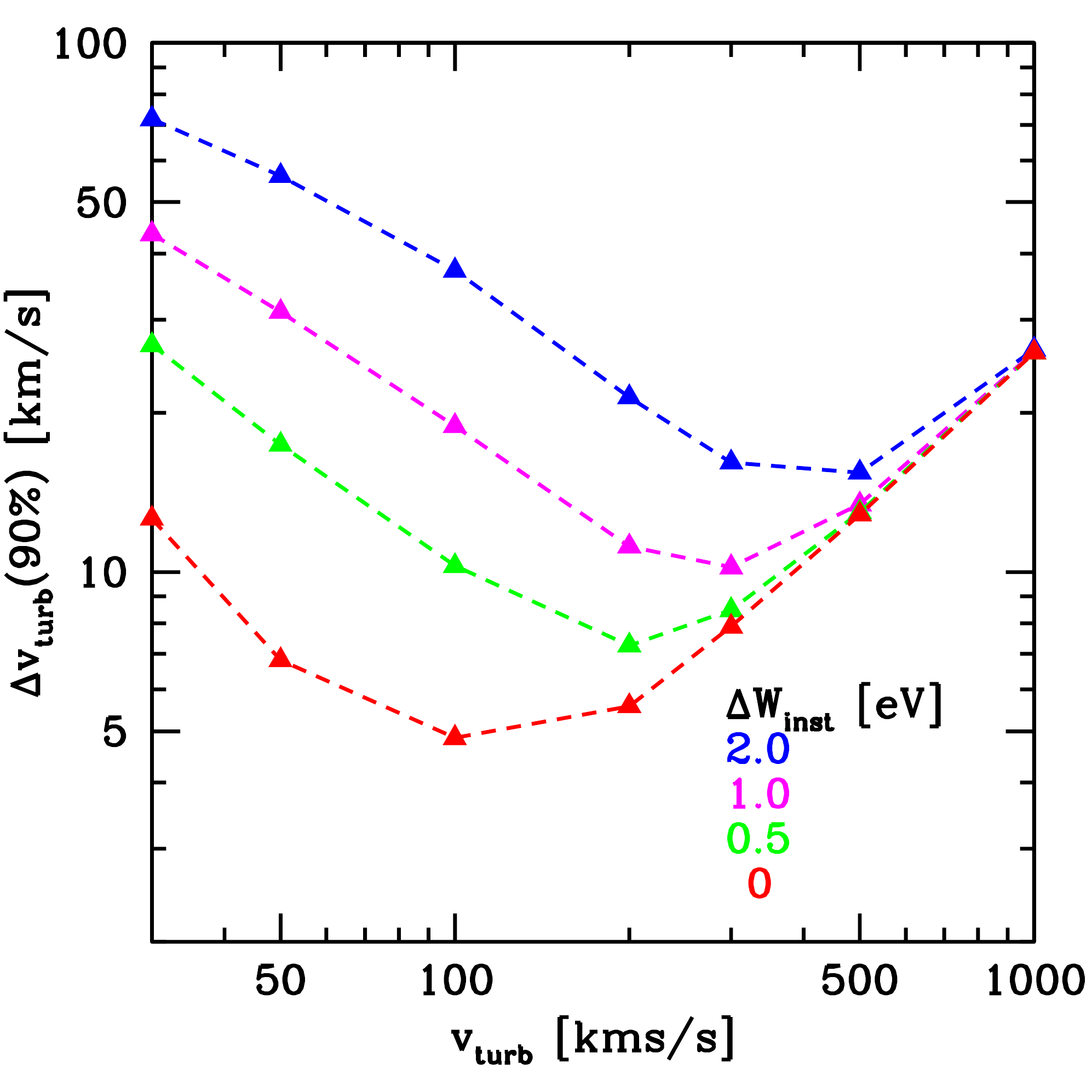}
\includegraphics[width=0.45\textwidth]{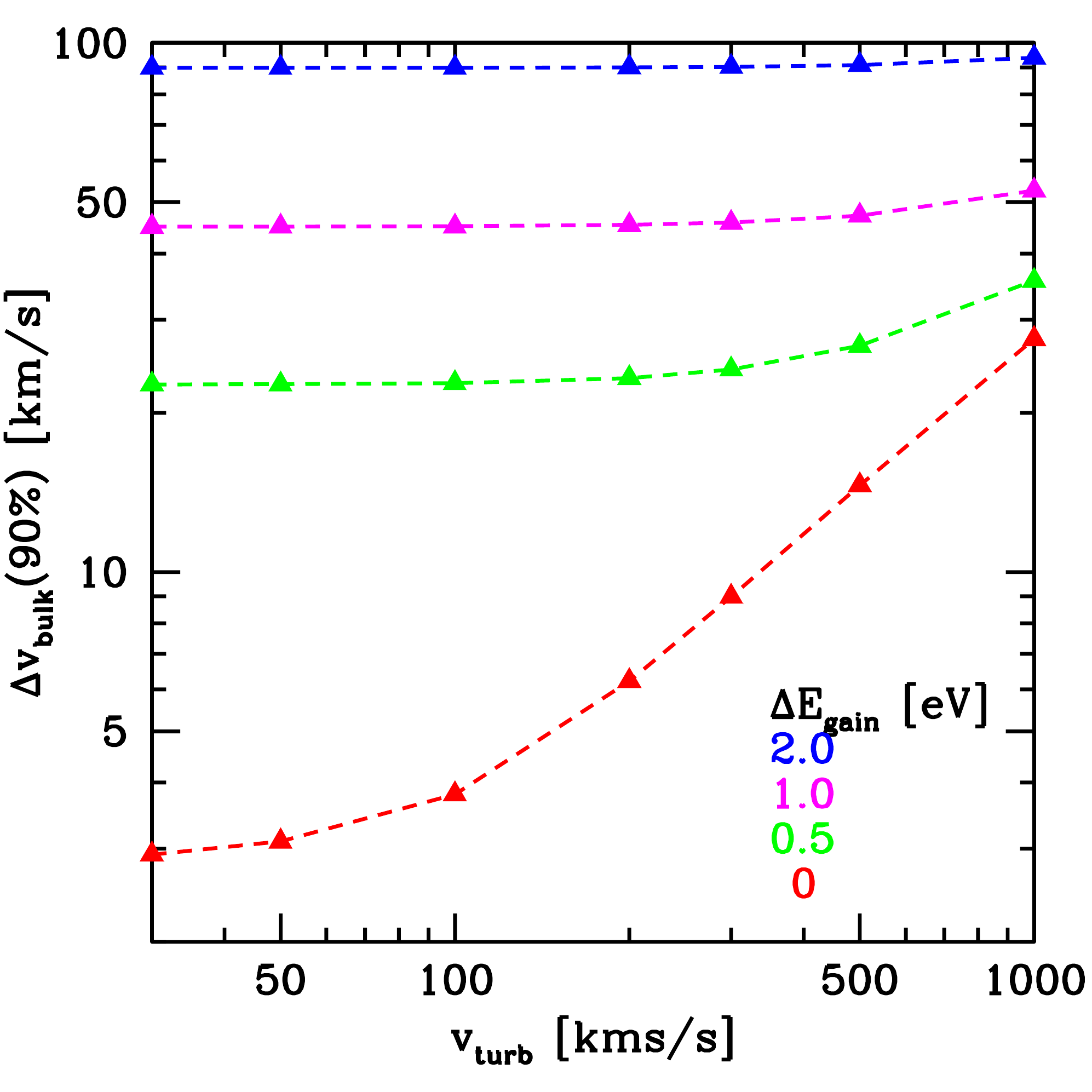}
\caption{Uncertainties of the measured $v_{\rm turb}$ (left panel) and
$v_{\rm bulk}$ (right panel) as a function of $v_{\rm turb}$ for
different uncertainties in the FWHM of the line spread function and the
gain. Note that the turbulent broadening adds to the observed line width
in quadrature and its systematic error, to the first order
approximation, is inversely proportional to the turbulent velocity. The
simulation assumes 11,000 counts in the 6.7~keV Fe-K lines.}
\label{fea:vturb-deltaz}
\end{center}\end{figure}


In bright, nearby galaxy clusters, such as the Perseus
Cluster, \astroh will determine the projected velocity $v_{\rm bulk}$
(line centroid) and the line-of-sight velocity dispersion $\sigma_{\rm
v}$ (line width) as a function of position, providing a measure of the
bulk and small scale velocities of the plasma.  $\sigma_{\rm v}$ is
essentially the 1D line-of-sight component of the characteristic
velocity of isotropic turbulence, and in this white paper we define
$v_{\rm turb}=\sigma_{\rm v}$.  The observed width of a spectral line
is $W_{\rm obs}=\sqrt{W_{\rm inst}^2+W_{\rm therm}^2+W_{\rm turb}^2}$,
where $W_{\rm inst}$ is the instrumental line broadening (expected to
have a FWHM of 5--7~eV), and $W_{\rm therm}=4.35$eV$\sqrt{kT/4{\rm
keV}}$ and $W_{\rm turb}=5.27{\rm eV}(v_{\rm turb}/100{\rm km/s})$
denote the FWHM of the \ion{Fe}{XXV} lines at 6.7~keV due to thermal
broadening and isotropic turbulence, respectively \citep[e.g. see
Table 2 in][and Appendix \ref{sec-sys}]{rebusco2008}.

\bigskip \noindent{\textit{Turbulent line broadening and bulk
velocities}}\\*[3pt] In our simulations, we assume turbulence with a
characteristic velocity of $v_{\rm turb}=300$~km~s$^{-1}$.  The
principal constraints on $v_{\rm bulk}$ and $v_{\rm turb}$ are provided
by the 6.7~keV Fe-K lines, which have the best combination of photon
statistics and spectral resolution $\Delta E/E$. To illustrate the power
of this spectral feature, in Table 1 and 2, we also present the expected
results for $v_{\rm bulk}$ and $v_{\rm turb}$ obtained by only fitting
the Fe-K lines.  The expected number of photons in the 6.7~keV Fe-K lines
in the central pointing is $\sim14000$, allowing us to measure $v_{\rm
bulk}$ and $v_{\rm turb}$ with statistical uncertainties of
$\sim6$~km~s$^{-1}$ (for an illustration of the power of the Fe-K lines
see Figure \ref{fea:core:fe-band}).  For the off-center pointings (with
the exception of the 300~ks pointing at $r_{2500}$), we aim to have at
least 250 counts in the 6.7~keV Fe-K lines in each studied spatial
region,
and measure $v_{\rm bulk}$ and $v_{\rm turb}$ with statistical
uncertainties smaller than 40~km~s$^{-1}$.
The 300~ks pointing at $r_{2500}$ will yield
$\sim100$ counts in the 6.7~keV lines.  As described in Section
\ref{sec-nonth}, observations of brightest, relaxed galaxy clusters,
will provide accurate constraints on turbulent pressure near $r_{2500}$
for each object.  An ensemble of these observations will together
provide a robust measurement of the average turbulent pressure support.

Degrading the energy resolution from 5~eV to~7eV, assuming $v_{\rm
turb}=300$~km~s$^{-1}$, the precision of the line broadening measurement
decreases by about 10--20\%.  For smaller $v_{\rm turb}$, the
degradation of energy resolution will result in larger decrease in the
precision of the line broadening measurements (e.g. for $v_{\rm
turb}=140$~km~s$^{-1}$ degrading the energy resolution from 5~eV to 7~eV
will result in a $\sim$40\% increase in $\Delta v_{\rm turb}$). Even
though numerical simulations predict $v_{\rm turb}\sim300$~km~s$^{-1}$,
little is known about the actual turbulence in clusters. The existence
of long filaments of cold/cool gas in the Perseus Cluster indicates that
at certain locations the turbulent velocities may be as low as 100
km~s$^{-1}$ \citep{fabian2011}. With 250 counts in the 6.7~keV Fe-K
lines, and 7eV energy resolution, we will still be able to measure
$v_{\rm turb}=100\pm30$~km~s$^{-1}$.

In 
deep observations, systematic errors due to
uncertainties in the gain and line-spread-function calibration are also
expected to be significant (for a more detailed discussion on the
calculation and impact of systematics see Appendix
\ref{sec-sys}). Figure \ref{fea:vturb-deltaz} shows the combined
statistical $+$ systematic uncertainties in $v_{\rm turb}$ (left panel)
and $v_{\rm bulk}$ (right panel) as a function of $v_{\rm turb}$ for
different systematic uncertainties in the FWHM of the
line-spread-function and the gain. The systematic uncertainties in the
measured velocity broadening are sensitive to the line spread function
calibration, which is expected to be better than 1.6~eV in FWHM. The
uncertainties in the measured $v_{\rm bulk}$ will be sensitive to the
gain calibration: for example a gain calibration systematic error of
1~eV, implies a systematic uncertainty $\Delta v_{\rm
bulk}\sim50$~km~s$^{-1}$. The uncertainty on the gain calibration is
expected to be smaller than 2~eV.

The systematic errors above are assumed to be uncorrelated. This is a
highly pessimistic assumption. In practice the different pixels are
expected to have correlated systematic errors, and observations taken
within a short period of time are expected to be affected by systematics
in a similar way. Such correlations will decrease the impact of
systematic uncertainties on some measurements (e.g. line broadening) and
can be marginalized over using standard Monte Carlo methods.

\begin{figure}\begin{center}
\includegraphics[width=0.5\textwidth]{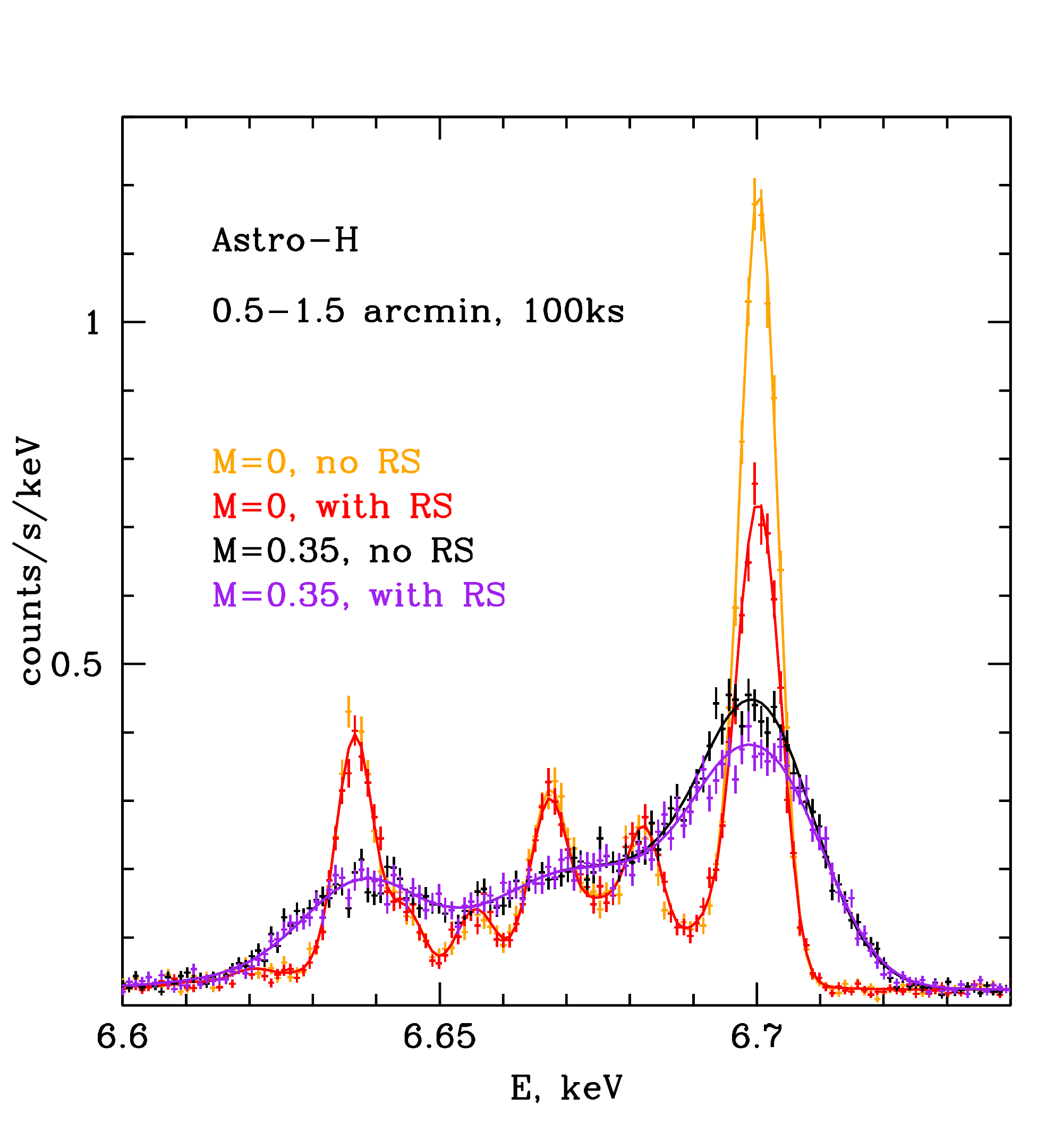}
\caption{Simulated Fe lines from the Perseus core affected by resonance
scattering (RS) for different turbulent velocities as a function of
rest-frame energies. The spectrum from the 100 ks central pointing
excludes the central
$1\times1$ arcmin region, where the contribution of the central AGN is
dominating the signal. Notice, that using a 100~ks \astroh
observation, we will be able to use the suppression of the resonance line
 to measure even relatively high line-of-sight turbulent velocity of
$v_{\rm turb}=360$~km~s$^{-1}$, corresponding to the Mach number $M_{\rm
1D}=0.35$ for $kT=4$~keV.}  \label{resscat}
\end{center}\end{figure}

\bigskip \noindent{\textit{Resonance scattering}}\\*[3pt] In rich,
massive, X-ray bright galaxy clusters resonance scattering is expected
to be particularly important for the He-like Fe line (1s2p) at 6.7 keV.
In the past, its effects were studied by comparing the flux from this
line with the flux of the nearby He-like Fe line (1s3p) at 7.9 keV
\citep{churazov2004}. With the spectral resolution of CCD type
detectors, however, the 7.9 keV Fe line is blended with the He-like Ni
line at 7.8 keV, which made the interpretation of results difficult and
ambiguous.

The much improved spectral resolution of the \astroh SXS (30 times
better than the energy resolution of the current X-ray CCDs around the
6.7 keV line), and new modeling and data analysis techniques that look
for resonance scattering in all resonant lines simultaneously
\citep[e.g.][]{zhuravleva2013}, will allow us to use resonance
scattering in the Perseus Cluster to obtain meaningful constraints on turbulence.

Figure \ref{resscat} shows the expected effects of resonance scattering
on the 6.7 keV He-like Fe-K line in the ICM for the central \astroh
pointing, assuming isotropic turbulence with a characteristic
line-of-sight velocity of $v_{\rm turb}=360$~km~s$^{-1}$, corresponding
to the Mach number $M_{\rm 1D} \equiv v_{\rm turb}/c_{\rm s} =0.35$,
where $c_{\rm s}$ is the sound speed.  The effects of resonance
scattering were calculated assuming a cluster atmosphere model (3D
distributions of density, temperature, and metallicity) derived from
spherically symmetric models fitted to \xmm and \chandra data
(Zhuravleva et al. 2013).

\subsubsection{The temperature structure of the ICM}

\noindent{\textit{Ion kinetic temperature}}\\*[3pt]
\label{fea:ion-t} Current X-ray detectors can only 
measure the temperature of the electrons in the ICM (from the
bremsstrahlung continuum and line ratios). The electron ion thermal
equilibration time-scales in the central regions of galaxy clusters are
small, and we expect the ion temperatures to equal the electron
temperatures. By measuring the thermal line broadening, deep
observations of the core of the Perseus Cluster with the \astroh SXS may
in principle 
verify this expectation and measure the ion temperature independently of
the electron temperature of the plasma.

Based on a single line profile, however, thermal line broadening cannot
be separated from broadening due to turbulence. Because $W_{\rm
therm}\propto E_{0}\sqrt{kT_{\rm e}/m_{\rm ion}}$ (where $E_{0}$ is the
line energy and $m_{\rm ion}$ is the ion mass), constraints on the ion
temperature require simultaneous measurement of the broadening of lines
from different ions with a large range of ion mass\footnote{We note that
the thermal broadening of low mass ions observed at lower energies is
small. E.g. the thermal widths for \ion{O}{VIII} (0.6~keV),
\ion{Mg}{XII} (1.47 keV), \ion{Si}{XIV} (2.01 keV), and \ion{Fe}{XXV}
(6.70 keV) are 0.79 eV, 1.46 eV, 1.84 eV, and 4.35 eV, respectively.}.
Under the simplifying assumptions of a single-temperature plasma of a
single characteristic velocity of isotropic turbulence of $v_{\rm
turb}=141$~km~s$^{-1}$, we simulated \astroh SXS spectra for a 100~ks
observation of the core of the Perseus Cluster. Fitting for the ion
temperature simultaneously with the other free parameters (see the
parameters in Table~\ref{simtable} and \ref{simtable7ev}) allows us to
measure $kT_{\rm ion}$ with statistical uncertainties of $kT_{\rm
ion}=3.43\pm0.49$~keV. If the turbulence is stronger, with a
characteristic velocity of $v_{\rm turb}=300$~km~s$^{-1}$, the
constraints on the ion temperature will be significantly weaker $kT_{\rm
ion}=3.43^{+1.7}_{-0.8}$~keV.

We note, however, that because the plasma in the cluster core is not single-temperature, and the gas motions in the cluster core can most likely not be characterized by isotropic turbulence with a single characteristic velocity, the measurements of the ion kinetic temperature will have significant systematic uncertainties.  

\begin{figure*}
\begin{center}
\includegraphics[width=0.8\textwidth]{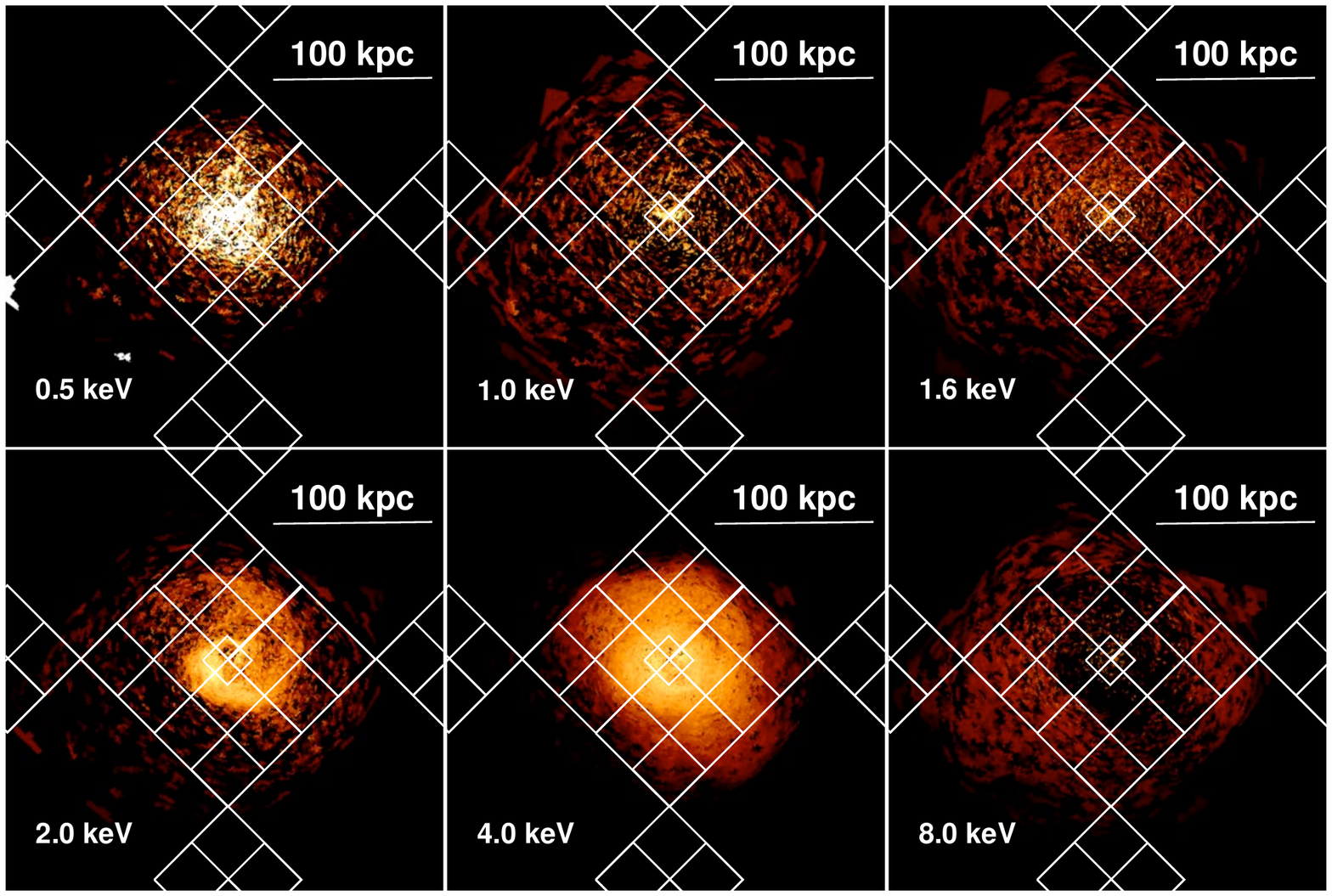}
\end{center}
\vspace{-0.3cm}
\caption{Results from a multi-temperature fit to the \chandra data of
the core of the Perseus Cluster. The images show the emission measure
per unit area of the different temperature components with the 
central \astroh pointing over-plotted. From \citet{sanders2007}.}
\label{multitemp}
\end{figure*}

\bigskip
\noindent{\textit{Multi-temperature structure and cooling}}\\*[3pt]
For central pointings we assume an absorbed multi-temperature
plasma model, the input parameters for which were determined using 2D
multi-temperature spectral maps produced by fitting deep \chandra data
of the core of the Perseus Cluster \citep[see
Figure~\ref{multitemp},][]{sanders2007}.  Cooling X-ray emitting gas (with
$kT\lesssim0.9$~keV) in the cluster center will be identified based on
the spectral lines of \ion{Fe}{XVII} and potentially by the presence of
the \ion{O}{VII} lines ($kT\lesssim0.35$~keV). While the \ion{Fe}{XVII}
lines have been detected in a number of cooling core clusters, including
the Perseus Cluster, the \ion{O}{VII} lines have never been seen in any
individual system - \ion{O}{VII} has only been detected in the stacked
archival RGS data of the coolest groups, clusters, and giant elliptical
galaxies \citep{sanders2011}.

Unfortunately, due to its decreased energy resolution ($\Delta E/E$) below 1~keV  and a smaller effective area, \astroh SXS is not expected to be significantly better at detecting cooling in cluster cores than \xmm RGS. The $\sim$0.5~keV phase that has been detected in the core of the Perseus Cluster with \xmm RGS and spatially mapped with \chandra will be detected with \astroh SXS at 6.1$\sigma$ significance (assuming an exposure of 100~ks). 
An important advantage, however, with respect to \xmm RGS is that
\astroh SXS will be providing high resolution spectra in well defined regions, also outside of the bright cluster core. 

Furthermore, deep \astroh observations of the Perseus Cluster may be used to search for departures from simple thermal Maxwellian electron distributions in the ICM. Non-thermal electrons, producing high-energy tails to the thermal Maxwellian electron distributions, may be accelerated in weak MHD shocks. The presence of such non-Maxwellian tails may be revealed by looking at the di-electronic satellite lines in the Fe-K band \citep[see][]{kaastra2009}.

\bigskip
\noindent{\textit{Multi-mission data analysis techniques}}\\*[3pt]
The combination of high resolution spectra from \astroh SXS and detailed 2D spectral information (with thousands of independent spatial bins in each of which we determine emission measure, temperature, metallicity, pressure, entropy) obtained in very deep observations with \chandra and \xmm, will provide a powerful tool to study the 3D thermal properties of the ICM. 
While the detailed spatially resolved spectral maps provide key information on the projected 2D temperature distribution, fitting multi-temperature differential emission measure models to high quality \astroh spectra will reveal the temperature structure along the line-of-sight.

\subsubsection{Chemical abundance measurements}
\label{sec-perseus-chem}

As shown in Table~\ref{simtable} and \ref{simtable7ev} we will be able
to accurately measure the chemical abundances of O, Ne, Mg, Si, S, Ar,
Ca, Fe, and Ni. The systematic uncertainties on the abundances will be
reduced by the dramatically improved spectral resolution of \astroh SXS.
Furthermore, the 
observations of this X-ray brightest galaxy
cluster will also 
permit measurement of
the abundances of rare elements
such as Na, Cl, Al, Cr, and Mn. The 100 ks pointing on the cluster core
will allow us to measure 
the abundances of Cr, Mn, and Al at a better
than $5\sigma$ significance, and the abundance of Na and Cl with a
significance better than $3\sigma$ (see Sec. \ref{sec-chem} for
details).  We will not be able to measure the abundance of N.

\subsubsection{The AGN emission}

The AGN emission of NGC~1275 has been studied in detail at all
wavelengths from radio to gamma rays all the way up to GeV
energies. However, detailed information in hard X-rays is relatively
scant. The data obtained by the Hard X-ray Imager (HXI) on \astroh will
allow us to study the spectral properties of the AGN and its time
variability up to energies of several tens of keV.

\subsubsection{Diffuse non-thermal emission}
The broad band X-ray data of the Perseus Cluster obtained with the wide
array of instruments on board \astroh, including the SXS, SXI, and the
HXI will also be used to search for non-thermal and very hot thermal
emission \citep[e.g.][]{Ota2008}. The hard X-ray data obtained in the
long exposures in combination with careful spectral modeling in
the soft X-rays will provide good constraints on diffuse hard X-ray
emission above the thermal emission (see Sec. \ref{sec-highe} for more
details).

\subsection{Remarks on sources of systematic uncertainty}

The Perseus Cluster, like all relaxed galaxy clusters, has a very peaked surface brightness distribution and the amount of light scattered from the bright cluster core into the off-center pointings will need to be carefuly evaluated (though the proximity of the target minimizes these concerns to the fullest extent possible). Based on the currently available simple azimuthally symmetric point-spread-function (PSF) model, in the radial range of 10--40~arcmin  the expected ratio of scattered light to signal is negligible. This means that our pointings planned at $r=240$~kpc and at $r=600$~kpc will not be affected by scattered light contamination. For the pointings at $r\sim140$~kpc, however, the preliminary, azimuthally symmetric PSF model predicts that up to 20\% of the photons will be scattered from the outside of the SXS field of view. However, according to our current understanding of the \astroh mirrors, the amount of light scattered from certain directions will be minimal - just like with the \suzaku mirrors - and we may be able to reduce the contamination by observing with one of the detector corners pointing towards the bright cluster center (this is the configuration shown in Figs~\ref{pointings2} and \ref{pointings1}). 
The exact strategy for minimizing scattered light contamination when observing the vicinity of the bright cluster core will need to be further explored once the Soft X-ray Telescope calibration files for the ray tracing simulator become available. 

The analysis of data from individual 1.5'$\times$1.5' resolution
elements will be challenging. With a half-power-diameter (HPD) for the
PSF of 1.3', such square shaped regions will contain only about 60\% of
the flux from the corresponding region on the sky - the rest of the
photons will be scattered out into the surrounding regions (this is an
estimate assuming a point source and even more photons will be scattered
out for a source with an extended surface brightness distribution). In
the same time photons from the surrounding regions will be scattered
into the region of interest. Spatially resolving the
velocity structure of the gas in the five pointings spanning the central $130\times130$~kpc$^{2}$ or in the
surrounding pointings at $r=140$~kpc will therefore require an excellent
understanding of the shape of the telescope PSF.

Finally, in order to take full advantage of the progress in X-ray instrumentation, innovation in the spectroscopic codes used for fitting the observed spectra will also be crucial. 
Updates in the ionization balance, implementation of a more extensive set of atomic parameters for Fe and Ni L-shell lines, and inclusion of rare elements that are currently not implemented in spectral fitting packages will be important. These updates will benefit the whole astrophysics community.





\bigskip 

\section{A Detailed View of AGN Feedback}
\label{sec-virgo}

\subsection*{Overview}

Active galactic nuclei (AGN) are the heating engines responsible for
preventing a cooling catastrophe in cluster cores. AGN inflate
bubbles filled with relativistic plasma which displace the ICM, drive
weak shocks, and drag filaments of cool gas out of cluster centers. One
of the best targets for studying AGN-ICM interaction is M87, the central
dominant galaxy of the Virgo Cluster. High-resolution SXS spectra for
this target will reveal the dynamics and microphysics of AGN feedback,
allowing us to map the line-of-sight velocity component of the gas in
the X-ray bright arms uplifted by the AGN, and measure the turbulence
induced by the rise of the radio bubbles and by the outer shock
front. We will place reliable constraints on the multi-phase structure
and distribution of cool gas in the cluster core, and accurately measure
the metallicity of the uplifted gas independently from that of the
surrounding ICM, robustly determining how gas motions induced by the AGN
spread out metals produced in the central galaxy.
\footnote{Coordinators of this section: A. Simionescu, K. Matsushita}

\subsection{Background and Previous Studies}

\subsubsection{The cooling flow problem and the necessity for AGN feedback}

If the energy radiated away at the centers of so-called `cool-core'
clusters of galaxies, which show sharp X-ray surface brightness peaks
and central temperature dips, came only from the thermal energy of the
hot, diffuse ICM, the ICM would cool and form stars at rates orders of
magnitude above what the observations suggest \citep[see][for a
review]{peterson2006}. It is currently believed that the energy which
offsets the cooling is provided primarily by interactions between active
galactic nuclei (AGN) in the central dominant galaxies of these clusters
and the surrounding ICM \citep[e.g.][]{churazov2000, Churazov01,
McNamara07}. Through a tight feedback loop, it is thought that the AGN
can provide the right amount of energy to prevent catastrophic cooling
and stem star formation.

Generally speaking, there are three main mechanisms by which the AGN-ICM
interactions take place. Cluster AGN have been observed to: 1) drive
weak shocks into the intracluster plasma, 2) inflate bubbles filled with
relativistic plasma, which then rise buoyantly, and 3) drag filaments of
cool gas out of cluster centers, in the wakes of the buoyantly rising
bubbles. It is not yet clear what determines the fraction of the total
energy imparted by the AGN that goes into each of these three processes
\citep{vr07,moj11}, nor do we understand in detail how the energy
associated with AGN-injected bubbles and uplifted gas filaments
eventually dissipates into heat.

While AGN-inflated bubbles and weak shocks are clearly seen in the innermost regions of the Perseus Cluster \citep[e.g.][]{bohringer1993,fabian2006}, the cool gas filaments uplifted by the radio lobes in this system are very faint in comparison with the rest of the ICM \citep[e.g.][]{sanders2007}, and the features related to the AGN feedback process will be difficult to resolve spatially with {\it ASTRO-H}. The core of M87 in the nearby Virgo Cluster therefore presents an ideal target for a more detailed look at the dynamics of AGN-ICM interaction.

\subsubsection{Previous observations of AGN feedback in M87}

The Virgo Cluster is the nearest galaxy cluster to us ($\sim16.7$ Mpc),
which allows us to resolve phenomena with exceptional spatial
resolution. Its central galaxy, M87 is among the brightest extragalactic
X-ray sources in the sky, providing a guarantee of excellent photon
statistics and accurate temperature and metal abundance
determinations. M87 is one of the best studied galaxies in the Universe
and its central region has been the target of very deep observations
with {\it XMM-Newton}, {\it Chandra}, {\it Suzaku}, and an array of
multiwavelength facilities.

The hot gas atmosphere of M87 shows striking signs of AGN-ICM interaction, including an AGN-driven classical shock \citep{Forman05, Forman06, Simionescu07, Million10}, X-ray cavities \citep{Forman05, Forman06, Million10}, and clear enhancements in X-ray surface brightness associated with the main radio lobes to the east and southwest of the core \citep{Feigelson87, Boehringer95, Belsole01, Young02, Forman05, Forman06, Werner10}. Initial {\it XMM-Newton} observations showed that multi-temperature models including cool gas components were needed to describe the spectra of these regions, known as the E and SW X-ray arms \citep{Belsole01, Molendi02}. 
 
With deeper {\it XMM-Newton} data, a correlation was found between the amount of cool gas and the metallicity in different regions along the arms, leading to the indirect conclusion \citep{simionescu2008a} that the cool, metal-rich gas is uplifted by the AGN from the central parts of the galaxy \citep[following the scenario proposed by][]{Churazov01} where metals are being produced by supernova explosions and stellar winds. \cite{Werner10} showed that the mass of gas in the arms is comparable to the gas mass in the approximately spherically symmetric 3.8 kpc core, demonstrating that the AGN has a profound effect on its immediate surroundings, effectively removing gas from the cluster center that would otherwise cool and form stars. 

The energetics and metal-transport mechanisms associated with the AGN activity in the central parts of M87 have therefore been the focus of much previous work, which has yielded important information about the physics of the AGN-ICM interaction. 

\subsection{Prospects and Strategy}


As discussed above, the uplifted gas in the X-ray bright arms, as well as in the core of M87, is multi-phase. Modeling the exact distribution of the temperature phases with low-resolution CCD spectra is very difficult and possibly subject to fitting biases. {\it XMM-Newton} RGS data have proven extremely useful for understanding the thermal state of the X-ray gas in the core of the galaxy and constraining the rate at which gas may be cooling out of the X-ray band \citep[below 0.5 keV, which is currently the lowest temperature detected in the M87 halo;][]{Werner06}. However, the extended nature of the X-ray bright arms has so far prevented high spectral resolution measurements of the properties of the uplifted gas with slitless grating spectrometers. These measurements are now enabled by the {\it ASTRO-H} SXS, and will be pivotal for furthering our understanding of AGN feedback physics.

\subsubsection{Dynamics of gas uplift}

From several arguments based on {\it Hubble Space Telescope}
observations of the superluminal motion in the jet of M87
\citep{biretta1999} as well as the geometry and degree of polarization
of the large-scale radio lobes, it is likely that the arms are oriented
at a relatively small angle with respect to the line of sight
\citep[][estimate this angle to be at most $35^\circ$]{Werner10}, and
therefore the {\it ASTRO-H} SXS should allow us for the first time to
measure the line-of-sight velocity of the low-entropy gas in the
arms. Detecting this outward motion of the cool X-ray gas will provide
the first direct proof for the scenario that the X-ray bright arms
indeed originate from gas uplift by the AGN. These observations will
moreover 
provide an estimate for the kinetic energy of the uplift, which
remains one of the main uncertainties in determining the total energy
associated with the AGN feedback.

\subsubsection{AGN-induced metal transport}

With lower-resolution spectra, it has so far not been possible to directly measure the metallicity of the uplifted gas independently of that of the hot ambient gas. When multi-temperature fits were performed, usually the abundances of the several phases had to be coupled to each other in order for the fit to be constrained. However, \cite{simionescu2008a} found a correlation between the amount of cool gas in a given spectral extraction region and the average metallicity of that region, which lead to the conclusion that the abundance of the cool gas must be roughly twice higher than that of the hot 2~keV phase. This is expected if the uplifted cool gas was dragged by the buoyantly rising radio lobes of the AGN from the cluster center, which is more metal-rich due to contributions from stellar winds and supernovae in M87. 

The abundance of the cool gas is a very important quantity for several reasons. Firstly, the emission from this cool 1~keV gas consists mainly of line emission; therefore, the metallicity and spectral normalization typically anticorrelate in low-resolution spectra, and in order to obtain an accurate estimate of the true emission measure, density, and mass of the uplifted cool gas, it is important to measure its metallicity and break this degeneracy. Secondly, the metallicity of the cool gas is important in order to understand the effect of the AGN on transporting metals produced by stars and supernovae in the central galaxy into the ICM. The distribution of metals in the centers of cool-core clusters of galaxies is much more broadly peaked than the distribution of optical light; if the stars produce the metals, then we must understand how the metals are spread out. \citet{Rebusco06} propose that AGN-ICM interaction plays a dominant role in this process, a role which can be quantitatively investigated for the first time using the proposed observations.

As we show in the next section, {\it ASTRO-H} will enable us to make the first accurate direct determination of the metallicity of the cool uplifted gas, and therefore measure the total gas mass and the mass of each metal that is being transported by the AGN in M87, and potentially test whether the chemical composition of the uplifted gas is different from that of the ambient gas at larger radii.

While M87 is the best system to study AGN-induced metal transport in detail, this phenomenon is wide-spread in cluster cores. Columns of relatively cool, metal-enriched gas lying along the edges of radio sources and X-ray cavities have been identified at the centers of other clusters and groups \citep[e.g. Hydra A,][]{simionescu2008b, Kirkpatrick11}, and the radial ranges of the metal-enriched outflows are found to correlate well with the jet power \citep{Kirkpatrick11}. Therefore our study of M87 will answer questions that are relevant to understanding AGN feedback in general, and will serve as a pathfinder for future observations of AGN-induced metal transport in other systems.

\subsubsection{Turbulent motions in the hot ICM}

With the observing strategy 
considered here, we will spatially map the turbulence in the hot
gas. With the excellent statistics offered by this nearby, bright
target, we will be able to take advantage of the {\it spatial}
resolution of the SXS, and measure the turbulent line broadening in 4--9
independent resolution elements, or ``spaxels'', \footnote{The term
``spaxel'' refers to a $1'\times 1'$ spatial resolution element of SXS.}
per pointing (see Section \ref{sys} for the discussion
of PSF scattering affecting the independence of such neighboring
regions).  This will allow us to compare, with high precision, the
turbulent velocities:

\begin{itemize}
\item in the cluster core, in the immediate vicinity of the AGN which is constantly injecting turbulence into the ICM;
\item within and immediately adjaecent to the X-ray bright arms, where the gas uplift by the buoyant radio lobes is driving the turbulence;
\item in a separate offset pointing targeting a more relaxed region far away from the direction of the uplift;
\item immediately inside and outside the AGN-driven shock at a radius of
      3 arcmin. While classical shocks are not expected to induce
      turbulence in the ICM, the SW arm appears very narrow inside the 3
      arcmin shock front and suddenly broadens beyond this radius
      \citep[e.g.][]{Werner10}. 
These observations will therefore help to elucidate the interplay
      between AGN uplift and shock dynamics in this region.
\end{itemize}

This will 
help us to understand the mechanisms which drive the
turbulence and to infer important details about the microphysics of the
ICM, which have so far remained elusive. In particular, we
may get a handle on the effective viscosity of the ICM,
which is also linked to the strength and geometry of the magnetic fields
which pervade the cluster.


In addition, as detailed in Section \ref{sec-chem}, these observations
will 
place important constraints on
types of supernova explosions that have contributed to the
chemical enrichment of the ICM over the life-time of the cluster.

\subsection{Targets and Feasibility}

\begin{figure*}
\begin{center}
\includegraphics[width=0.49\textwidth, bb=0 0 541 534]{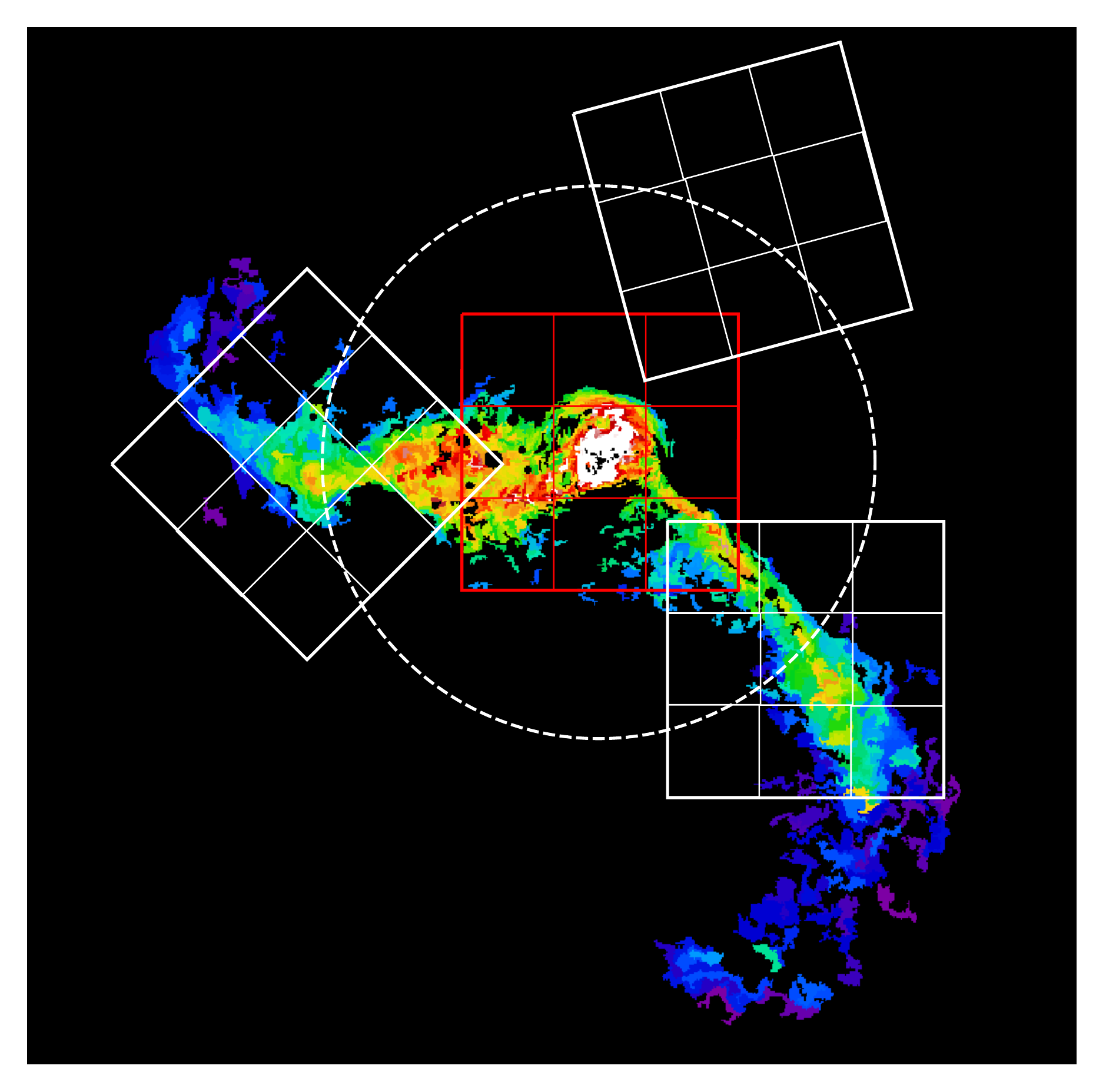}
\includegraphics[width=0.49\textwidth, bb=0 0 541 534]{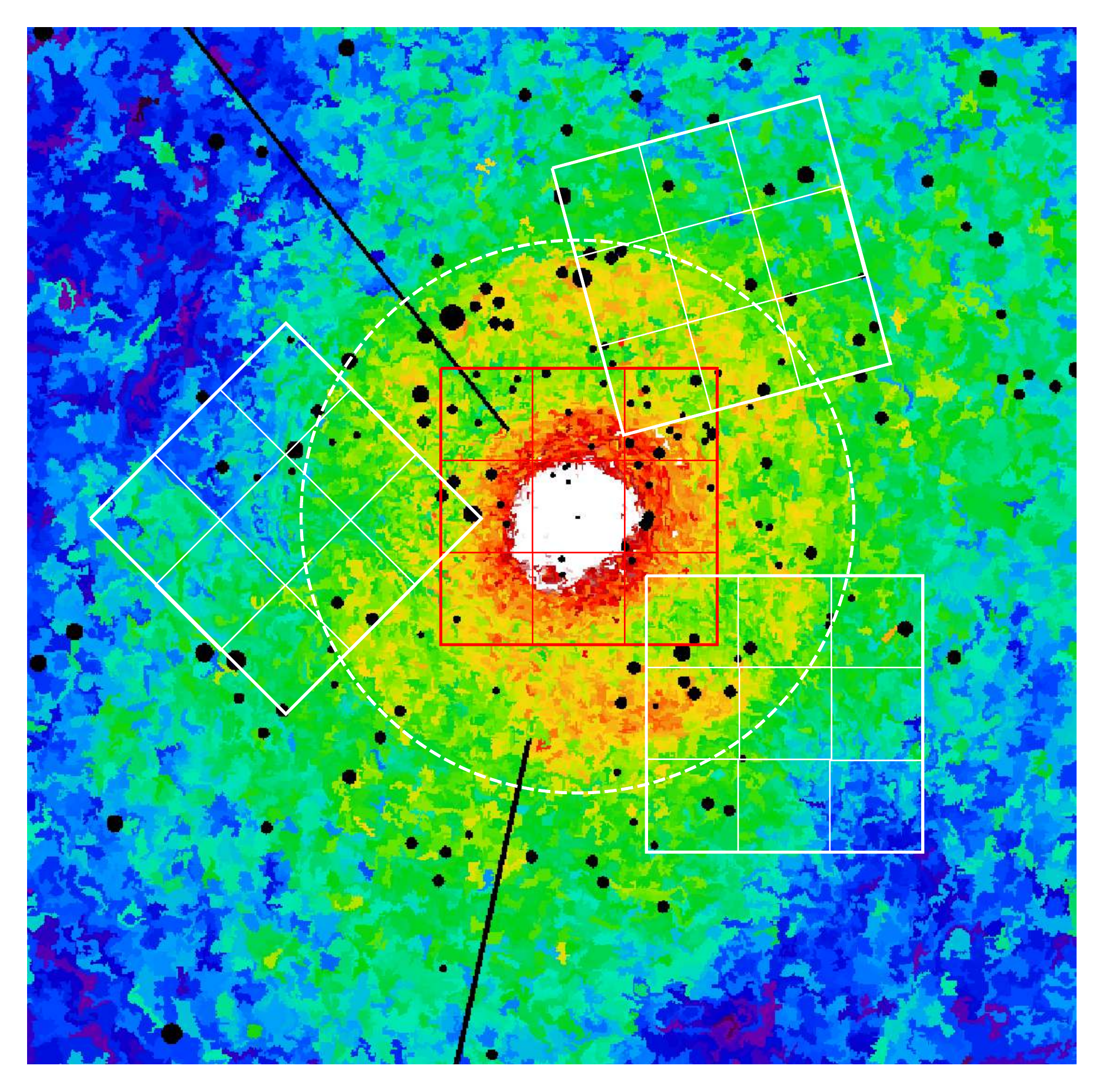}
\end{center}
\caption{{\it Left:} Spectral normalization of the uplifted cool gas from {\it Chandra} spectral fitting \citep{Werner10}. The proposed pointings (shown as boxes, each divided into nine spaxels) will probe the center, the E and SW arms, as well as one off-arm pointing where no (or significantly less) disturbance is expected due to the uplift. 
{\it Right:} Average pressure map from {\it Chandra} spectral fitting \citep{Million10}, showing evidence for a weak shock front with a radius of 3 arcmin (marked with a dashed white circle). The proposed pointings will probe the turbulence in the hot gas both inside and outside this shock front. 
At the distance of M87, each $3\times3^\prime$ box corresponds to a
 $14\times14$ kpc field of view.}
\label{pointings}
\end{figure*}

The goals stated above can be achieved with four pointings, each 100~ks
long (for a total of 400~ks). The locations of these pointings are shown
in Figure \ref{pointings}. One pointing will be centered on the
cluster core, while three others will be offset by approximately 3
arcminutes towards the E, SW, and NW. The E and SW pointings will sample
the two X-ray bright arms, which are believed to have been produced as
the AGN in M 87 inflated bubbles filled with relativistic plasma which
rose buoyantly through the cluster atmosphere, dragging filaments of
cool gas out of the cluster center in their wake. The NW pointing, in
turn, will probe a relatively undisturbed off-axis region. Because the
target is bright and nearby, the systematic uncertainties due to the
cosmic and instrumental backgrounds are minimal, and we will obtain
excellent statistics as demonstrated here.

\subsubsection{The central pointing}

The central pointing will allow us to measure accurately the properties
of the cool gas in the approximately spherically symmetric 3.8 kpc core
\citep[][and Figure \ref{pointings}]{Werner10}, and determine the level
of turbulent motions in the hot, 2~keV gas, in the immediate vicinity of
the AGN that is generating this turbulence. 
%

For a feasibility study, we extracted {\it XMM-Newton} MOS spectra of three annular regions (0.0'-0.5', 0.5'--1.5', and 1.5'--2.5') centered on M~87, and fitted each spectrum with a two-temperature vAPEC model.
We then used simx-1.3.1 to simulate the SXS spectra for the central pointing with a 5 eV response file, assuming exposure times of 50 and 100 ks.

\begin{figure}
\begin{center}
 \includegraphics[width=0.49\textwidth, bb=0 0 792 612]{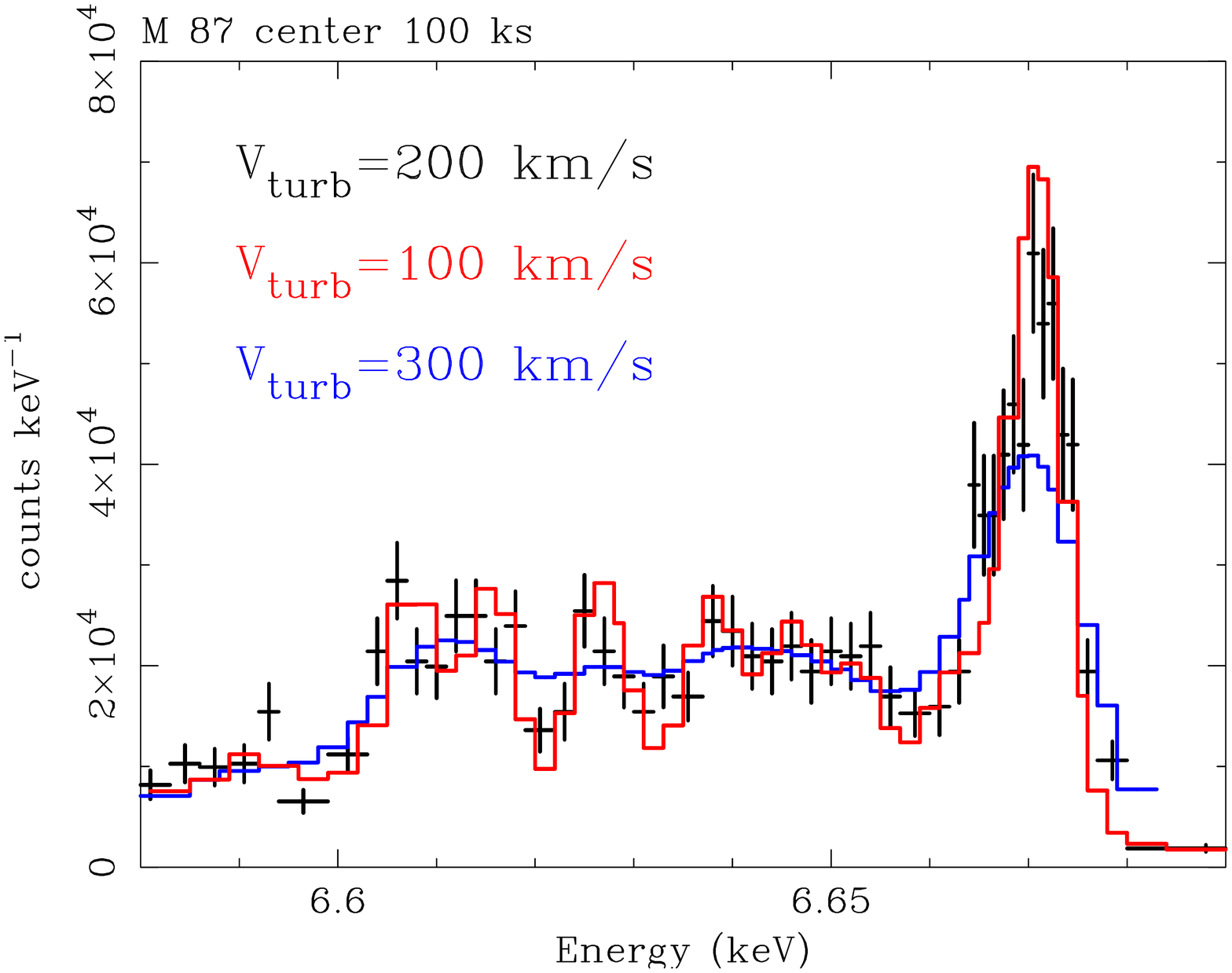}
 \includegraphics[width=0.49\textwidth, bb=0 0 792 612]{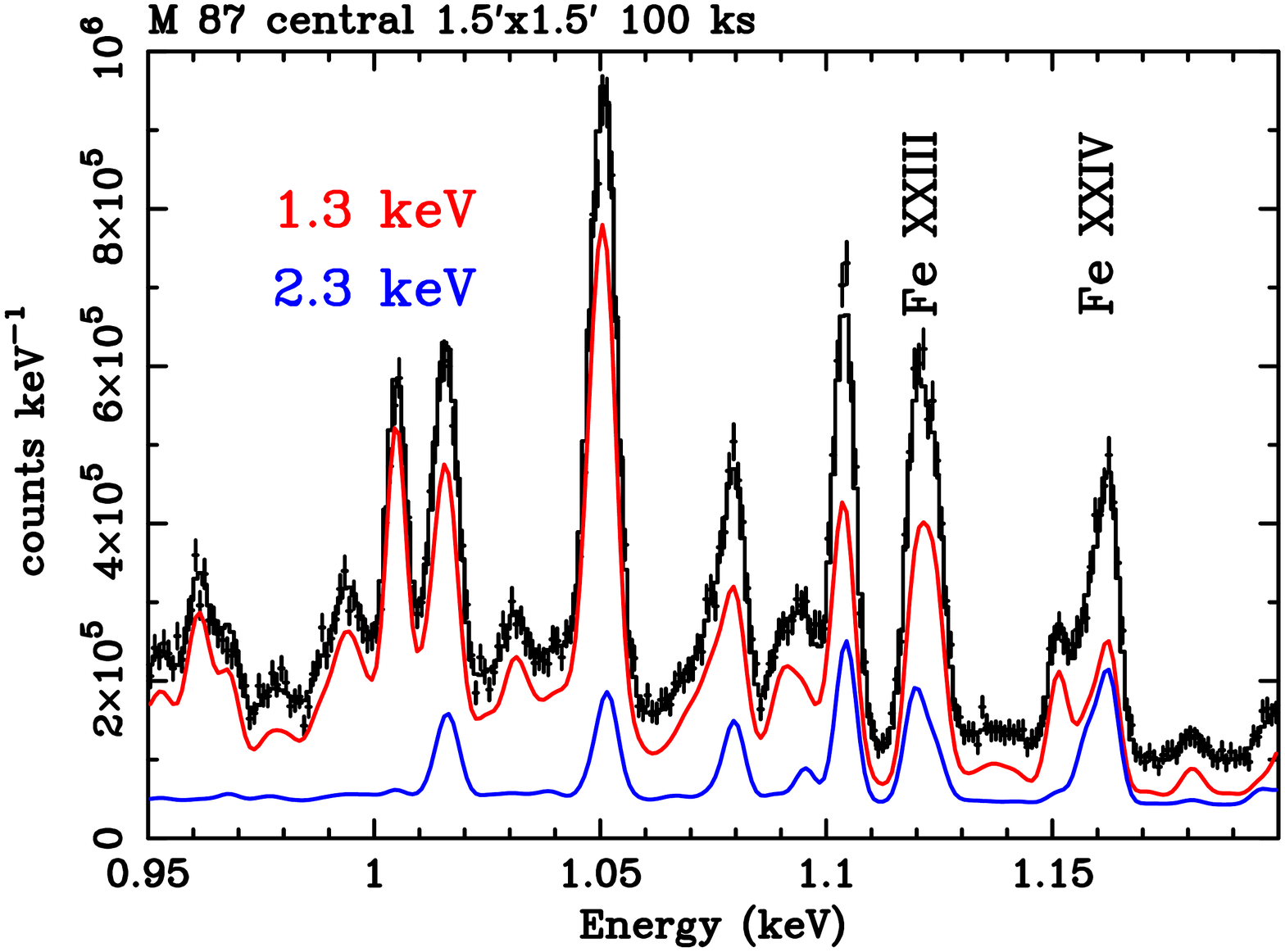}

\caption{ (left) The simulated SXS spectrum for the central 3'x3' of M
87 around the 6.7~keV Fe-K lines with a 100 ks exposure.  (right) The
same as the left panel but for the central 1.5'x1.5' of M 87 around the
Fe L lines. The Fe L lines with optical depth $\sim$ 1 are labeled. The
red and blue lines show the contributions of the two temperature
components.  } \label{m87turb}
\end{center}
\end{figure}

As shown in the left panel of Figure \ref{m87turb}, the 6.7 keV Fe line is dominated by the hotter component. Therefore, using this line alone to determine the turbulent velocity of the $\sim2$~keV gas provides an alternative method which is entirely independent of the properties of the cool component.
For example, assuming a 
$v_{\rm turb}=$200 km/s turbulence,
fits of the simulated SXS spectrum around the 6.7 keV line (6.0--7.5 keV) give turbulent velocities of 170 $\pm$ 20 km/s and 200 $\pm$ 13 km/s with 50 ks and 100 ks exposures, respectively, considering only $1\sigma$ statistical errors.

The study of the effect of the resonant line scattering of the 6.7 keV
line is complementary with direct measurements of the turbulence using
line broadening.  In the absence of turbulent motions, the expected
optical depth of the 6.7 keV line of the core of M 87 is 1.4
\citep{Churazov10}, and therefore we would expect this line to be
suppressed. This suppression is reduced by the presence of random gas
motions.  Just as in the case of Perseus discussed in Section
\ref{sec-perseus}, the 
ratio of the 6.7~keV (1s2p) line strength to that of the 7.9~keV (1s3p)
He-like Fe line, the latter of which is expected to be optically thin,
can be used as an indicator of the effect of scattering, which in turn
depends on the level of turbulence in the ICM.

However, with CCD detectors, the He-like Ni line at 7.8 keV and the
He-like Fe line at 7.9 keV are blended into a single bump, and thus the
inferred level of the resonant scattering in the 6.7 keV line couples
with the Ni/Fe ratio. The improved spectral resolution of the {\it
ASTRO-H} SXS will
refine such measurements of the resonant line scattering.

Unfortunately, because of the low average temperature of the ICM in M87,
the number of photons in the He-like Fe line at 7.9 keV over the
central SXS field of view is expected to be only 60 counts with a 100 ks
exposure, therefore we must use either a significantly longer
observation, or use other pairs of lines in order to measure the effects
of resonant scattering. One possibility is to compare the line strengths
of the resonance line at 6.7 keV to other lines within the 6.7 keV line
complex.
With a 100 ks exposure, the expected total counts in the 6.7 keV line 
complex
and the resonance line at the center of M 87 with SXS are
about 1700 and 460 counts, respectively, which provides sufficient
statistics for constraining the optical depth of the resonance line.
However, the flux ratios of lines in the 
complex
depend also on the ICM temperature, and there may be systematic
uncertainties due to the atomic data. The temperature structure in the
ICM can be constrained accurately, using the bremsstrahlung continuum as
well as other line ratios (Si, S, Ar and Ca).  To constrain the
systematic uncertainties in the plasma codes, we can use the spatial
variation of the contribution of the resonance line to the other lines
in the 6.7 keV line 
complex.  
In the cluster core, the optical depth of the resonance line is expected
to be the highest, while this line should be practically optically thin
at larger radii. In the central 1.5'x1.5' region, we can use 40\% of the
total counts for the central 3'x3' region, 
and test for this spatial trend (though this measurement
will be affected by PSF scattering and require accurate modeling).

Moreover, the expected optical depths of two Fe-L lines (Fe XXIII at 1.129 keV and Fe XXIV at 1.168 keV) are about unity \citep{Churazov10}.
As shown in the right panel in Figure \ref{m87turb}, these two lines mostly come from the 1 keV component. 
Therefore, we can use these two lines to constrain the turbulent motions of the cool 1 keV component in the cluster core. Because the lines from this cool component are seen at low energies, where the spectral resolution $\Delta E/E$ is not as high as at 6.7~keV, resonant line scattering may offer the only viable method to measure the turbulent velocities in the cool gas. Again, this will require systematic uncertainties in the atomic data to be taken into account.

\subsubsection{The offset pointings}

We have extracted spectra of the off-center regions shown in Figure \ref{pointings} using a deep {\it Chandra} observation and applying the latest calibration. Using the SPEX spectral analysis package, the spectra for the E and SW arms were fit with a two-temperature model, while the spectra for the NW offset pointing were fit with a single temperature model. Following the conclusions of \cite{simionescu2008a}, we fixed the abundance of the cool 1~keV gas to be twice that of the hot 2~keV ambient phase. 
The best-fit abundances of Si, S, Ar, Ca, and Fe for the hot gas were consistent with 1 solar in the units of \cite{grevesse1998}, and were therefore fixed to this value. For O, Mg, and Ne, where {\it Chandra} spectra do not allow an accurate determination of the metal abundance, we have assumed their ratios with respect to Fe to be 0.5, 0.6, and 0.8, respectively, as determined with RGS \citep{Werner06} and converted to the solar units assumed here.

Using the best-fit parameters from the {\it Chandra} fits, we then
assumed a line-of-sight velocity of the cool gas of $v_{\rm bulk}^{\rm
cool}=300$
km/s (roughly half of the sound speed in the ambient gas) and a
turbulent velocity in hot gas of
$v_{\rm turb}^{\rm hot}=$141 km/s, and simulated the expected spectra for a
100~ks observation with the SXS, using the predicted 5eV response files.
We have included the cosmic and instrumental background models, however we find these to be an insignificant contribution to the expected emission from the source.

We find that, ignoring systematic uncertainties and using a full band fit in the 0.3--10 keV range, we can measure parameters of interest with the statistical precisions shown in Table \ref{basesim}. We show simulations of the full field of view for the E and SW pointings, as well as one simulation of a typical 1x1 arcmin spaxel in a two-temperature region (the center of the SW arm) and a typical 1x1 arcmin spaxel in a single temperature region (the center of the NW pointing). We note that the C-stat space in our simulations is complex, with more than one local minimum. This is reflected by several highly asymmetric formal $\Delta C=1$ error bars in Table \ref{basesim}. Monte Carlo simulations will be needed to assess the error bars more robustly. 



\begin{table}[t]
\caption{Simulation of the offset pointings. Statistical error bars are given at the $\Delta$C=1 level. Detections below the $4\sigma$ level are not reported.}\label{basesim}
\begin{footnotesize}
\begin{center}
\begin{tabular}{lccccc}
\hline
Pointing position & E & SW & SW & SW  & NW  \\
 & entire pointing & entire pointing & central spaxel & central spaxel &
		     central spaxel \\
Exposure time (ks) & 100 & 100 & 100 & 50 & 100\\ 
\hline 
$kT$ cool (keV) & $1.14_{-0.01}^{+0.01}$ & $1.05_{-0.01}^{+0.01}$ & $1.13_{-0.02}^{+0.01}$ & $1.09_{-0.01}^{+0.03}$ & - \\
$Y$ cool ($10^{64}$ cm$^{-3}$) & $0.31_{-0.03}^{+0.02}$ & $0.18_{-0.01}^{+0.005}$ & $0.086_{-0.004}^{+0.008}$ & $0.073_{-0.003}^{+0.006}$ & - \\
$kT$ hot (keV) & $2.11_{-0.01}^{+0.01}$ & $2.14_{-0.01}^{+0.01}$ & $1.94_{-0.03}^{+0.02}$ & $1.92_{-0.02}^{+0.05}$ & $2.40_{-0.03}^{+0.03}$ \\
$Y$ hot ($10^{64}$ cm$^{-3}$) & $4.51_{-0.03}^{+0.04}$  &$5.27_{-0.03}^{+0.03}$ & $0.61_{-0.02}^{+0.01}$ & $0.63_{-0.02}^{+0.01}$ & $0.503_{-0.006}^{+0.006}$ \\
\hline
$v_{\rm bulk}$ cool (km/s) & $311_{-23}^{+33}$ & $303_{-18}^{+14}$ & $280_{-33}^{+13}$ & $256_{-28}^{+60}$ & - \\
$v_{\rm turb}$ hot (km/s) & $135_{-8}^{+8}$ & $153_{-8}^{+7}$ & $142_{-27}^{+28}$  & $134_{-30}^{+30}$ & $138_{-26}^{+24}$ \\
\hline
O cool & $0.98_{-0.13}^{+0.14}$ $^\dagger$& $0.82_{-0.22}^{+0.26}$ $^\dagger$ & $0.96_{-0.22}^{+0.22}$ $^\dagger$ & - & - \\
Si cool &$1.97_{-0.17}^{+0.22}$& $2.85_{-0.40}^{+0.44}$ & $2.02_{-0.47}^{+0.50}$ & - & - \\
S cool &$2.17_{-0.30}^{+0.53}$ & $3.13_{-0.53}^{+0.76}$ & $1.94_{-0.37}^{+0.54}$  & - & - \\
Fe cool &$2.08_{-0.10}^{+0.09}$ & $2.28_{-0.11}^{+0.14}$& $1.93_{-0.16}^{+0.08}$ &  $1.72_{-0.15}^{+0.19}$ & - \\
\hline
O hot & - & - & - & - & $0.52_{-0.06}^{+0.07}$ \\
Ne hot &  $0.82_{-0.03}^{+0.02}$ & $0.83_{-0.02}^{+0.02}$ & $0.76_{-0.07}^{+0.08}$ & $0.89_{-0.08}^{+0.10}$ & $0.89_{-0.08}^{+0.08}$ \\
Mg hot &$0.61_{-0.02}^{+0.02}$ & $0.58_{-0.02}^{+0.02}$ &  $0.52_{-0.06}^{+0.06}$ & $0.66_{-0.08}^{+0.08}$ & $0.63_{-0.07}^{+0.07}$ \\
Si hot & $0.97_{-0.03}^{+0.02}$  & $0.98_{-0.02}^{+0.02}$ & $0.96_{-0.08}^{+0.08}$ & $0.83_{-0.09}^{+0.10}$ & $0.94_{-0.06}^{+0.07}$ \\
S hot & $0.95_{-0.03}^{+0.03}$  & $0.96_{-0.03}^{+0.03}$ & $0.98_{-0.09}^{+0.08}$  &$1.01_{-0.16}^{+0.14}$& $0.93_{-0.08}^{+0.08}$ \\
Ar hot & $0.98_{-0.07}^{+0.07}$ & $0.96_{-0.06}^{+0.07}$  & $1.12_{-0.20}^{+0.21}$ & $1.27_{-0.27}^{+0.35}$  &$1.18_{-0.23}^{+0.25}$ \\
Ca hot & $1.05_{-0.08}^{+0.08}$ &  $0.96_{-0.07}^{+0.07}$ & $1.33_{-0.24}^{+0.27}$ & - &$0.83_{-0.20}^{+0.25}$ \\
Fe hot & $0.98_{-0.01}^{+0.01}$ & $1.01_{-0.01}^{+0.01}$ & $0.94_{-0.05}^{+0.04}$ & $0.99_{-0.06}^{+0.06}$ &$1.03_{-0.04}^{+0.04}$ \\
Ni hot & $0.95_{-0.05}^{+0.06}$ &  $1.00_{-0.05}^{+0.05}$ & $1.02_{-0.15}^{+0.14}$ & $0.92_{-0.22}^{+0.19}$& $1.22_{-0.18}^{+0.19}$ \\
\hline
\end{tabular}
\end{center}
$^\dagger$ ${\rm O^{hot}}$ fixed with respect to ${\rm Fe^{hot}}$ by the ratio measured in the relaxed off-center pointing.
\end{footnotesize}
\end{table}


Our simulations use exposure times that provide robust measurements of
the Fe abundance and line-of-sight velocity of the uplifted gas, ${\rm
Fe^{\rm cool}}$ and $v_{\rm bulk}^{\rm cool}$, as well as the level of
turbulence in the ambient 2~keV gas.
These are the quantities which could not be constrained before {\it ASTRO-H}.
These exposure times 
will allow us to spatially map the gas motions by determining
$v_{\rm bulk}^{\rm cool}$ with a 10\% precision and obtaining a
$>5\sigma$ detection of the turbulent velocity broadening in each 1x1
arcmin spaxel. Using the full field of view of each offset pointing, we
will obtain ${\rm Fe^{cool}}$ with an exquisite statistical precision of
5\%, while $v_{\rm bulk}^{\rm cool}$ can be measured to within $\pm 20$
km/s and
$v_{\rm turb}^{\rm hot}$ to within $\pm10$ km/s. Assuming a 7~eV
response instead, we can measure ${\rm Fe^{cool}}$ to 7.5\% precision
per full field of view and 
$v_{\rm turb}^{\rm hot}$ to within $\pm15$ km/s (a 50\%
increase of the error bars compared to 5~eV), while the uncertainty on
$v_{\rm bulk}^{\rm cool}$ 
remains unchanged.

The statistical errors given in Table \ref{basesim} can be considered as a lower limit on the precision that we expect to achieve in our observations. Most of our results will likely be systematics dominated, as explained in more detail in Section \ref{sys}. The exposure times that we have used for these simulations are designed such that, per full field of view, we will still obtain significant detections of the three quantities of interest discussed above, even with a lower spectral resolution or if a possibly significant fraction of counts is scattered outside the detector footprint due to the PSF. Moreover, we have required a sufficient number of counts in order to measure the centroid of at least three Fe-L lines with a statistical precision of around 0.1 eV (30 km/s at 1~keV). A zoom-in on the expected spectrum of the Fe-L complex for the E arm is shown in Figure \ref{specmodel}. We clearly detect lines emitted only by the cool gas, only by the hot gas, and by both phases. The {\it relative} velocity shift between the lines from the cool and the hot gas should be much less subject to systematic biases than an absolute measurement of the bulk velocity of the gas, which is limited by the nominal 1--2~eV gain calibration uncertainty.


The metal abundances of O, Ne, Mg, Si, S, Ar, Ca, Fe, Ni in the hot gas are all determined with statistical errors of less than 10\%. The errors on the abundances of other metals in the cool gas, e.g. Si and S, are relatively large. To measure the chemical composition of the uplifted gas, therefore, we can hope to combine the SW and E pointings for an improved accuracy. Because only one line from O is seen in the spectrum, the abundances of O in the hot and cold gas can not be fit independently. We can estimate the O abundance in the cool gas by fixing the ratio of O/Fe in the hot gas to the value determined in the single-temperature regions. We obtain in that case a similar accuracy for O as that for Si and S (15--30\%).


\subsection{Remarks on sources of systematic uncertainty}
\label{sys}

Of potential concern in terms of the systematic uncertainties for these
measurements are PSF scattering from the bright center
of M87 outwards and scattering between the neighboring 1'-1.5' regions,
and gain calibration.
We have modeled the spectra from $r<0.5^\prime$ and 0.5-1.5$^\prime$
from the center of M87 obtained with {\it XMM-Newton} and, using a
radially symmetric model for the PSF, predicted the amount of PSF
scattering expected at 3 arcmin off axis towards the north, for the same
distance from the center where our offset pointings are located.  The
result is shown in Figure \ref{stray} and demonstrates that the
scattered light contribution from the center of M87 should be small.

\begin{figure}
\begin{center}
\includegraphics[width=0.6\textwidth, bb=0 0 574 574]{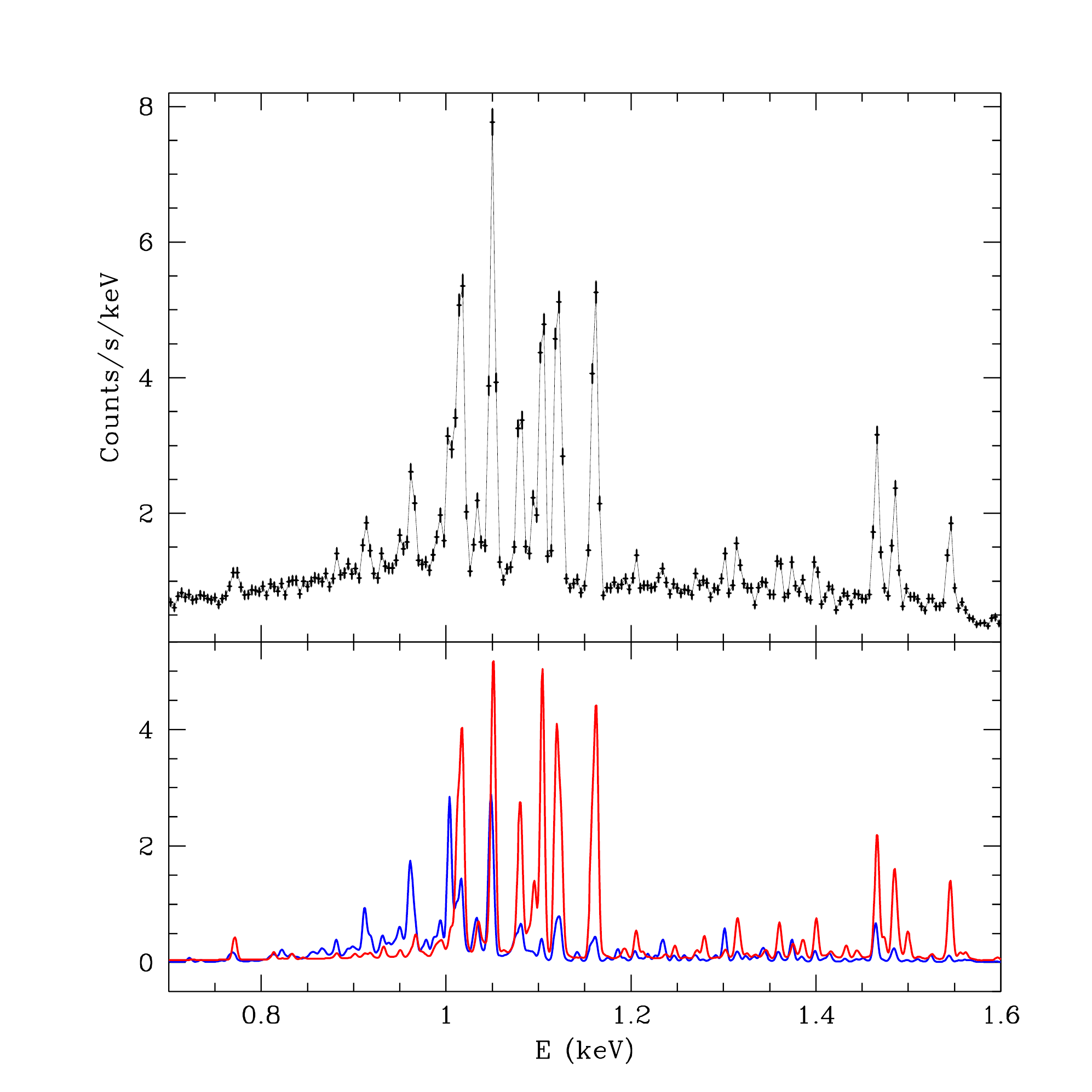}
\end{center}
\caption{Zoom in on the Fe-L complex in the X-ray bright arms. The top panel shows the expected spectrum for the E pointing with {\it ASTRO-H}, while the bottom panel shows the line contributions to the model from the cool 1~keV component (in blue) and the hot 2~keV component (in red).}\label{specmodel}
\end{figure}


Moreover, it will be important to understand how many counts are lost from a given spatial extraction region due to the PSF and, in turn, how many photons from immediately outside the footprint of the
detector are scattered into the field of view, as well as how many
photons are scattered from spaxel to spaxel. Million et al. (2010) show
that the M87 atmosphere outside the arms is well stratified, and
isothermal within a given radial bin, such that scattering in and out of
the field of view in the tangential direction should not influence the
results on the cool gas or the temperature of the hot ICM; however, the
metallicity and turbulent velocity of the hot gas may be subject to an
increased systematic uncertainty. In terms of the measurements per
spaxel, with a 1.3$^\prime$ HPD PSF, a $1\times1^\prime$ spaxel is only
expected to contain about 44\% of the flux of a point source located at
its center - therefore, mixing between different spatial regions will be
very important, and understanding its effects will require a precise
calibration of the PSF and detailed 
modeling using the
existing high-spatial resolution {\it Chandra} data.

\begin{figure}
\begin{center}
\includegraphics[width=0.65\textwidth, bb=0 0 792 612]{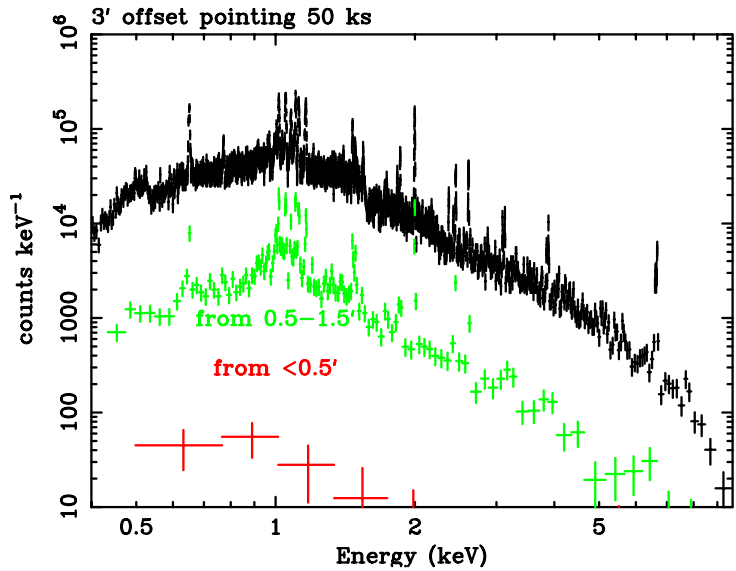}
\end{center}
\caption{Expected PSF scattering
contribution from the central $r<0.5^\prime$ (in red) and
0.5-1.5$^\prime$ (in green) at our proposed offset pointings centered
$3^\prime$ off axis, compared to the expected source spectrum (shown in
black). We assumed a radially symmetric PSF.}\label{stray}
\end{figure}

Gain calibration uncertainties can be mitigated by measuring the relative redshifts between lines emitted (primarily) by the cool vs. hot gas, and which are close in energy (see Figure \ref{specmodel}). With a 100~ks observation, we expect to be able to measure the centroid of most of the Fe-L lines in Figure \ref{specmodel} with a statistical precision of around 0.1 eV, and therefore determine the relative velocity of the cool gas with respect to the hot gas with a precision much exceeding the nominal 1~eV gain calibration uncertainty.
 
 \subsection{Beyond Feasibility}
 
The expected optical depths of two Fe-L lines which originate mostly from the 1 keV component are approximately unity in the core of M87. Provided that the atomic physics of these transitions is calibrated accurately enough, we may be able to constrain the turbulent motions of the cool gas in the cluster center, and compare this result to the turbulence in the hot gas. Because the lines from this cool component are seen at low energies, where the spectral resolution $\Delta E/E$ is not as high as at 6.7~keV, resonant line scattering offers the only viable method to measure the turbulent velocities in the cool gas. 

With deeper exposures, we will be able to measure any potential
differences in chemical composition between the different gas
phases. Currently, only the Fe abundance of the gas uplifted by the AGN
can be measured accurately, independently from the Fe abundance of the
hot gas. The Si and S abundances of the cold gas are measured with a
relatively large statistical error of 30\%, which probably will not
be sufficient for drawing any statistically significant
conclusions regarding differences in Si/Fe and S/Fe between the uplifted
and ambient gas.






\bigskip

\section{Plasma Kinematics and Cluster Masses}
\label{sec-nonth}

\subsection*{Overview}

One the most significant unanswered questions in both cluster
astrophysics and cluster cosmology concerns the role of bulk motions and
turbulence in supporting the intracluster medium (ICM) against gravity.
Simulations predict that, even in relaxed clusters, bulk velocities will
vary with radius and reach an appreciable fraction of the sound speed,
thus providing significant non-thermal pressure support. \astroh is the
first X-ray observatory capable of testing these predictions and
quantifying this support.  In this section we consider how best to use
spatially resolved measurements of Fe K line centroids and widths as a
function of radius to constrain non-thermal pressure support in bright,
relaxed clusters. These observations will improve our understanding of
cluster masses, and, ultimately, of cluster cosmology.  The most
promising targets are Abell 2029, 2199, 1795, 478, 496, 426 (Perseus)
and PKS0745.  We simulate observations of the first four of these
objects, which span the redshift range $0.03 < z < 0.10$, to explore the
cluster distance and flux limits at which such measurements are feasible
with the SXS (Perseus is considered in detail in
Sec. \ref{sec-perseus}).  We find that interesting constraints can be
placed on the run of intra-cluster plasma motion with radius
in each of these objects,
with a precision of $\sim 100$
km s$^{-1}$ for the mean line-of-sight velocity and $\sim$ 100--150
km$^{-1}$ for the RMS velocity dispersion.
A most compelling observation 
would be a set of pointings at $0.8 \times r_{2500}$ in Abell 2029,
apparently the most relaxed cluster in the local Universe.
These must be supplemented by short pointings at
smaller radii, 
to bound systematic errors due to
scattering from the bright cluster core. 
If ICM motion is not detected, these observations will provide a
systematics-limited upper bound on random motions of $v_{\rm turb} < 100$ km
s$^{-1}$ and an upper limit on the non-thermal pressure fraction $< 1$\%
at a radius of interest for cosmological studies. Our second 
goal is to extend similar but somewhat less precise measurements to a
small sample of nearby, relaxed clusters to constrain the scatter in
non-thermal pressure support and mass.
\footnote{Coordinators of this section: N. Ota, M. Bautz}

\subsection{Background and Previous Studies}
Galaxy clusters mark a cross-roads of astrophysics and cosmology. Other sections of 
this white paper discuss a variety of ways in which \astroh will advance our 
understanding of astrophysical processes shaping individual clusters.  
The cluster population comprises the largest and  therefore the youngest
virialized  structures in 
the
Universe, and its evolution
has already provided significant constraints on cosmological parameters. 
Here we discuss that \astroh observations that will open a new path to profound advances in 
the power of cluster cosmology.

Cluster studies can constrain cosmology in two independent ways.  One
method relies on the direct connection between the growth rate of cosmic
density inhomogenities and the cosmic expansion history. Since clusters
are just such density inhomogenities, the evolution of the cluster mass
function (the number of clusters per unit volume and per unit cluster
mass, as a function of cluster mass) can be related to the cosmic
expansion history \citep{Vikhlinin09b}.  Moreover, the mass we infer for
a cluster in general depends on the cluster distance, which in turn must
be inferred from an assumed redshift distance relation; the latter also
depends on the cosmological model.  Taken together, these dependencies
allow us to infer the cosmic expansion history (and in particular, the
role of dark energy in the cosmic expansion) from the evolution of the
cluster mass function. \cite{Vikhlinin09b} also point out that, since the
local cluster mass function can be directly related to the current
amplitude of the power spectrum of density inhomogenities, one can
derive an independent cosmological constraint by comparing the local
power spectrum measurement from clusters to the high-redshift power
spectrum derived from the anisotropies of the cosmic microwave
background (CMB).  This approach is often termed the
``growth-of-structure'' technique.

A second approach to cluster cosmology is based on the (very reasonable) assumption that the most massive clusters 
are large enough to be  ``fair samples'' of the cosmic matter distribution \citep{white93,sasaki96,allen2008}. In this case,   the
fraction of cluster mass which is baryonic can be expected to be independent of redshift. Because  X-ray measurements of the cluster baryon fraction depend sensitively on the angular diameter distance to a cluster, it is
possible to  obtain  independent, geometrical constraints on the cosmic
expansion history by  measuring the apparent baryon fraction as a function of redshift. In
recent years both this ``baryon-fraction'' technique and  the growth-of-structure method described above 
have yielded significant cosmological constraints  \citep[][and references therein]{Allen11,Vikhlinin09b}. 

Both of these techniques require that cluster masses can be  accurately characterized for substantial numbers of 
clusters. In practice, modern cosmological studies rely on scaling relations between cluster mass and
observable ``mass proxies'' which can more readily be measured than cluster masses themselves. Of course, these  
scaling relations must ultimately be calibrated by means of direct cluster mass measurements.  
In fact, accurate cosmological constraints require that 
both  the mean mass-observable relation and the scatter in mass about that relation are known as functions of the 
observable. 

In principle, X-ray measurements can yield accurate masses 
provided that the X-ray emitting cluster plasma is in hydrostatic
equilibrium (HSE), and that thermal pressure alone supports the plasma
against the cluster's gravity.   If these assumptions are correct, and if
the three-dimensional shape of the cluster is known, then X-ray
measurements of the spatial distribution of  plasma  density and
temperature provide direct measurements of the cluster mass profile.  If, on
the other hand, significant non-thermal pressure supports the plasma, then
HSE measurements will, in general, underestimate the true cluster mass. 
To date it has not been possible to test the HSE assumption directly.
For this reason, significant uncertainties must be associated with  HSE masses,
uncertainties which limit their power in constraining cosmology. 

Theoretically, it has been recognized for some time
\citep{Evrard90,Norman99} that macroscopic motions of the intra-cluster
plasma may indeed provide significant non-thermal pressure support.
Subsonic plasma flows may be produced by infalling material, including
dark matter, subclusters and galaxies, and could be manifest as
organized or random motions on a variety of spatial scales.  Modern
N-body-hydrodynamic simulations \citep[e.g.,][]{Lau09,Rasia12} present a
reasonably consistent picture of these motions with the following
characteristics.  The magnitude of the typical mean bulk velocity,
$\bar{v}$, and of the random motions about the mean,
$\sigma_{v}$, each vary systematically with radius within a cluster, in
the sense that the motions are smallest near cluster core and rise with
radius.  The motions are found to be larger in unrelaxed clusters than
in relaxed clusters. The magnitudes of these motions, after scaling by a
characteristic velocity (e.g. $v_{500} \equiv \sqrt{GM_{500}/r_{500}}$;
see Table \ref{tab:targets} below) seem to be nearly independent of
cluster mass, at least for massive ($M_{500} \grtsim 3 \times 10^{14}
M_{\odot}$) objects including both relaxed and unrelaxed systems
\citep{Lau09}.  \footnote{$M_{500}$ and $r_{500}$ are the mass and
radius associated with the volume within which the mean cluster density
is $500 \times$ the critical cosmological matter density.}

The motions are expected to be large enough to be significant for
cosmological studies.  For example, \citet{Lau09} predict, for the
relaxed clusters used to calibrate scaling relations, bulk motions 
of $\bar{v}/v_{500} \approx 0.2$ and one-dimensional velocity
dispersons of $\sigma_{v}/v_{500} \approx 0.10$ at $r = r_{2500}$, so
the velocities are typically 100--200 km s$^{-1}$.  Such motions should
bias hydrostatic mass estimates for relaxed clusters low by typically
$<10$\% at $r_{2500}$ and $<15$\% at $r_{500}$. This is a significant
fraction of the expected systematic uncertainty in the normalization of
state-of-the art mass scaling relations
\citep[e.g.,][]{Benson13}. Indeed, recent results from the Planck
satellite, which imply some tension between the cosmological parameters
determined from the growth of structure method and the CMB data,
have suggested that biases in HSE masses may be even larger
\citep{planckXX,linden2014}.

Some observational efforts have been made to detect cluster plasma
motions. Although \cite{Tamura11} used the \suzaku XIS to measure the
1500 km s$^{-1}$ infall speed of a subcluster in the well-known
merging system Abell 2256, for cosmologically important relaxed
clusters only upper limits have been established to date.  CCD
spectrometers do not have the resolution required to detect motions at
the level expected in these systems.  For example, \cite{Ota07} used
the \suzaku XIS to place an upper limit of $\sim 1400$ km s$^{-1}$ on
bulk motions in the Centaurus cluster.  \cite{sanders2011turb} used the \xmm
RGS to place upper limits on velocity broadening, generally $< 1000$
km s$^{-1}$, in a number of clusters. These results pertain only to
the cores (with $r \lesssim 0.5^{\prime}$) of cool-core clusters, and
are somewhat dependent on the unknown spatial variation of the
spectra. Model-dependent upper limits on broadening of 100--300 km
s$^{-1}$ are claimed for the cores of Abell 496 and Abell 1795
\citep{sanders2011turb}, for example. These limits are consistent with the
predictions noted above.

We show here that  \astroh
SXS spectroscopy can reveal the expected motions and the associated
non-thermal pressure support. An empirical constraint on the resulting bias
in scaling relations  between X-ray observables and  'true' 
masses determined from weak lensing, would  clearly be an important contribution to cluster cosmology. 
We note also that simulations are frequently relied on
for estimates of the scatter in scaling relations
\cite[e.g.,][]{Kravtsov06}, so any  empirical test of the validity of the
simulations could provide an indirect assessment of the assumed scatter. 
Finally, if measurements of non-thermal pressure support could eventually 
be performed  on a sufficiently large sample of objects, it might be possible 
to use this information  to discover selection criteria with which to assemble 
a sub-sample of objects (presumably those with the least non-thermal support) 
for which especially low-scatter mass estimates can be made. Such a sub-sample would in turn permit exceptionally precise cosmological constraints.

Such measurements would also improve our understanding of cluster
astrophysics. For example, the simulations referred to above all require
assumptions about transport processes (e.g., viscosity and thermal
conduction) in the intracluster plasma, about which little is known.
Direct measurements of the plasma velocities could test some of these
assumptions,  and  perhaps even improve our knowledge of the allowed range
for the transport coefficients. More fundamentally, there is evidence that
the  two approaches to  hydrodynamical simulation (smoothed-particle
hydrodynamics, or SPH, and Eulerian) may not produce consistent results in
detail \citep{Rasia12}, so tests of their basic predictions may be helpful in
judging the reliability these  techniques.

\subsection{Prospects and Strategy}

The \astroh SXS, with its superior spectral resolution, will enable
accurate measurements of line broadening and shifts for the He-like
iron-K emission. For some objects, measurements of the H-like iron-K line 
may also be useful. We aim at revealing the
dynamical states of the cluster gas and improving the cluster mass model
by incorporating the pressure support due to gas motions. Our strategy is to observe the 
most relaxed clusters for which SXS observations are practical, and 
to map them to a radius as close to  $\sim r_{\rm 2500}$ as possible. 
Relaxed  clusters are the most useful  for cosmological work because they are expected to obey 
the most accurate (least biased) and lowest-scatter mass-observable relations.  
Several factors motivate us to map to radii approaching 
 $\sim r_{\rm 2500}$.  As noted above, the magnitude of non-thermal pressure support is expected to increase with 
radius. Moreover, as discussed below, the contaminating effects of scattering from the bright cores of these relaxed objects should generally decrease with increasing radius.  Finally, since the most accurate mass measurements are ultimately expected to be obtained from weak lensing observations of an ensemble of 
clusters, we wish to measure the non-thermal pressure support to a radius at which 
weak lensing measurements can be made. At present weak lensing mass measurements 
can be made to radii as small as $\sim r_{\rm 2500}$; at smaller radii, confusion 
between cluster galaxies and (lensed) background objects becomes problematic. As weak lensing 
techniques improve, we expect that the reliable mass measurements may be made at radii 
as small as $0.75 r_{\rm 2500}$. 

Measurements of line broadening will yield the line-of-sight velocity
dispersion that can arise from turbulence in the
ICM. If the turbulence is isotropic, it will provide the pressure
support $p_{\rm turb}=\rho_{\rm gas} v_{\rm turb}^2$ of the amount 
\begin{eqnarray}
\frac{p_{\rm turb}}{p_{\rm therm}} 
\simeq 1.3 \times 10^{-2} 
\left(\frac{v_{\rm turb}}{100 \mbox{ km s$^{-1}$}}\right)^2
\left(\frac{\mu}{0.6}\right)
\left(\frac{k T}{5 \mbox{ keV}}\right)^{-1}. 
\label{eq-pturb}
\end{eqnarray}
where $v_{\rm turb}^2$ is the one-dimensional velocity dispersion
of turbulence, $\rho_{\rm gas}$ is the gas mass density, $p_{\rm
therm}=\rho_{\rm gas}k T /\mu m_{\rm p}$ is the thermal pressure, $\mu$
is the mean molecular weight, and $m_{\rm p}$ is the proton mass.
These pressure terms contribute to the hydrostatic mass via
\begin{eqnarray}
M(<r) = - \frac{r^2}{G\rho_{\rm gas}}
\frac{d (p_{\rm therm} + p_{\rm turb})}{d r}.  
\label{eq-hydrostatic}
\end{eqnarray}
Our primary goal is hence to measure for the first time the contribution
of turbulence to the total pressure support with a percent order
accuracy and to infer its impact on the mass estimates of galaxy
clusters.  The measurements around the bright cluster center will be
limited by systematic errors on the width of the line spread function
and thermal broadening for which a conservative estimation gives the sum
of $\Delta v_{\rm turb} \sim 100$ km s$^{-1}$ (90\%), whereas the extent to
which the measurements can be done within feasible observing time is
controlled by photon statistics. This level of accuracy is also
necessary for unveiling or giving meaningful constraints on relatively
mild gas motions inferred from numerical simulations near the cluster
core \citep[e.g.,][]{Lau09}. Equation (\ref{eq-hydrostatic}) further
implies that an accurate measurement of the radial gradient of
$v_{\rm turb}^2$ is crucial for mass reconstruction.

In addition, measuring the shifts of a line centroid will directly yield
the line-of-sight bulk velocity $v_{\rm bulk}$ of the amount
\begin{eqnarray}
\frac{v_{\rm bulk}}{v_{\rm sound}}
= 8.7 \times 10^{-2} \left(\frac{v_{\rm bulk}}{100 \mbox{ km s$^{-1}$}}\right)
\left(\frac{\mu}{0.6}\right)^{1/2}
\left(\frac{k T}{5 \mbox{ keV}}\right)^{-1/2},
\end{eqnarray}
where $v_{\rm sound}=\sqrt{5 k T/3 \mu m_{\rm p}}$ is the sound
speed. By mapping $v_{\rm bulk}$ over an appreciable range of radius, we
can also perform a direct test on the validity of hydrostatic
equilibrium and explore the nature of any coherent motions such as
rotation of the ICM. The measurements will also be limited by systematic
errors on the energy scale near the center and by statistical errors at
the outermost observable regions. It is expected that an accuracy of
$\Delta v_{\rm bulk} < 100$ km s$^{-1}$ is achievable including systematics,
which improves the current accuracy by more than a factor of $\sim 5$
\citep[][for the merging cluster A2256]{Tamura11} and allows the first
measurements of the gas bulk motion in relaxed clusters.

Another important and nontrivial systematic effect on this study arises
from scattering of photons by relatively broad wings of the Point Spread
Function (PSF); the sharply peaked X-ray emission toward the cluster
center may contaminate the velocity measurements at fainter off-center
regions. Further exacerbating this problem are the
centrally peaked abundance profiles in relaxed, cool core clusters,
which result in even greater contamination of the Fe line flux in the
outer regions compared to the continuum flux. As demonstrated by our
detailed simulations presented in the following sections, this effect
can be adequately accounted for
by performing a series of pointings along the radius from the center of
a cluster and solving for intrinsic distributions of the line width and
shifts (plus temperature and metallicity if necessary). Existing high
spatial resolution data from \chandra and \xmm will play a crucial role
in modeling accurately the scattered contributions of the continuum
emission.

\begin{table}[t]
\begin{center}
\caption{Summary of parameters for a sample of relaxed clusters.}
\label{tab:targets}
\begin{threeparttable}
\begin{tabular}{l|cccccc}
\hline
\hline
Cluster   & $z$\tnote{a}      &  $kT$\tnote{a} & $r_{2500}$\tnote{b}  &
 $r_{500}$\tnote{b} & $v_{500}$
& ref.\tnote{b}\\
          &                   &   [keV]  &  [Mpc, arcmin] & [Mpc,
		 arcmin] & [km s$^{-1}$] & \\
\hline
A2029     & 0.077   &  8.5 & 0.66, 7.6  & 1.36, 15.7 & 1510 & V \\
A2199     & 0.030   &  4.0 & 0.41, 11.5 & 1.02, 28.7 & 1130 & R \\
A1795     & 0.062   &  6.1 & 0.50, 7.0  & 1.24, 17.5 & 1370 & V \\
A478      & 0.088   &  7.3 & 
0.65, 6.6 & 1.34, 14.6 & 1480 & V \\
\hline                                                                 
A496      & 0.033   &  4.1 & 0.36, 10.1 & 0.89, 25.2 & 990  & R \\
PKS0745   & 0.103   &  8.0 & 0.52, 4.8  & 1.31, 12.0 & 1450 & R \\
\hline                                                            
\hline
\end{tabular}
\begin{tablenotes}
\item[a]{\footnotesize From the X-ray Galaxy Clusters Database (http://bax.ast.obs-mip.fr).}
\item[b]{\footnotesize Scale radii are from \citet[][`R']{reiprich2002} and
     \citet[][`V']{vikh2006}.}
\end{tablenotes}
\end{threeparttable}
\end{center}
\end{table}

\begin{figure}[t]
\centering
\begin{minipage}[t]{.49\textwidth}
\centering
\includegraphics[width=0.9\hsize]{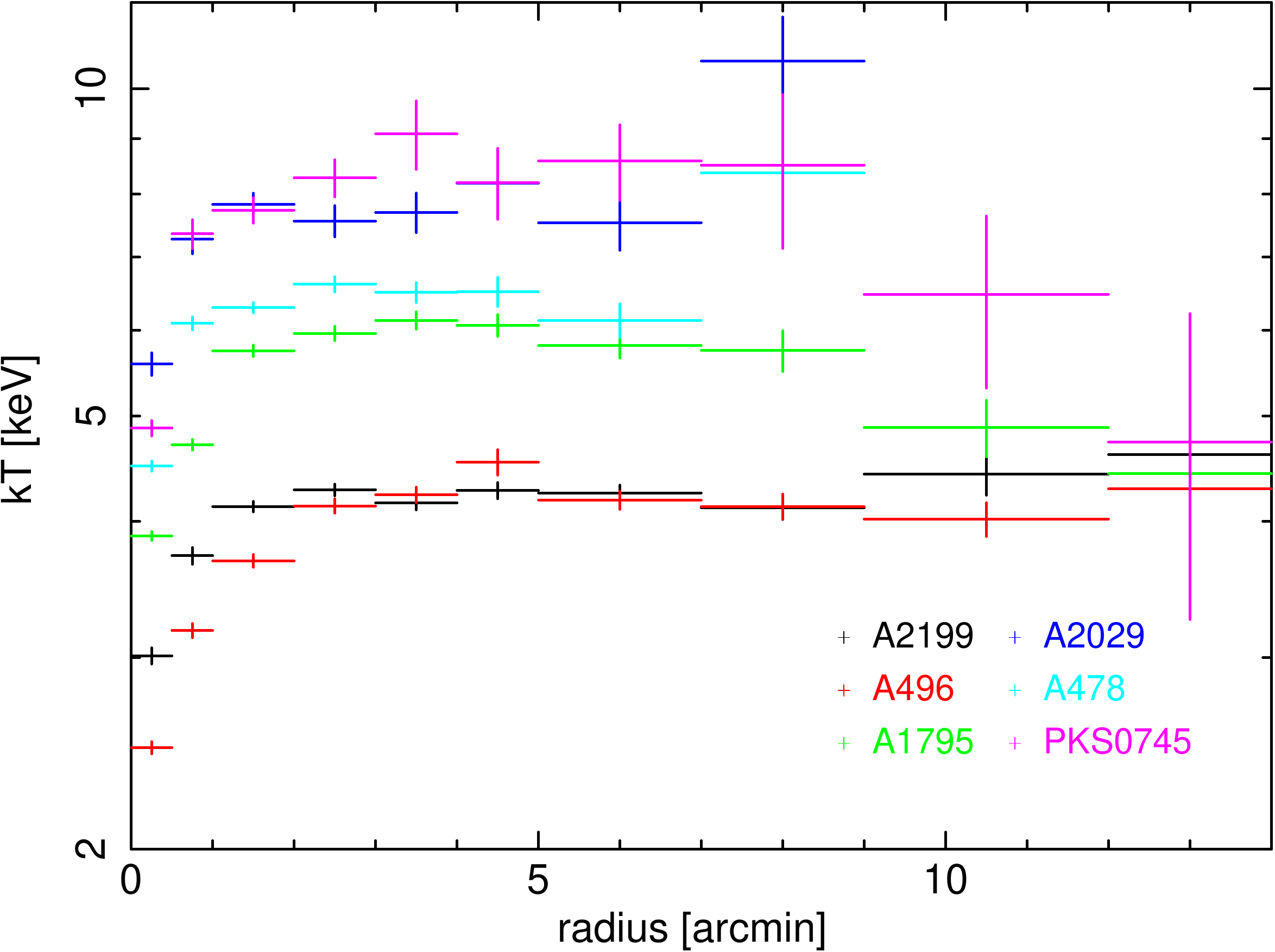}
\end{minipage}
\begin{minipage}[t]{.49\textwidth}
\centering
\includegraphics[width=0.9\hsize]{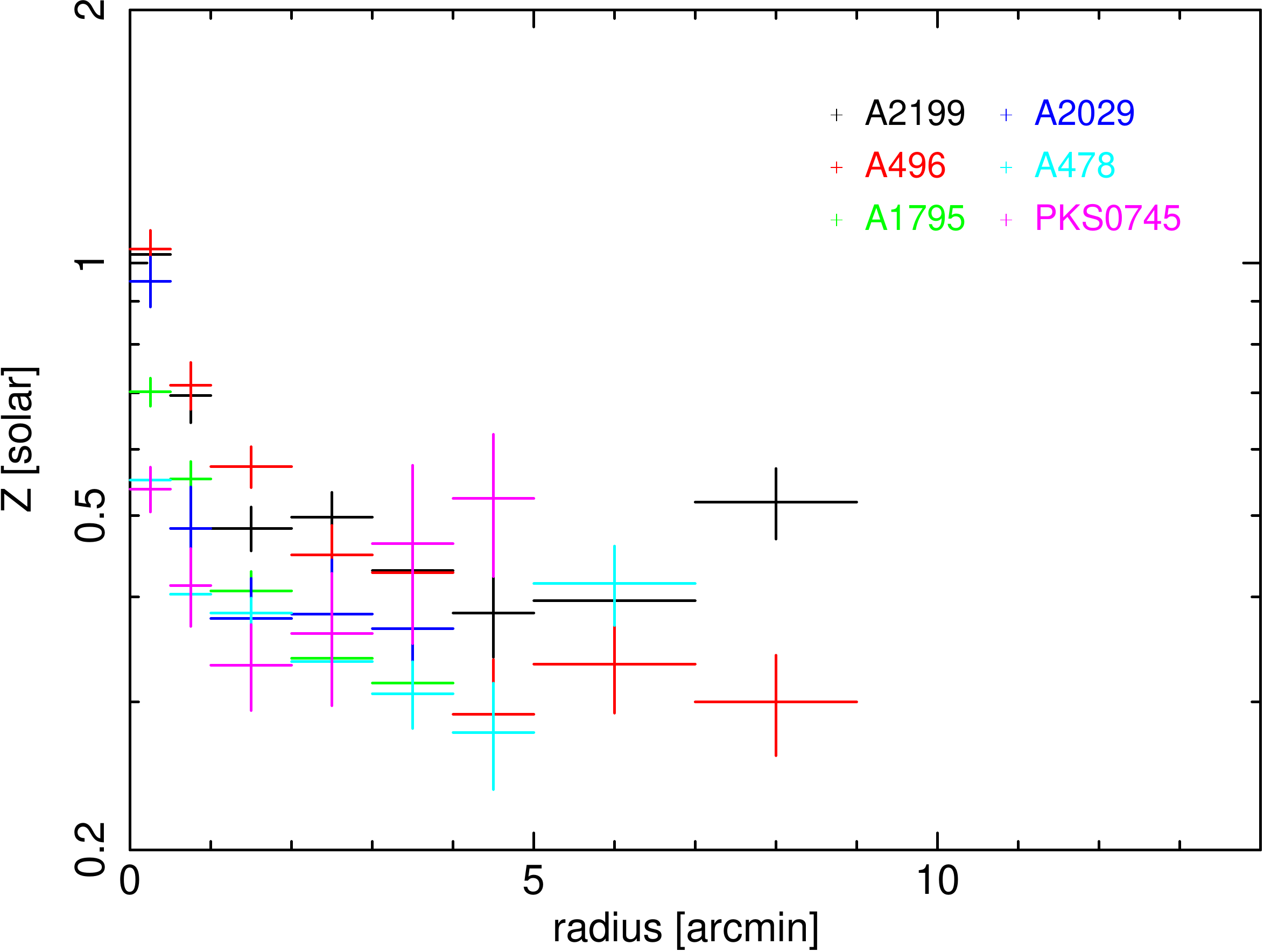}
\end{minipage}
\begin{minipage}[b]{.49\textwidth}
\centering
\includegraphics[width=0.9\hsize]{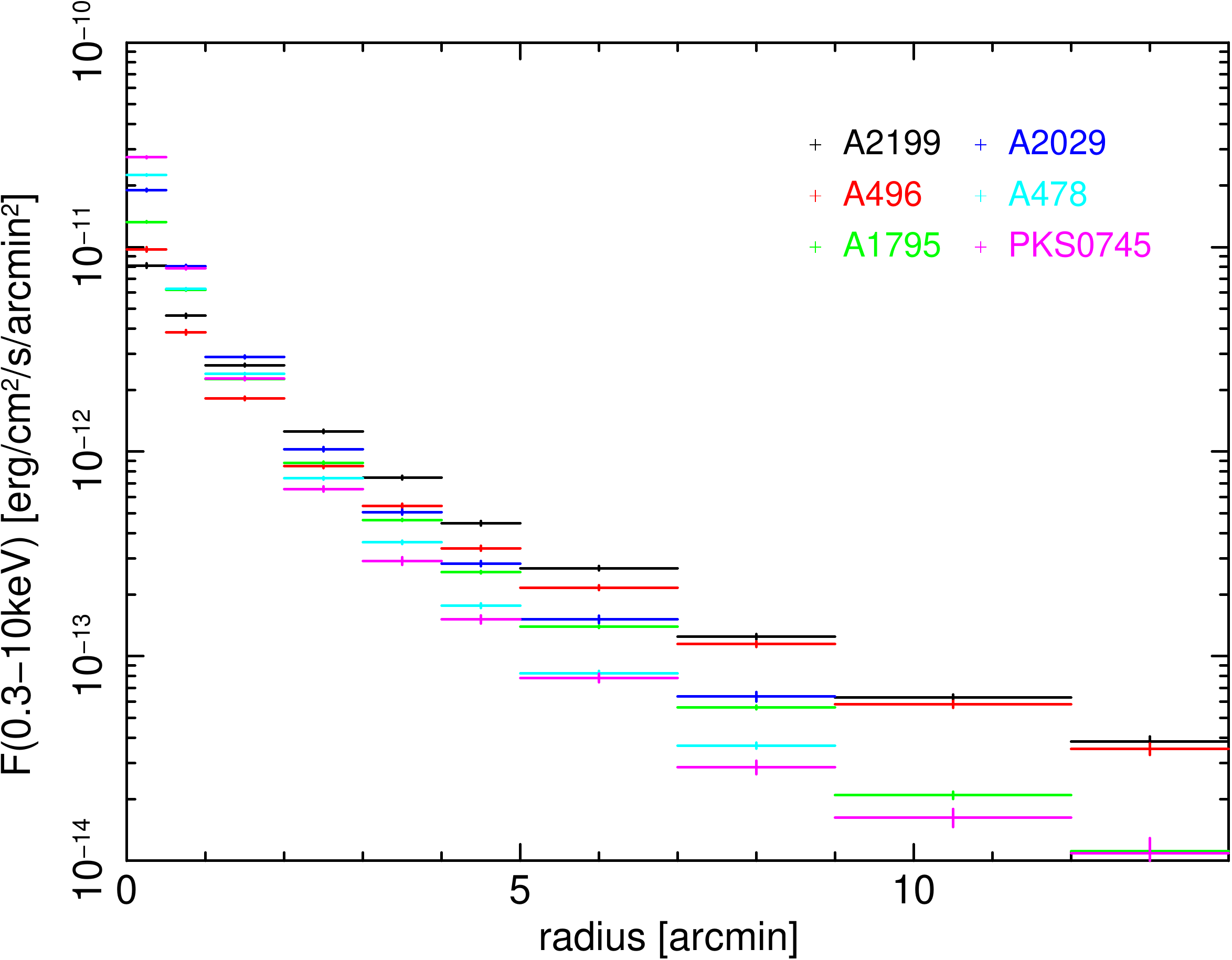}
\end{minipage}
\caption{Temperature, metal abundance, and surface brightness profiles for the
cluster sample, from \citet{Snowdenetal2008}.}
\label{fig:profiles}
\end{figure}

\begin{table}[t]
\begin{center}
\caption{Radial positions of simulated regions in A2029, A2199, A1795,
 and A478.}  \label{tab:regions}
\begin{tabular}{c c c}
\hline
\hline
Region & Center & Range between boundaries\\
number & [arcmin] & [arcmin] \\
\hline
1 & 0 & -1.5 -- 1.5 \\
2 & 3 & 1.5 -- 4.5 \\
3 & 6 & 4.5 -- 7.5 \\
4 & 9 & 7.5 -- 10.5 \\
5 & 12 & 10.5 -- 13.5 \\
\hline
\hline
\end{tabular}
\end{center}
\end{table}


\subsection{Targets and Feasibility}

The foregoing considerations imply that viable targets for our study are
bright, relaxed clusters with sufficient radial extent to minimize the
PSF scattering effect, the electron temperature above $\sim$ 4 keV to
produce prominent Fe-K line emission, and existing high quality \chandra
and \xmm data. The best targets include Abell 2029, 2199, 1795, 478, 496,
and PKS0745; the parameters of these clusters are listed in Table
\ref{tab:targets}, and temperature, metal abundance, and surface
brightness profiles are shown in Figure \ref{fig:profiles}.  In the
following, we present our simulation results for the first four
clusters, which cover a range of redshifts, mass (mean
temperature), and different levels of the impact of the PSF effect
mentioned above.

Abell 2029, though relatively distant, 
appears to be the most relaxed cluster in the redshift range we
consider (Mantz et al., in preparation), and thus the best choice for
contributing to cluster cosmology.  Abell 2199 at $z=0.030$ serves as an
ideal example of a very extended cluster with a relatively mild central
emission peak for which the PSF effect is likely to be minimal. A1795 at
$z=0.062$ is a more representative case of a larger sample of clusters
that can be resolved by \astroh with the stronger PSF effect. Finally,
A478 at $z=0.088$ has the higher mass but is amongst the most distant
clusters resolvable by {\it ASTRO-H}.

\begin{figure}[t]
\begin{center}
\rotatebox{0}{\scalebox{0.35}{\includegraphics{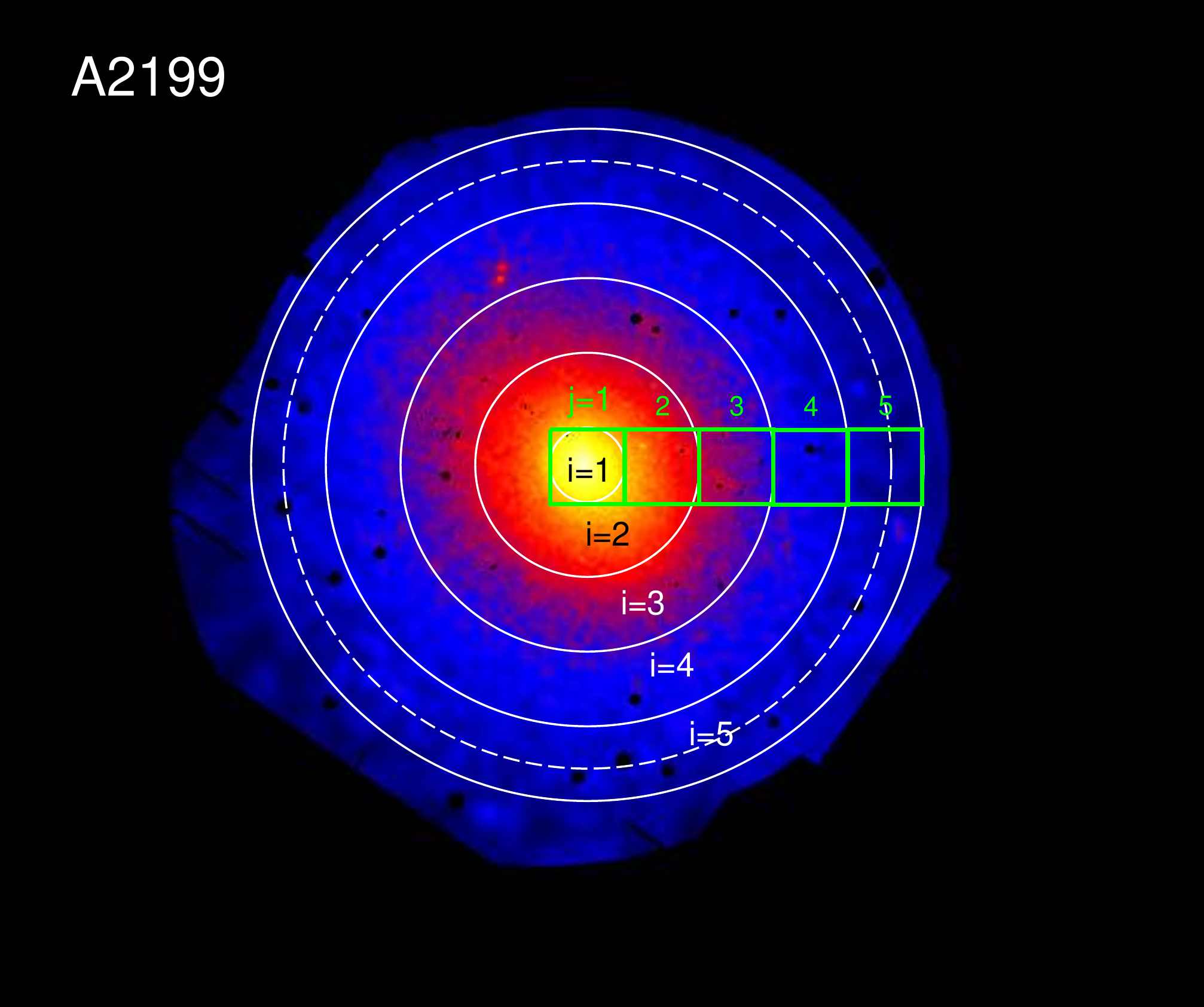}}}
\rotatebox{0}{\scalebox{0.35}{\includegraphics{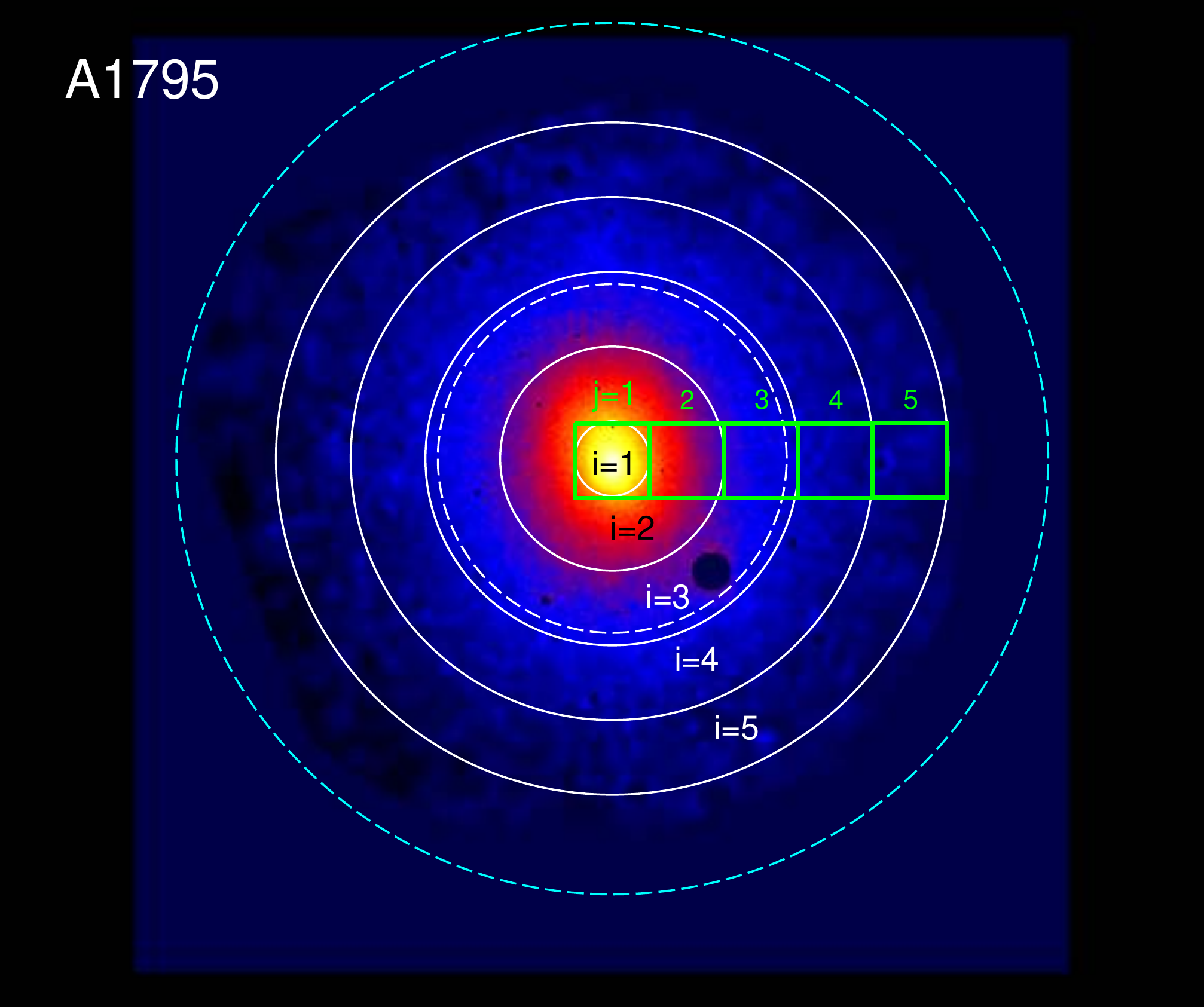}}}
\rotatebox{0}{\scalebox{0.35}{\includegraphics{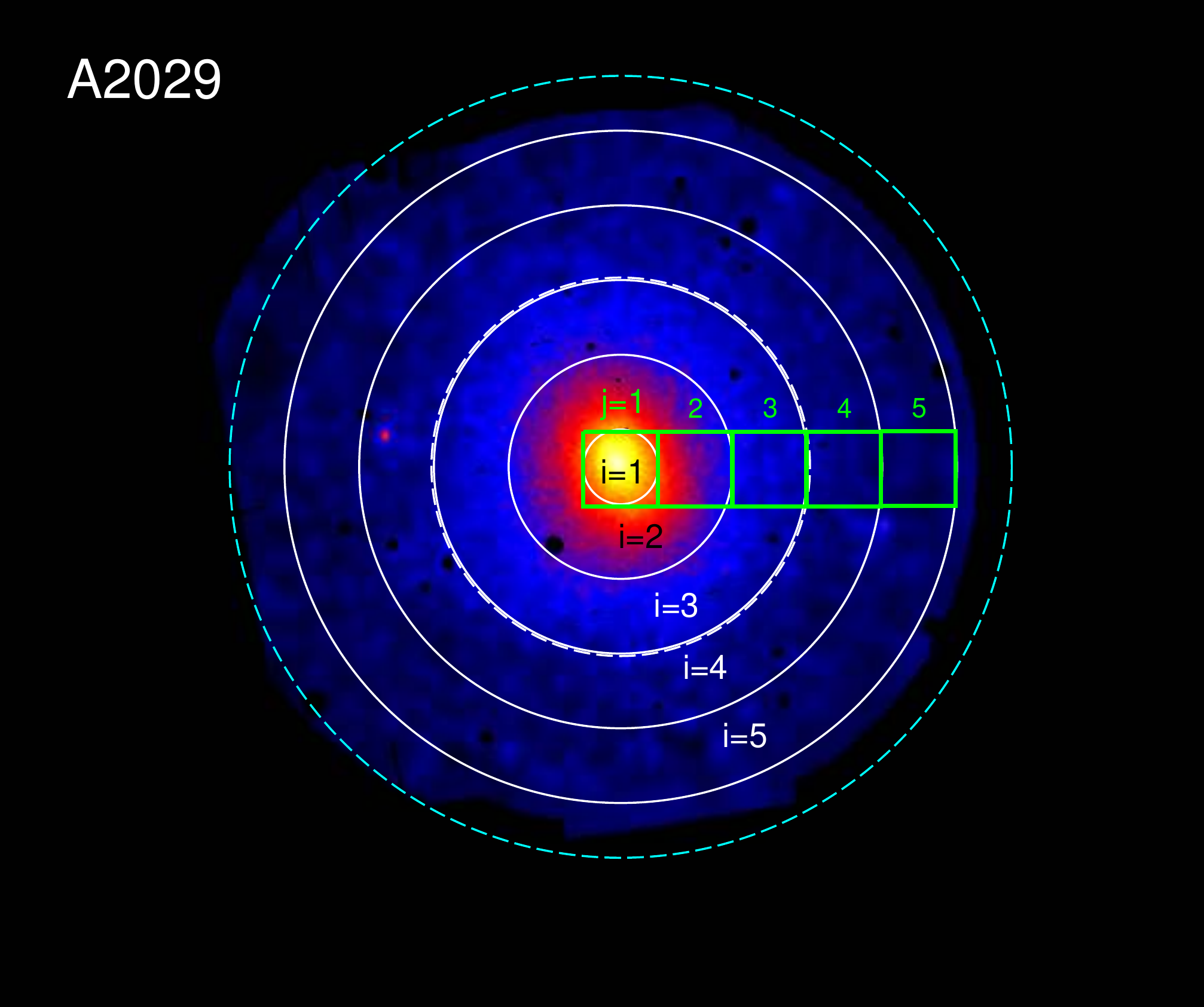}}}
\rotatebox{0}{\scalebox{0.35}{\includegraphics{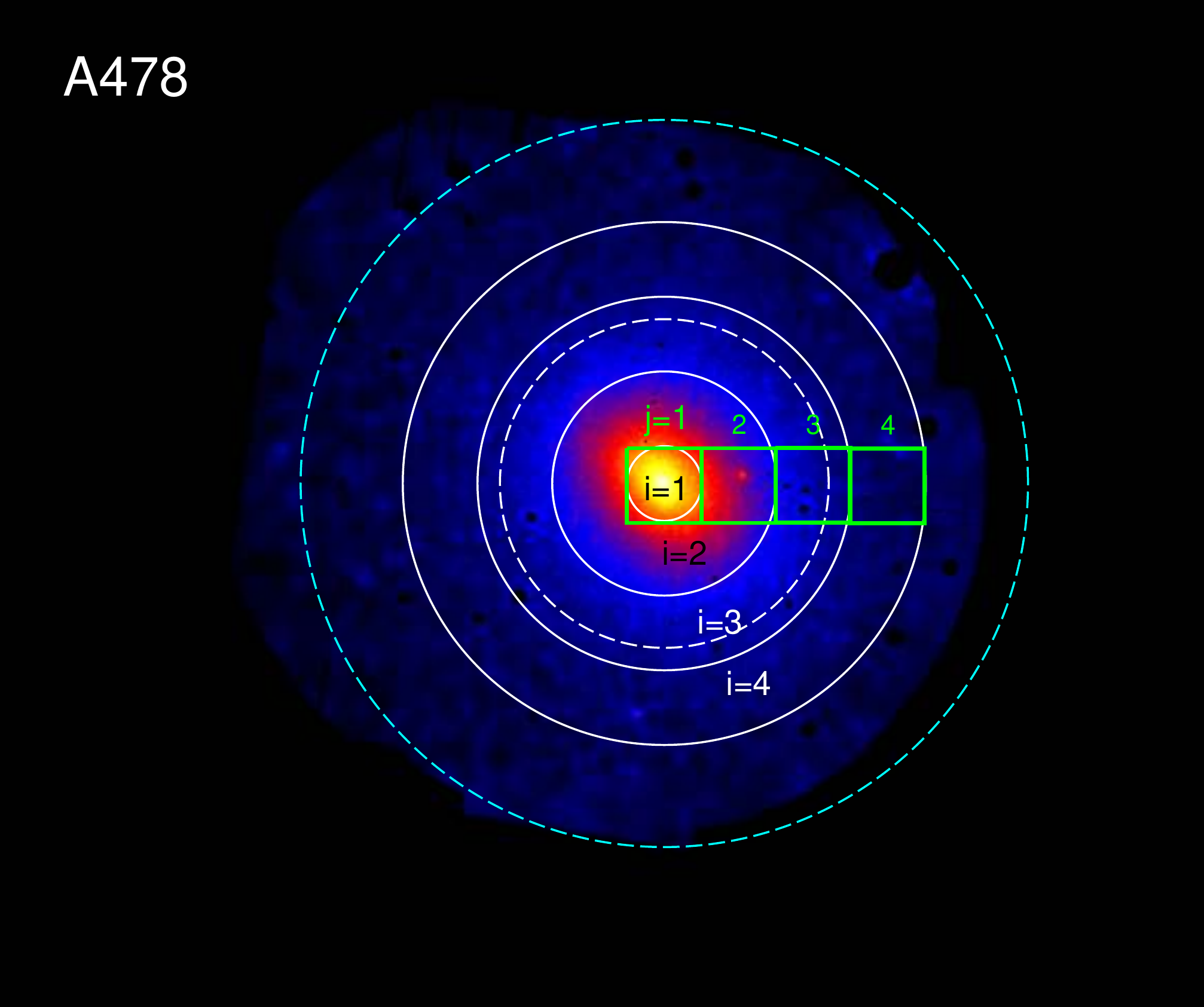}}}
\caption{\xmm images of A2199 ($z=0.030$;
top left), A1795 ($z=0.062$; top right), A2029($z=0.077$; bottom left)
and A478($z=0.088$; bottom right).  The annuli ($i=1,2,..$) and the
$3'\times 3'$ pointing regions ($j=1,2,..$) used in our simulations are
indicated by white circles and green boxes, respectively.  The scale
radii of $r_{2500}$ and $r_{500}$ are shown with the inner and and outer
dashed circles. For A2199, $r_{500}$ is outside the
image.}\label{fig:xmm}
\end{center}
\end{figure}

In order to constrain the contribution of non-thermal pressure to the
hydrostatic condition, we aim to measure the line broadening and shift
of He-like Fe-K line at 6.7 keV with an accuracy better than $\sim
100~{\rm km\,s^{-1}}$ in these nearby relaxed clusters. We consider
one-dimensional SXS mapping of bright clusters and perform feasibility
studies to estimate the necessary exposure time based on detailed
spectral simulations including the PSF scattering effect.  The
simulations are performed for a series of pointings on annuli shown in
Figure~\ref{fig:xmm} with boundaries given in Table~\ref{tab:regions}.
To take into account large uncertainties in the radial distribution
of turbulence, we investigate the following three cases; 
(a) $v_{\rm turb}$ is
constant over radius, (b) $v_{\rm turb}$ 
increases with radius, and (c) $v_{\rm turb}$ decreases sharply as
$v_{\rm turb} = 500~{\rm km\,s^{-1}}$ inside the core and $v_{\rm turb}
= 0~{\rm km\,s^{-1}}$ elsewhere.  Cases (a) and (b) are consistent with
the current upper limits on turbulence in the cluster cores
\citep{sanders2011turb}, whereas case (c) is intended to exhibit a limit
in which the PSF scattering from the bright core has the largest impact
on velocity measurements at larger radii.

\subsubsection{Abell 2029}

\begin{figure}
\begin{center}
     \includegraphics[width=.75\textwidth]{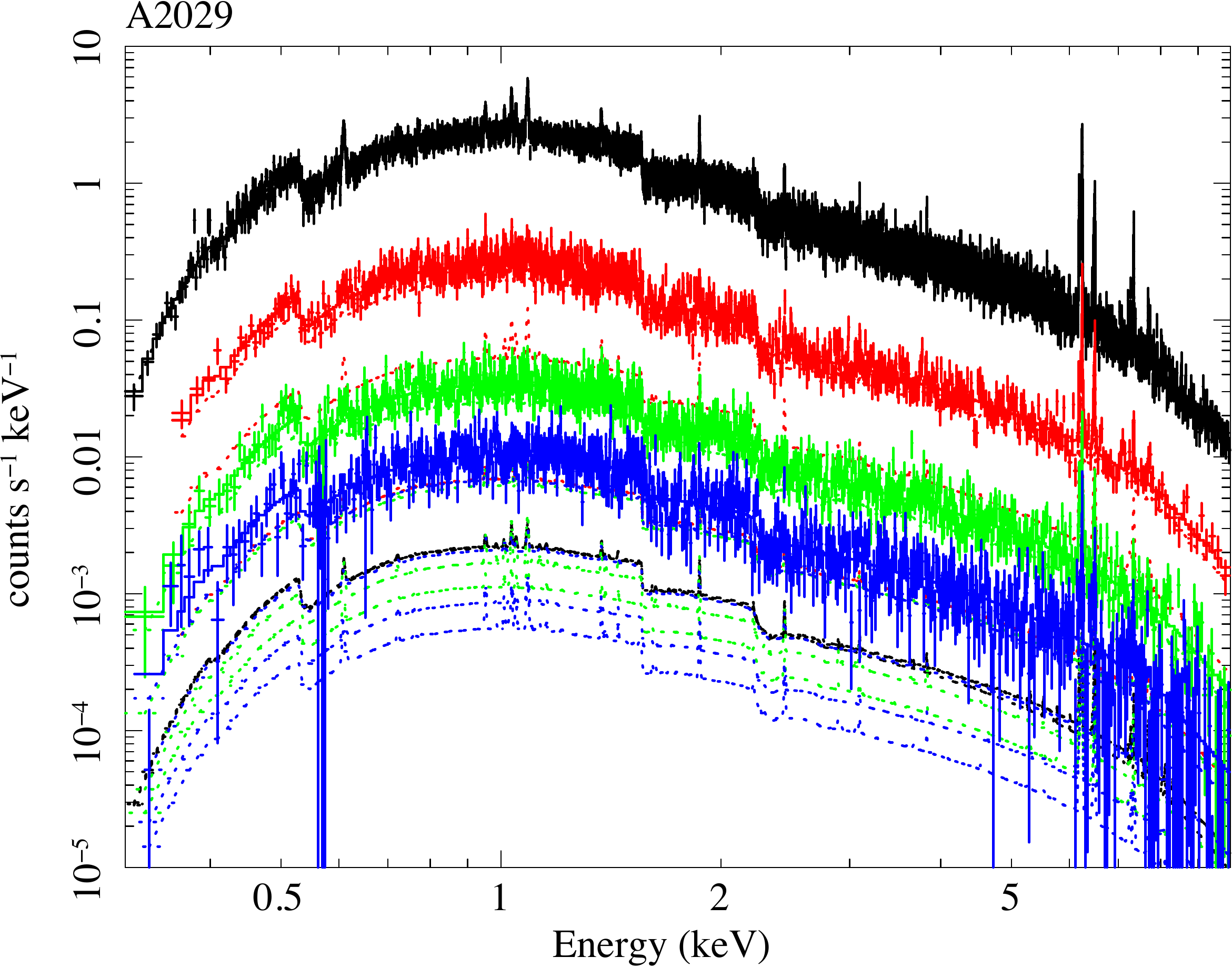}
     \includegraphics[width=.75\textwidth]{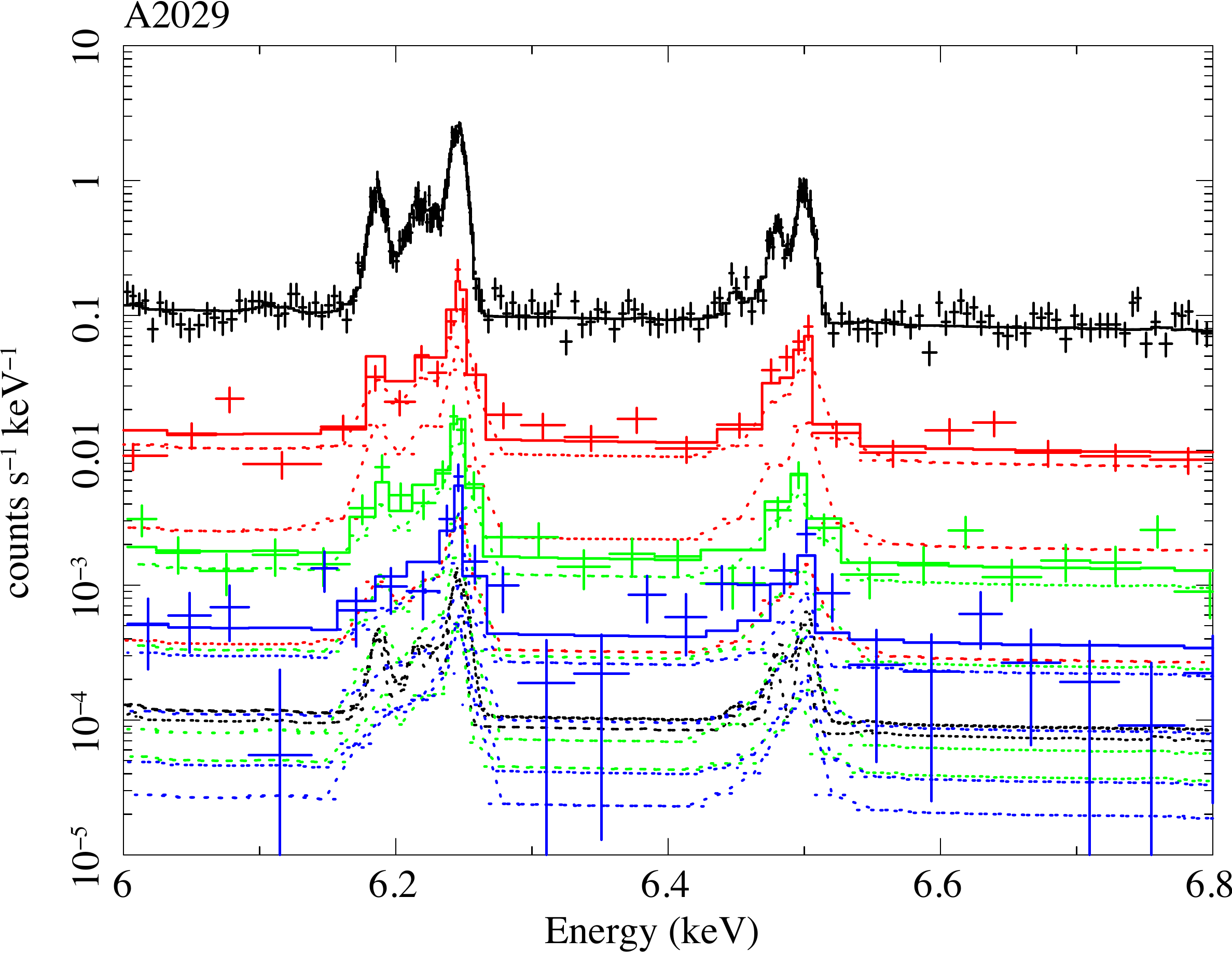}
     \caption{Top: Simulated SXS spectra of A2029 for case
     (a). The data for the regions $j=$ 1/2/3/4 are shown with the
     black/red/green/blue crosses, respectively and the scattered
     components are indicated with the dotted lines. Bottom: Blow-up of
     the top panel.}\label{fig:a2029sxs}
\end{center}
\end{figure}

\begin{figure}
\begin{center}
\begin{minipage}{0.46\textwidth}
\rotatebox{0}{\scalebox{0.28}{\includegraphics{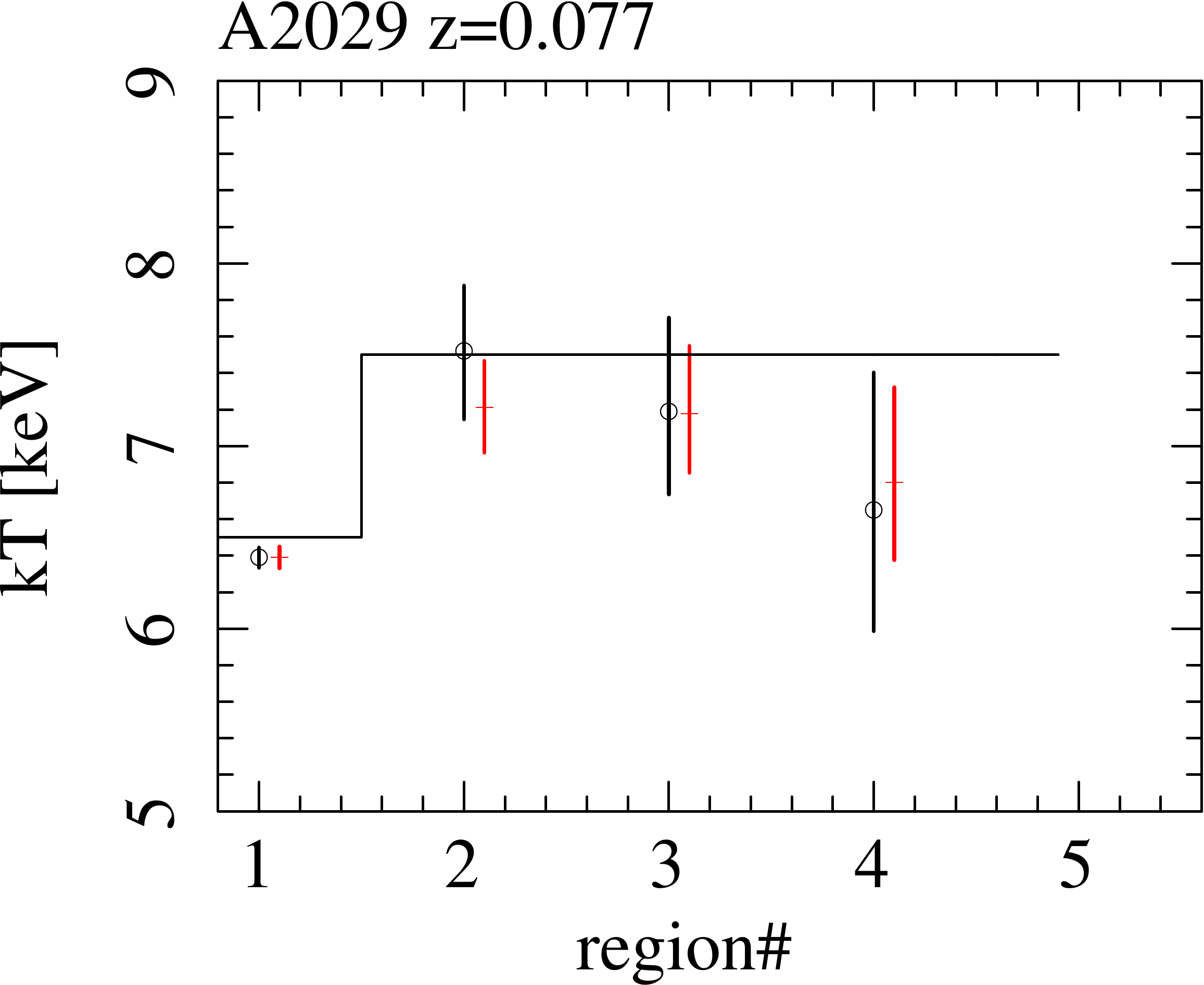}}}
\rotatebox{0}{\scalebox{0.28}{\includegraphics{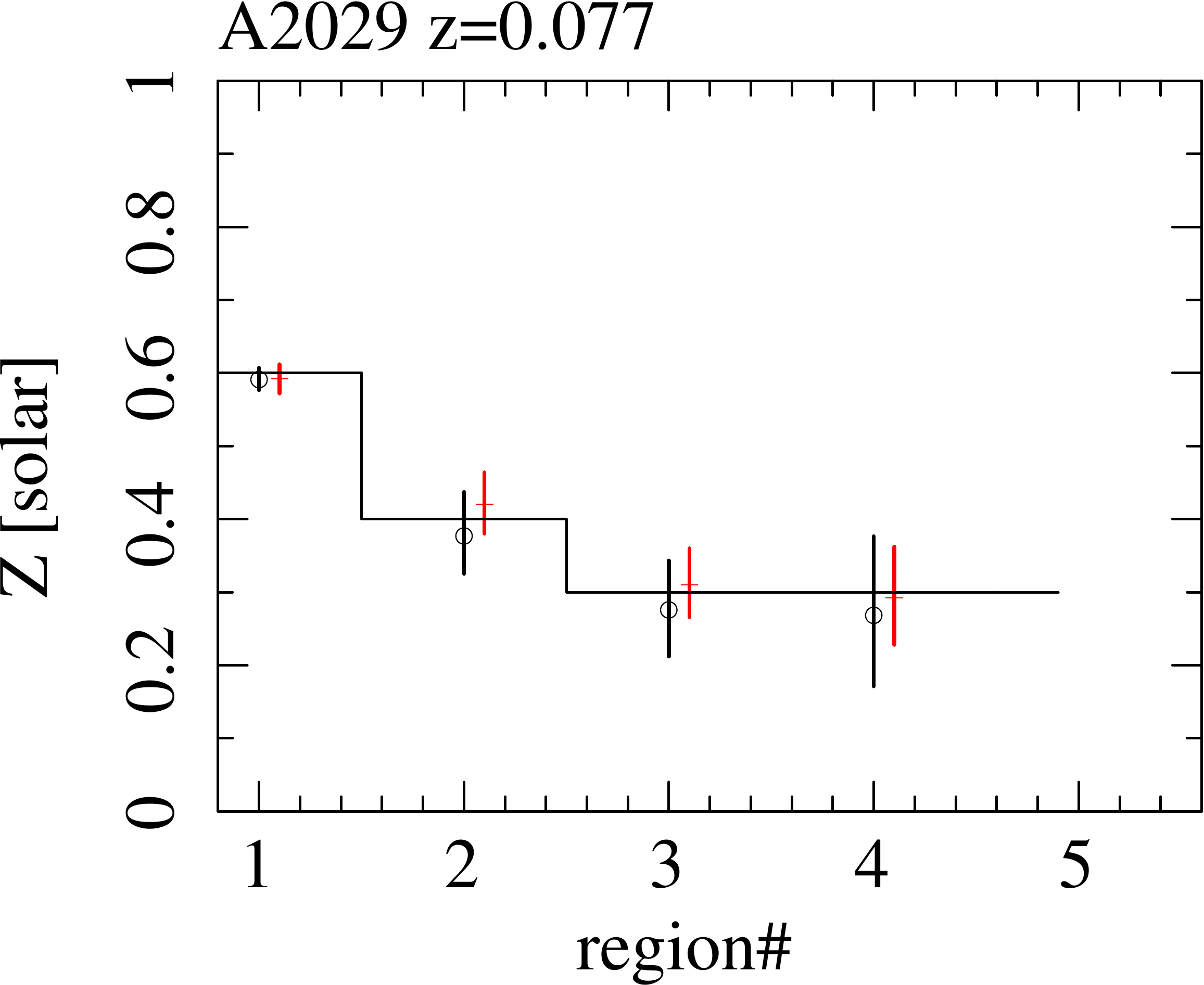}}}
\rotatebox{0}{\scalebox{0.28}{\includegraphics{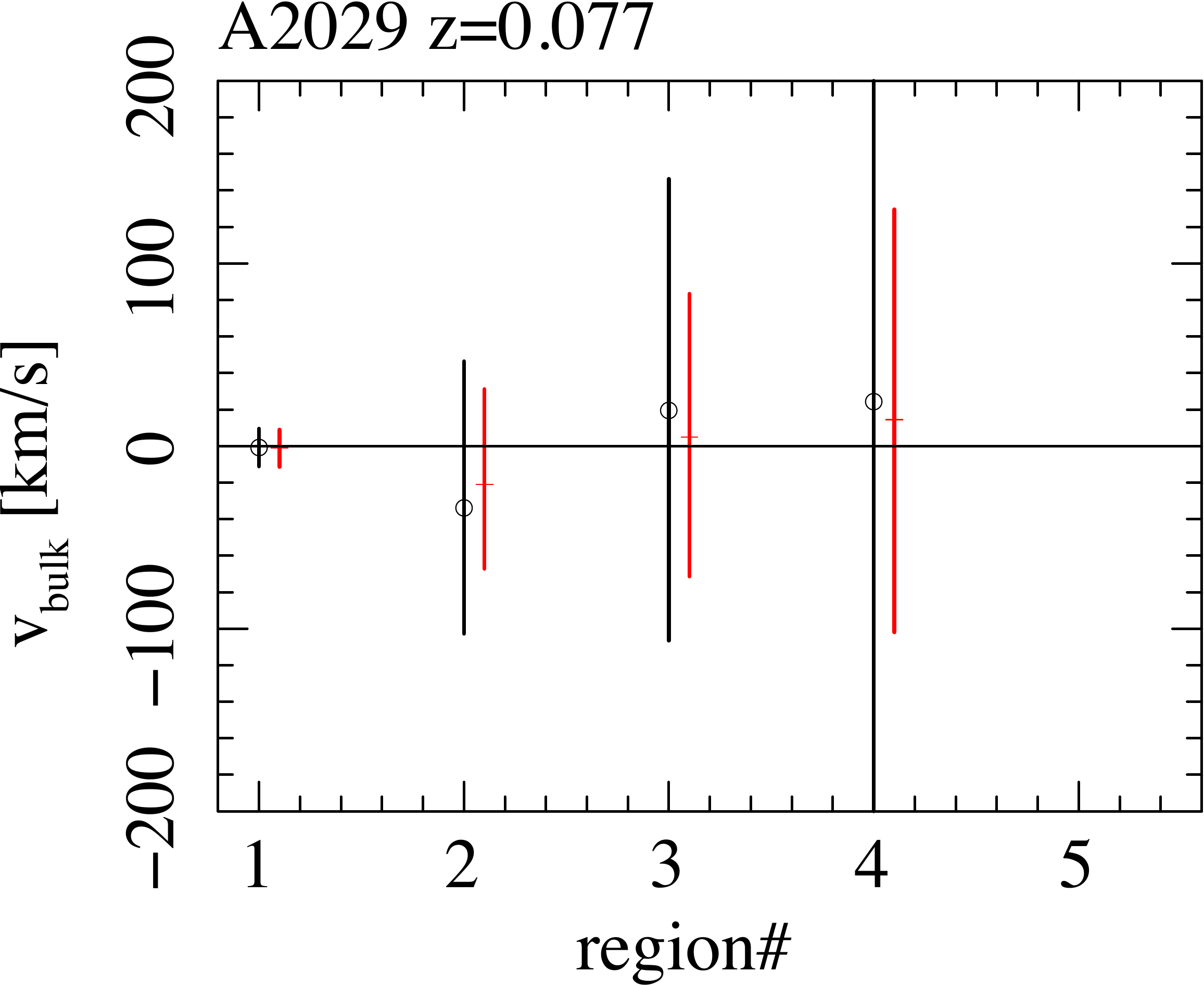}}}
\rotatebox{0}{\scalebox{0.28}{\includegraphics{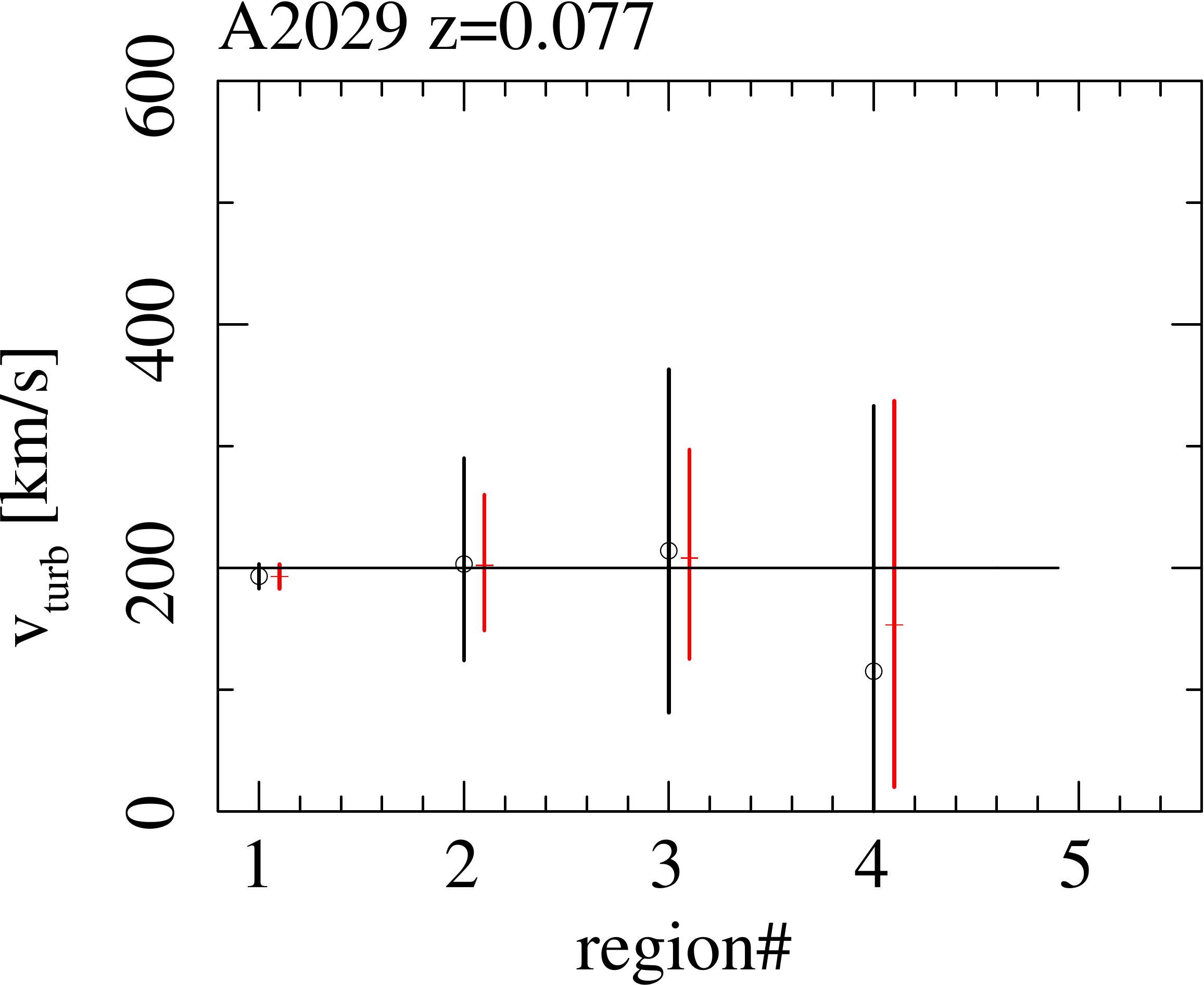}}}
\caption{Results of simultaneously fitting the spectra of all annuli
(black circles) or separately fitting the spectrum of each annulus (red
crosses) in A2029 for case (a): $v_{\rm turb}$ is constant over radius.
The solid line indicates the input parameter values of the
simulations. Errors are at the $90\%$ confidence.}  \label{fig:a2029con}
\end{minipage}~~
\begin{minipage}{0.46\textwidth}
\rotatebox{0}{\scalebox{0.28}{\includegraphics{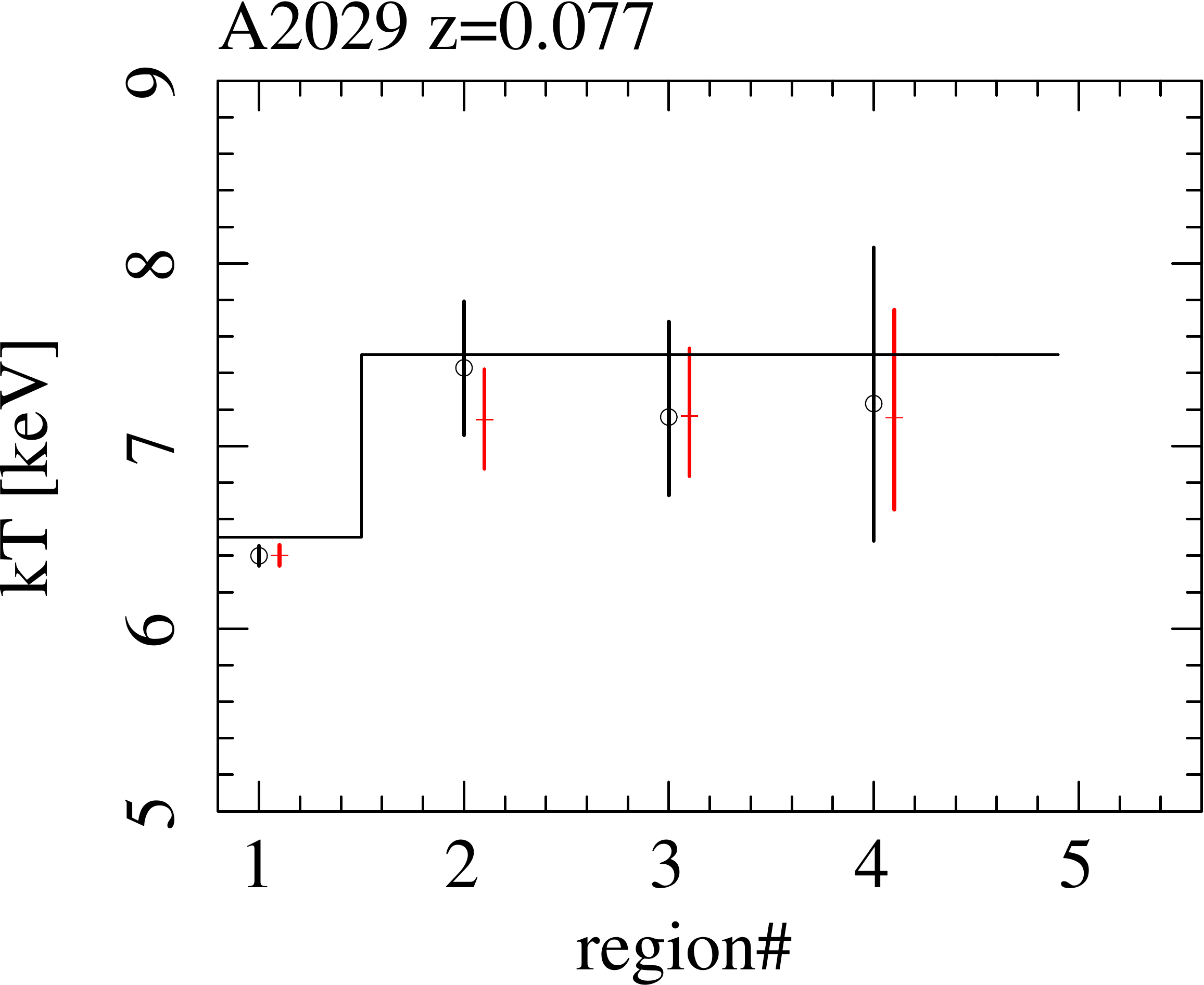}}}
\rotatebox{0}{\scalebox{0.28}{\includegraphics{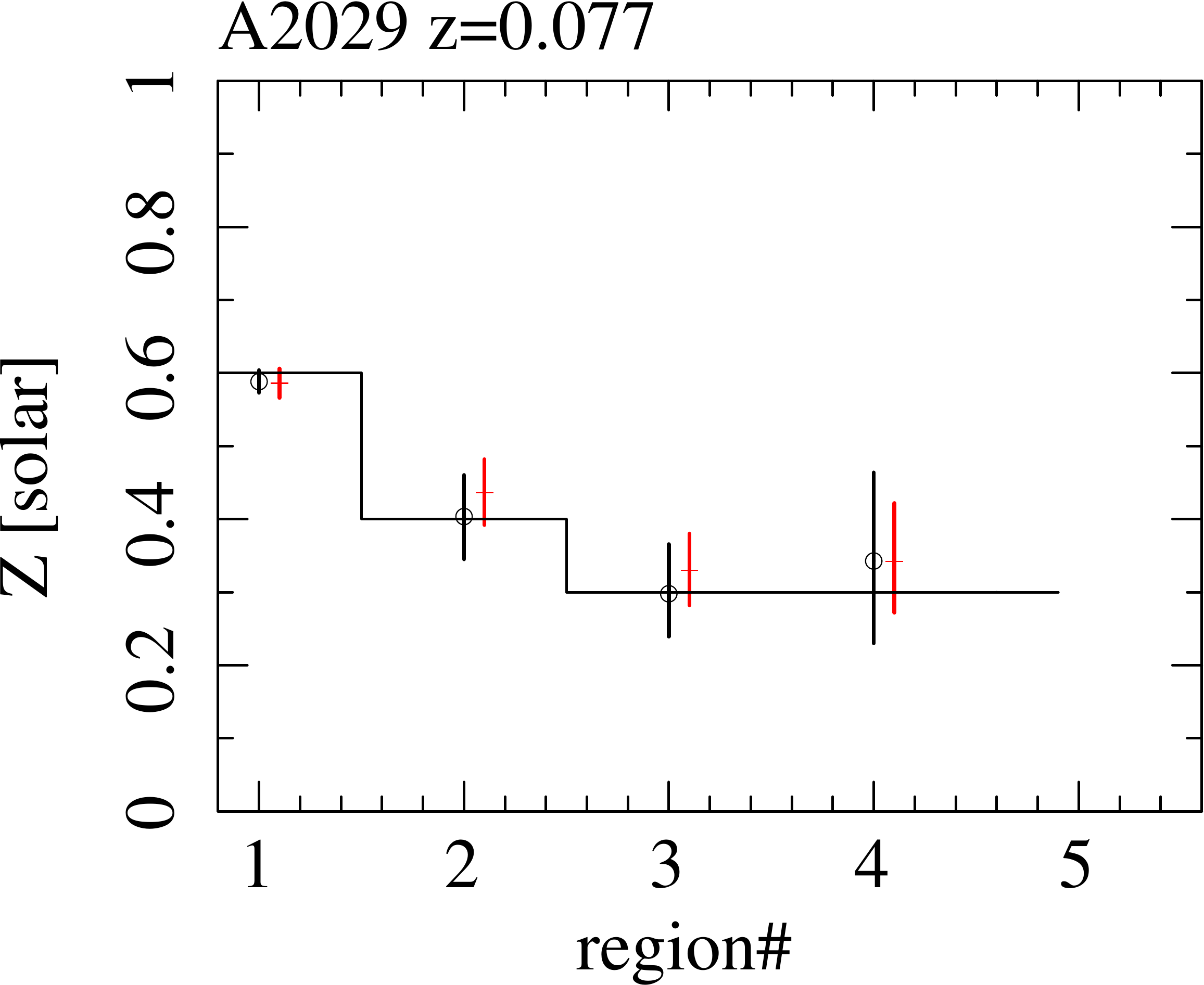}}}
\rotatebox{0}{\scalebox{0.28}{\includegraphics{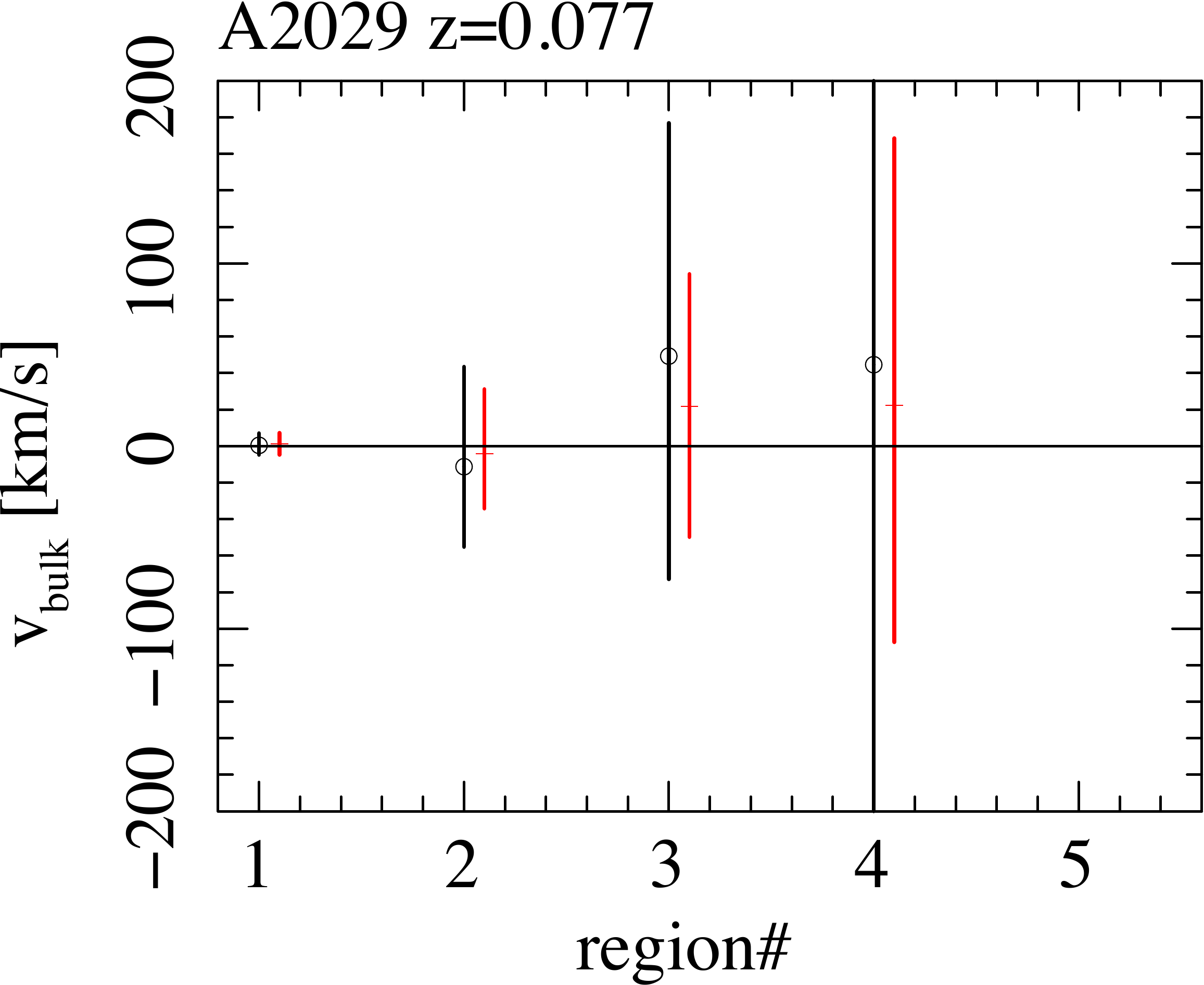}}}
\rotatebox{0}{\scalebox{0.28}{\includegraphics{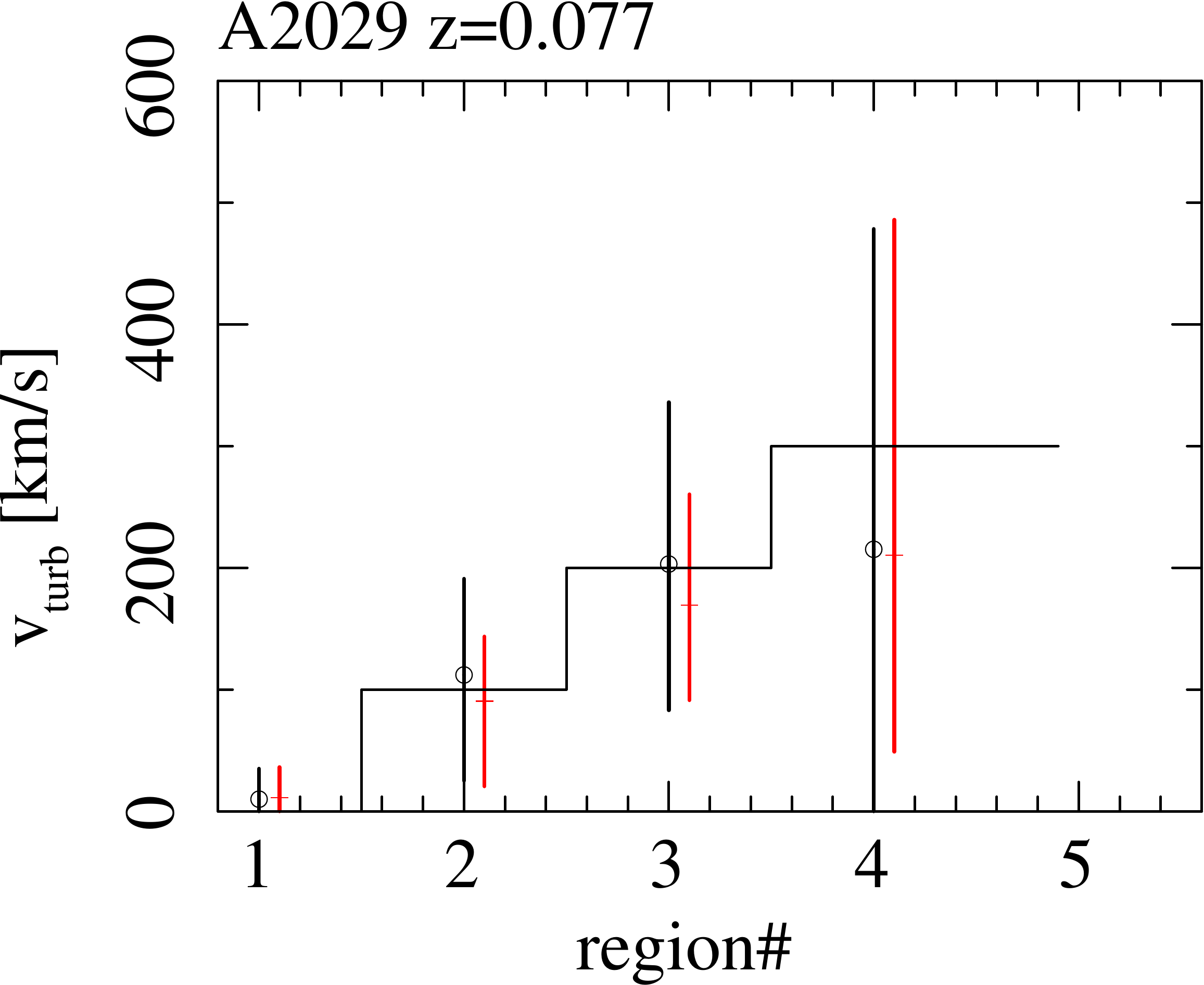}}}
\caption{Same as Figure~\ref{fig:a2029con}, but for case (b): $v_{\rm
turb}$ increases with radius.}
\label{fig:a2029pos}
\vspace{1.8cm}
\end{minipage}\end{center}
\end{figure}

\begin{figure}
\begin{minipage}{0.46\textwidth}
\rotatebox{0}{\scalebox{0.28}{\includegraphics{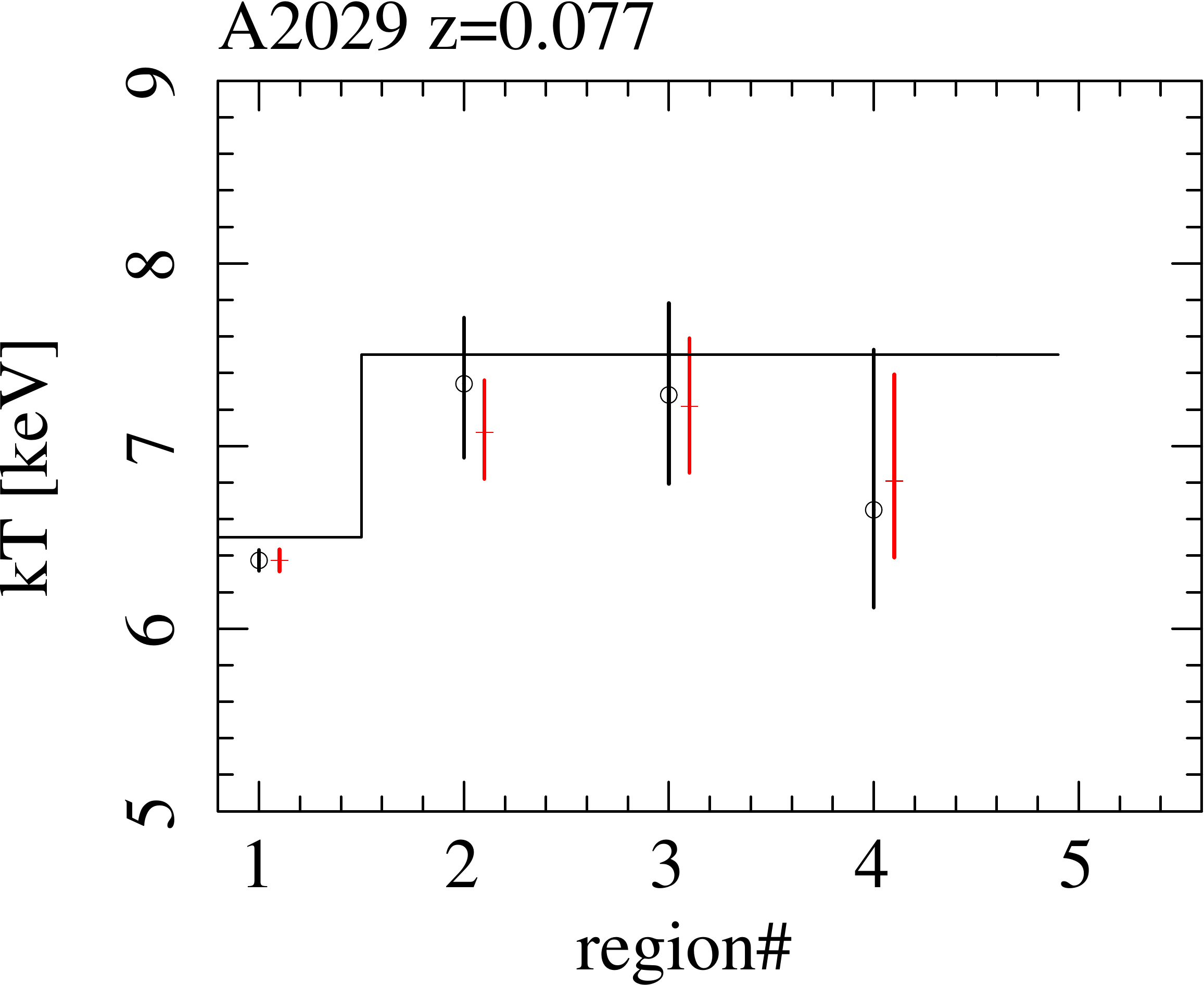}}}
\rotatebox{0}{\scalebox{0.28}{\includegraphics{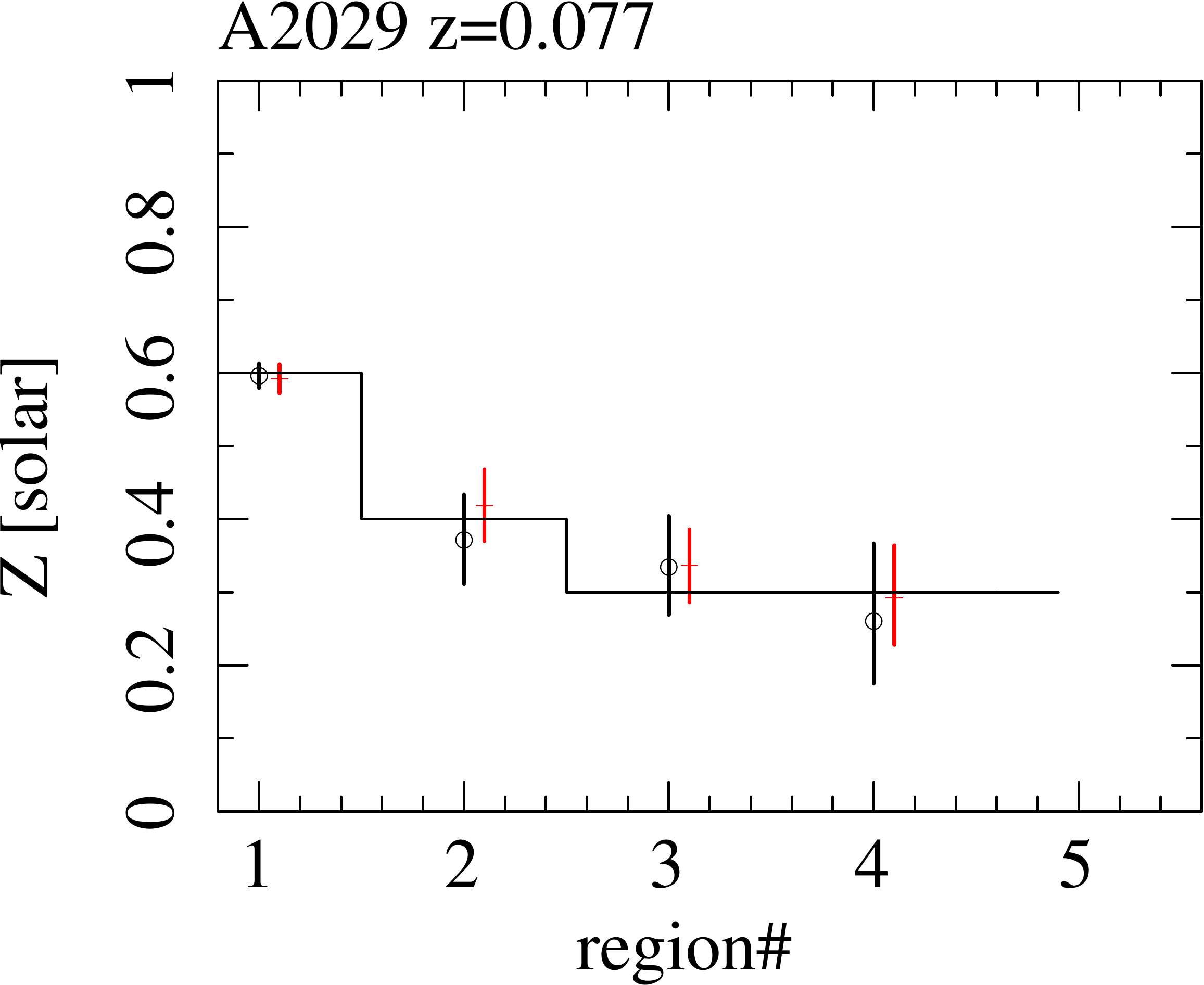}}}
\rotatebox{0}{\scalebox{0.28}{\includegraphics{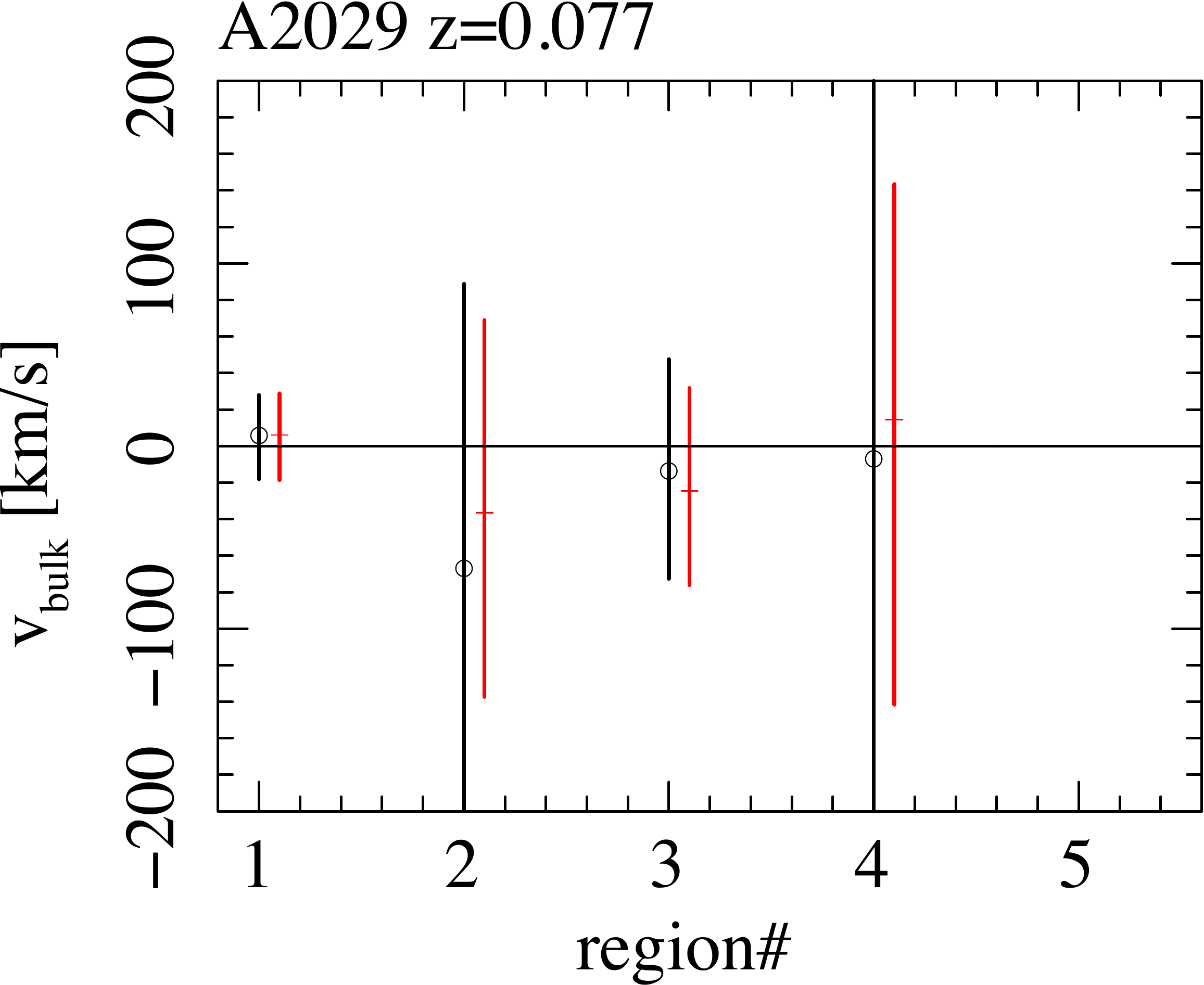}}}
\rotatebox{0}{\scalebox{0.28}{\includegraphics{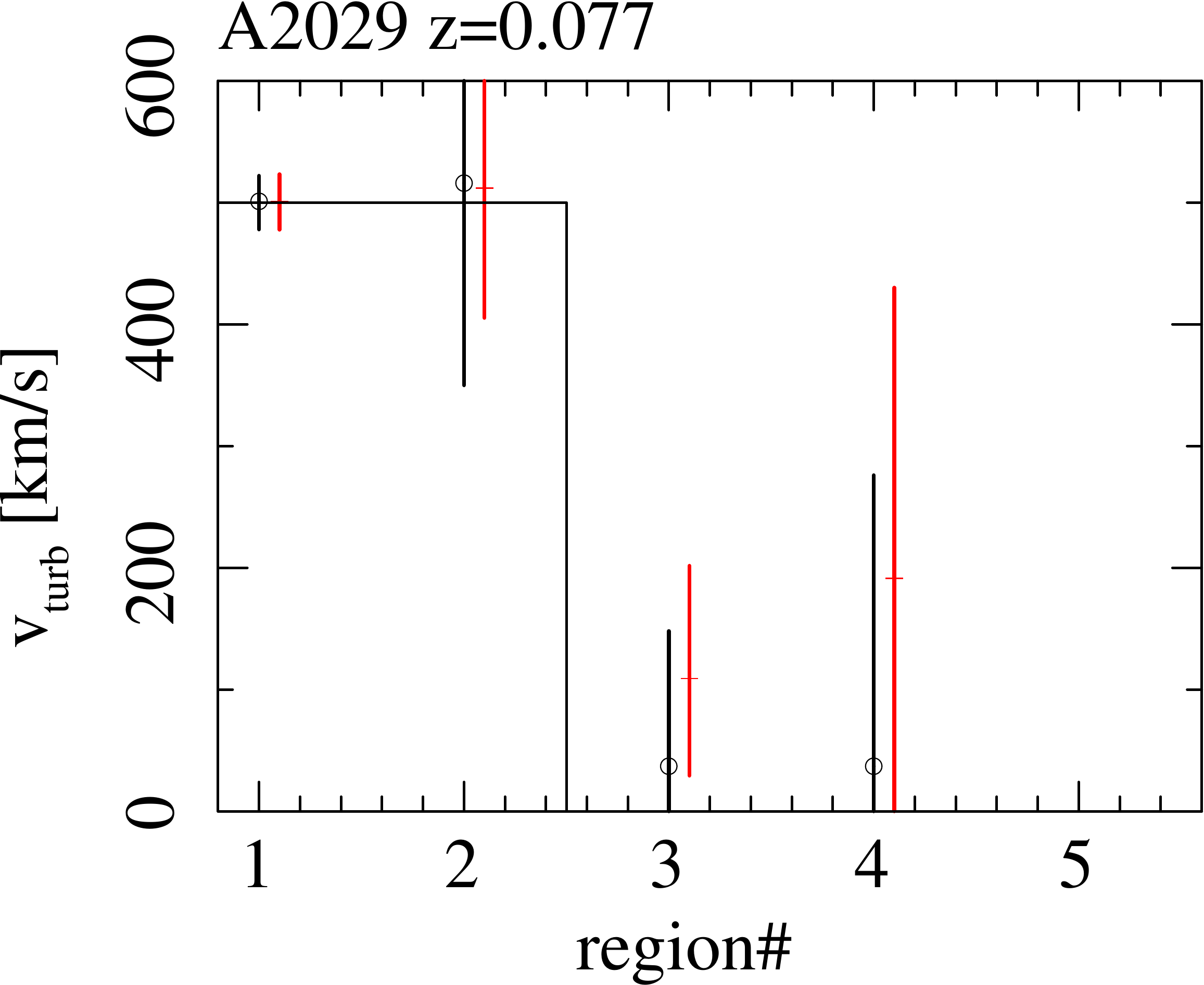}}}
\caption{Same as Figure~\ref{fig:a2029con}, but for case
(c): $v_{\rm turb}$ decreases with radius.}
\label{fig:a2029neg}
\end{minipage}~~
\begin{minipage}{0.46\textwidth}
\rotatebox{0}{\scalebox{0.28}{\includegraphics{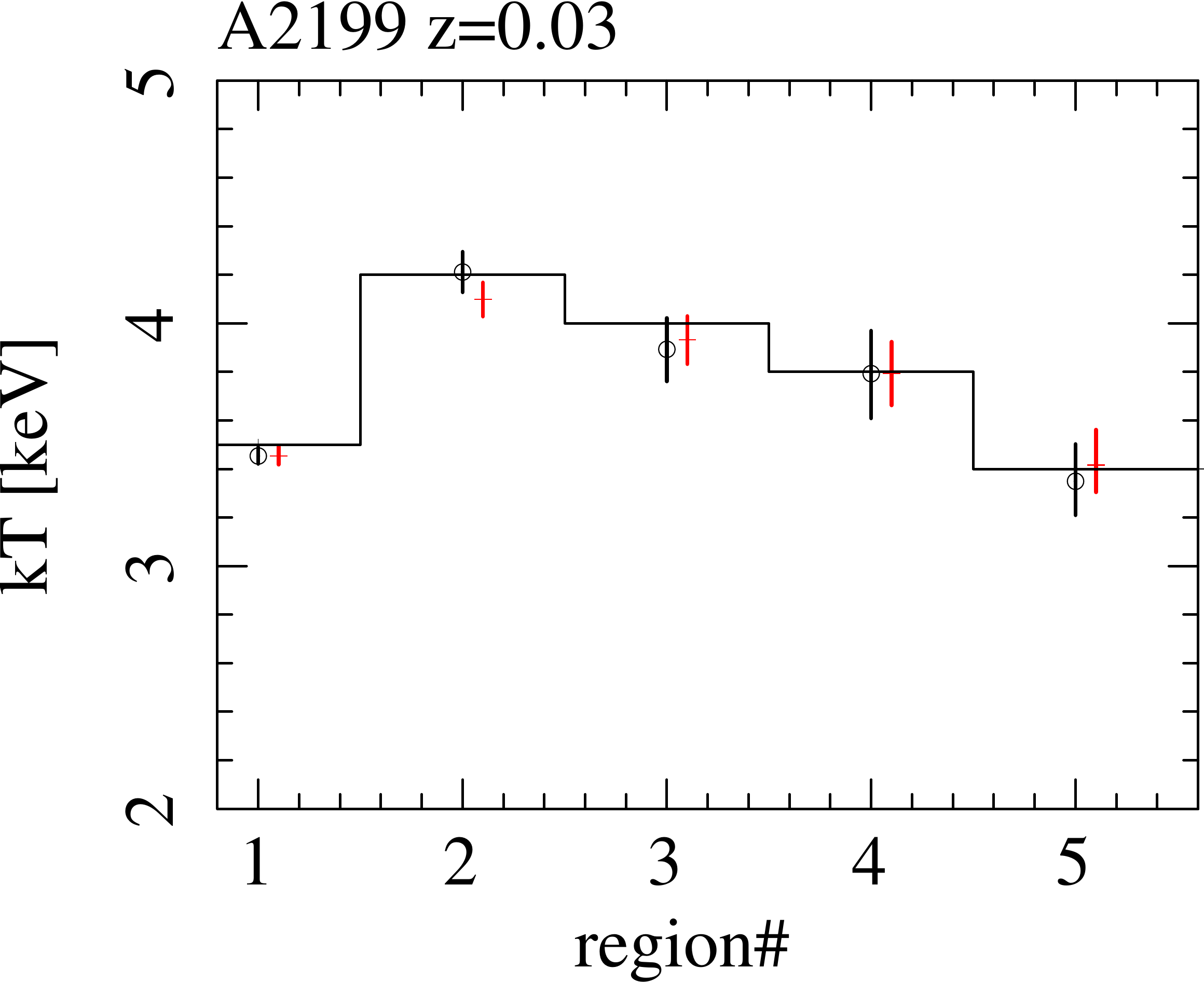}}}
\rotatebox{0}{\scalebox{0.28}{\includegraphics{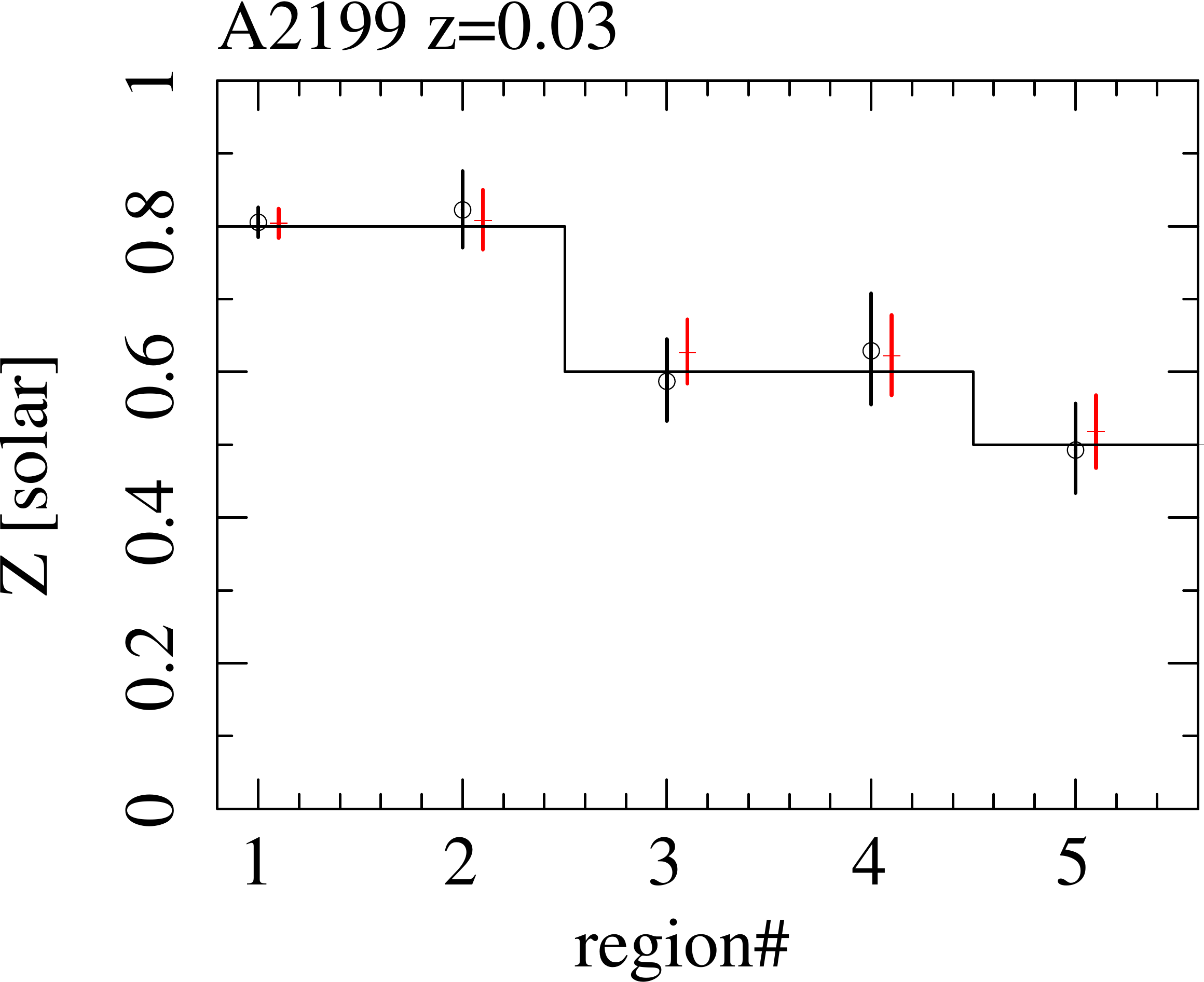}}}
\rotatebox{0}{\scalebox{0.28}{\includegraphics{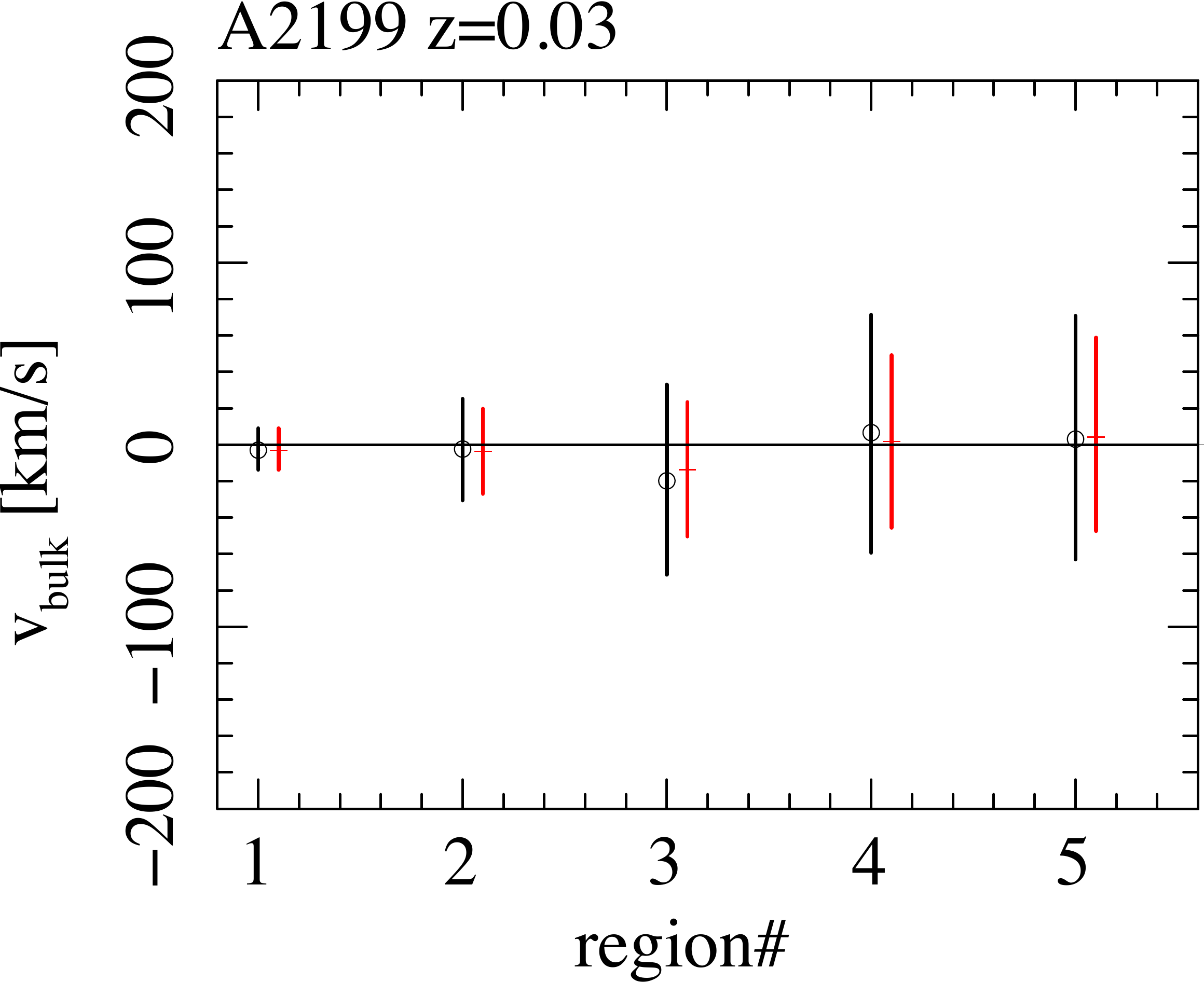}}}
\rotatebox{0}{\scalebox{0.28}{\includegraphics{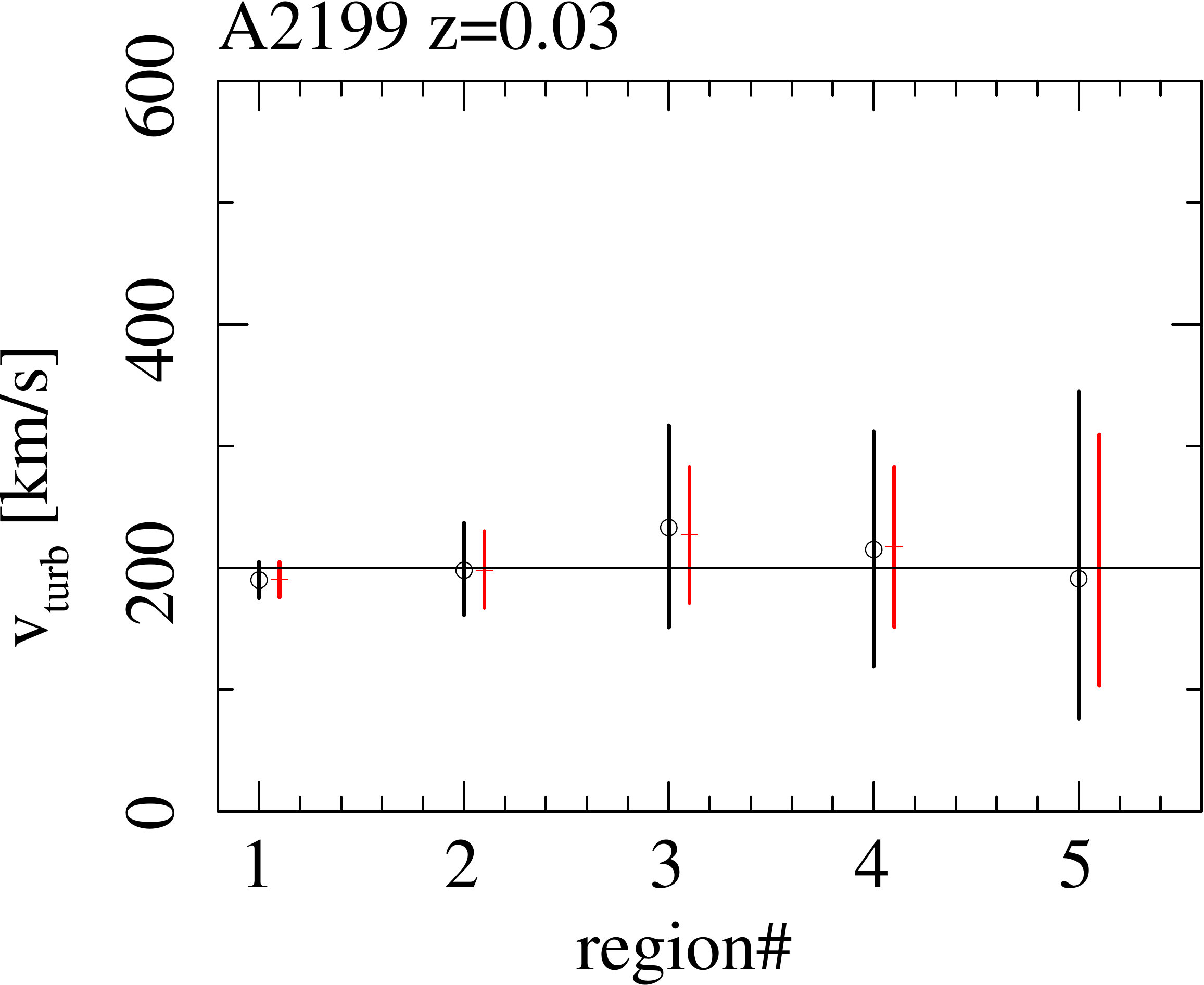}}}
\caption{Same as Figure~\ref{fig:a2029con}, but for A2199 and case (a).}
\label{fig:a2199con}
\end{minipage}
\end{figure}

We simulate SXS observations of Abell~2029 for the pointing regions
$j=1-4$; region 3 is centered at about $0.8 \times r_{2500}$ (540 kpc,
6.5 arcmin), and its outer boundary is very close to
$r_{2500}$. The assumed spectral parameters are listed
in Table~\ref{tab:a2029spec}. The effect of PSF scattering is estimated
by convolving the \xmm surface brightness distribution in the
0.3--10~keV band \citep[Table~5 in][]{Snowdenetal2008} with the 
SXT PSF model of January 2013.
Table~\ref{tab:a2029frac} gives the fraction of photons in the $j$-th
SXS pointing region, originating from the $i$-th annular image. Here the
vignetting factor is taken into account by referring to
``SXT\_VIG\_to110119.fits''. Using the calculated photon distributions
and the SXS response file with FWHM = 5~eV \footnote{The detector
response: ah\_sxs\_5ev\_basefilt\_20100712.rmf (the baseline version)
and the telescope arf: sxt-s\_120210\_ts02um\_intallpxl.arf are used in
the spectral simulation.}, we have created the SXS mock spectra for four
pointing regions, $j=1$--4, in the cluster (Figure~\ref{fig:a2029sxs}).
The parameters, $kT$, $Z$, $z$, $v_{\rm turb}$, and the spectral
normalization are allowed to vary and fitted in each annulus
simultaneously, while the fractions of scattered photons (both continuum
and the 6.7 keV Fe line) are fixed at the values given in Table
\ref{tab:a2029frac}.  The four spectra are fitted simultaneously and
Figures~\ref{fig:a2029con}--\ref{fig:a2029neg} show the results for the
cases (a)--(c), respectively. The quoted errors are at the 90\%
confidence. \footnote{Calibration errors of the PSF shape lead to the
systematic uncertainty in measuring line profiles particularly in the
outer regions of clusters.
Our earlier anlysis suggested that a PSF calibration
accuracy of 10\% or better would be sufficient for mapping nearby
clusters.}
%
%

\begin{table}[t]
\begin{center}
\caption{
Input parameters assumed in the spectral simulations for A2029.  Three
models are considered for the radial distribution of turbulence: (a)
$v_{\rm turb}$ is constant over radius, (b) $v_{\rm turb}$ increases
with radius, and (c) $v_{\rm turb}$ decreases sharply outside the core.
The hydrogen column density is fixed at $N_{\rm H} = 3.0\times 10^{20}$
cm$^{-2}$.}\label{tab:a2029spec}
\begin{tabular}{cccccccc} \hline\hline
Reg & $kT$~[keV] & $Z$~[solar]  & Flux $[{\rm erg\,s^{-1}\,cm^{-2}}]$&  \multicolumn{3}{c}{$v_{\rm turb}~{\rm [km\, s^{-1}]}$\tnote{a}}  & Exposure~[ks]\\ 
     &      & & (0.3--10 keV)       & (a) & (b) & (c)&  \\\hline
1  & 6.5 & 0.6 & $6.5\times10^{-11}$ &200 & 0 & 500 & 50 \\
2  & 7.5 & 0.4 & $7.5\times10^{-12}$ &200 & 100 & 500 & 50\\
3  & 7.5 & 0.3 & $1.0\times10^{-12}$ &200 & 200 & 0 & 300\\
4  & 7.5 & 0.3 & $3.0\times10^{-13}$ &200 & 300 & 0 & 600\\ \hline
\end{tabular}
%
\bigskip \caption{Fraction of photons at 0.3--10 keV in the $j$-th
simulated regions originating from the $i$-th annuli for A2029
(note the scattering fractions of the Fe line flux are
greater, because of the peaked abundance profile given in Table
\ref{tab:a2029spec}).  }\label{tab:a2029frac}
\begin{tabular}{lllll}\hline \hline
   & $i=1$ & $i=2$ & $i=3$ & $i=4$ \\ \hline
$j=1$ & {\bf 1.000} & 0.000 & 0.000 & 0.000 \\ 
$j=2$ & 0.201 & {\bf 0.772} &	0.027 & 0.000 \\
$j=3$ & 0.048 	& 0.185 &	{\bf 0.734} & 0.033 \\
$j=4$ & 0.053 	& 0.086 &	0.203 & {\bf 0.658} \\ \hline
\end{tabular}
\end{center}
\end{table}%

With the total exposure of about 400~ks, the turbulent velocity can be
measured with a statistical accuracy of $\lesssim100~{\rm km\,s^{-1}}$
out to region $j=3$, corresponding to about $0.8\,r_{2500}$ for
this cluster. As expected, the PSF effect has the largest impact in case
(c) and can bias the measured values of $v_{\rm turb}$ outside the
cluster core.



\begin{figure}
\begin{minipage}{0.46\textwidth}
\rotatebox{0}{\scalebox{0.28}{\includegraphics{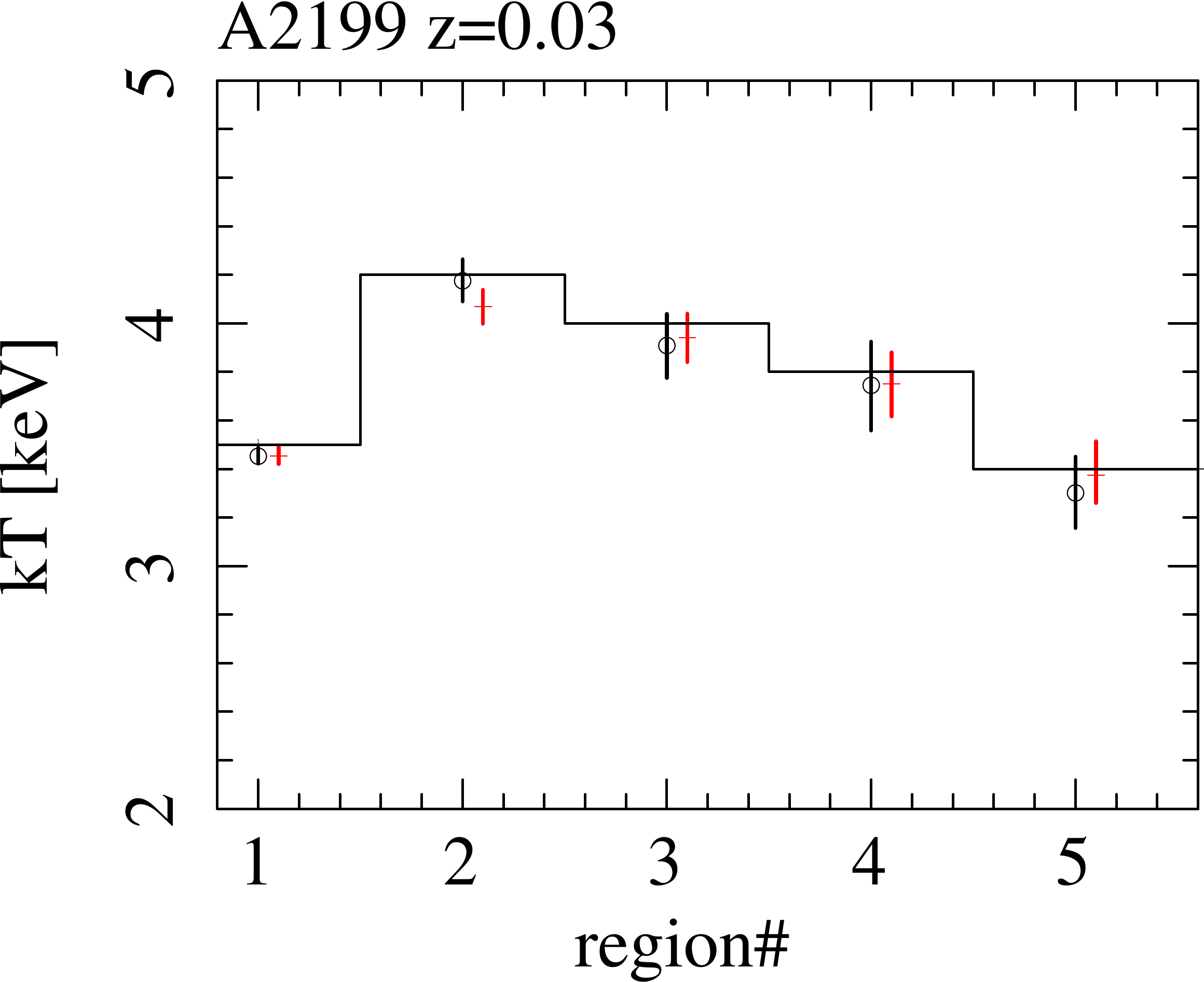}}}
\rotatebox{0}{\scalebox{0.28}{\includegraphics{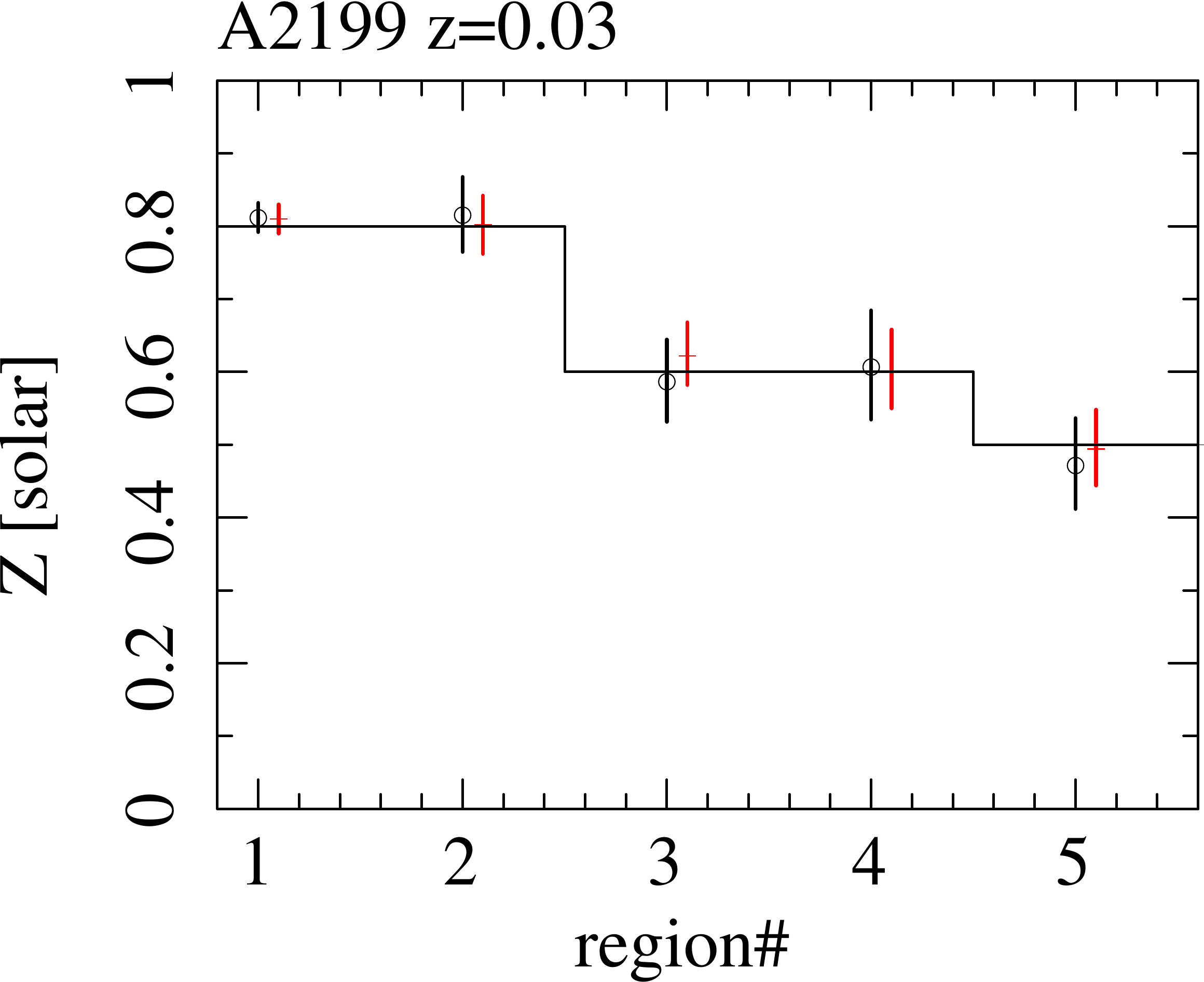}}}
\rotatebox{0}{\scalebox{0.28}{\includegraphics{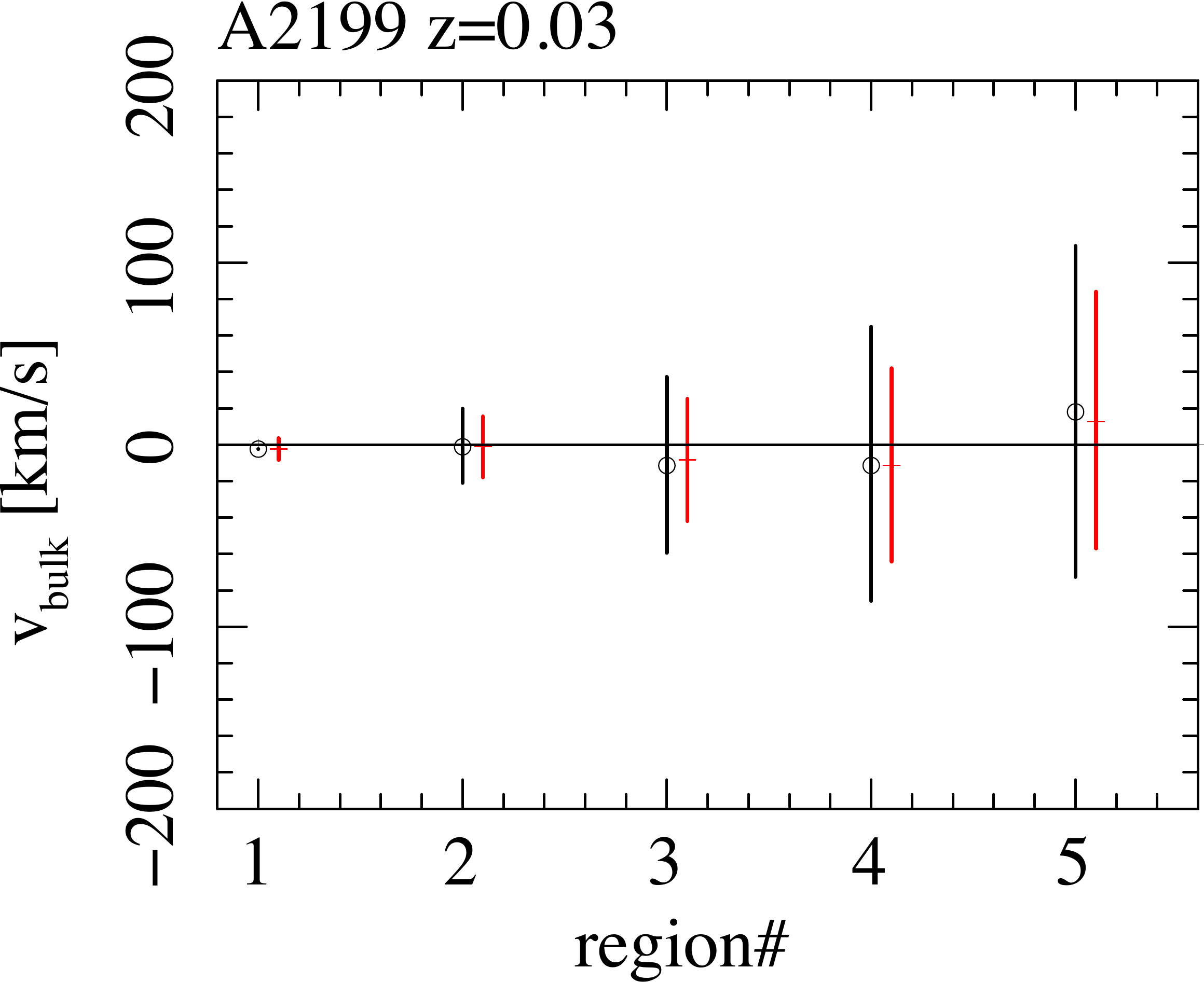}}}
\rotatebox{0}{\scalebox{0.28}{\includegraphics{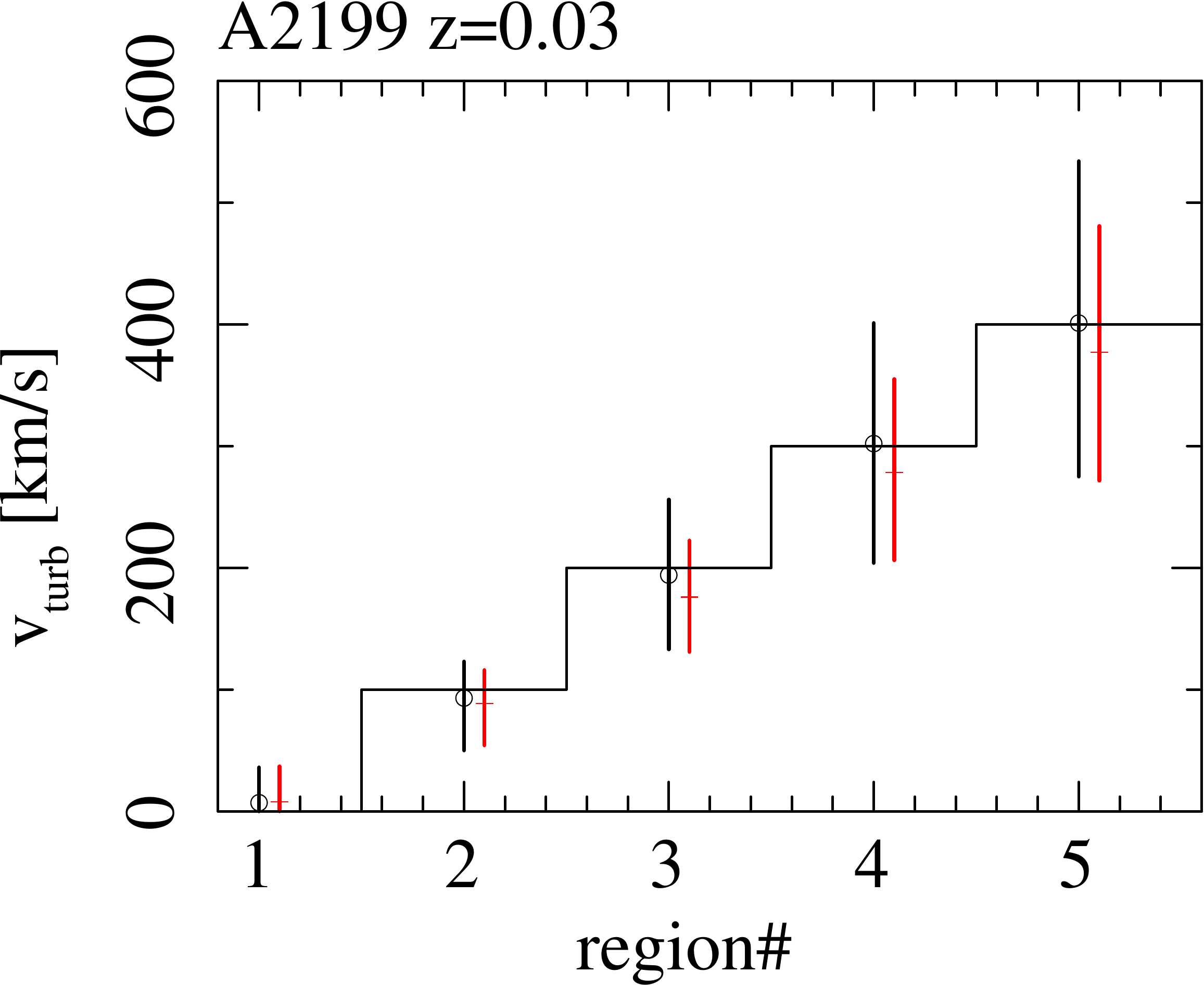}}}
\caption{Same as Figure~\ref{fig:a2029con}, but for A2199 and case (b).}
\label{fig:a2199pos}
\vspace{1.8cm}
\end{minipage}~~
\begin{minipage}{0.46\textwidth}
\rotatebox{0}{\scalebox{0.28}{\includegraphics{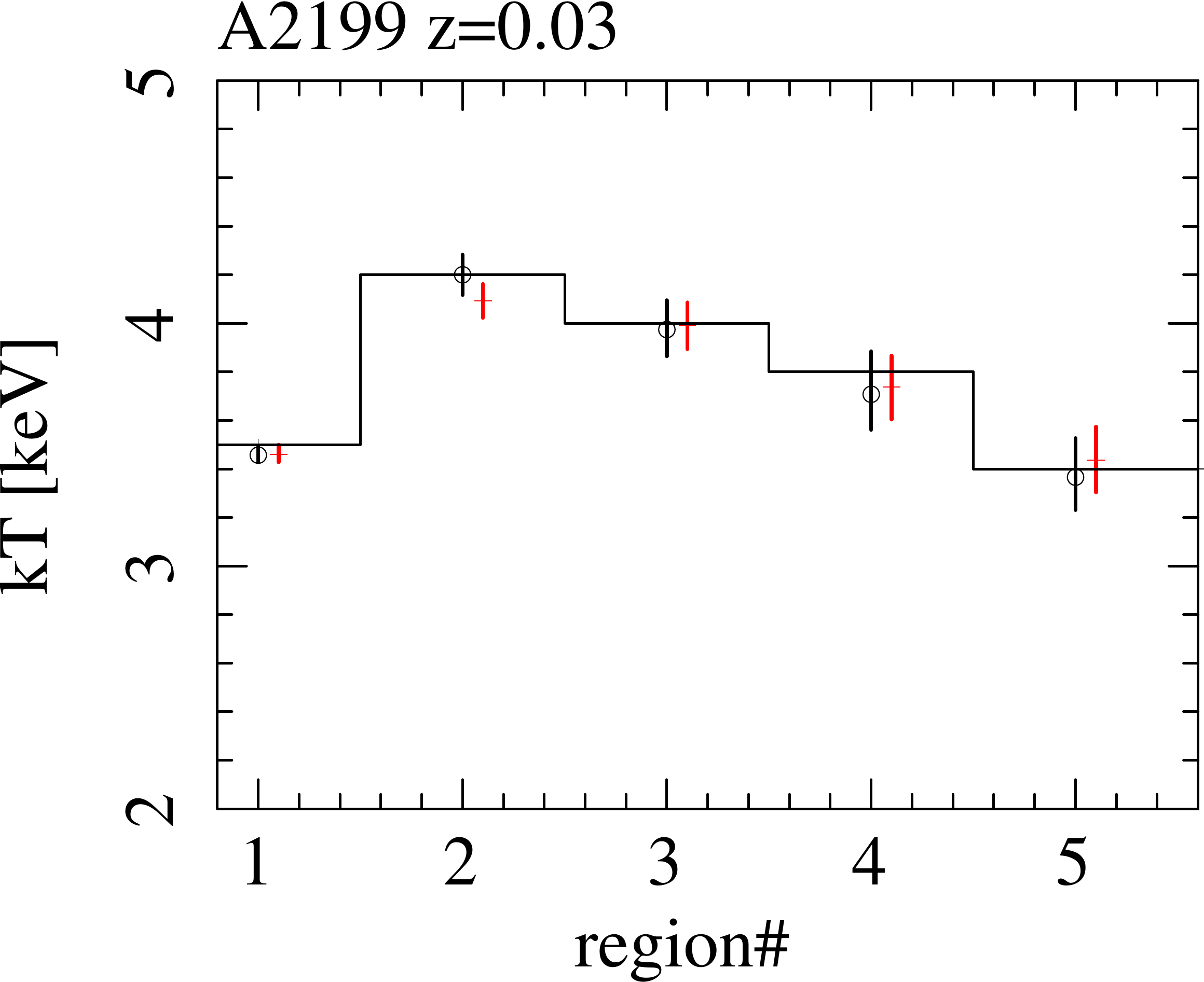}}}
\rotatebox{0}{\scalebox{0.28}{\includegraphics{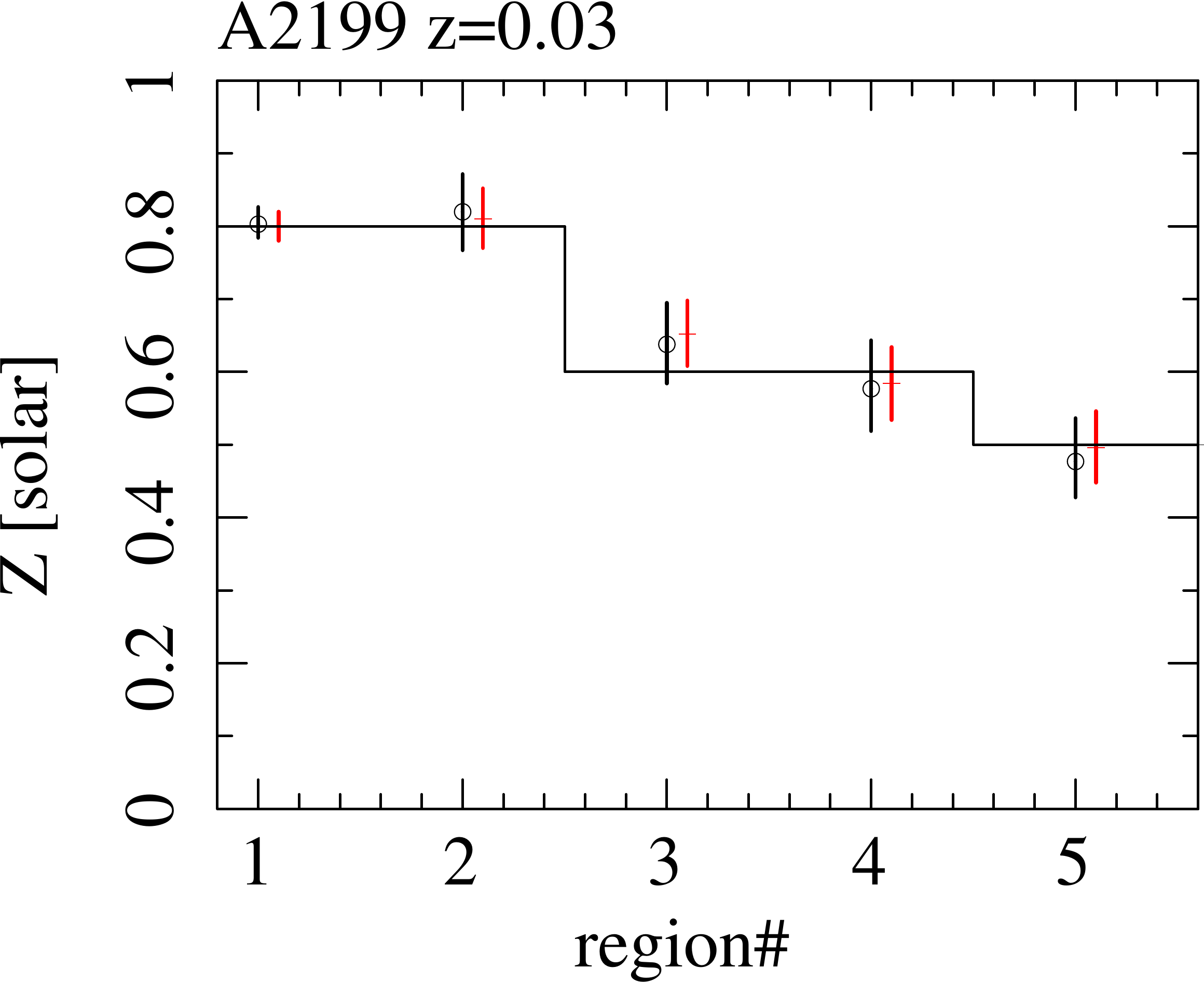}}}
\rotatebox{0}{\scalebox{0.28}{\includegraphics{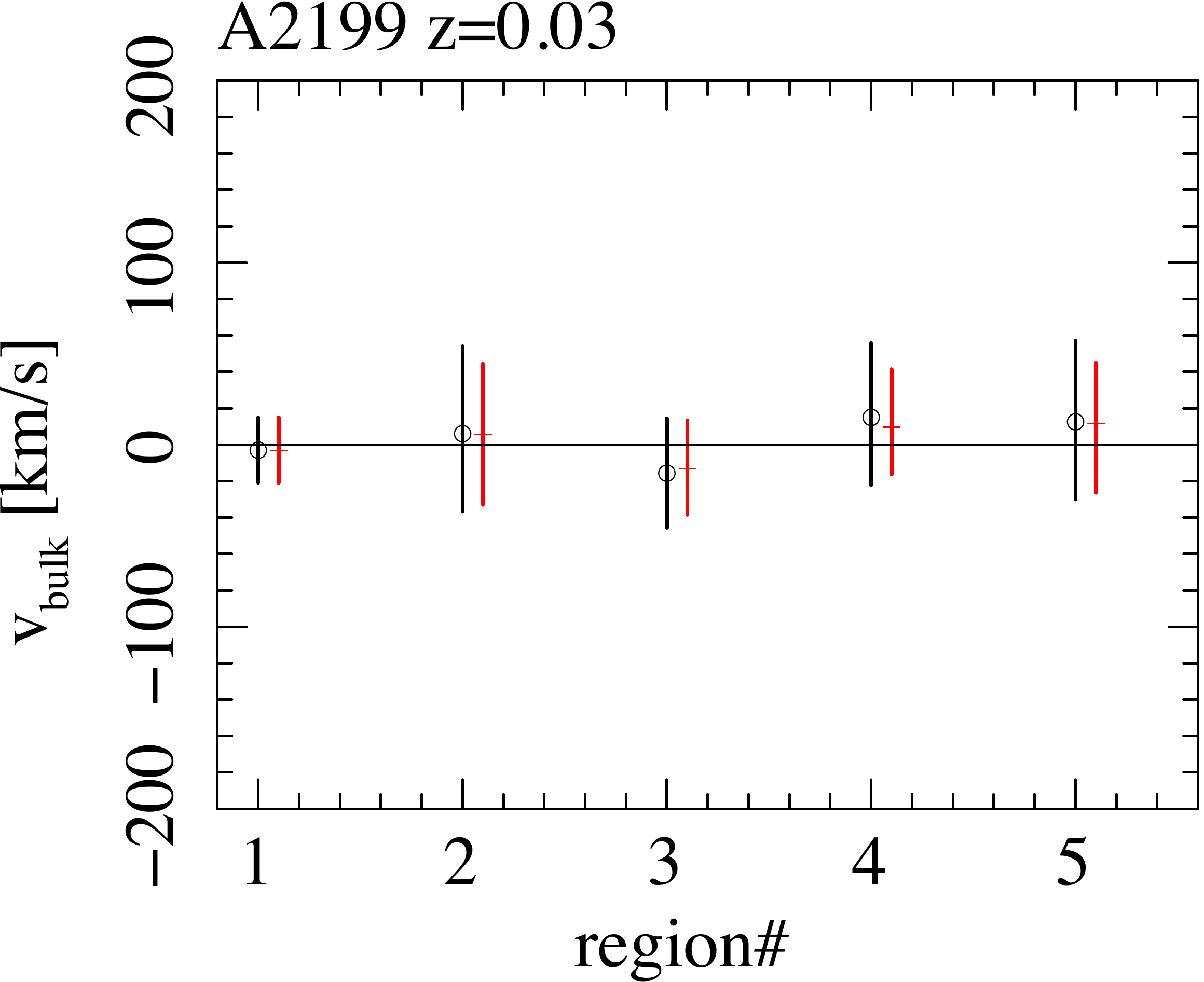}}}
\rotatebox{0}{\scalebox{0.28}{\includegraphics{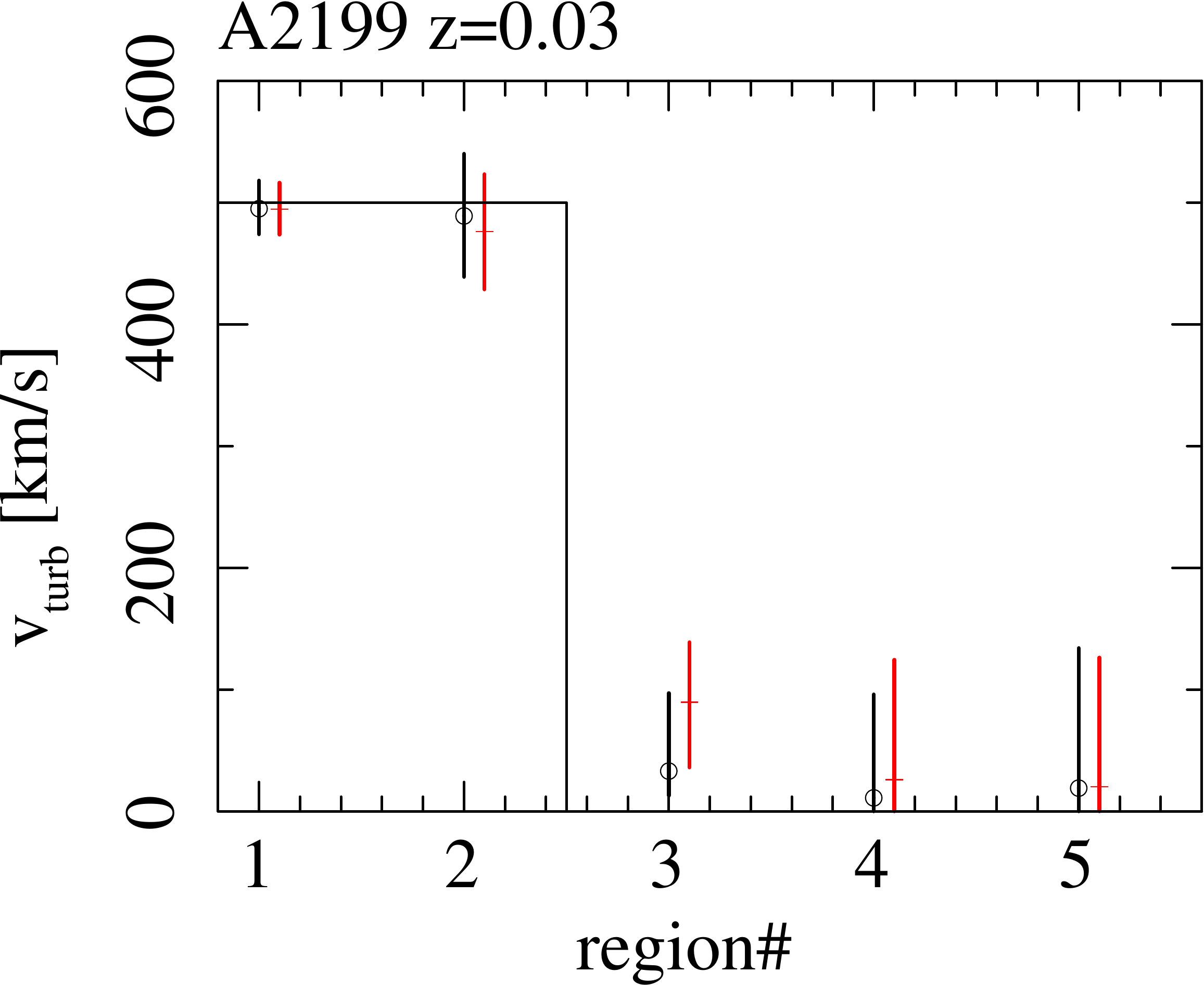}}}
\caption{Same as Figure~\ref{fig:a2029con}, but for A2199 and case (c).}
\label{fig:a2199neg}
\vspace{1.8cm}
\end{minipage}
\end{figure}

\subsubsection{Abell 2199}

We simulate SXS observations of Abell 2199 for the five SXS pointing
regions $j = 1 - 5$, assuming the spectral parameters in
Table~\ref{tab:a2199spec}, and the scattered light contribution given in
Table~\ref{tab:a2199frac}.  These pointings cover the cluster to a
maximum radius of about 450 kpc, with the outermost pointing centered
just beyond $r_{2500}$ (see Table~\ref{tab:targets}).  Using the SXS
response files noted in the previous subsection, we have created mock
SXS spectra for the five regions. 
The five spectra are fitted simultaneously over the 0.3 -- 10 keV
band. Figures~\ref{fig:a2199con}--\ref{fig:a2199neg} show the results
for the cases (a)--(c), respectively.

\begin{table}[t]
\begin{center}
\caption{Same as Table \ref{tab:a2029spec}, but for A2199 and 
$N_{\rm H} = 0.88\times 10^{20}$ cm$^{-2}$.}
\label{tab:a2199spec}
\begin{tabular}{cccccccc} \hline\hline
Reg & $kT$~[keV] & $Z$~[solar]  & Flux $[{\rm erg\,s^{-1}\,cm^{-2}}]$ &  \multicolumn{3}{c}{$v_{\rm turb}~{\rm [km\, s^{-1}]}$}  & Exposure~[ks]\\ 
     &      &       & (0.3--10~keV) & (a) & (b) & (c) & \\\hline
1  & 3.5 & 0.8 & $3\times10^{-11}$ &200 & 0 & 500 & 50 \\
2  & 4.2 & 0.8 & $9\times10^{-12}$ &200 & 100 & 500 & 50\\
3  & 4.0 & 0.6 & $2\times10^{-12}$ &200 & 200 & 0 & 150\\
4  & 3.8 & 0.6 & $7\times10^{-13}$ &200 & 300 & 0 & 300\\ 
5  & 3.4 & 0.5 & $4\times10^{-13}$ &200 & 400 & 0 & 450\\ \hline
\end{tabular}
\bigskip 
\caption{Same as Table \ref{tab:a2029frac}, but for A2199.}
\label{tab:a2199frac}
\begin{tabular}{llllll}\hline \hline
   & $i=1$ & $i=2$ & $i=3$ & $i=4$ & $i=5$ \\ \hline
$j=1$ & {\bf 1.000} & 0.000 & 0.000 & 0.000 & 0.000 \\ 
$j=2$ & 0.125 & {\bf 0.834} &	0.040 & 0.000 & 0.000 \\
$j=3$ & 0.016 	& 0.156 &	{\bf 0.784} & 0.044 & 0.000 \\
$j=4$ & 0.012 	& 0.040 &	0.141 & {\bf 0.747} & 0.059 \\
$j=5$ & 0.008 	& 0.025 &	0.028 & 0.110 & {\bf 0.829} \\ \hline
\end{tabular}
\end{center}
\end{table}%

With the total exposure time of about 550~ks (1~Ms), the turbulent
velocity can be measured to the statistical accuracy of
$\lesssim100~{\rm km\,s^{-1}}$ out to region $j=4$ ($j=5$),
corresponding to about $0.85\,r_{2500}$ ($1.1\,r_{2500}$) in this
cluster.


\begin{table}[ht]
\begin{center}
\caption{Same as Table \ref{tab:a2029spec}, but for A1795 and 
$N_{\rm H} = 1.2\times 10^{20}$ cm$^{-2}$.}
\label{tab:a1795spec}
\begin{tabular}{cccccccc} \hline\hline
Reg & $kT$~[keV] & $Z$~[solar]  & Flux $[{\rm erg\,s^{-1}\,cm^{-2}}]$ &  \multicolumn{3}{c}{$v_{\rm turb}~{\rm [km\, s^{-1}]}$}  & Exposure~[ks]\\ 
     &      &       & (0.3--10~keV) & (a) & (b) & (c) & \\\hline
1  & 4.0 & 0.60 & $3.7\times10^{-11}$ &200 & 0   & 500 & 50 \\
2  & 6.0 & 0.35 & $8.4\times10^{-12}$ &200 & 100 & 500 & 50\\
3  & 6.0 & 0.30 & $1.3\times10^{-12}$ &200 & 300 & 0   & 200\\
4  & 5.0 & 0.30 & $4.3\times10^{-13}$ &200 & 400 & 0   & 400\\ \hline
\end{tabular}
\bigskip 
\caption{Same as Table \ref{tab:a2029frac}, but for A1795.}
\label{tab:a1795frac}
\begin{tabular}{lllll}\hline \hline
   & $i=1$ & $i=2$ & $i=3$ & $i=4$ \\ \hline
$j=1$ & {\bf 1.000} & 0.000 & 0.000 & 0.000 \\ 
$j=2$ & 0.184 & {\bf 0.790} &	0.026 & 0.000 \\
$j=3$ & 0.040 	& 0.184 &	{\bf 0.741} & 0.035 \\
$j=4$ & 0.034 	& 0.065 &	0.158 & {\bf 0.701} \\ \hline
\end{tabular}
\end{center}
\end{table}%

\subsubsection{Abell 1795}

\begin{figure}
\begin{minipage}{0.46\textwidth}
\rotatebox{270}{\scalebox{0.28}{\includegraphics{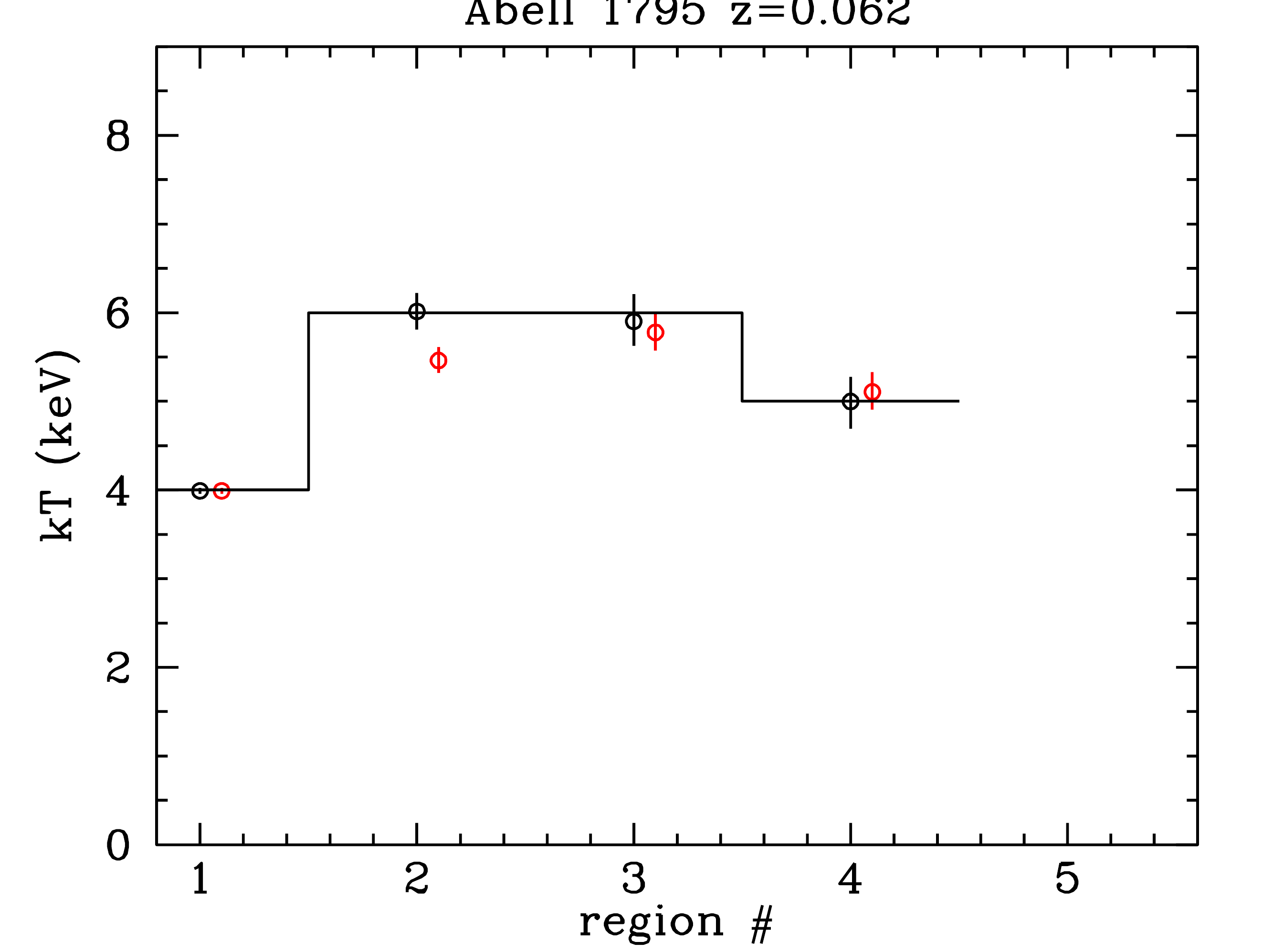}}}
\rotatebox{270}{\scalebox{0.28}{\includegraphics{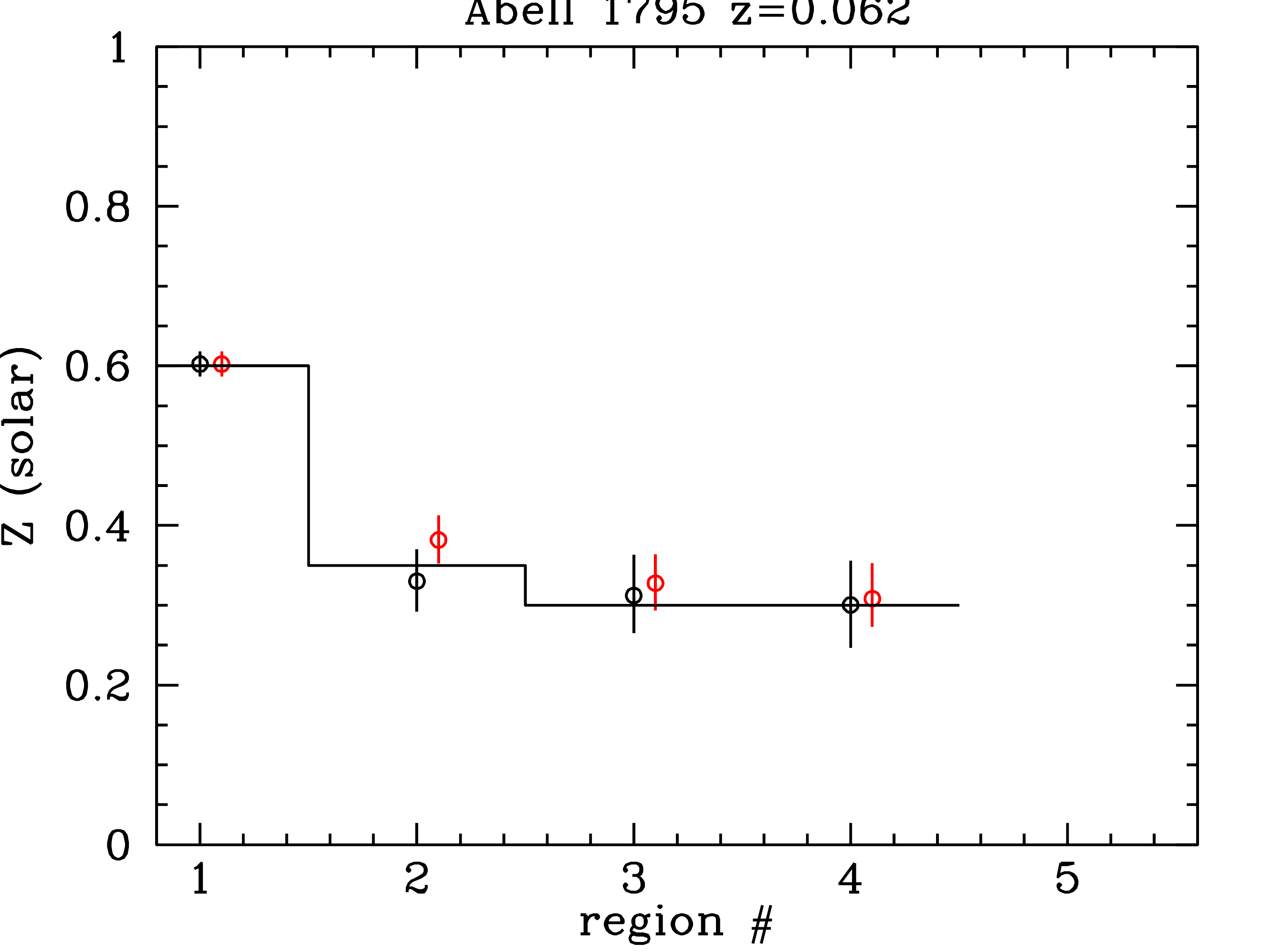}}}
\rotatebox{270}{\scalebox{0.28}{\includegraphics{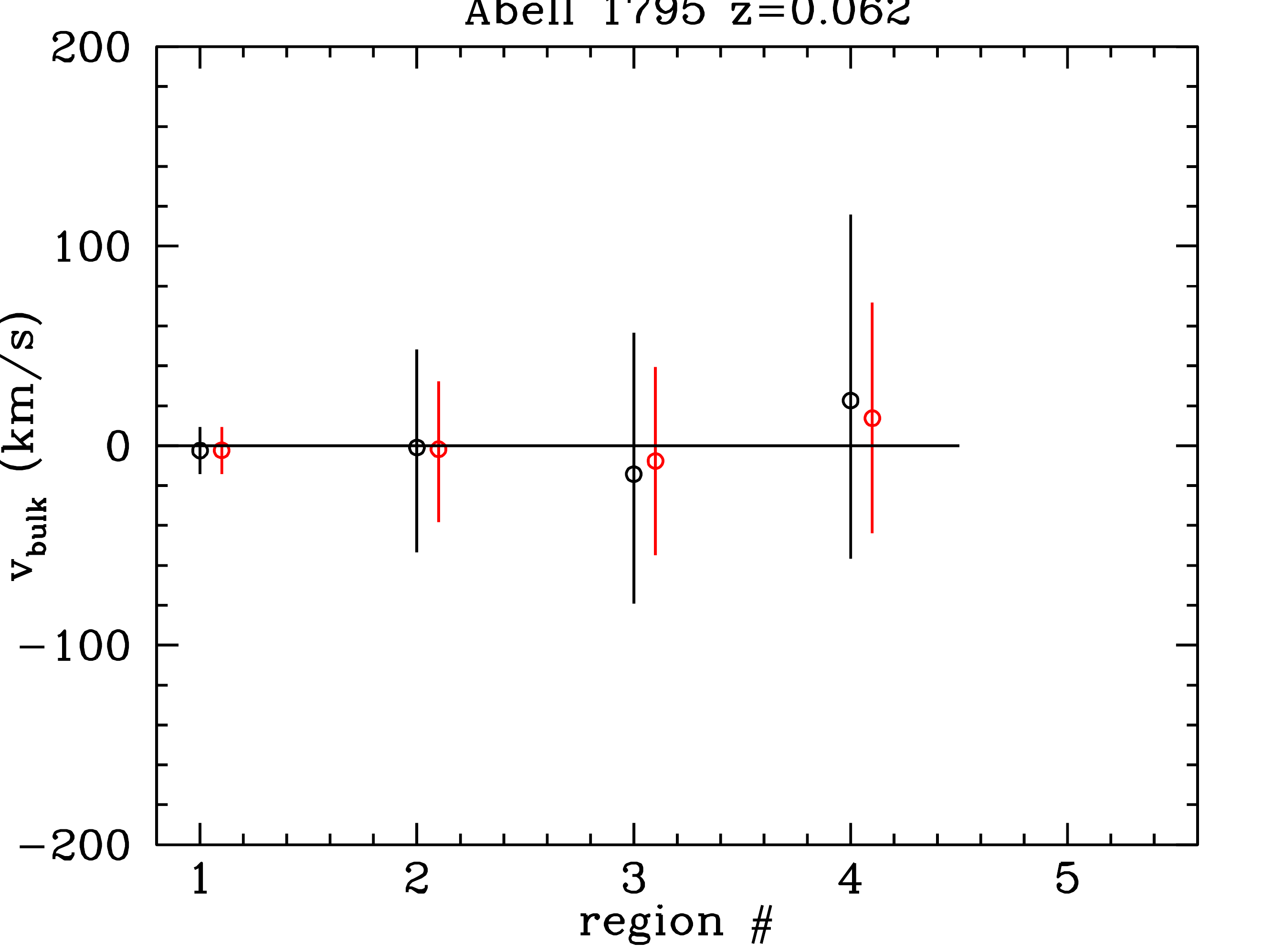}}}
\rotatebox{270}{\scalebox{0.28}{\includegraphics{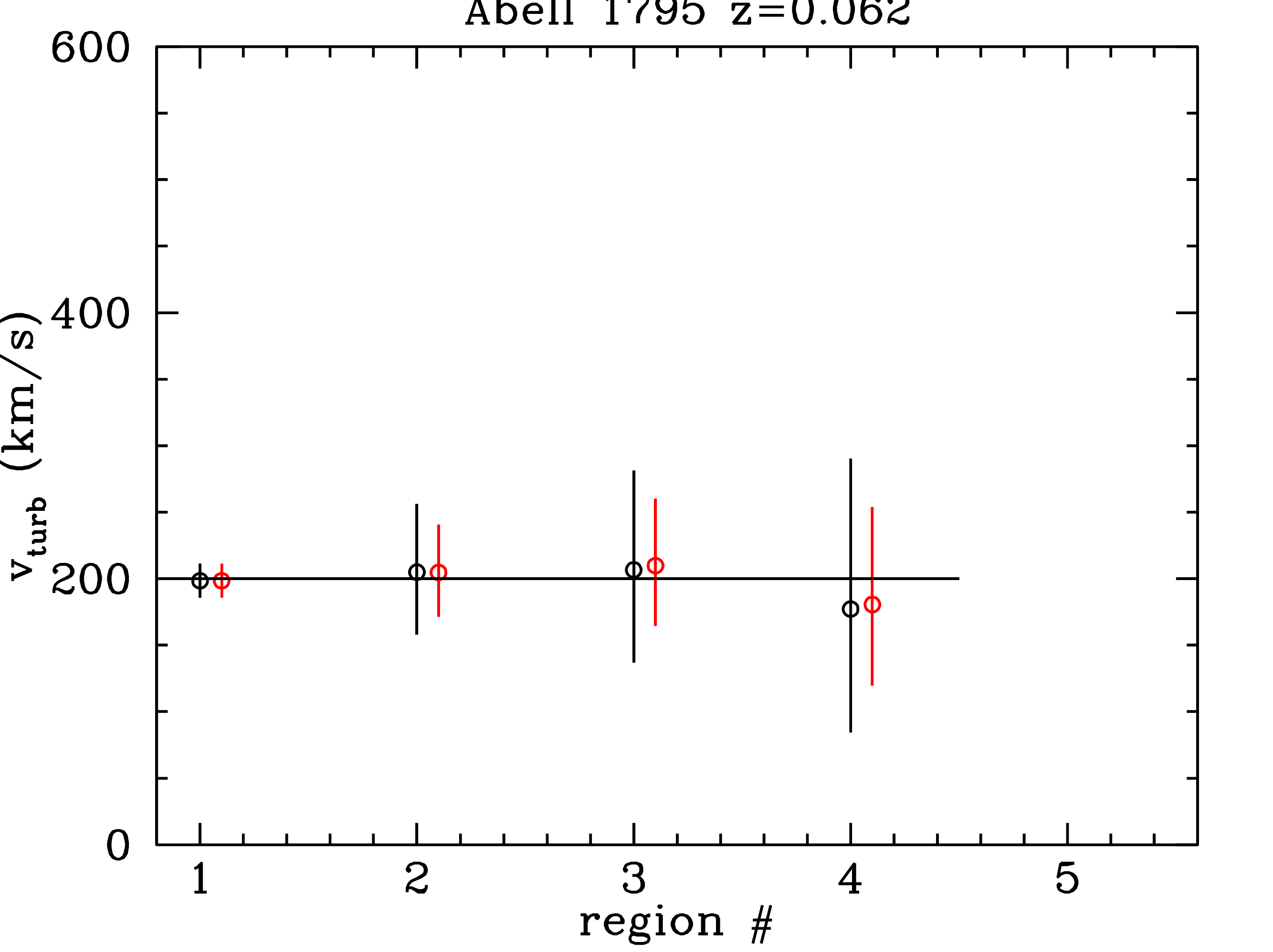}}}
\caption{Same as Figure~\ref{fig:a1795con}, but for A1795 and case (a).}
\label{fig:a1795con}
\end{minipage}~~
\begin{minipage}{0.46\textwidth}
\rotatebox{270}{\scalebox{0.28}{\includegraphics{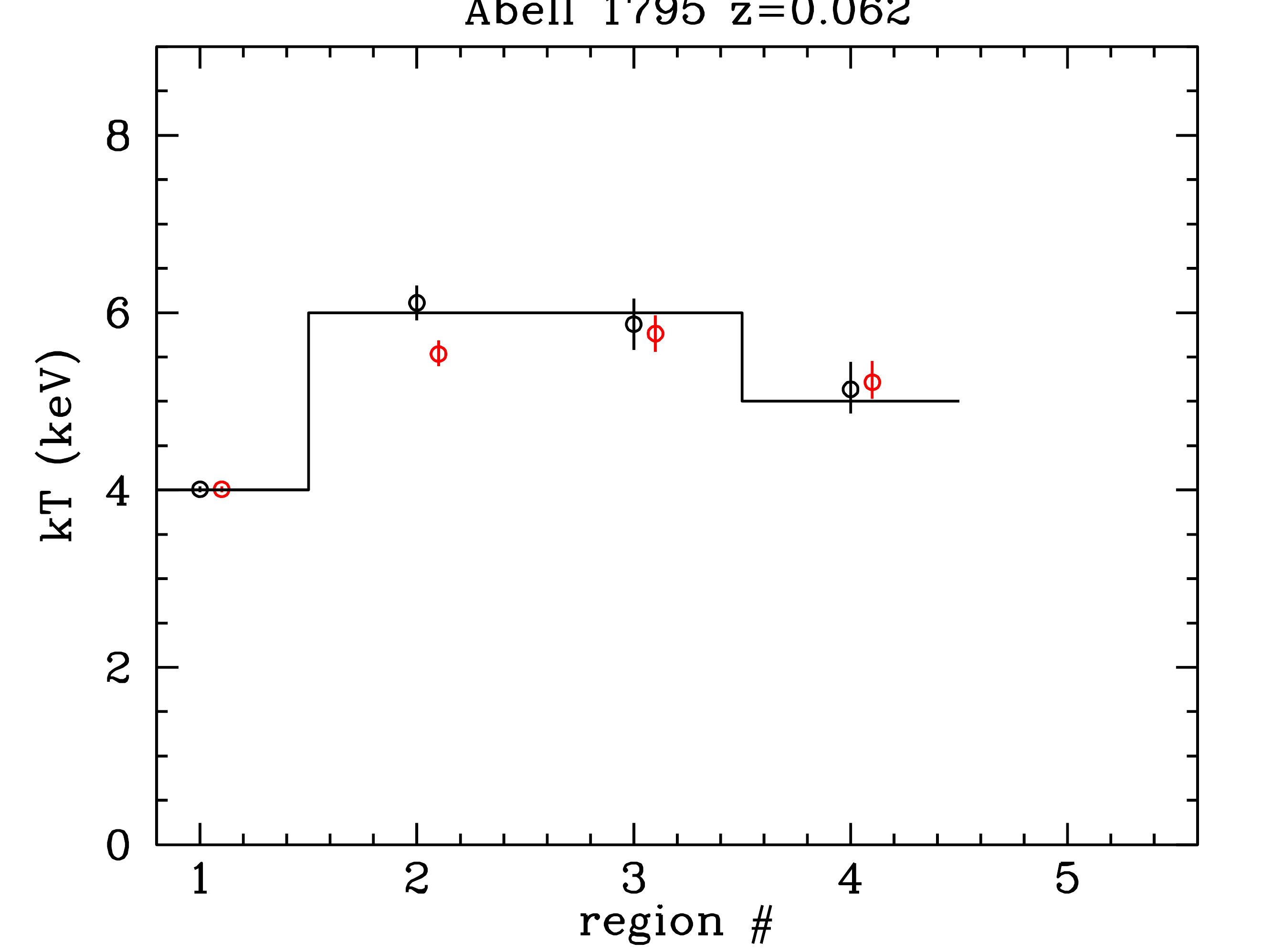}}}
\rotatebox{270}{\scalebox{0.28}{\includegraphics{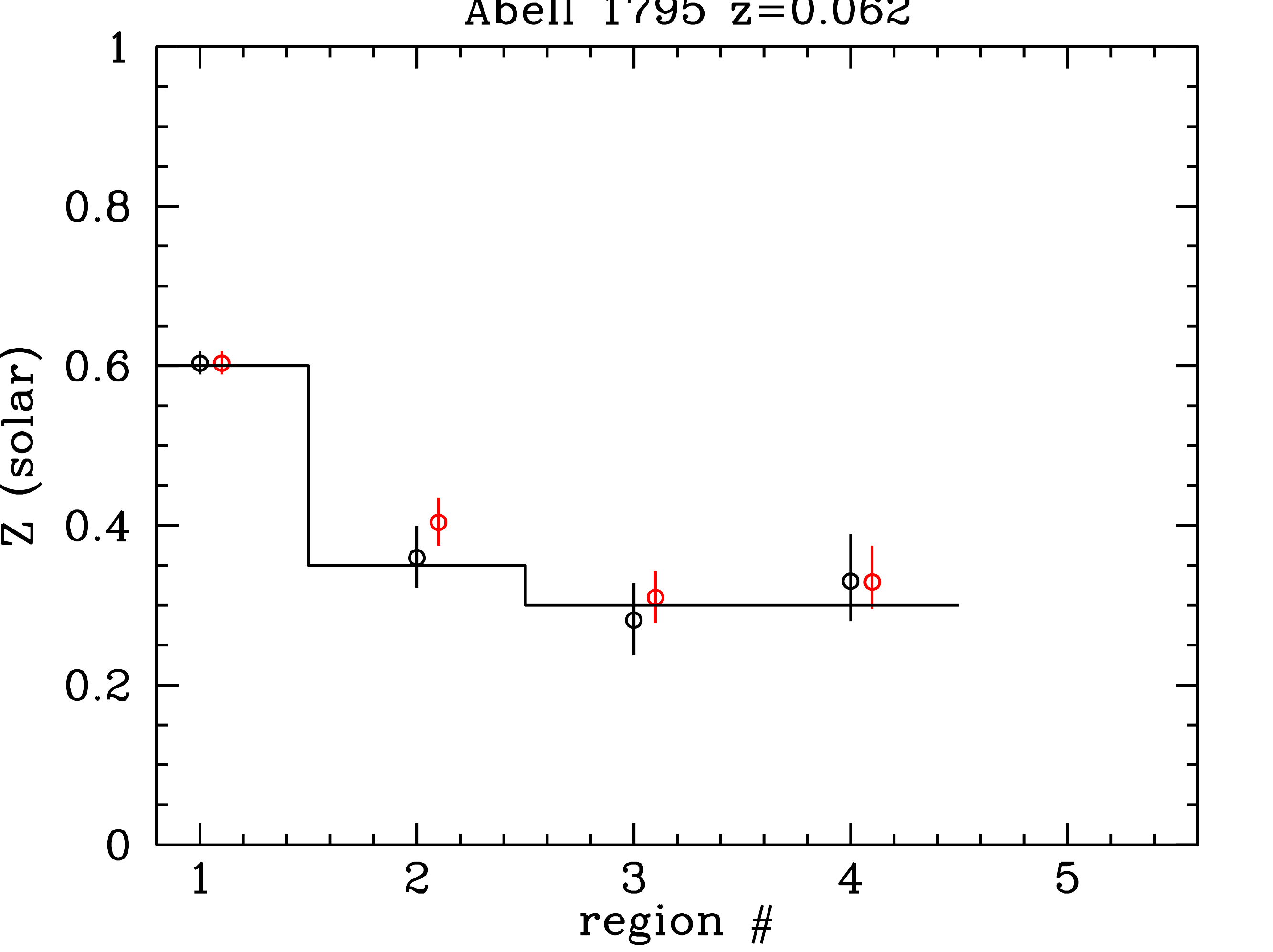}}}
\rotatebox{270}{\scalebox{0.28}{\includegraphics{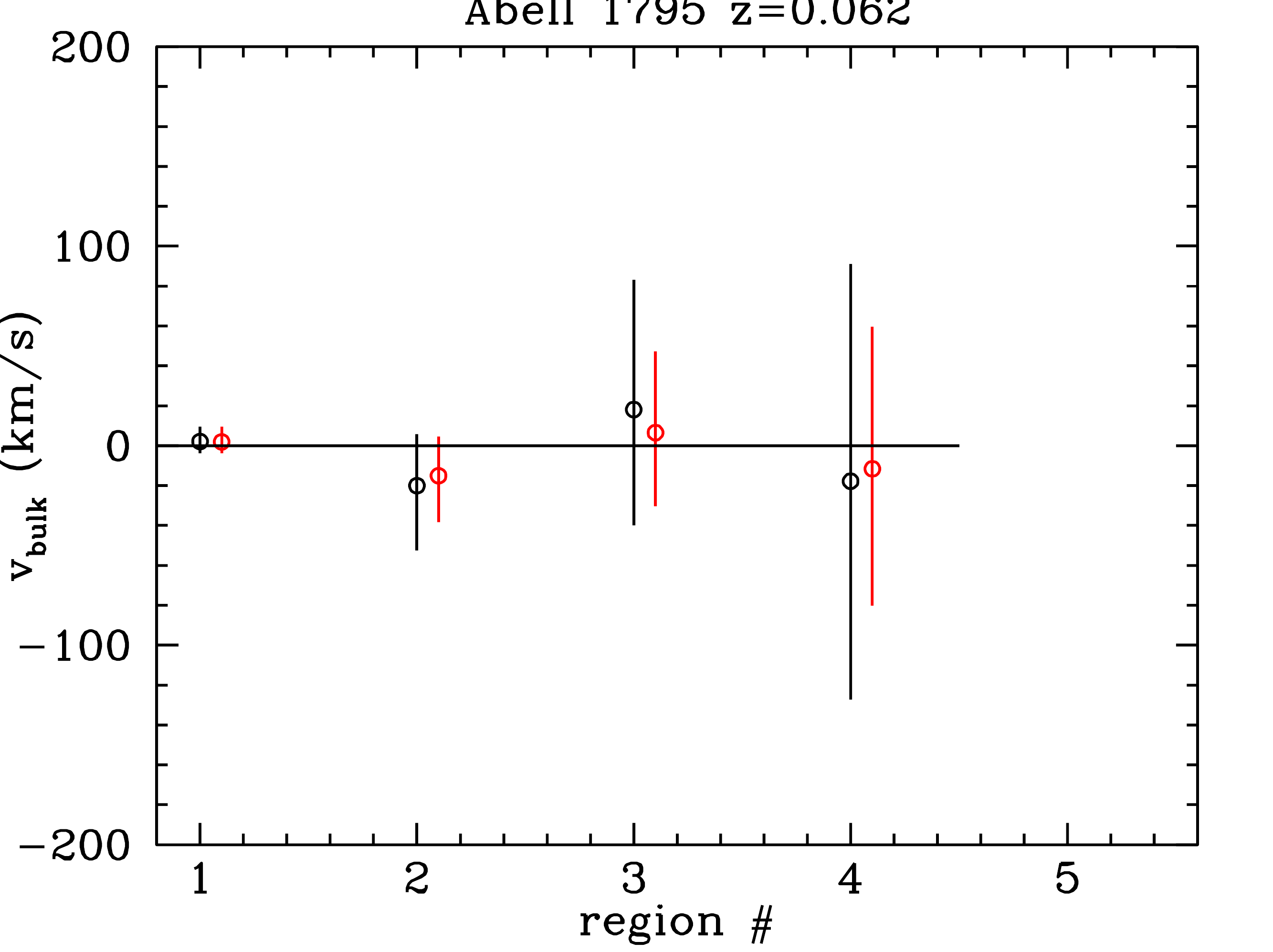}}}
\rotatebox{270}{\scalebox{0.28}{\includegraphics{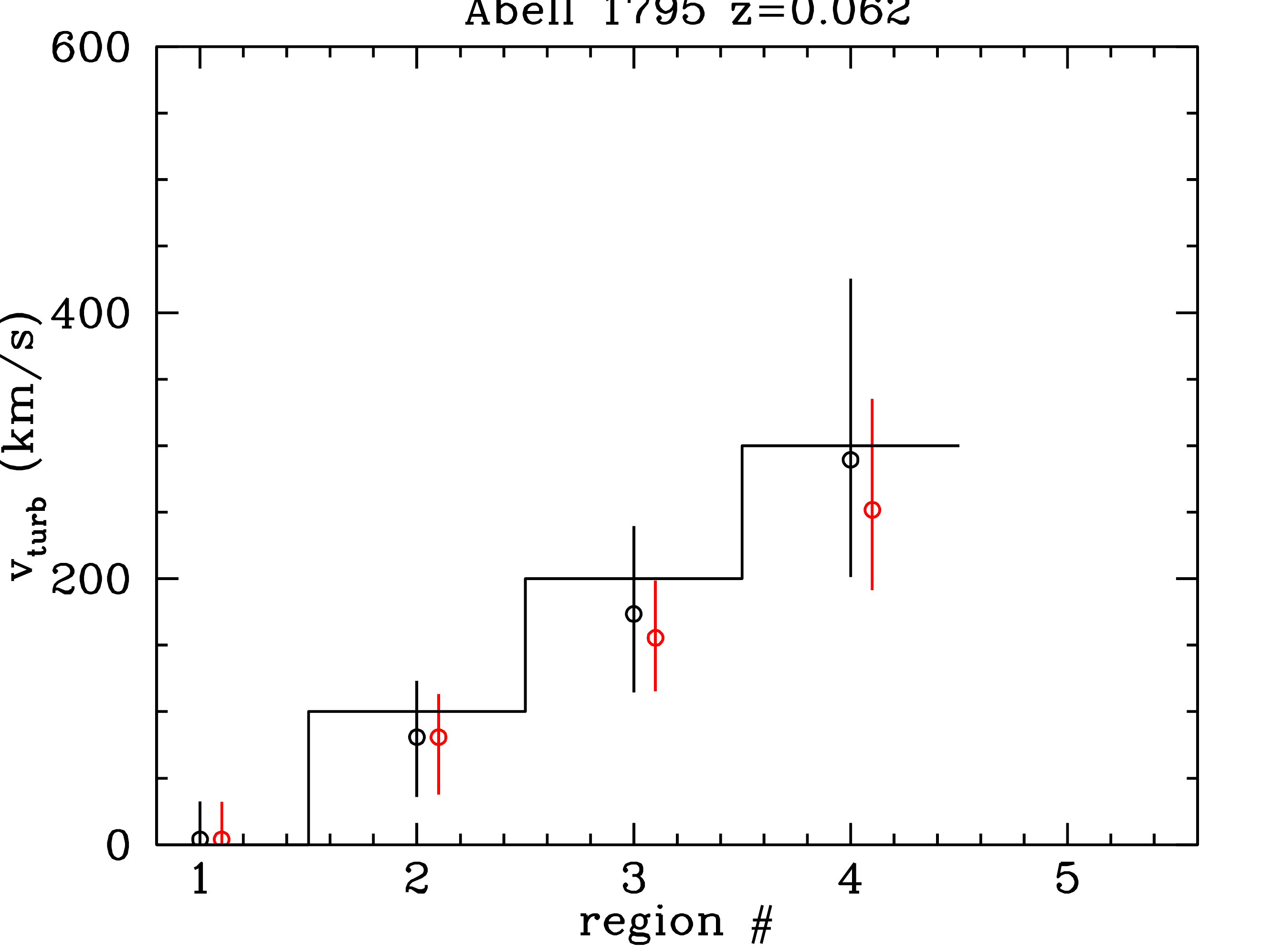}}}
\caption{Same as Figure~\ref{fig:a2029con}, but for A1795 and case (b).}
\label{fig:a1795pos}
\end{minipage}
\end{figure}

\begin{figure}
\begin{minipage}{0.46\textwidth}
\rotatebox{270}{\scalebox{0.28}{\includegraphics{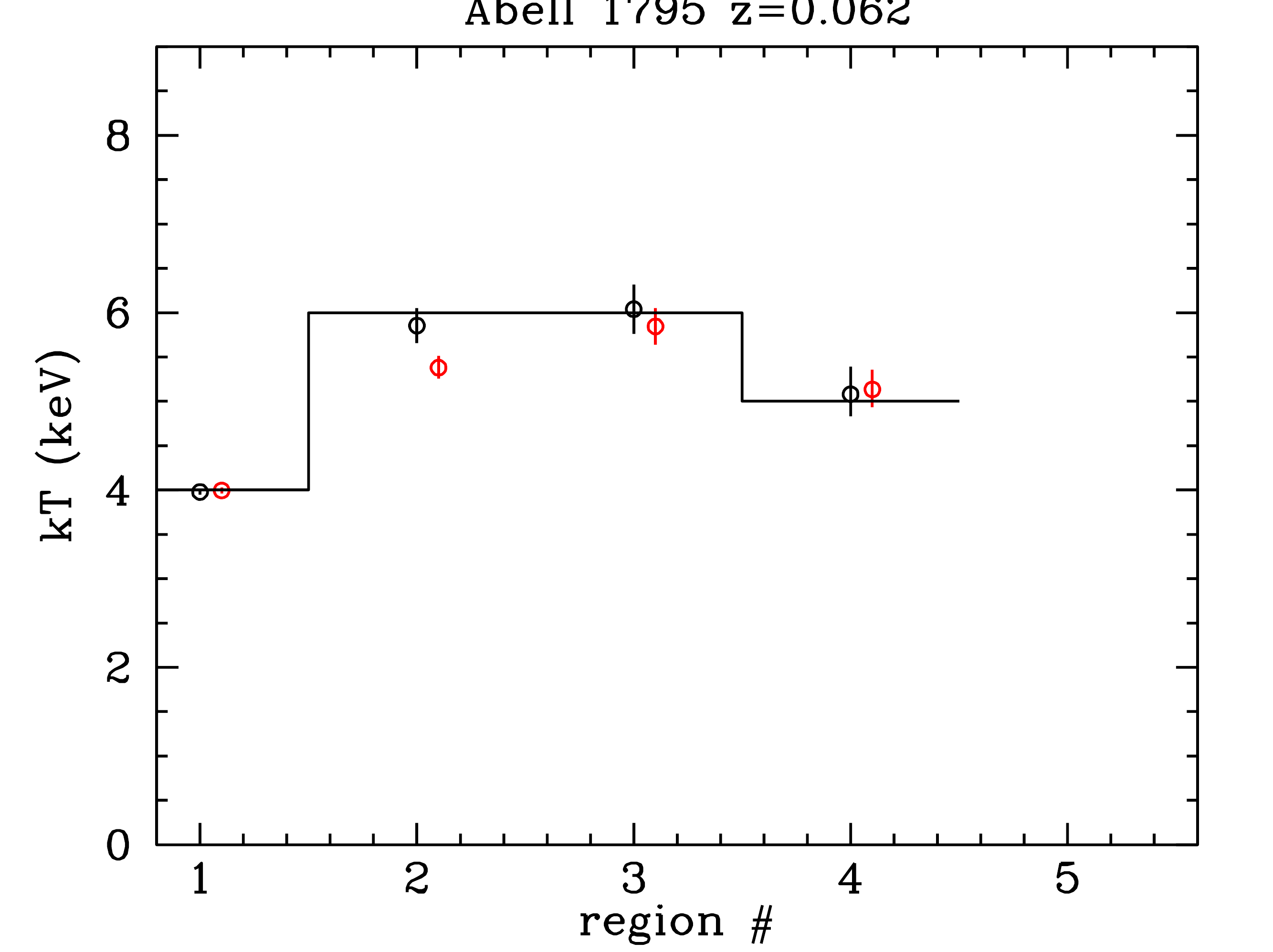}}}
\rotatebox{270}{\scalebox{0.28}{\includegraphics{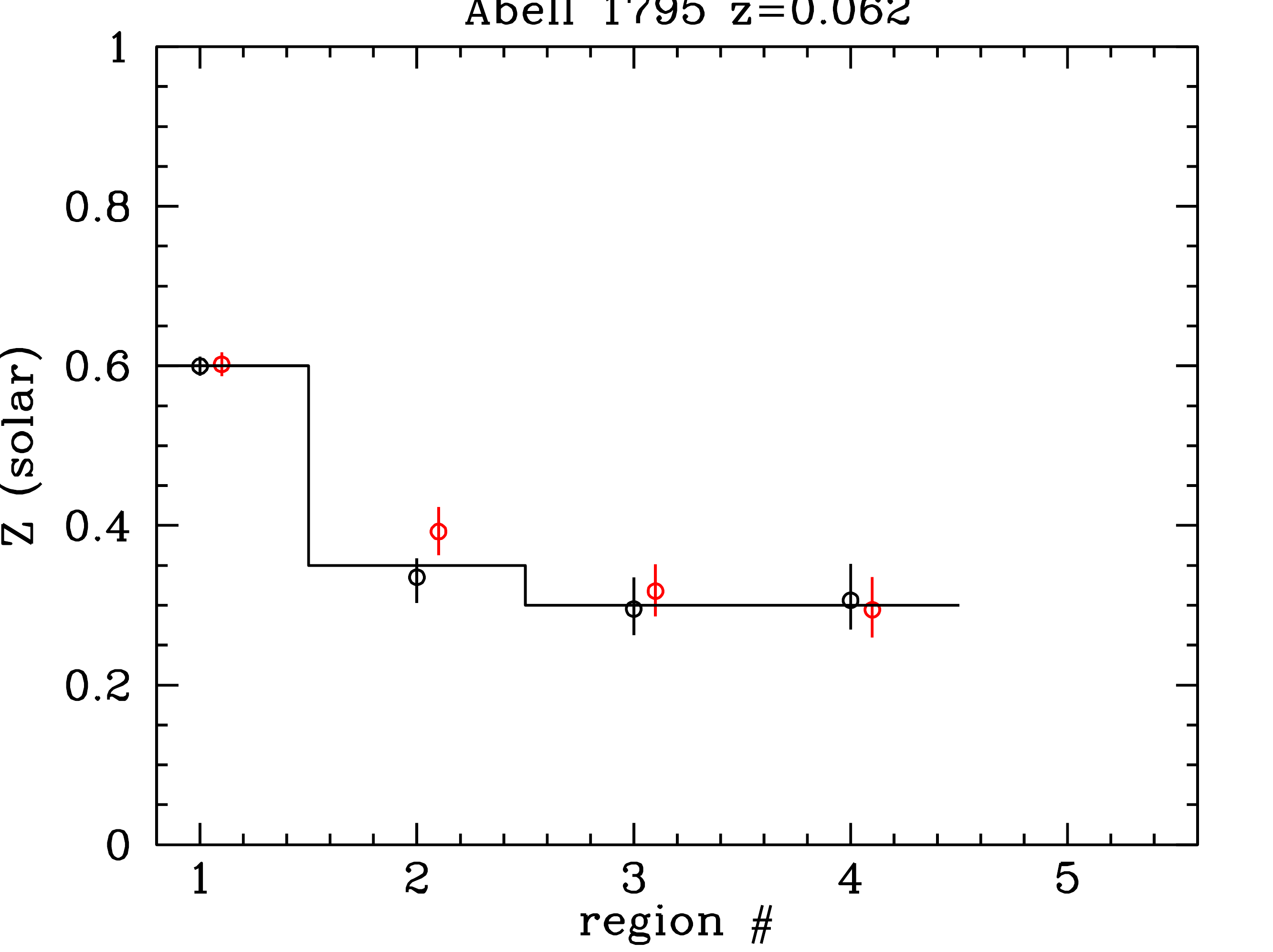}}}
\rotatebox{270}{\scalebox{0.28}{\includegraphics{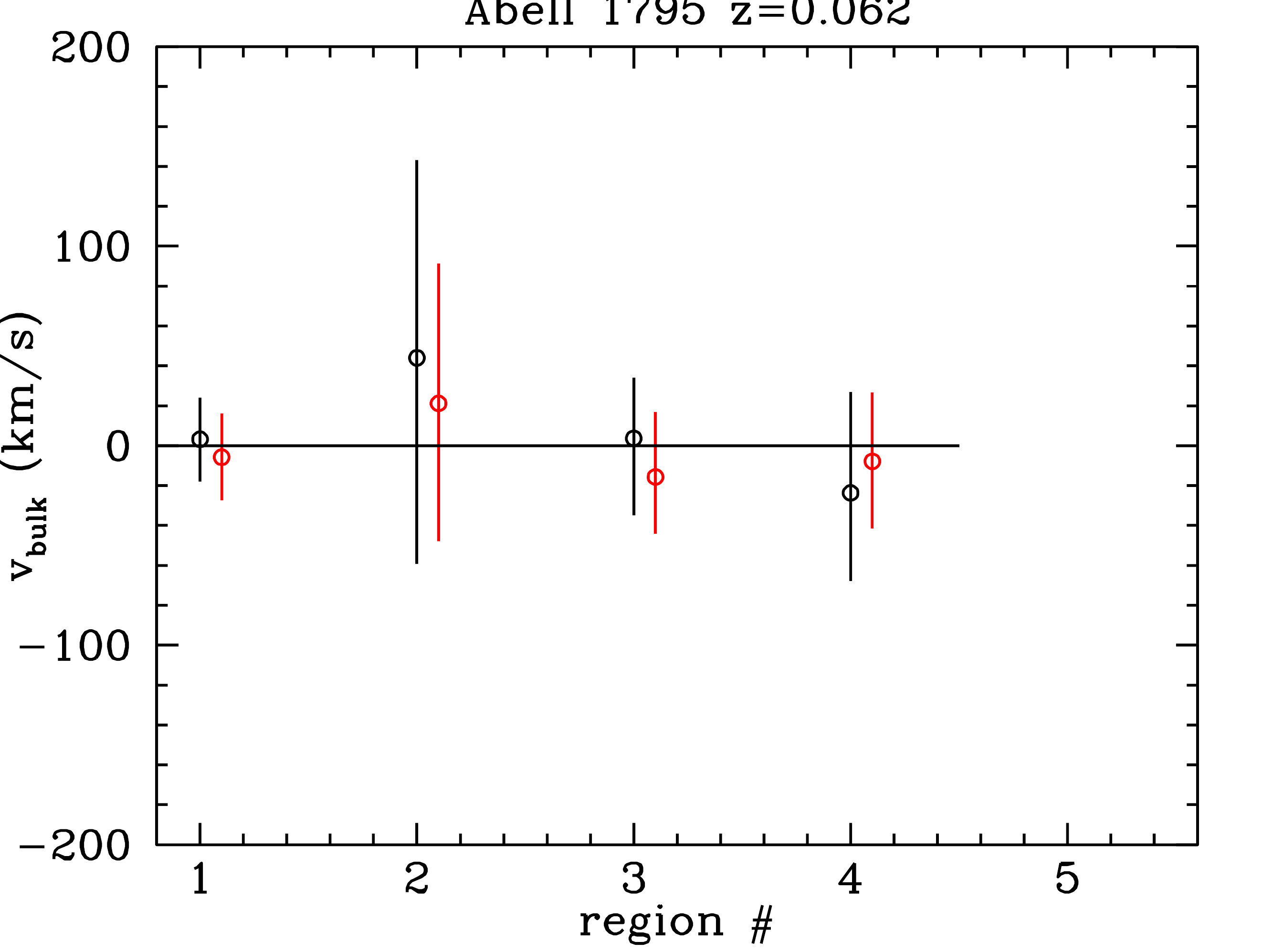}}}
\rotatebox{270}{\scalebox{0.28}{\includegraphics{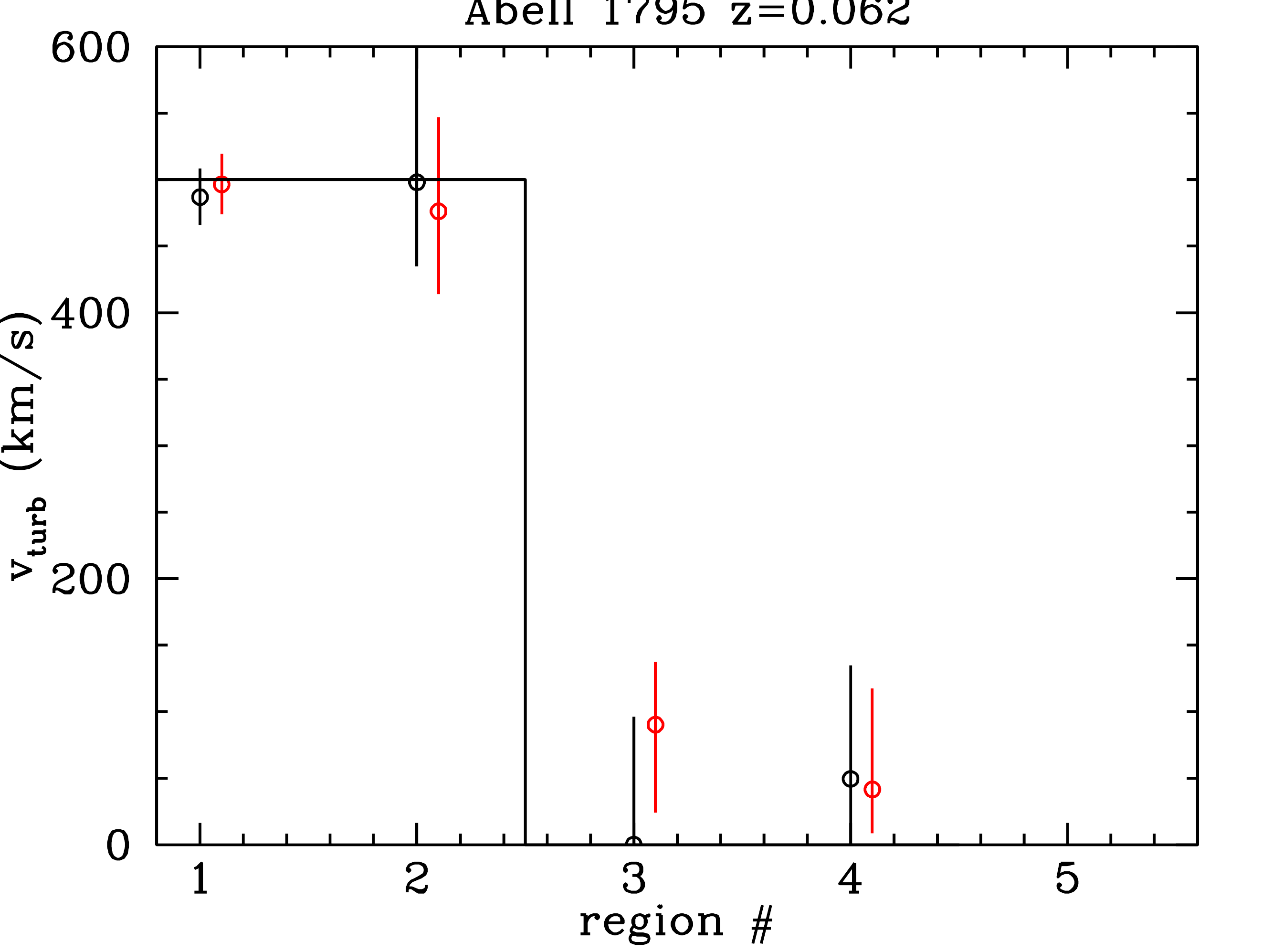}}}
\caption{Same as Figure~\ref{fig:a2029con}, but for A1795 and case (c).}
\label{fig:a1795neg}
\end{minipage}~~
\begin{minipage}{0.46\textwidth}
\rotatebox{0}{\scalebox{0.28}{\includegraphics{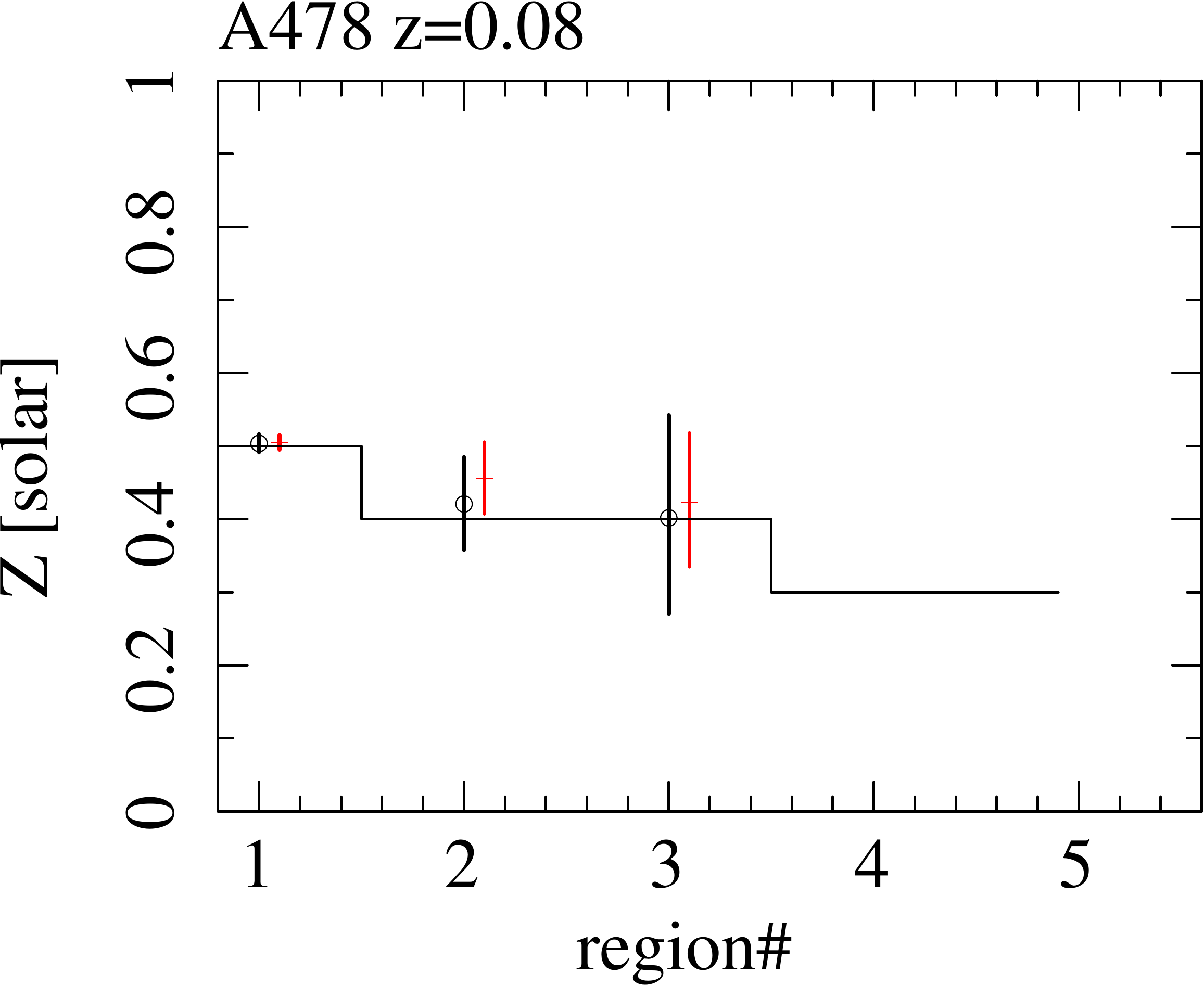}}}
\rotatebox{0}{\scalebox{0.28}{\includegraphics{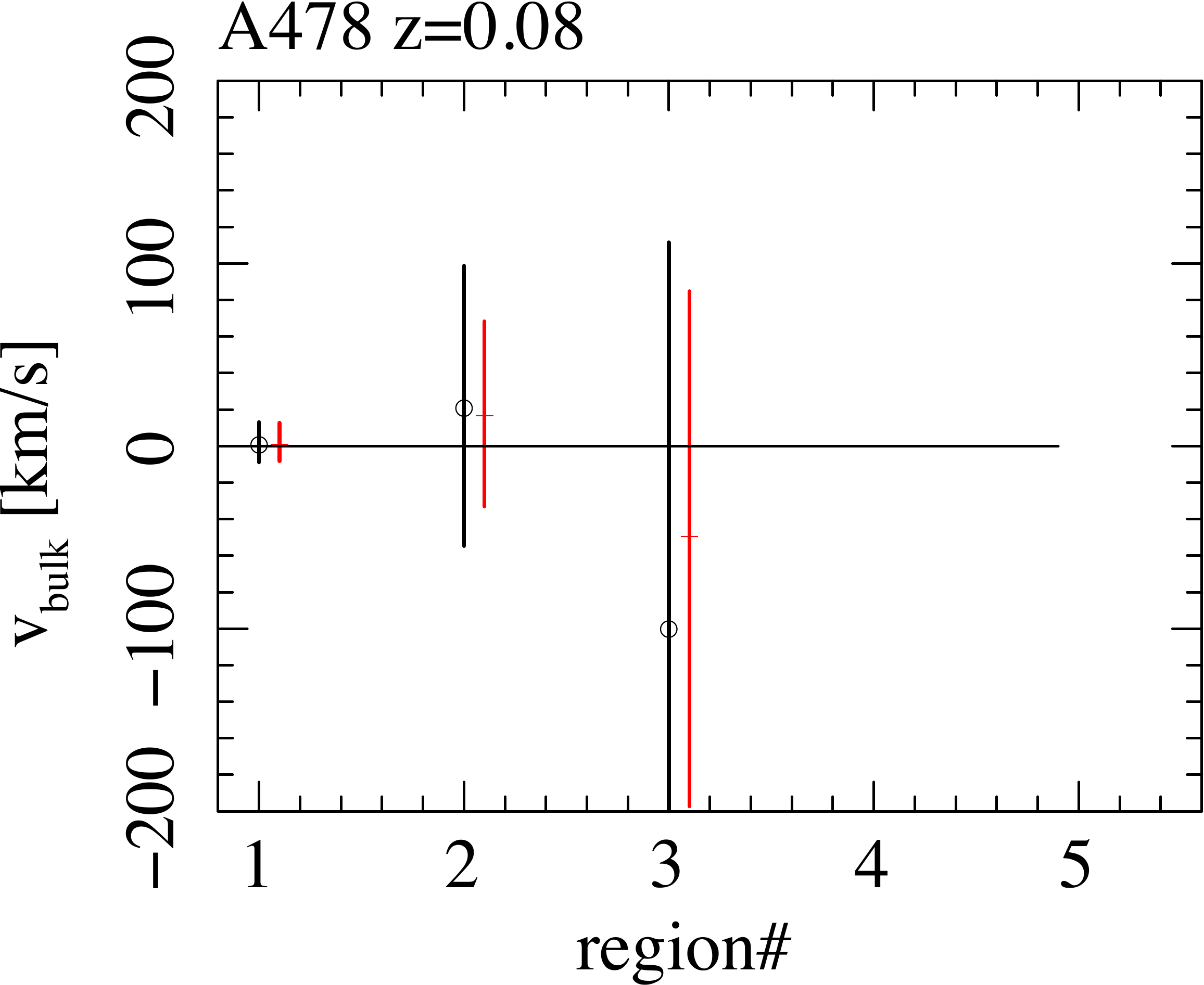}}}
\rotatebox{0}{\scalebox{0.28}{\includegraphics{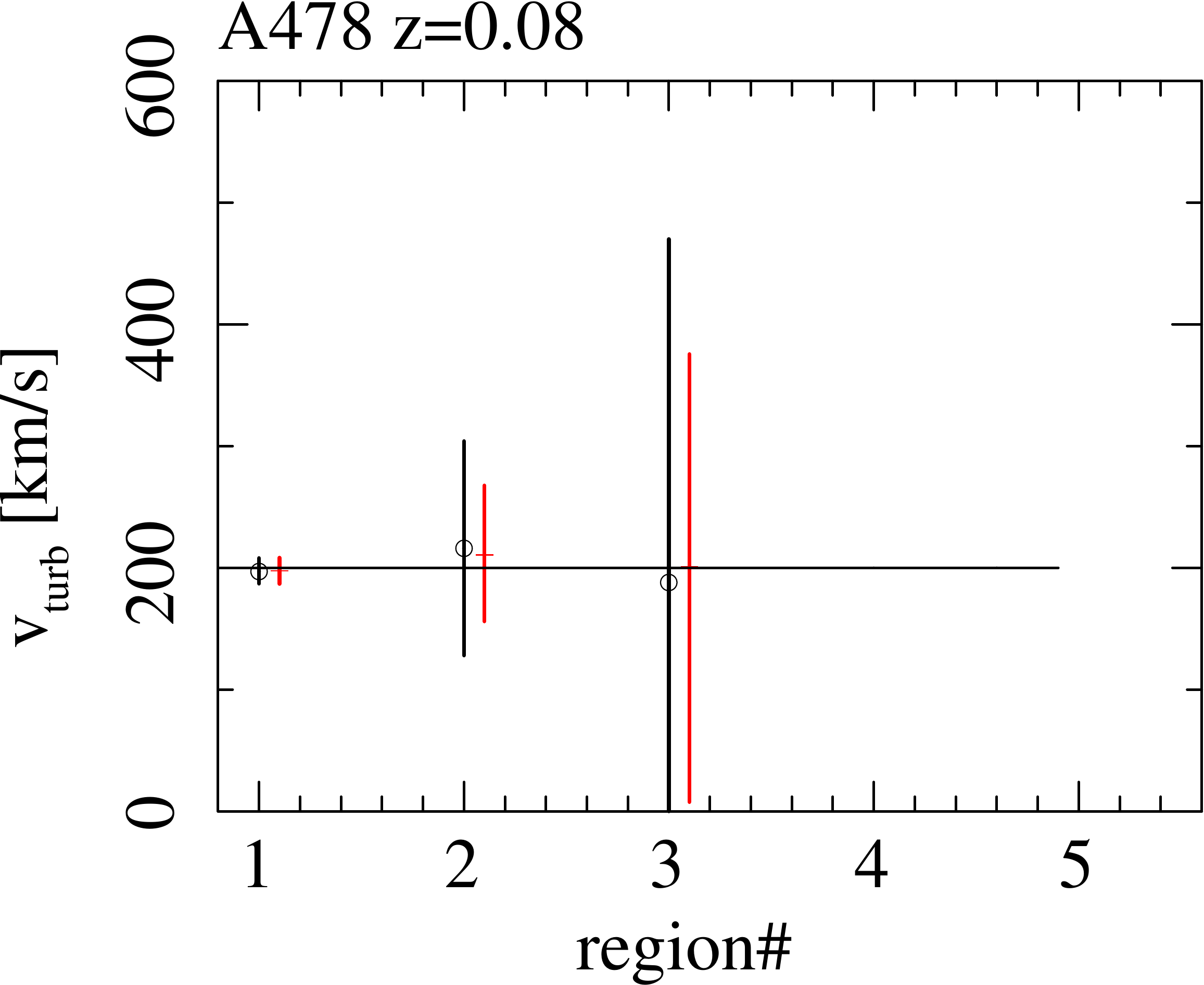}}}
\caption{Same as Figure~\ref{fig:a2029con}, but for A478 and case (a). As
described in the main text, temperature is fixed at the input values
listed in Table \ref{tab:a478spec} for this cluster.}
\label{fig:a478con}
\end{minipage}
\end{figure}

We simulate observations of Abell 1795 in a similar way to Abell 2029
and Abell 2199, for the input parameters listed in Table
\ref{tab:a1795spec}. We use the PSF model to estimate the scattered
light contribution from each annulus into each SXS field, as shown in
Table \ref{tab:a1795frac}.  In this case, we simulate four SXS
pointings, extending to 10.5 arcmin.  The outer edge of Region 3 reaches
just past $r_{2500}$ (500kpc, 7.0 arcmin). We simulate case (a): a
constant $v_{\rm turb}$, case (b): a rising $v_{\rm turb}$ profile, and case (c): $v_{\rm turb}$ decreases sharply.
As with Abell 2199, the parameters $kT$, $Z$, $z$, the turbulent
velocity $v_{\rm turb}$, and normalization are allowed to vary, while
the fractions of scattered photons are fixed to those given by the
image simulations.  The best-fit parameter constraints are shown in
Figures~\ref{fig:a1795con}--\ref{fig:a1795neg} for three cases. The quoted and plotted errors
are at 90\% confidence.

With a total exposure time of 400~ks, the turbulent velocity
can be measured to a statistical accuracy of 
$\lesssim100~{\rm km\,s^{-1}}$ out to region $j=3$, corresponding
to about $r_{2500}$.  The additional pointing for region $j=4$
($0.6\,r_{500}$) requires a large amount of time; 400 ks will enable
interesting constraints only if $v_{\rm turb}$ is quite large, as in case
(b).

\subsubsection{Abell 478}

\begin{figure}
\begin{center}
\begin{minipage}{0.46\textwidth}
\rotatebox{0}{\scalebox{0.28}{\includegraphics{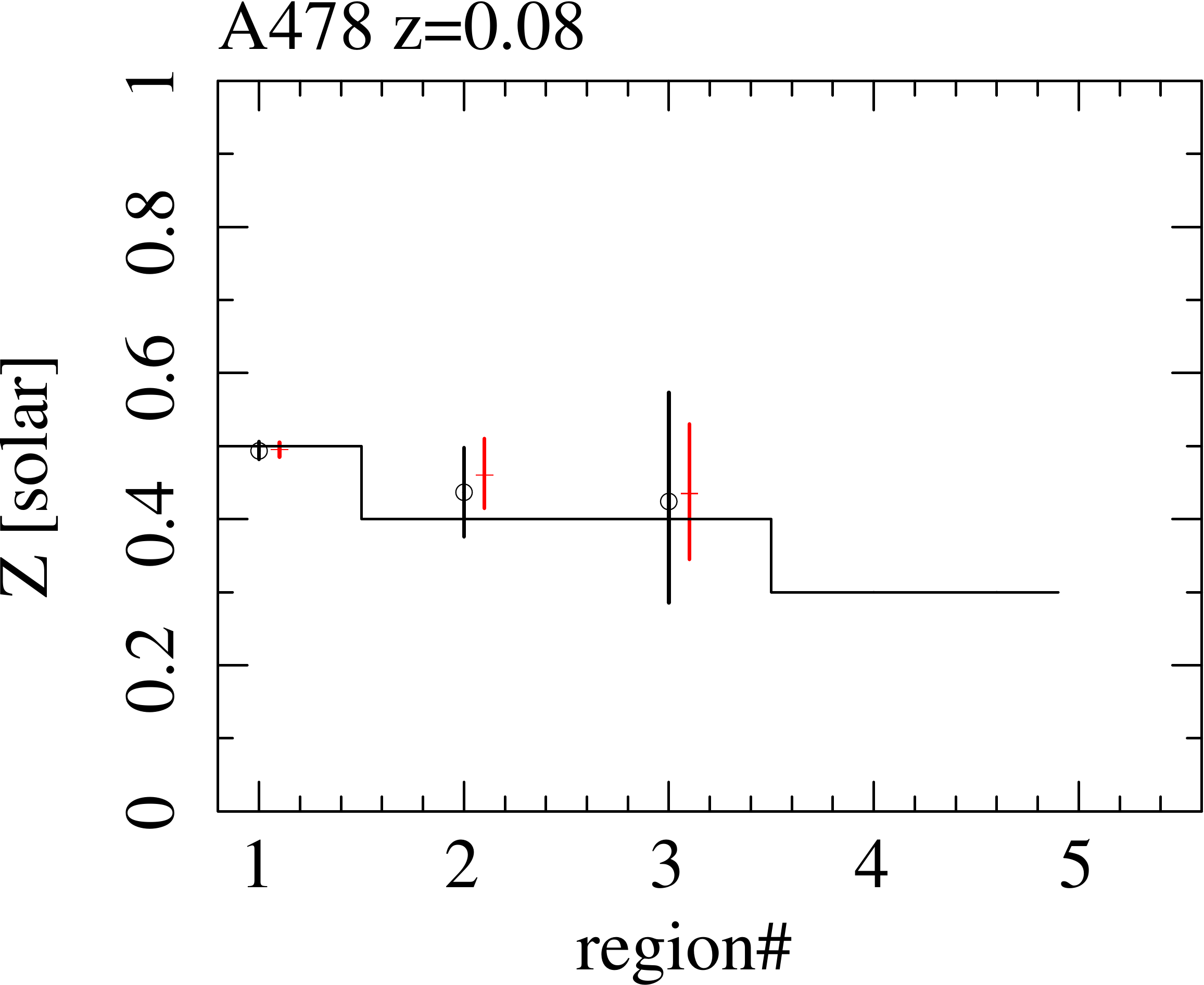}}}
\rotatebox{0}{\scalebox{0.28}{\includegraphics{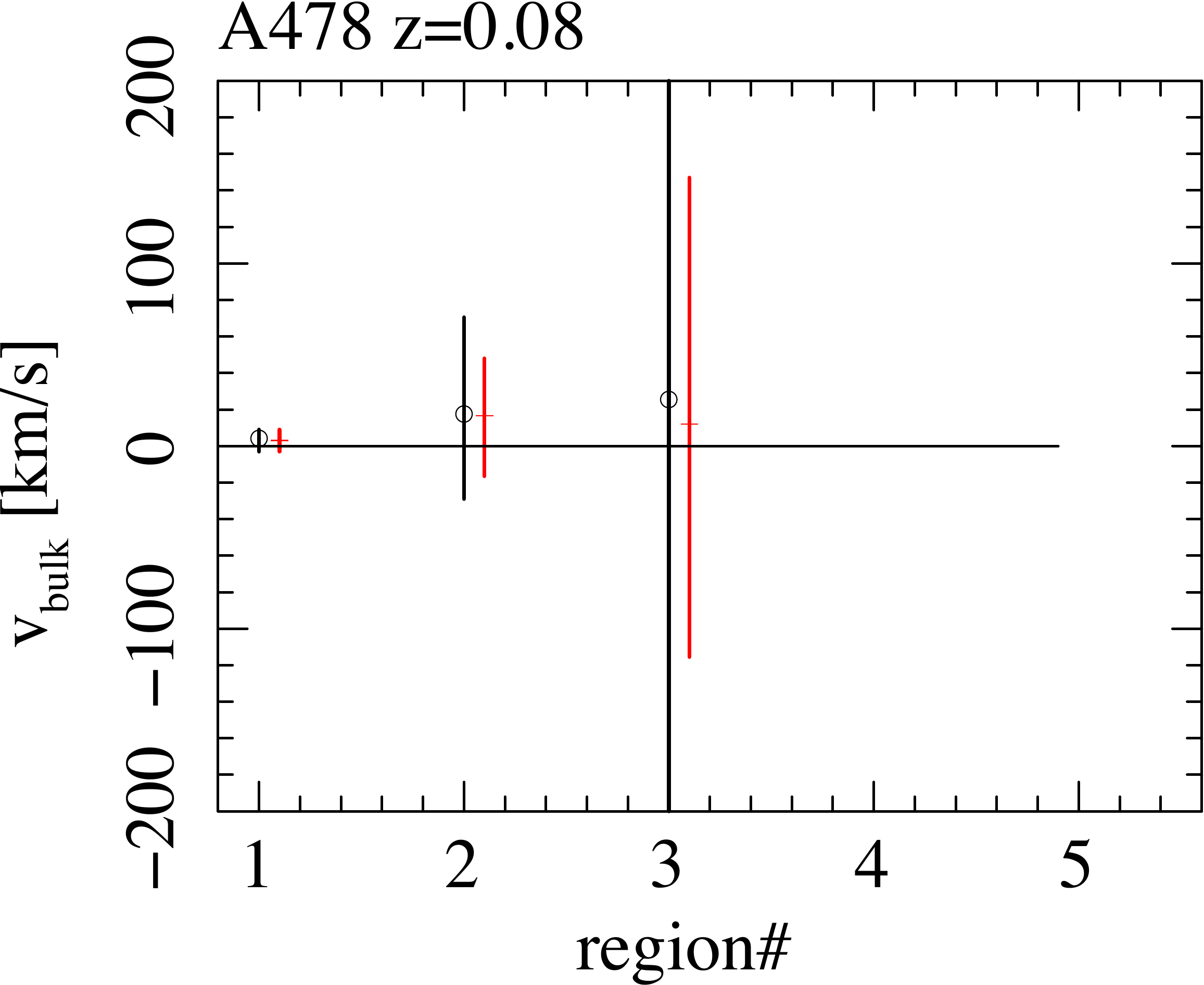}}}
\rotatebox{0}{\scalebox{0.28}{\includegraphics{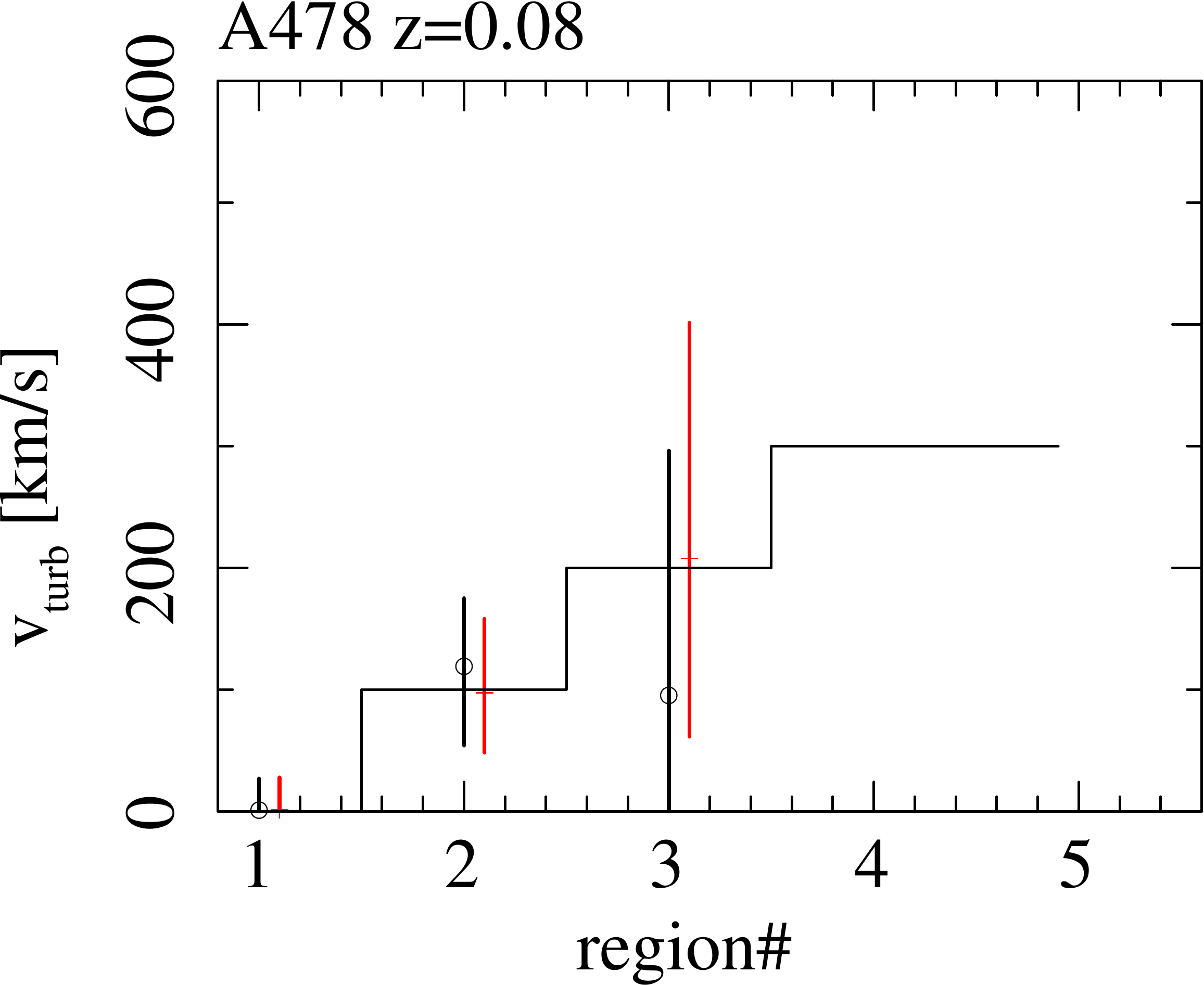}}}
\caption{Same as Figure~\ref{fig:a2029con}, but for A478 and case (b).}
\label{fig:a478pos}
\end{minipage}~~
\begin{minipage}{0.46\textwidth}
\rotatebox{0}{\scalebox{0.28}{\includegraphics{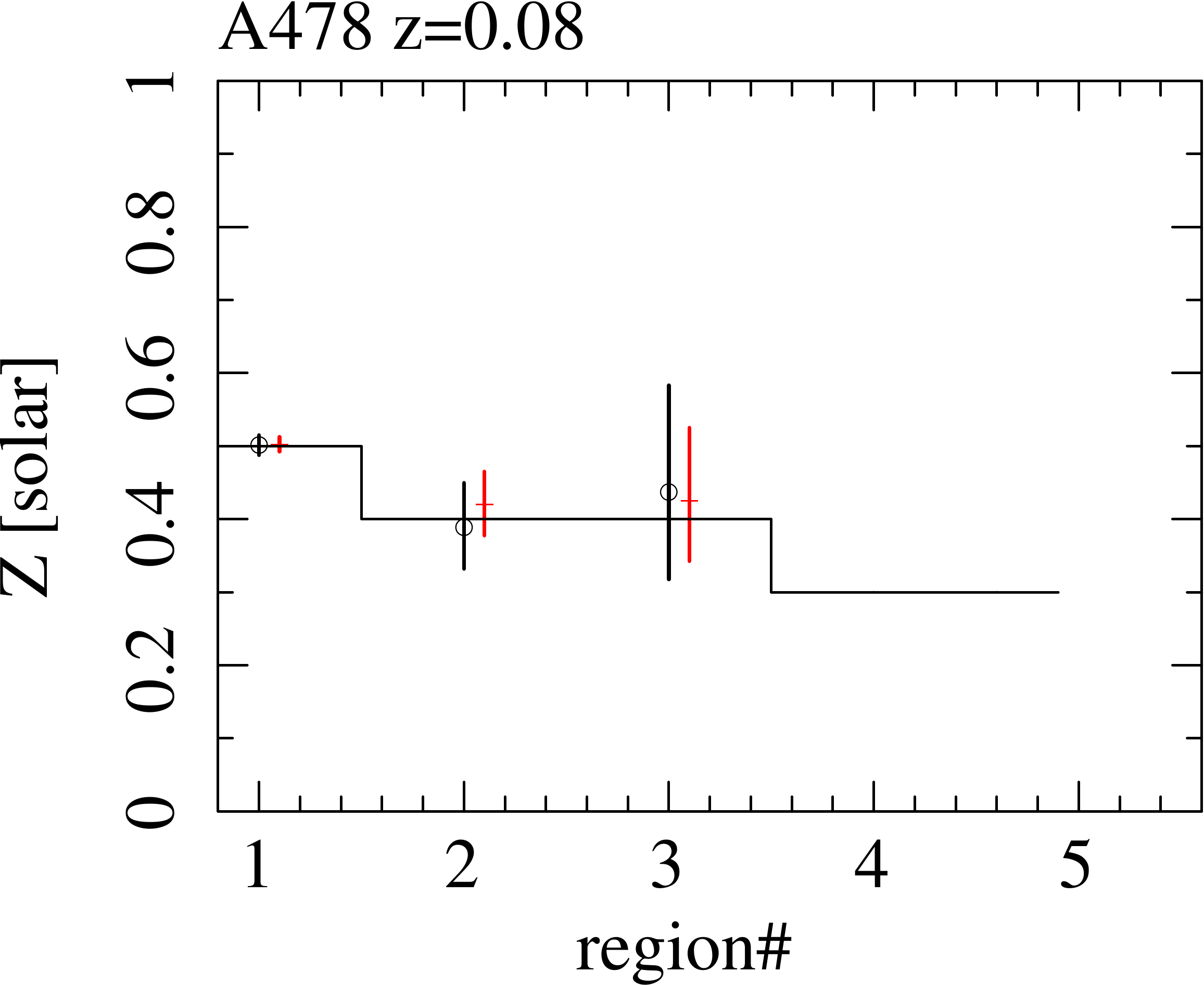}}}
\rotatebox{0}{\scalebox{0.28}{\includegraphics{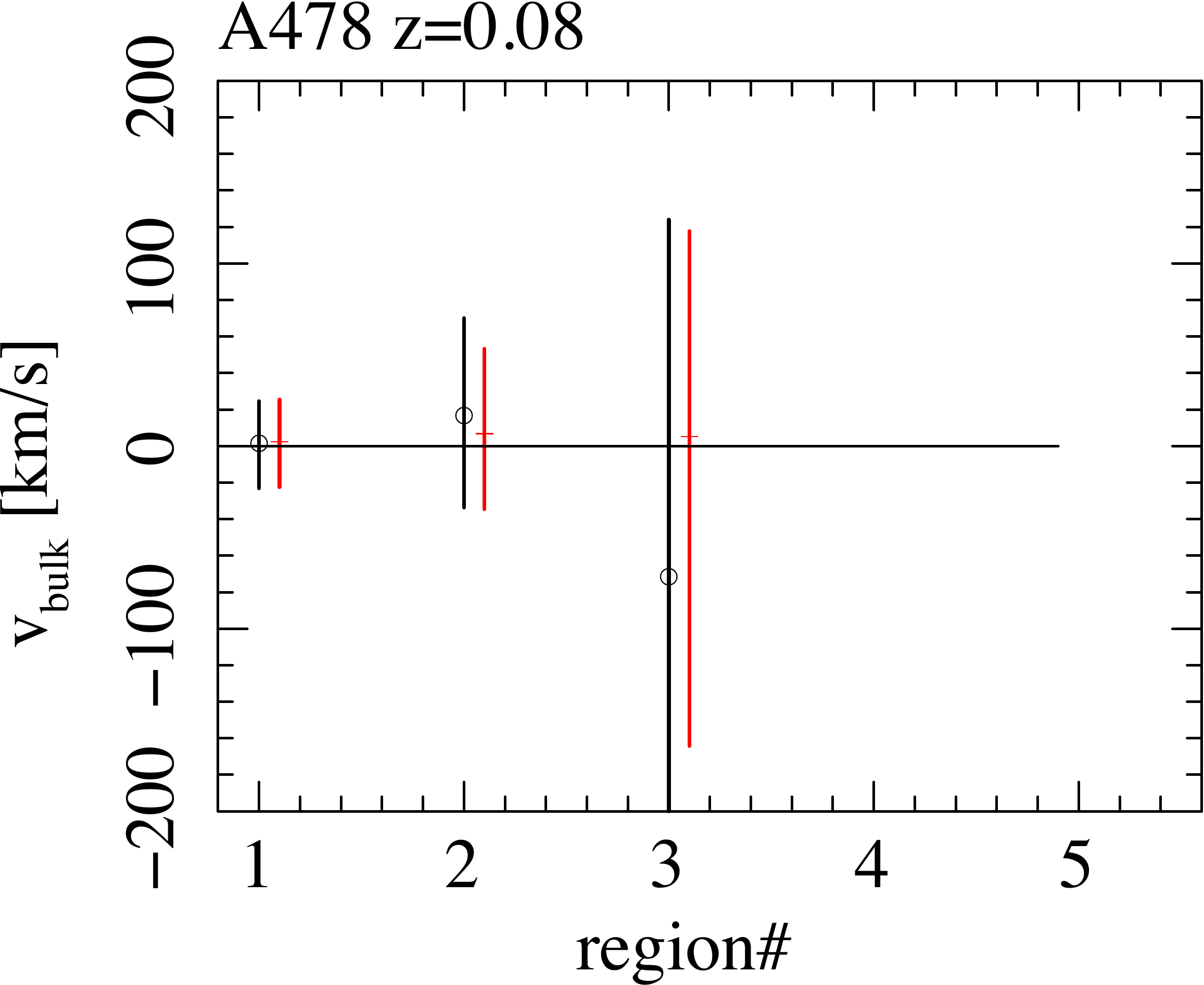}}}
\rotatebox{0}{\scalebox{0.28}{\includegraphics{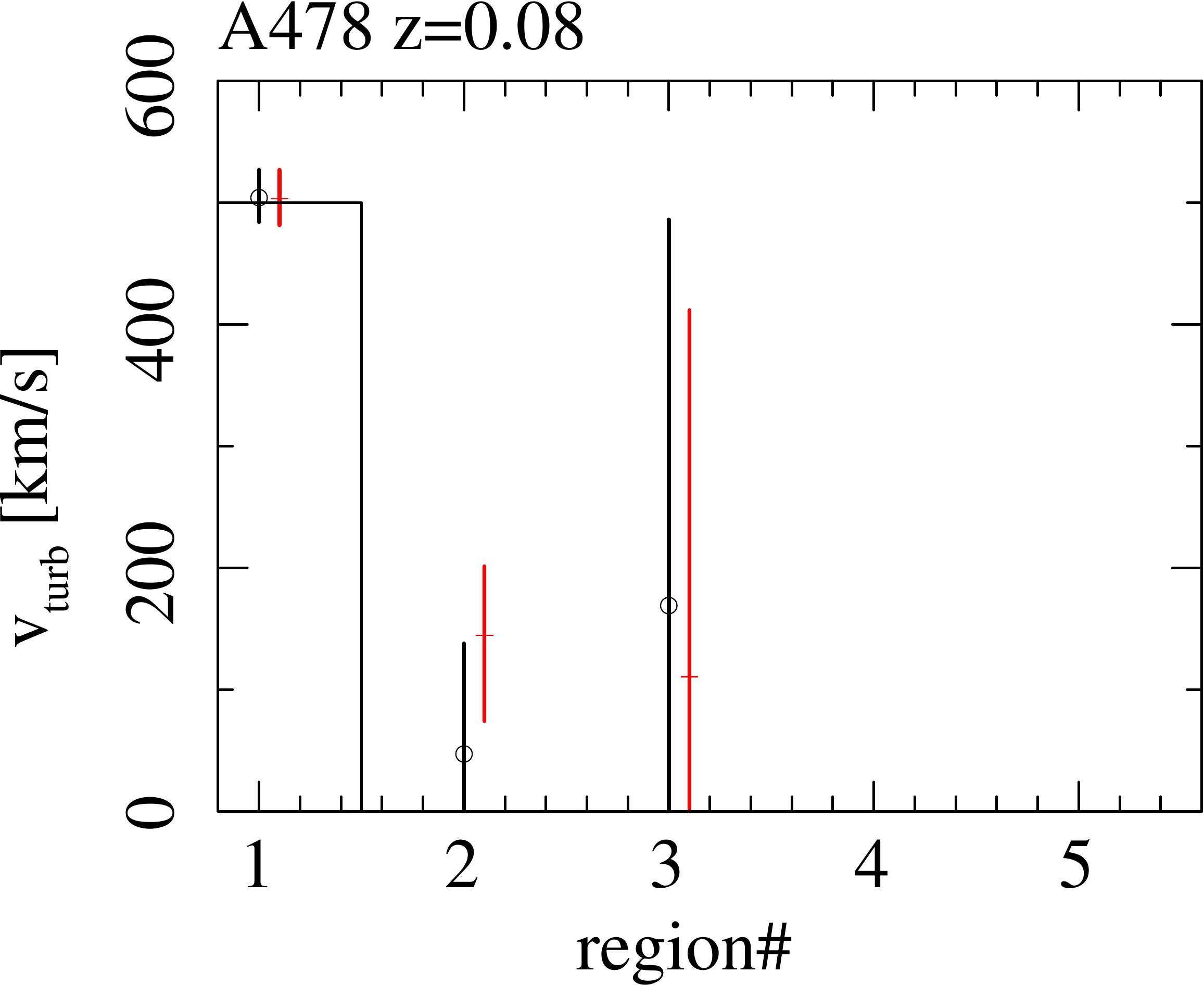}}}
\caption{Same as Figure~\ref{fig:a2029con}, but for A478 and case (c).}
\label{fig:a478neg}
\end{minipage}
\end{center}
\end{figure}

Simulated SXS spectra for A478 are created assuming the
spectral parameters given in Table~\ref{tab:a478spec}, using the methods
described above for the previous three clusters. The scattered fractions
of the photons are shown in Table \ref{tab:a478}. Region 3 is centered
just beyond $r_{2500}$ (5.8 arcmin). 
Given large fractions of scattered photons in this cluster, we fix the
temperature of each region to the input value and fit simultaneously the
spectra of regions $j=1$--3 to constrain $v_{\rm turb}$, $v_{\rm bulk}$,
and $Z$ ($j=4$ is not used due to large statistical errors). The errors
are at the 90\% confidence. Figures~\ref{fig:a478con}--\ref{fig:a478neg}
show the fitting results for the cases (a)--(c), respectively.

\begin{table}[ht]
\begin{center}
\caption{Same as Table \ref{tab:a2029spec}, but for A478 
and $N_{\rm H} = 1.35 \times 10^{21}$ cm$^{-2}$.} 
\label{tab:a478spec}
\begin{tabular}{cccccccc} \hline\hline
Reg & $kT$~[keV] & $Z$~[solar]  & Flux $[{\rm erg\,s^{-1}\,cm^{-2}}]$ &  \multicolumn{3}{c}{$v_{\rm turb}~{\rm [km\, s^{-1}]}$}  & Exposure~[ks]\\ 
     &      &       & (0.3--10 keV)& (a) & (b) & (c) & \\\hline
1  & 5.0 & 0.5 & $5.8\times10^{-11}$ &200 & 0 & 500 & 50 \\
2  & 6.5 & 0.4 & $5.8\times10^{-12}$ &200 & 100 & 0 & 50\\
3  & 7.5 & 0.4 & $8.0\times10^{-13}$ &200 & 200 & 0 & 150\\ 
4  & 8.0 & 0.3 & $2.5\times10^{-13}$ &200 & 300 & 0 & 300\\ \hline
\end{tabular}
\bigskip 
\caption{Same as Table \ref{tab:a2029frac}, but for A478.}
\begin{tabular}{lllll}\hline \hline
   & $i=1$ & $i=2$ & $i=3$ & $i=4$  \\ \hline
$j=1$ & {\bf 1.000} & 0.000 & 0.000 & 0.000 \\ 
$j=2$ & 0.230 	& {\bf 0.748} &	0.021 &	0.000   \\
$j=3$ & 0.072 	& 0.211 &	{\bf 0.684} &	0.032   \\ 
$j=4$ & 0.078 	&0.107 &	0.189 & {\bf 0.626} 	  \\ \hline
\end{tabular}
\label{tab:a478}
\end{center}
\end{table}%

Abell 478 is the most challenging cluster among those discussed in
detail here. With a total of 250 ks exposure, one obtains useful
constraints on the bulk velocity in three innermost bins, but the
turbulent velocity dispersion ($v_{\rm turb}$) is well-constrained only
in the central two pointings, and this result is achieved only if the
plasma temperature is known a priori.  We have not yet determined
whether temperature information from CCD-resolution spectra is adequate
to allow the turbulent velocities to be constrained in the manner we
have simulated here.  We note that, if so, the two central pointings,
requiring a total of 100 ks, can constrain $v_{\rm turb}$ at $r <
r_{2500}$ in this cluster.



\subsection{Discussion}
The tradeoff between cluster distance, measurement radius, and detected
iron-line flux for our sample is summarized in
Table~\ref{tab:linecounts}. The table compares the iron-line counting
rates, for both He-like and H-like ionization states, for Abell 2199,
our nearest and least massive cluster, and Abell 2029, one of our most
distant and most massive objects, at several radii.  The Perseus cluster
is also included for reference. The tabulated line counting rates are
obtained assuming zero line broadening. The values in the table do not
account for vignetting, which will reduce the line counting rates by
about 15\%.

\begin{table}[t]
\begin{center}
\caption{Summary of parameters for A2029, A2199, and Perseus.}
\label{tab:linecounts}
\begin{threeparttable}
\begin{tabular}{l c c c  c c c c}
\hline
\hline
Cluster & $z$ &  Region & $r_{\rm reg}/r_{2500}$\tnote{a} & $kT$    &$Z$
 & Flux \tnote{b} & Fe-K~(He,H) line\tnote{c} \\
        &   &  number &                    & [keV] & [solar] & [erg
			 s$^{-1}$ cm$^{-2}$] & [cts/100ks/SXS] \\
\hline
A2029   & 0.077 & 1   & 0    & 6.5 & 0.6 & $6.5 \times 10^{-11}$ & 5107, 2049 \\
A2029   & 0.077 & 2   & 0.39 & 7.5 & 0.4 & $7.5 \times 10^{-12}$ & 413, 165 \\
A2029   & 0.077 & 3   & 0.79 & 7.5 & 0.3 & $1 \times 10^{-12}$ & 40, 16 \\
A2029   & 0.077 & ``3.5''\tnote{d}   & 1.0 & 7.5 & 0.3 & $5.8 \times 10^{-13}$ & 25, 9 \\

\\
A2199   & 0.030 & 4   & 0.78 & 3.8 & 0.3 & $7 \times 10^{-13}$ & 35, 2 \\
A2199   & 0.030 & 5  & 0.96  & 3.4 & 0.3 & $4 \times 10^{-13}$ & 19, $<1$ \\
\hline
Perseus & 0.018  & $--$ & 1.0 & 5.0 & 0.3 & $8 \times 10^{-13}$ & 47, 7 \\
\hline
\hline
\end{tabular}
\begin{tablenotes}
\item[a]{\footnotesize Geometrical center (not emission weighted) of
simulated regions.}
\item[b]{\footnotesize Flux at 0.3--10 keV within the SXS
field of view.}
\item[c]{\footnotesize Photon counts per 100~ks at
$(6.59-6.74)/(1+z)$ keV and $(6.9-7.0)/(1+z)$ keV for the He-like and
H-like Fe-K lines, respectively. No turbulent broadening is assumed. To
take into account vignetting,
the flux and line counts shown here should be multiplied by $\sim 0.85$.
}
\item[d]{\footnotesize Midpoint between regions 3 and 4, which lies
at $r_{2500}$.}
\end{tablenotes}
\end{threeparttable}
\end{center}
\end{table}

Table~\ref{tab:linecounts} shows that reducing the measurement radius
from $r_{2500}$ to $0.8~r_{2500}$ can raise the counting rate by a
factor of $1.5-2$.  It also shows that the hotter ICM in A2029 provides
an additional bonus of 40\% more counts from the H-like line, compared
to that in A2199.  The photons from the H-like line not only provide
statistically stronger constraints on line broadening; they also provide
a valuable consistency check, and allow an independent measurement of
the plasma temperature as well. These results, taken together with other
data that indicate no cluster closer than Abell 2029 is more relaxed
(Mantz et al., in preparation), imply that an observation of Abell
2029 at an emission-weighted radius of $0.8 \times r_{2500}$ is the
optimal choice.

Of course, it is ultimately necessary to investigate non-thermal
pressure support in more than one relaxed cluster. Besides testing
whether or not A2029 is representative, observations of a set of such
objects would constrain the scatter in non-thermal pressure support,
mass and apparent baryon fraction to be expected amongst relaxed
clusters. It would also give some indication of the variation of
non-thermal pressure support with cluster mass.  Having already
established a (nearly) systematics-limited constraint with A2029, we
would observe other 
objects with shorter exposure times 
that are sufficient to provide an interesting upper limit on scatter in non-thermal support. With current upper limits  
at  about  5\% \citep{Allen11}, exposures of  250 ks (or less) per object, with formal limits on 
non-thermal pressure support of 1-2\% per object, would be adequate for
this purpose. 
Specific candidates for these observations include Abell 2199,
which is relatively nearby and somewhat less massive than A2029, and
Abell 1795, which may be more typical of objects used for cosmology than
the extremely relaxed Abell 2029. 
Perseus, PKS0745, A478, and A496 are also plausible candidates.

\subsection{Beyond Feasibility}

The observations outlined above will provide sensitive first searches for
dynamically important motions in the intracluster medium.  As the \astroh
mission progresses, with sufficiently long exposures, we may be able to
address scientific questions not discussed above.  For example, simulations
predict that plasma motions near the cluster core may be largely tangential
\citep{Lau09}, with systematic motions dominating random ones. This
prediction could be tested by measuring velocity characteristics along
multiple directions from the cluster center. Another possibility is that,
with sufficiently deep observations one may be able to distinguish
multiple temperature components in the plasma  outside the complicated,
cooling cluster cores. This could be important for understanding thermal
conduction, and might also have significance for cluster mass estimates,
since temperature (and therefore mass) measurements of multiphase gas
with low-resolution (CCD) may be biased low \citep{Rasia12}. Finally, a
sufficiently deep and spatially dense sampling of the velocity fields in
relaxed clusters like those discussed here could extend our knowledge of
turbulent dissipation beyond that obtained from larger, brighter, but
perhaps unique objects (Perseus, Virgo, and Coma) discussed elsewhere in
this whitepaper.


\bigskip




\section{Mapping Gas Flows and Turbulence in Merging Galaxy Clusters}
\label{sec-coma}

\subsection*{Overview}

Mergers of galaxy clusters are the most energetic events in the Universe
since the Big Bang. They dissipate the vast kinetic energy of a cluster
collision into thermal energy of the ICM via shocks and turbulence,
while channeling some of it into magnetic fields and ultrarelativistic
particles. While we do observe cluster collisions and their end result
--- heated ICM and synchrotron emission from cosmic-ray electrons ---
and have even imaged shock fronts in a handful of clusters, so far it
has not been possible to study probably the most prevalent energy
conversion mechanism in the ICM, which consists in the bulk flows and turbulence that
they generate. The SXS will provide the first opportunity to detect and
map the bulk and turbulent velocities in the ICM of nearby merging
systems. The generation and dissipation of turbulence depends on a
number of unknown microphysical properties of the ICM (e.g., viscosity
and magnetic field structure). Mapping the bulk and turbulent velocities
in well-resolved nearby mergers at different stages of unrest (e.g.,
Coma, A754) and across the prominent ``cold fronts'' in some merging
clusters (A3667, A2319), combined with radio observations and numerical
MHD simulations, will help to devise a detailed physical model of the
ICM. This is a critically important piece for cluster precision
cosmology, one to which \astroh can contribute.\footnote{Coordinators of
this section: M.~Markevitch, H.~Akamatsu}






\subsection{Background and Previous Studies}

Clusters of galaxies form and grow via gravitational infall and accretion of
surrounding matter --- including smaller clusters. Cluster mergers are the
most energetic events in the Universe since the Big Bang, with the total
kinetic energy of the colliding subclusters reaching $10^{65}$ ergs. In the
course of a merger, a significant portion of this energy, that carried by
the gas, is dissipated (on a Gyr timescale) via shocks and turbulence in the
intracluster medium (ICM).  Eventually, the gas heats to a temperature that
approximately corresponds to the depth of the newly formed gravitational
potential well. The exact details of energy dissipation are not well
understood, because they depend on complex microphysics of the intracluster
plasma. This plasma is permeated by weak, tangled magnetic fields; although
the magnetic pressure is of order a percent of thermal pressure and thus
unlikely to be dynamically important in most of the cluster volume, such a
field is more than sufficient to make the plasma collisionless and
qualitatively alter its physics.

To complicate the picture, many merging clusters host giant diffuse
radio halos, produced by synchrotron radiation of ultra-relativistic
electrons (the Lorentz factor of $\gamma\sim 10^4$) in the cluster
magnetic field (e.g., Feretti et al.\ 2012). These rapidly cooling (the
radiative cooling time of $t_{\rm cool}\sim 10^{7-8}$ yr) relativistic
particles are mixed with thermal plasma and have to be constantly
accelerated in-situ to explain the extent of the radio halos.  The
current thinking is that they are accelerated by scattering on
magnetosonic waves generated by turbulence (e.g., Brunetti \& Jones
2014). The turbulence also reorders and amplifies the intracluster
magnetic field. Some MHD simulations suggest that during the violent
stage of a merger, energy densities in turbulence, magnetic field and
relativistic electrons may reach the same order of magnitude (e.g.,
ZuHone et al.\ 2013). Thus, complex non-hydrodynamic phenomena arising
from the nature of the ICM as collisionless, magnetized plasma may have
significant effects on the cluster physics and energy budget, and thus
on relating the various ICM observables to the cluster total masses. The
cluster masses must be known to high accuracy in order to fully realize
the potential of clusters for precision cosmology (see also
Sec. \ref{sec-nonth} for details).

Temperatures and densities of the thermal ICM component are readily derived
from the X-ray spectro-imaging. With the existing and past X-ray
instruments, we've observed many cluster collisions and their end result ---
the irregularly distributed, nonuniformly heated intracluster
medium. With recent high-resolution X-ray imaging, we even caught
sight of a few shock fronts that quickly pass across the cluster collision site (e.g.,
in the Bullet cluster, A520, A754, A2146 and a few others). Detailed studies of
shock fronts provide a merger velocity in the plane of the sky, and for
well-studied shocks, constrained the electron-proton equilibration timescale
(Markevitch 2006; Russell et al.\ 2012). In many mergers, we've also
observed ``cold fronts,'' or sharp contact discontinuities. They are
unresolved even with {\it Chandra}, providing proof that the intracluster
plasma is indeed collisionless. They also appear stable, potentially
indirectly constraining the effective ICM viscosity (Markevitch \& Vikhlinin
2007).

A critical piece of the puzzle that has been missing so far is direct
observations of the gas motions, both streaming and turbulent. Both types of
motions are expected to exist (and often coexist) at various stages of
cluster mergers. The line-of-sight bulk (or streaming) gas velocity would
provide the missing ``third dimension'' for understanding the geometry and
velocity of a cluster merger. Turbulence should cascade from the large
linear scales of the merger down to the unknown damping scale, which is
determined by the ICM microphysics. This is where {\it ASTRO-H}\/ can make
some of its most important scientific contributions.

\subsection{Prospects and Strategy}

By characterizing the bulk velocities and turbulence in merging clusters
and using these measurements, in combination with the X-ray imaging and
radio data, as input for detailed numerical MHD simulations, we hope to
greatly improve our understanding of the various physical processes in
the ICM.  The superb energy resolution of SXS, and its moderate angular
resolution, will enable accurate, spatially resolved measurements of
line broadening and shifts for the 6.7 keV Fe-K line for many nearby
clusters. Line shifts reveal variations of the line-of-sight bulk
velocities of the ICM, while line broadening measures turbulence on
linear scales smaller than the beam size.  Strictly speaking, lines can
be broadened by any line-of-sight velocity difference within the beam
size. ``True'' turbulence, with random motions on all linear scales
following a certain power spectrum, can be distinguished from, e.g., a
superposition of oppositely directed streams by the shape of the
broadened line. When this distinction is not clear-cut, we will have to
rely on reasonable assumptions and hydrodynamic simulations.  For
example, if we {\em don't}\/ detect turbulence (or random motions) on
small scales in clusters, but do detect line-of-sight (LOS) velocity
gradients on larger scales, this would immediately tell us that plasma
viscosity should be high.  For another example, the strength of
turbulence in the area of the cluster radio halos will constrain
theories of cosmic-ray acceleration. If high levels of turbulence (e.g.,
prevalent near-sonic velocity dispersions) are detected, it would mean
that mechanical energy is a significant fraction of the total energy
budget. This would provide important input for cosmological (as well as
idealized) cluster simulations.

These measurements will be expensive in terms of the exposures required.
Merging clusters usually do not have cool cores, and their surface
brightness is relatively low even at their centers, typically requiring over
100 ks per pointing to accumulate enough photons for interesting line
constraints (as we will see below). Furthermore, given the $1.1'-1.3'$ HPD
angular resolution, the $3'\times3'$ SXS FOV contains only $\sim 2-4$
independent imaging pixels, requiring mosaics to cover an interesting range
of angular scales for velocity mapping.

These observations have two broad goals. The easier one is to map the
line-of-sight (LOS) gas velocities in a merger, which is critical for the
reconstruction of the geometry and dynamical stage of a particular merger,
in combination with all other existing data (X-ray, optical, radio) and
hydro simulations. A more difficult task is to characterize the power
spectrum of turbulence, which could be used as input for high-precision MHD
simulations to constrain the ICM microphysics.  For the latter goal, one may
prefer clusters without cool cores, despite their lower surface brightness,
because cool cores always come with physical complications such as steep
entropy gradients (which are very convectively stable and thus make the
turbulence develop preferentially in radial shells instead of isotropically), 
and moving bubbles from the central AGN activity.  Because 
the range of scales between the size of the PSF and the size of the bright
cluster region accessible for measurements is small even for the nearest
systems, it may be best to select the least contaminated examples; Coma may
be the best candidate.

For mergers at a more violent stage, it would be extremely interesting to
observe A3667, A754, and possibly A2256. All three are famous mergers with
many past attempts to reconstruct their merger geometry --- on which we will
build using the SXS velocity data. These clusters are expected to have large
gradients of LOS velocity as well as high levels of turbulence.

In addition, A3667 has a spectacularly sharp ``cold front'' (Vikhlinin et
al.\ 2001), a contact discontinuity in the gas that enables some unique ICM
physics tests on its own (Markevitch \& Vikhlinin 2007). The gas density
jump is unresolved even with the {\it Chandra}\/ $1'' \approx 1$ kpc
resolution, while the shape of the front hints at the onset of hydrodynamic
instabilities.  Another such high-contrast cold front is found in A2319.
Cold fronts arise either from subcluster stripping by ram pressure, or from
``sloshing'' of the low-entropy gas in a disturbed gravitational potential
well (Markevitch \& Vikhlinin 2007). The fronts in A3667 and A2319 can
be of either origin, although
the initial interpretation for A3667 was stripping. In the stripping
scenario, bulk motions of the subclusters are mostly in the plane of the
sky, while in the sloshing case, we may see a near-sonic flow along the LOS
inside the front.  SXS measurements would resolve this ambiguity, providing
solid basis for the derivation of physical constraints from these cold
fronts.

Among the clusters that we consider, Coma, A754 and A2319 have giant radio
halos (while A2256 has a faint halo and a bright ``relic'' near its core,
and A3667 has a pair of bright peripheral relics). Simultaneous observations with HXI
may yield interesting constraints on the cluster nonthermal matter
components. The relativistic electrons responsible for the synchrotron radio
halos are also expected to produce inverse Compton emission at high X-ray
energies (upscattering the CMB photons); their detection at both wavebands
would allow one to disentangle the magnetic field strength and the density
of the relativistic electrons (the radio observations only give a product of
the electron density and $B^{\,2}$, where $B$\/ is the highly uncertain
magnetic field strength; for more details, see Sec. \ref{sec-highe}). 

\begin{table}[thb]
\caption{\normalsize Cluster Sample}  
\label{table-targets}
\begin{center}
\begin{tabular}{lcccccc}
\hline
\hline
Name   &$z$ & $kT$ [keV] & kpc/$3'$ ($h=0.7$)\\
\hline
Coma      & 0.023 & 8 & 84 \\
A3667     & 0.053 & 7 & 185 \\
A754      & 0.054 & 9 & 190 \\
A2319     & 0.056 & 9 & 194 \\
A2256     & 0.058 & 7 & 202 \\
\hline                                         
\end{tabular}
\end{center}
\end{table}

\subsection{Targets and Feasibility}

For mapping the LOS velocity and turbulent broadening, we will mosaic
the clusters and derive spectra in $2'\times2'$ or $3'\times3'$
regions. For a relatively flat brightness distribution such as that
in the proposed clusters (except for cold fronts in A3667 and A2319 and
the shock front in A754, see below), such a pixel size allows us to
disregard PSF scattering -- the spectra will be largely uncorrelated.

\begin{figure}
\begin{center}
\rotatebox{0}{\includegraphics[width=7cm]{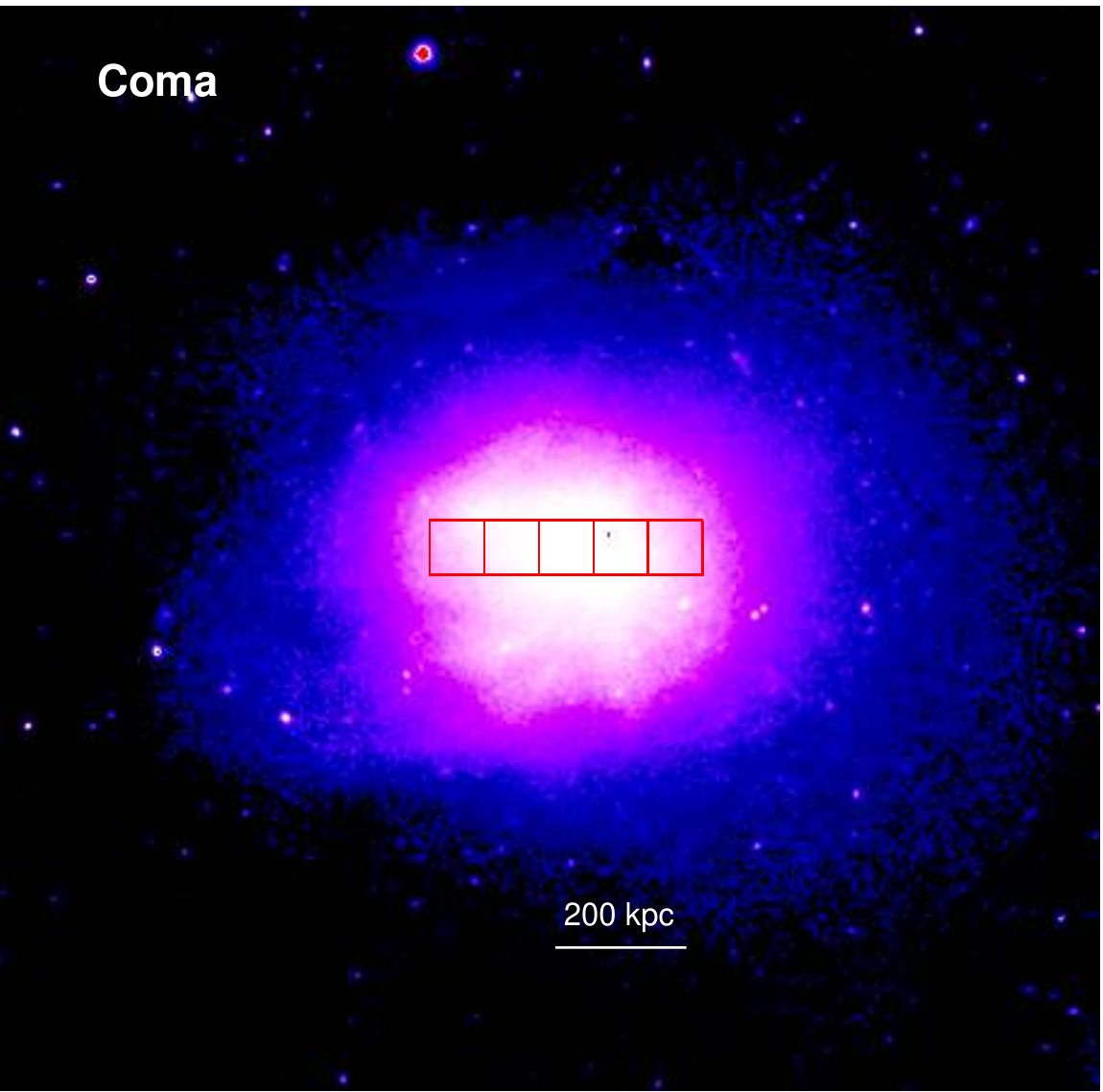}}
\hspace{5mm}
\rotatebox{0}{\includegraphics[width=7cm]{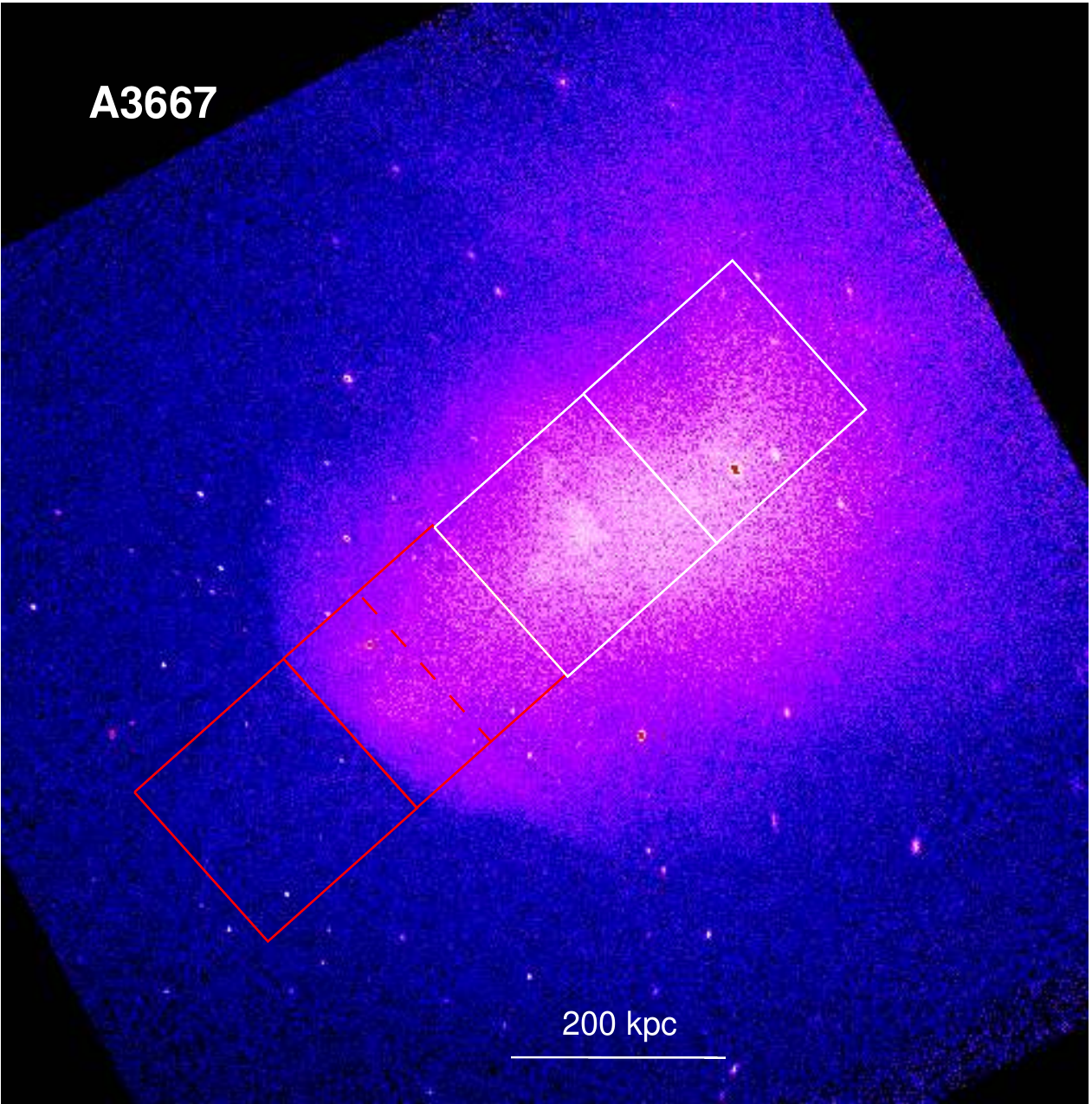}}
\vspace{5mm}
\rotatebox{0}{\includegraphics[width=7cm]{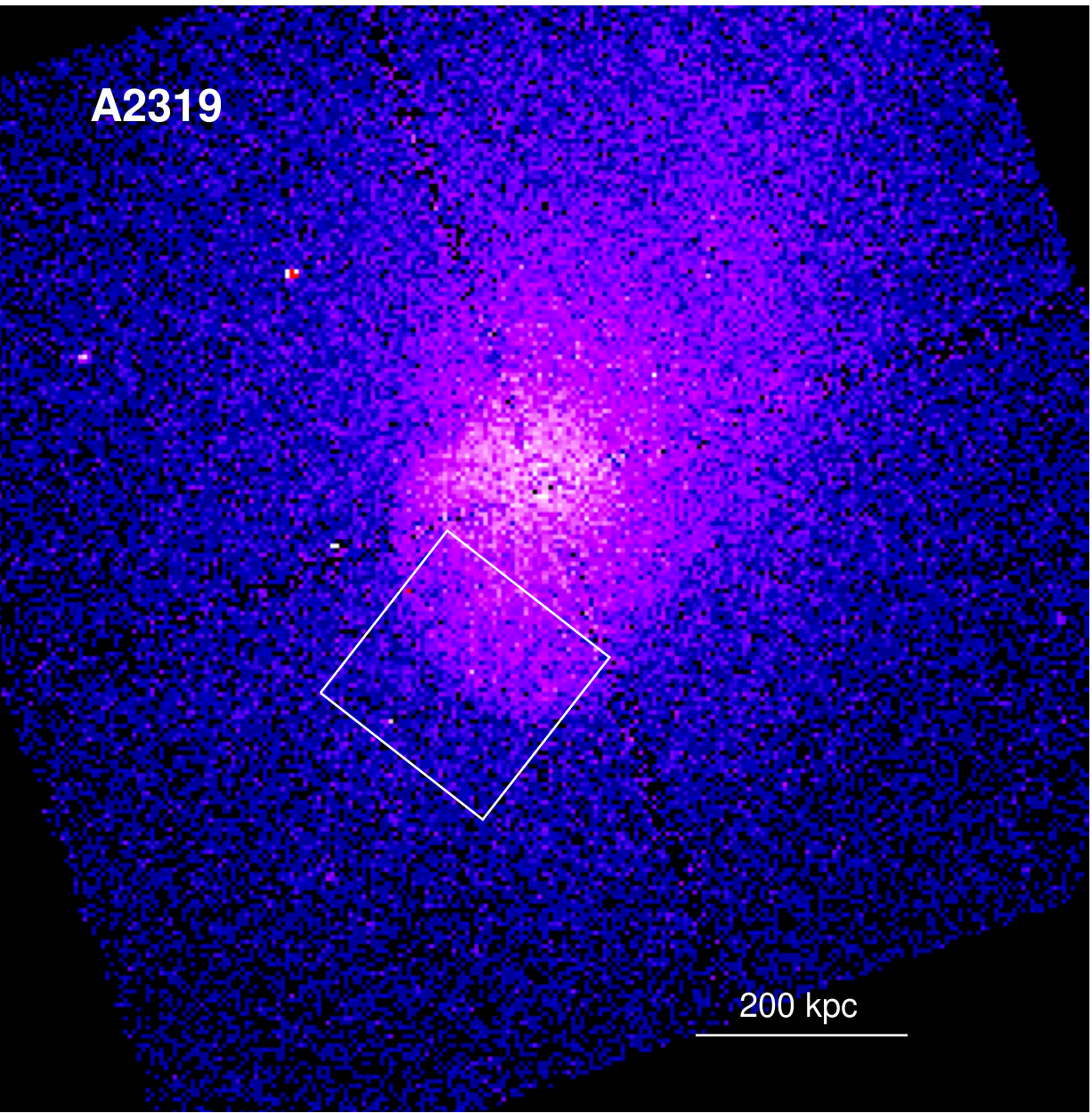}}
\hspace{5mm}
\rotatebox{0}{\includegraphics[width=7cm]{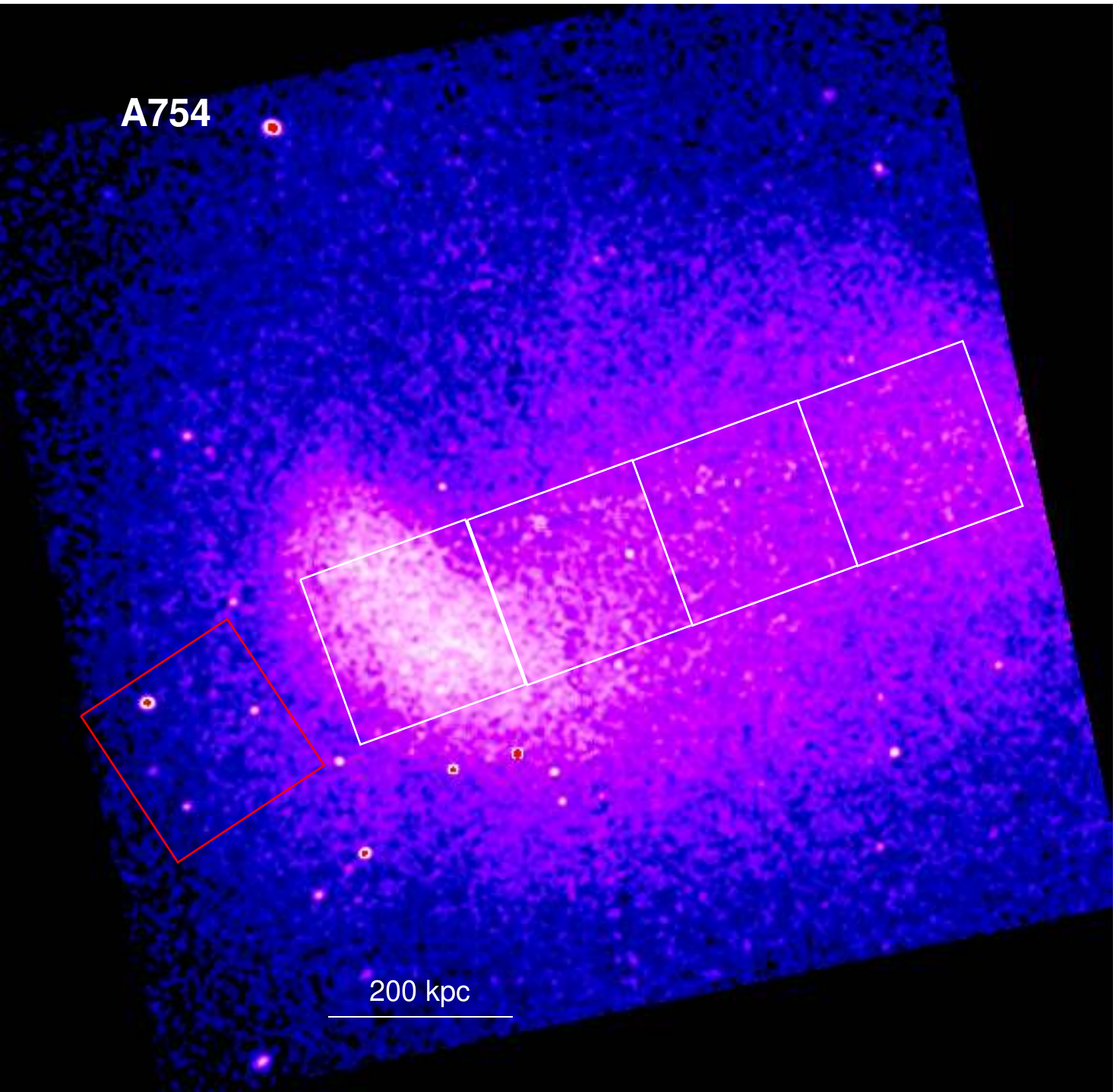}}
\caption{X-ray images of Coma, A754, A3667 and A2319 with the proposed
  $3'\times3'$ regions overlaid. The Coma image is from {\it XMM-Newton} and the
  rest is from {\it Chandra}. Red squares in the A3667 panel show the
  pointings inside and outside the prominent cold front, which are
  higher-priority than the rest for this cluster (white regions). Red square
  in A754 panel shows a post-shock region, which is the only one that can be
  resolved with SXS, but requires an unrealistically long exposure. White
  regions are higher-priority pointings for each cluster.}
\label{fig-merger_images}
\end{center}
\end{figure}

We aim at a minimum of 500 counts in the 6.7~keV Fe-K line for each
spectrum. The statistical accuracies of the line shift and turbulent
broadening strongly depend on the line width
(Figure \ref{fea:vturb-deltaz}). For the expected high values of
turbulence with the LOS velocity dispersion of $v_{\rm turb} \sim
300-1000$\ \kms (corresponding to the turbulence Mach number $M_{\rm 1D}
\equiv v_{\rm turb}/c_{\rm s} \sim 0.2-0.6$, where $c_{\rm s}$ is the
sound speed), we will obtain 90\% statistical uncertainties on line
position of 50 (160)\ \kms, and uncertainties on $v_{\rm turb}$ of 60
(170)\ \kms, for $v_{\rm turb}=$ 300 (1000)\ \kms, respectively, by
fitting both parameters and the line normalization simultaneously. These
constraints conservatively come from the He-like line alone, while in
these hot clusters the H-like line will slightly improve the
accuracy. With such statistical accuracy, the line shift measurement
will be dominated by the systematic uncertainty of SXS gain stability
(2.0 eV or 100 \kms\ at 90\%;
see also Figure \ref{fea:vturb-deltaz}) at $v_{\rm turb} \lax 500$
\kms.  At these levels of turbulence, the line broadening measurement
will be governed by statistics --- the expected uncertainty on the
line response is negligible at $v_{\rm turb} \gax 100$ \kms 
(Appendix \ref{sec-sys}).


The parameters of clusters that we find promising for the velocity and
turbulent mapping are given in Table\ \ref{table-targets}.

\subsubsection{Coma}
\label{subsec-coma}

Coma is the nearest bright cluster that appears to have undergone a merger
fairly recently. It offers the best chance to characterize the turbulence on
a range of scales, free of complications related to AGN activity. Coma has a
large, near-constant density core free of strong entropy gradients and
stratification, where turbulence should develop in 3D in an isotropic,
qualitatively simple way. On the other hand, the constant core also
unavoidably means strong LOS projection effects, which will have to be
addressed in a statistical manner using hydro simulations. One
possible observational setup is a ``transect'' of 5 contiguous 
pointings (Figure\ \ref{fig-merger_images}) spanning the cluster
core. It will allow measurements of velocities, line broadening, and their spatial correlations
on scales $2'-15'$ ($50-400$ kpc). The largest offsets will constrain the
cluster-scale rotation, which is a very interesting quantity. The
differences in velocities as a function of distance may be quantified
by a ``structure function'',
$SF(r)\equiv \langle(v(x+r)-v(x))^2\rangle$, where the
brackets denote averaging over pairs of directions separated by
$r$\/ in the sky (e.g., Zhuravleva et al.\ 2012). The structure
function has a one-to-one correspondence to a 3D power spectrum of
turbulence.  

\begin{figure}
\begin{center}



\includegraphics[width=\hsize]{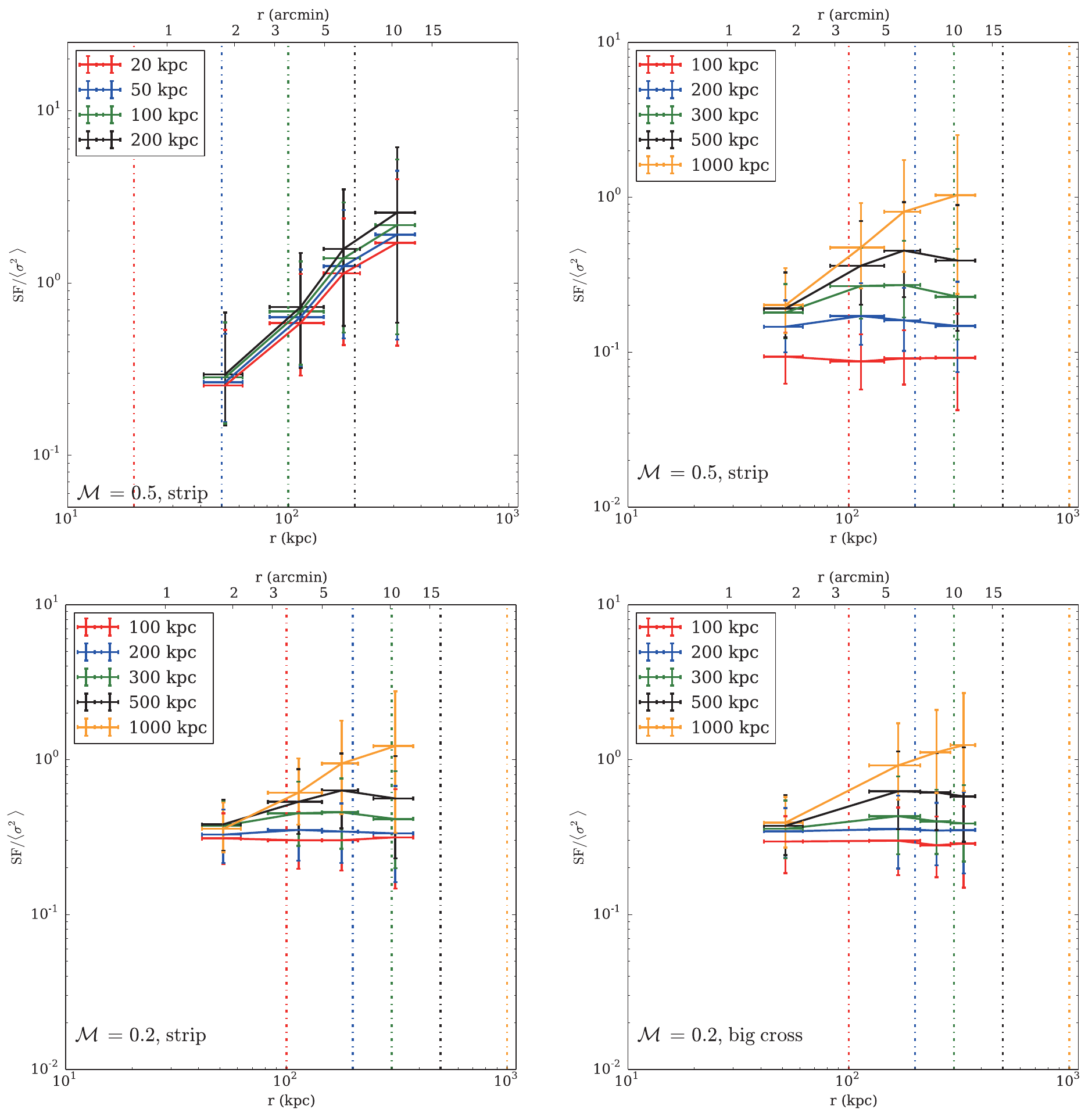}

\caption{Simulated structure functions for Coma, normalized to the
  average velocity dispersion. The measurement errors on velocities
  (statistical and systematic) are included. Panels labeled ``Strip''
  are for the pointing setup shown in the Coma panel of Figure\
  \ref{fig-merger_images}, while that marked ``Big cross'' is for a
  setup in which the small offsets are replaced with $6'$ offsets in the
  perpendicular direction. Errors are 68\%. The turbulence Mach number
  ($M_{\rm 1D}=0.2$ or 0.5) is given in labels.  The upper-left panel is for a
  Kolmogorov power spectrum with an exponential cutoff at small linear
  scales (different scales are shown by different colors), while other
  panels are for the cutoff at large linear scales. A small-scale cutoff
  is not detectable because of the dominance of the statistical
  uncertainties of the velocities, while the presence or absence of
  large-scale turbulent eddies can be detected.}  \label{fig-coma_sf}
\end{center}
\end{figure}

A simulation that includes the statistical and systematic errors on
velocities from the expected 500 line counts per $3'\times 3'$ pointing
is shown in Figure\ \ref{fig-coma_sf} (ZuHone \& Markevitch 2014, in
prep.). The simulation assumed an adiabatic, beta-model cluster
gas distribution, and considered several values of the turbulence Mach
number as well as various deviations of the turbulent power spectrum
from the pure Kolmogorov power law. In particular, we considered an
exponential cutoff
at various
small linear scales (representing
a dissipation scale of the turbulent cascade) and a cutoff at various large
scales (representing a ``driving scale'' of the cascade). The SF
curves in Figure\ \ref{fig-coma_sf} are normalized by the average
velocity dispersion, in order to show what 
extra information can be extracted from the Coma offset pointings once 
the central pointing is observed and the presence of the
turbulence and its strength are established.

In the core of a typical hot, non-cool-core cluster such as Coma, the
ICM collisional mean free path is of order 10 kpc; if turbulence
dissipates at this scale, its power spectrum will start turning over on
scales several times that, which for Coma is resolvable with {\it
ASTRO-H}.  However, the upper-left panel of Figure\ \ref{fig-coma_sf}
shows that we do not expect to be able to detect the presence of a
dissipation scale (unless it is implausibly big). Without the inclusion
of the random velocity measurement errors, the SF curves for different
dissipation cutoffs do differ significantly (ZuHone \& Markevitch 2014,
in prep.), but once those errors are included, they dominate the
small-scale power and overwhelm the intrinsic power spectrum
differences. However, the SF curves for different large-scale cutoffs
(the turbulence driving scales) can easily be distinguished, even for
moderate $M_{\rm 1D}\approx 0.2$. The lower-right panel (labeled ``big
cross'') shows the measured SF curves for a different pointing setup,
where the two $3'$ offsets were replaced with two extra $6'$ offsets to
form a cross. The accuracy of the large-scale variations is improved (at
the expense of the intermediate scales; the smallest-scale data points
come from the differences measured within each $3'\times 3'$ FOV); this
may be a better setup for studying the shape of the power spectrum on
large scales.

Exposure times required to collect 500 line counts from each
$3'\times3'$ pointing in the ``strip'' setup are (left to right) 150,
110, 100, 100, 150 ks, for a total of 610 ks.  These early measurements,
depending on their results, will provide a pilot dataset of a possible
future ``key project'' to map the whole Coma core, because the accuracy
of the power spectrum constraints increases proportionally to the number
of pairs of pointings at a given angular separation.

%

\subsubsection{A3667}

\begin{figure}[t]
\begin{center}
\rotatebox{0}{\includegraphics[width=0.6\hsize]{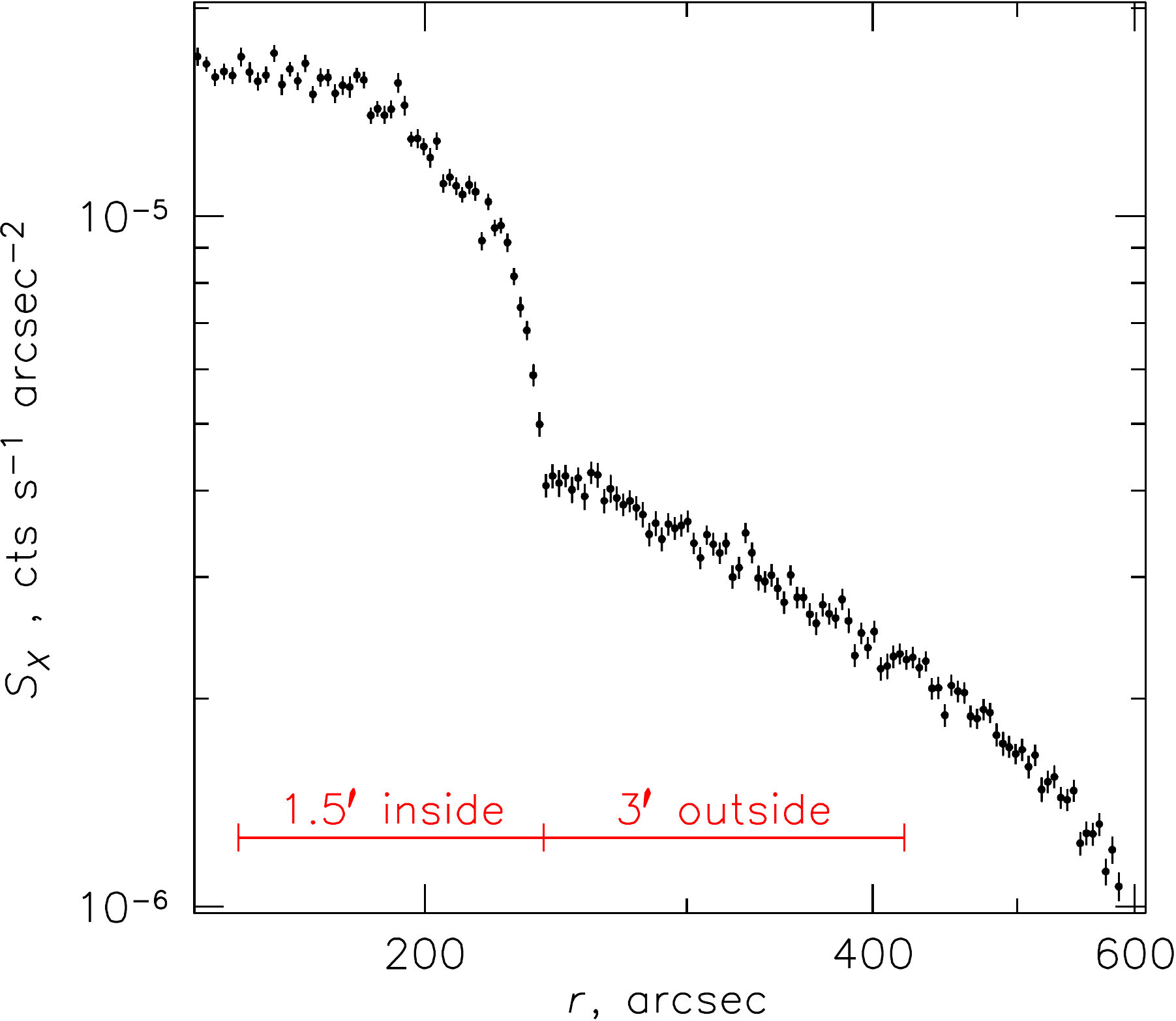}}
\caption{{\it Chandra}\/ X-ray brightness profile across the sharp cold
  front in A3667. The proposed observations will probe the gas properties
  across the front. Simulated spectra discussed in Tables
  \ref{table-a3667params} and \ref{table-a3667fits} correspond the to radial
  ranges shown in red.}
\label{fig-a3667briprof}
\end{center}
\end{figure}

A3667 is a classic cold front cluster.  Large-scale velocity mapping in
regions shown in Figure\ \ref{fig-merger_images} will require 200, 180, 100, and 120
ks, for a total of 600 ks. The two pointings shown in red will allow a
detailed study of the cold front (a strikingly sharp edge in X-ray
brightness shown in Figure\ \ref{fig-a3667briprof}), which we simulated in
detail.

\begin{table}[t]
\begin{center}
\caption{\normalsize A3667. Gas properties for the cold front based on {\it Chandra}\/ and {\it XMM-Newton}.}
\vspace{2mm}
\begin{tabular}{lcccc}
\hline
\hline
Region & $kT$ [keV] & Abundance [solar] & Flux$^a$ [erg cm$^{-2}$
 s$^{-1}$]& Fe line photons$^b$\\ 
\hline
Inside  & 4.5 & 0.55 & $2.50 \times 10^{-13}$& 600\\
Outside & 7.7 & 0.30 & $0.45 \times 10^{-13}$& 180\\
\hline
\label{table-a3667params}
\end{tabular}
\begin{minipage}{0.8\textwidth}
\footnotesize
$^a$ 0.5--2.0 keV \\
$^b$ $6.25-6.4$ keV counts in $1.5'\times 3'$ inside region and $3'\times
  3'$ outside region; 200 ks exposure; not including PSF scattering
\end{minipage}
\end{center}
\end{table}

For a high-contrast feature such as this cold front, PSF scattering will
have a significant effect. Based on the {\it Chandra}\/ image (and using the
PSF version from January 2013), we estimated the scattered contribution from
the bright $1.5'\times 3'$ region inside the cold front (half of the red
square shown in Figure\ \ref{fig-merger_images} adjacent to the front; see also Figure\ 
\ref{fig-a3667briprof}) into the $3'\times 3'$ region outside the front (the
leftmost red square in Figure\ \ref{fig-merger_images}) to be 30\% of the total line
photons in that spectrum. For error estimates, it was included as an
additional thermal component in the fit for the outside region.

Table \ref{table-a3667params} summarizes the X-ray quantities around the
cold front, for the ``inside'' and ``outside'' regions described above. For the
gas outside the front, we have assumed (reasonably conservatively) a
turbulent broadening of $v_{\rm turb}=500$ km~s$^{-1}$.  For the gas {\em inside}\/ the
front, we tried two scenarios --- (a) that gas has no LOS velocity
difference from the gas outside the front, but it does have a velocity
dispersion (hereafter called ``turbulence'' for simplicity) of
$v_{\rm turb}=1000$ km~s$^{-1}$.  This scenario represents a ``stripping'' cold front,
where this broadening is due to the gas inside the front flowing in opposite
directions along the LOS. Scenario (b) is for the gas on the inside to have
no turbulence, but to move along the LOS with $v=1000$ km~s$^{-1}$ -- this
represents a ``sloshing'' front geometry. (In reality, the latter scenario
can also have turbulence inside the front; as long as we can detect the LOS
velocity difference for the two regions, this would still point to a
``sloshing'' geometry.)

\begin{figure}[t]
\begin{center}
\includegraphics[width=0.4\hsize]{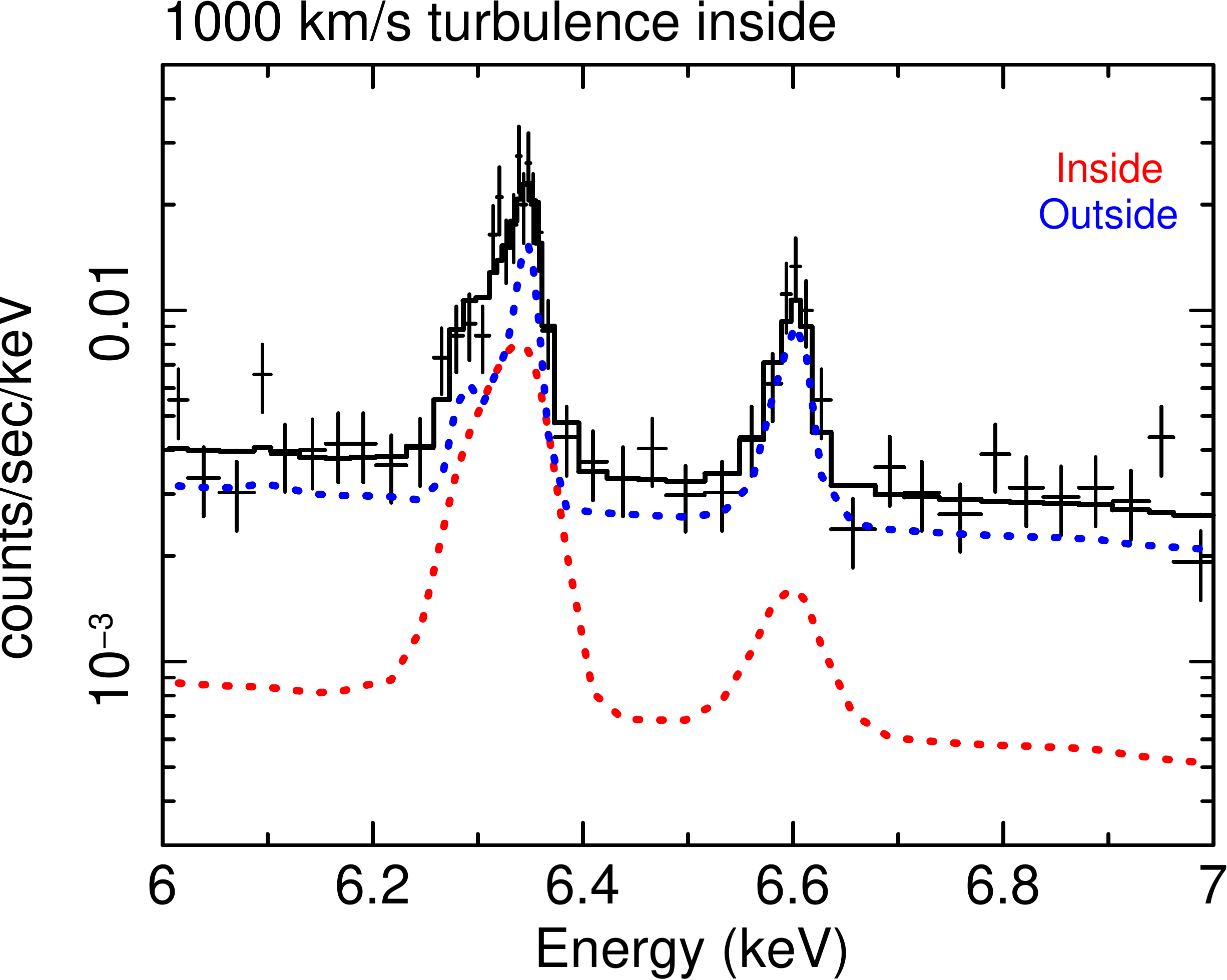}
\hspace{5mm}
\includegraphics[width=0.4\hsize]{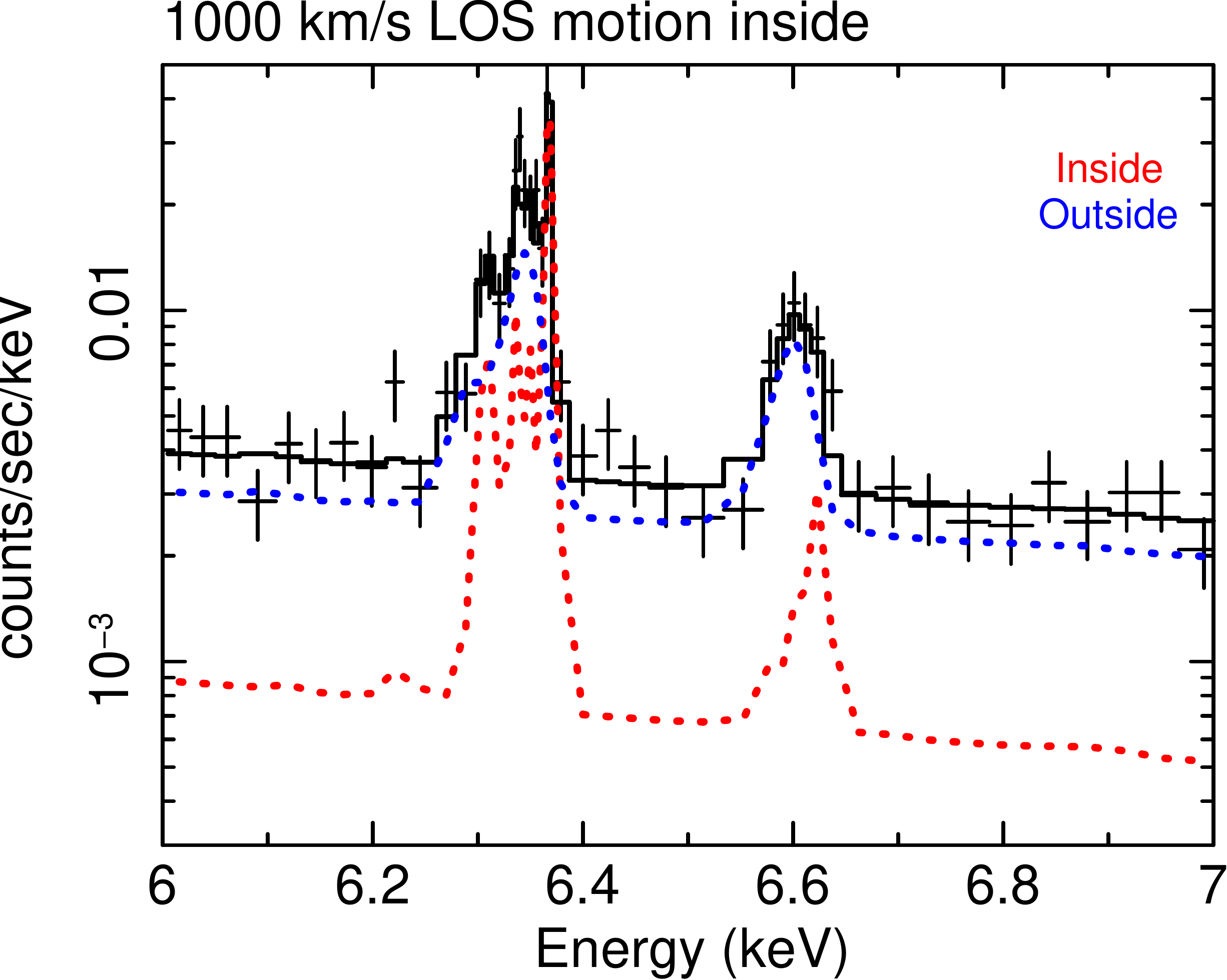}

\caption{Simulated spectra from the $3'\times 3'$ region {\em outside}\/ the
  cold 
  front in A3667 (see Figure\ \ref{fig-merger_images}), for a 200 ks exposure. Red
  lines show contamination from the gas inside the front due to PSF
  scattering; blue line shows the contribution from the underlying region in
  the sky. Two panes correspond to two assumptions about the gas inside the
  front (see text).}
\label{fig-a3667spec}
\end{center}
\end{figure}

Figure \ref{fig-a3667spec} shows the simulated spectra for the {\em
  outside}\/ region for these two scenarios, illustrating the PSF-scattered
contribution from the brighter side of the front. Table
\ref{table-a3667fits} shows the resulting LOS velocity and broadening
accuracies from two 200 ks exposures straddling the front, for the above two
assumptions.  For these fits, the whole spectrum was used, that is, both the
He and H-like lines contributed to the constraints; all parameters,
including both temperatures and abundances, were free.  Systematic
uncertainties on gain accuracy are similar to the statistical uncertainties
for LOS velocities for both regions; for line broadening, the statistical
uncertainties dominate over both the systematic resolution uncertainty and
the $T_{\rm ion}$ uncertainty. With two 200 ks exposures, we can obtain
interesting constraints on velocities and even constrain turbulence outside
the front.

\begin{table}[t]
\begin{center}
\caption{\normalsize A3667. Statistical accuracy ($1\sigma$) for
  velocities and velocity dispersions from two 200 ks
  front pointings, for scenarios (a) and (b) for the state of gas
  inside the front. In both cases, the gas outside has turbulent
  broadening with $v_{\rm turb}=500$ km~s$^{-1}$.}
\vspace{2mm}
\begin{tabular}{lcc}
\hline
\hline
Region                  & $v_{\rm bulk}$ [\kms] & $v_{\rm turb}$ [\kms] \\
\hline
Constraints for gas inside: & & \\
(a) $v_{\rm bulk}=0$, $v_{\rm turb}^{\rm inside}=1000$ \kms &   ...      & $1005\pm23$ \\ 
(b) $v_{\rm bulk}=1000$ \kms,  $v_{\rm turb}^{\rm inside}=0$ & $\pm43$ &    ...         \\ 
\hline
Constraints for gas outside: & & \\
(a) $v_{\rm bulk}=0$,  $v_{\rm turb}^{\rm inside}=1000$ \kms & $\pm100$ & $420\pm150$ \\
(b) $v_{\rm bulk}=1000$ \kms,  $v_{\rm turb}^{\rm inside}=0$ & $\pm136$ & $571\pm143$ \\
\hline
\label{table-a3667fits}
\end{tabular}
\end{center}
\end{table}

\begin{figure}[t]
\begin{center}
\rotatebox{0}{\includegraphics[width=11cm]{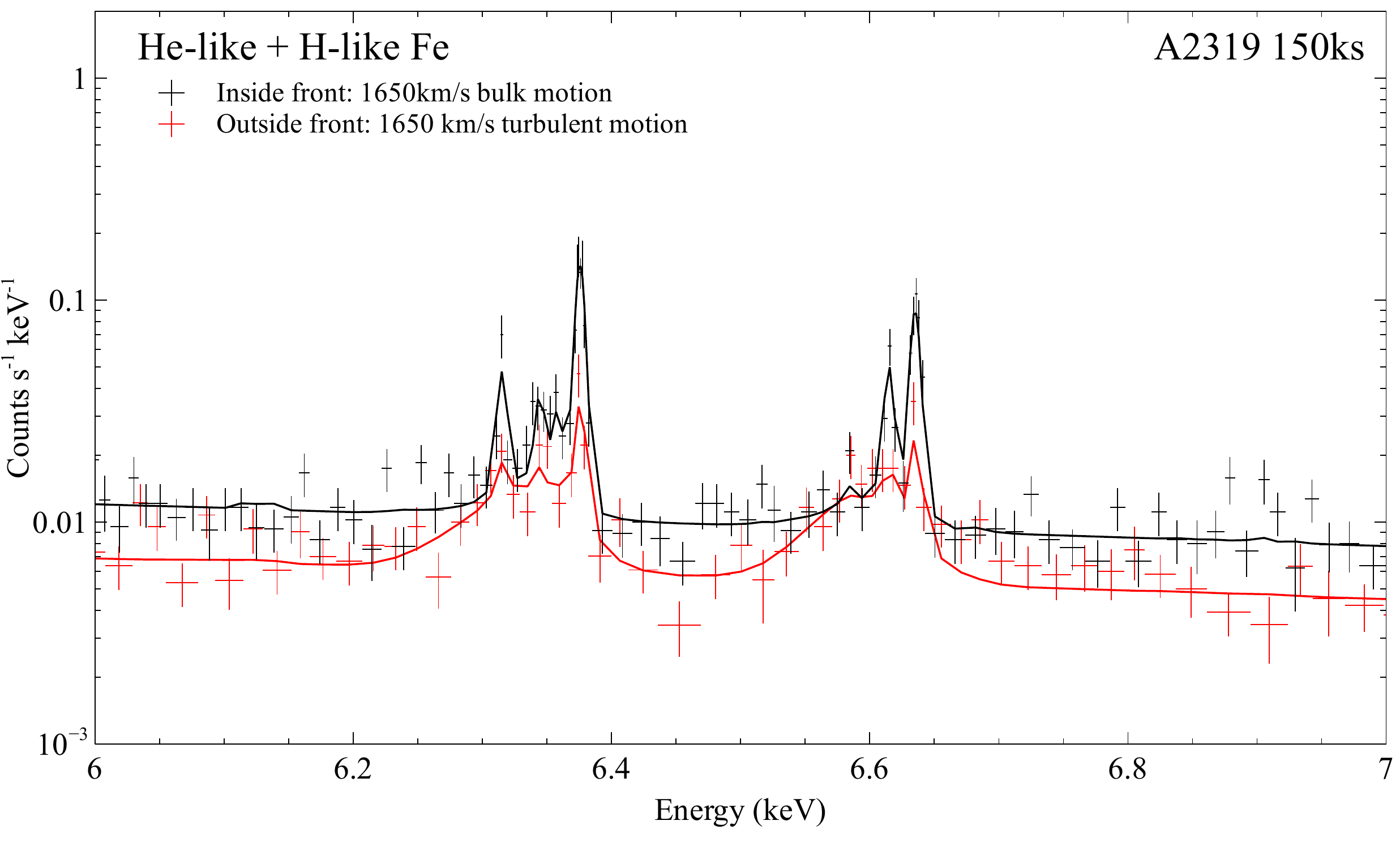}}

\caption{Simulated spectra from the $1.5'\times 3'$ regions inside and
  outside the cold front in A2319 (two halves of the SXS FOV shown in Figure\ 
  \ref{fig-merger_images}), for a 150 ks exposure and scenario (b) with
  turbulent broadening outside the cold front. Black and red lines show the
  inside and outside regions, respectively, and include both the intrinsic
  and PSF-scattered components.}
\label{fig-a2319spec}
\end{center}
\end{figure}

\subsubsection{A2319}

A2319 features a prominent cold front with as high a gas density jump as in
A3667, but located in a brighter region of the cluster. We simulated
the velocity and line broadening constraints that can be obtained from one
150 ks exposure straddling the front (see Figure\ \ref{fig-merger_images}). PSF
scattering was included based on {\it Chandra}\/ image. The contaminant was
included in the fit for the outside region as an additional thermal
component; it contributes 35\% of the total emission in the outside
spectrum.  Parameters for both components were varied to determine the
uncertainties for the outside region. For this simulation, we assumed a
sonic relative LOS velocity between the two components (1650 \kms) and a rather
extreme, $M_{\rm 1D}\approx 1$ turbulent broadening for the gas inside the
front in scenario (a) or outside the front in scenario (b). The
resulting constraints are given in Table\  
\ref{table-a2319fits}, and simulated spectra for scenario (b) are shown in
Figure\ \ref{fig-a2319spec}. The constraints are interesting and can clearly
distinguish these two extreme cases, though, as expected from such small
numbers of line photons (Table\ 
\ref{table-a2319fits}), the errors are large and less-extreme motions
will not be easy to constrain. This underscores the need to expose
well to collect at least $\sim 500$ line counts for robust
measurements of gas dynamics. 

\begin{table}[t]
\begin{minipage}{\textwidth}
\caption{\normalsize A2319. Constraints on velocities for $1.5'\times3'$
  regions and exposure time 150 ks.  Errors are $1\sigma$. Gain
  stability for systematic uncertainty is 1 eV. Line counts ($6.3-6.4$
  keV) include continuum but exclude PSF scattered component. Scenario
  (a) assumes turbulence inside the cold front and none outside, whereas
  (b) assumes turbulence outside. In both models, there is a LOS
  velocity difference of the gas inside the front in the reference frame
  of the gas outside.}
\vspace{2mm}
\begin{center}
\begin{tabular}{l c c c c}
\hline
\hline
 & \multicolumn{2}{c}{scenario (a)} & \multicolumn{2}{c}{scenario (b)} \\
 & Inside & Outside & Inside & Outside \\
\hline
$v_{\mathrm{bulk}}$ [\kms] & $1640^{+24}_{-21}$ & $28\pm160$ & $1635\pm10$ & $140^{+320}_{-290}$ \\
$v_{\mathrm{bulk}}$ sys. err. [\kms] & 45 & 45 & 45 & 45\\
$v_{\rm turb}$ [\kms] & $1320^{+140}_{-180}$ & $<47$ & $<29$ & $1500^{+280}_{-240}$ \\
$v_{\rm turb}$ sys. err. [\kms] & 1 & 60 & 60 & 1 \\
He-like Fe counts & 325 & 160 & 400 & 120 \\
\hline
\end{tabular}
\end{center}
\label{table-a2319fits}
\end{minipage}
\end{table}

\subsubsection{A754}

This is a well-studied merger with a $M=1.6$ shock front (Macario et al.\ 
2010). Our estimate shows that it is not feasible to study the post-shock
gas (see \S\ref{sec:beyond}). From detailed examination of the available
X-ray data, one may expect the bright, cool elongated ``core'' to have a
different LOS velocity from the rest of the cluster (it appears to be a
``sloshing'' core disrupted by a violent cluster collision). A transect
shown by white squares in Figure\ \ref{fig-merger_images} will require the following
exposures: 100, 170, 220, 240 ks, for a total of 730 ks.

\subsubsection{A2256}

A2256 is another famous merger studied by all the X-ray missions. Tamura
et al.\ (2011) reported a LOS velocity gradient so large that it was
detectable with {\it Suzaku} even with a CCD resolution. A2256 would
thus be a very promising target for velocity mapping. However, its peak
surface brightness is half that of Coma, so exposures will have to be
very long ($\sim 200$ ks in the center and several times longer in the
region of the large velocity differences).

\subsection{Beyond Feasibility}
\label{sec:beyond}

Shock fronts are unique tools for studying the ICM microphysics. However,
none of the known shocks will be resolved by {\it ASTRO-H}. The most
prominent one in the Bullet cluster is only $30''$ away from the bright cool
bullet; A520 and A2146 have similarly small separations from their bright
driver subclusters.

A754 has a shock front observed by {\it ROSAT}\/ and {\it Chandra}\/
(Krivonos et al.\ 2003; Macario et al.\ 2011), which is better separated
from the bright core than the more famous ones.  However, our estimate
showed that even in this nearby cluster, PSF scattering from the bright
cluster core $2-3'$ away will dominate the flux in the post-shock region.
For example, in a 150 ks exposure shown by red square in Figure\ 
\ref{fig-merger_images}, we will collect a total of 190 line counts, of which only
50 counts originate in the underlying post-shock region and the rest is
scattered from the outside.  Such a large scattered contribution combined
with low brightness makes it impossible to derive any constraints for that
region.








\bigskip

\section{High-energy Processes}
\label{sec-highe}

\subsection*{Overview}

The Hard X-ray Imager (HXI) on board {\it ASTRO-H}, together with the
foregoing \nustar mission, will comprise the first generation of hard
X-ray instruments with imaging capability in the 5--80 keV energy
band. The prime targets of HXI include the inverse-Compton radiation by
non-thermal electrons in galaxy clusters and the thermal emission from
very hot ($kT\gg 10$ keV) intracluster gas. The former is of particular
importance for revealing the origin of the
ultrarelativistic particles, which are presently observed only in the
radio bands; it will also provide information on the strength and
structure of the intracluster magnetic fields. The latter is critical
for detailed understanding of the heating mechanism
of the intracluster plasma as well as assessing correctly the total
pressure support in clusters. A major advantage of {\it ASTRO-H} will be
its wide spectral coverage, which plays a key role in disentangling a
number of emission components including backgrounds; the HXI data will
be obtained simultaneously with those of the Soft X-ray Spectrometer
(SXS) and the Soft X-ray Imager (SXI) at $0.3-12$ keV. We also address
meaningful lessons from early operations of {\it NuSTAR}.\footnote{Coordinators of this section: M. Kawaharada, G. Madejski}



\subsection{Background and Previous Studies}

Besides being luminous sources of thermal X-ray emission, galaxy
clusters often harbor faint diffuse radio sources. "Giant radio halos"
and "minihalos" are unpolarized sources found in merging clusters and in
cluster cool cores, respectively. Some clusters also exhibit "radio
relics" -- elongated, strongly polarized sources in the periphery (for a
review see Feretti et al.\ 2012).
This strongly suggests that there is a substantial population of
relativistic particles. However, the details of the energy distribution
of such particles are not fully understood, since there is an inherent
degeneracy with the strength of the magnetic fields and their
distribution, when made on the basis of the radio data alone.
Furthermore, the process of electrons being accelerated to the
relativistic energies required to account for the radio emission is far
from clear.  Peripheral relics can plausibly be produced
by Fermi acceleration on merger shocks, but the mechanism responsible
for the giant halos and minihalos has to act simultaneously throughout
the cluster, because the radiative cooling time of such energetic
electrons ($10^{7-8}$ yr) is much shorter than their diffusion or
advection time across the halo (e.g., Petrosian 2001), so they have to
be produced in-situ. Several physical mechanisms have been proposed (for
a review see Brunetti \& Jones 2014).

The very same electrons that generate the radio emission must also
inverse-Compton scatter any ambient radiation. This
provides an interesting opportunity to determine the magnetic field
strength and the number density of relativistic electrons separately
(Harris \& Romanishin 1974), taking advantage of the fact that the radio
synchrotron emission is proportional to the cosmic ray density times the
energy density of the magnetic field, while inverse Compton emission is
independent on the magnetic field and is instead proportional to the
energy density of the background radiation.
There are two, possibly three sources of such ambient photons with
significant energy density.  One is the Cosmic Microwave Background,
with well-determined energy density.  
Another is the cluster's own thermal X-ray emission, which
is well-measured in clusters from sensitive soft X-ray data, with
temperatures of a few up to $\sim 15$ keV (and possibly even more; e.g.,
Markevitch et al. 2006).  These two photon fields should
be Compton-upscattered by the ultrarelativistic electrons to the hard
X-ray and $\gamma$-ray bands, respectively.
(The third source is starlight from the member galaxies, but
while within the member galaxies it might be appreciable, its energy
density in clusters is significantly lower.)

Since these two sources of photon fields in clusters are reasonably well
determined (both the co-moving CMB intensity, and cluster X-ray
luminosity are robustly measured), the detection of inverse Compton
radiation should break the degeneracy between the distribution of
relativistic electrons and the strength of the magnetic field, allowing
the determination of both quantities separately, and possibly even
measuring the spatial structure of the field.  It is important to note
that even a sensitive {\it non-detection} has a profound impact on
cluster studies, since at the sensitivity of the current imaging hard
X-ray instruments (\astroh HXI, but also \nustar), those upper limits on
non-thermal flux correspond to {\it lower} limits on the strength of
magnetic field (e.g. Nakazawa et al. 2009).  
Furthermore,
the robust determination of the distribution of relativistic particles
is important in determining the cluster masses, which are essential when
one wishes to use clusters as tools for precision
cosmology. This is because the non-thermal particles provide an
additional source of pressure, which needs to be considered in
derivation of cluster masses, traditionally relying on an assumption of
hydrostatic equilibrium. In fact, a deviation from hydrostatic
equilibrium has been reported in some clusters (e.g.  Kawaharada et
al. 2010; Ichikawa et al. 2013; van der Linden et al. 2014), by
comparing the thermal pressure and the total pressure derived from
lensing analysis; the relativistic components, including
the magnetic fields and cosmic rays, are possible reasons, along with
turbulence and the effects of asphericity.

Gamma-ray measurements so far have not revealed any emission from
clusters.  This is possibly due to a limited sensitivity of instruments,
but also due to the fact that the inverse Compton scattering of soft
X-rays operates at least partially in the Klein-Nishina regime.  The
best limits so far were from the {\it Fermi} LAT instrument (Ackermann et
al. 2010).  The hard X-rays offer a promising probe of the
inverse-Compton radiation by relativistic electrons.
The non-thermal emission is expected to have a power-law shape, as
inferred from the radio spectra.  However, the {\it thermal} cluster
emission is so dominant that the {\it non-thermal} emission is very
difficult to disentangle (e.g., Million \&
Allen 2009; Wik et al.\ 2014).
Further complication arises from the fact that the clusters' X-ray
emission cannot be described by a single temperature (see, e.g.,
Peterson and Fabian 2006; Anderson et al. 2009).  Such multi-temperature
structure can very easily mask the power-law component potentially
present in addition to the thermal emission {\it within} the soft X-ray
band.

The presence of very hot ($kT \gg 10$ keV) gas has in fact been inferred
in some merging clusters such as RX J1347.5--1145 (Kitayama et al. 2004;
Ota et al. 2008) and A3667 (Nakazawa et al. 2009), whereas definite
distinction from the non-thermal emission is yet to be
made. Its emission measure and spatial distribution are
uncertain, and it may significantly contribute to the total ICM
pressure, while largely evading detection by previous X-ray instruments
that only cover energies below 10 keV.
A wider spectral coverage is crucial for disentangling
this component from the non-thermal emission and various background
components.
Hard X-ray imaging should also play a complementary role in revealing
the nature of the hot gas to spatially resolved Sunyaev-Zel'dovich
effect observations (for a review see Kitayama 2014).

\subsection{Prospects and Strategy}

The hard X-ray band is suitable for detecting the inverse-Compton
radiation as well as the thermal emission from gas with $kT\gg
10$ keV.
The caveats are much higher levels of contamination from the cosmic
X-ray background (CXB) and non-X-ray (or particle) background (NXB) than
the soft X-ray band.
In fact, some of the previous claims of detection of the
cluster nonthermal emission from collimated hard X-ray instruments
(workhorses for the hard X-ray astronomy before the advent of focusing
hard X-ray telescopes) 
have been controversial (e.g., Fusco-Femiano et
al. 2000; Rephaeli et al. 2006; Petrosian et al. 2006; Ajello et
al. 2010).  Convincing and conclusive hard X-ray measurements
require both imaging capability and an improved control of the
backgrounds.

\astroh is one of the very first observatories featuring such focusing
hard X-ray instruments. The Hard X-ray Imager (HXI) is sensitive up to
$\sim 80$ keV, well beyond the tail of clusters' thermal spectrum.
Since the HXI is co-aligned with the SXS, the data will be taken by both
instruments and, fortuitously, there is a strong synergy between the
SXS, SXI and HXI regarding cluster observations.  One of the key
requirements is accurate cross-calibration between the soft and hard
X-ray instruments, as is also the case
for many other joint soft and hard X-ray observations.


There are a few important cases where the non-thermal emission in a
cluster arises in regions with relatively weak thermal soft X-ray
emission.  The striking radio relic at the northwestern
periphery
of A3667 shows relatively strong radio emission, but weak soft X-ray
emission.  Its radio emission exhibits a sharp linear
edge, which would make the detection of the accompanying nonthermal
X-ray emission easier,
since it allows for the search via positional coincidence.  In the
central regions of A3667, the very hot gas could be detected with the
combined analysis of HXI and SXI.  The radio halo in the Coma cluster is
another major object from which non-thermal emission is expected. These
would be observations 
for which the HXI science provides
the main driver in the early phase of \astroh observations.


The scientific goals addressed via the HXI observations are also 
complementary to those discussed in Section
\ref{sec-nonth}. There, the main tool used for studies of the
non-thermal pressure is the high resolution X-ray spectroscopy with the
SXS, while here, the expected impact of the HXI observations is
discussed.  However, ultimately, it is the joint evidence from both
instruments that will be most valuable in studying the departure from
a simple 
thermal picture of clusters.  One example regards the effect of
the population of non-thermal electrons on the line emission in the soft
X-ray band
(e.g. Kaastra et al. 2009).  Those joint studies should address one
important and overarching goal: how the non-thermal / high energy
processes in clusters affect the reliability of determination of cluster
mass, and thus applicability of X-ray observations of clusters to
cosmology.

\begin{figure}[t]
 \begin{center}
  \includegraphics[width=0.37\textwidth,clip,angle=0]{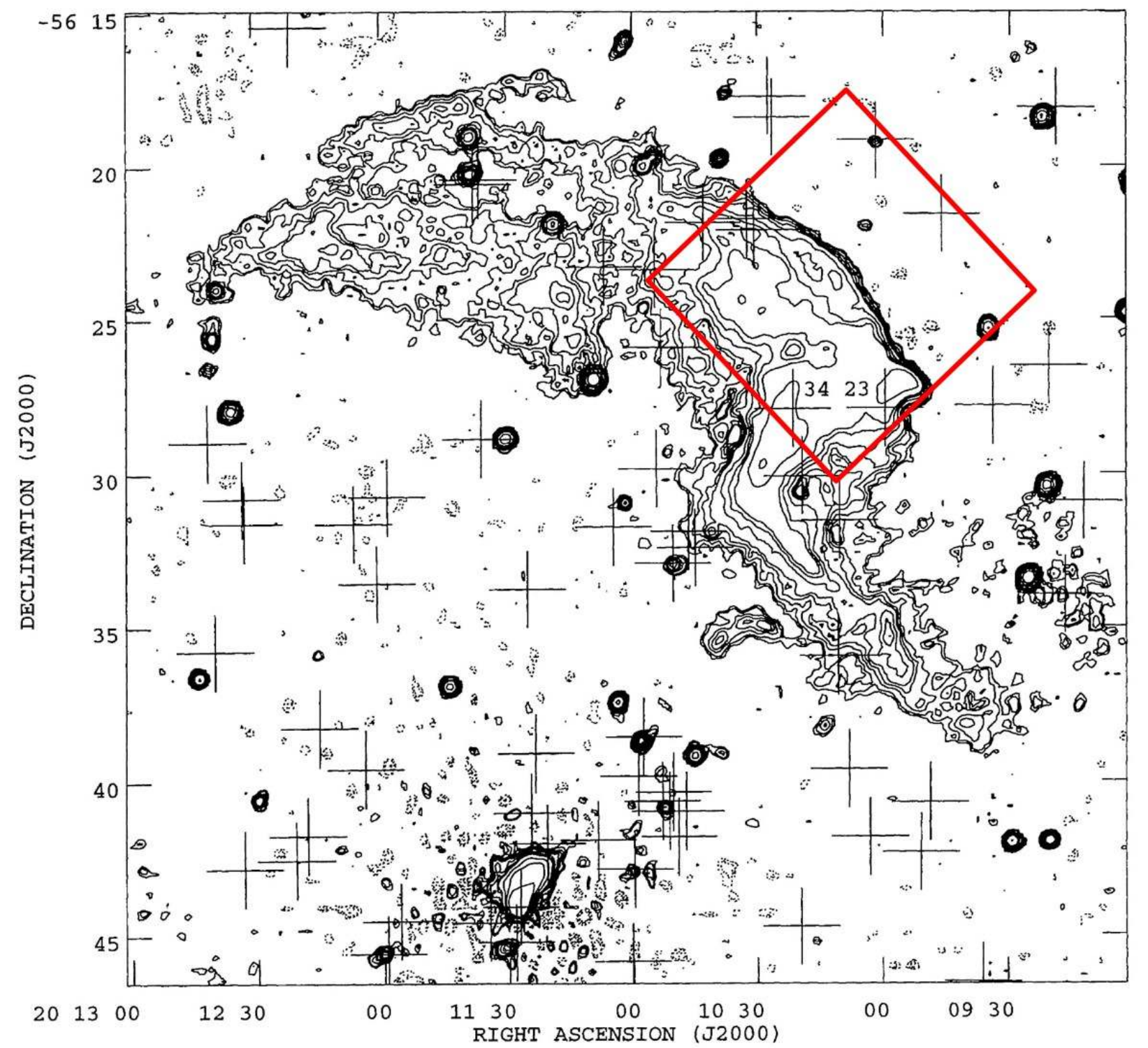}
  \includegraphics[width=0.38\textwidth,clip,angle=0]{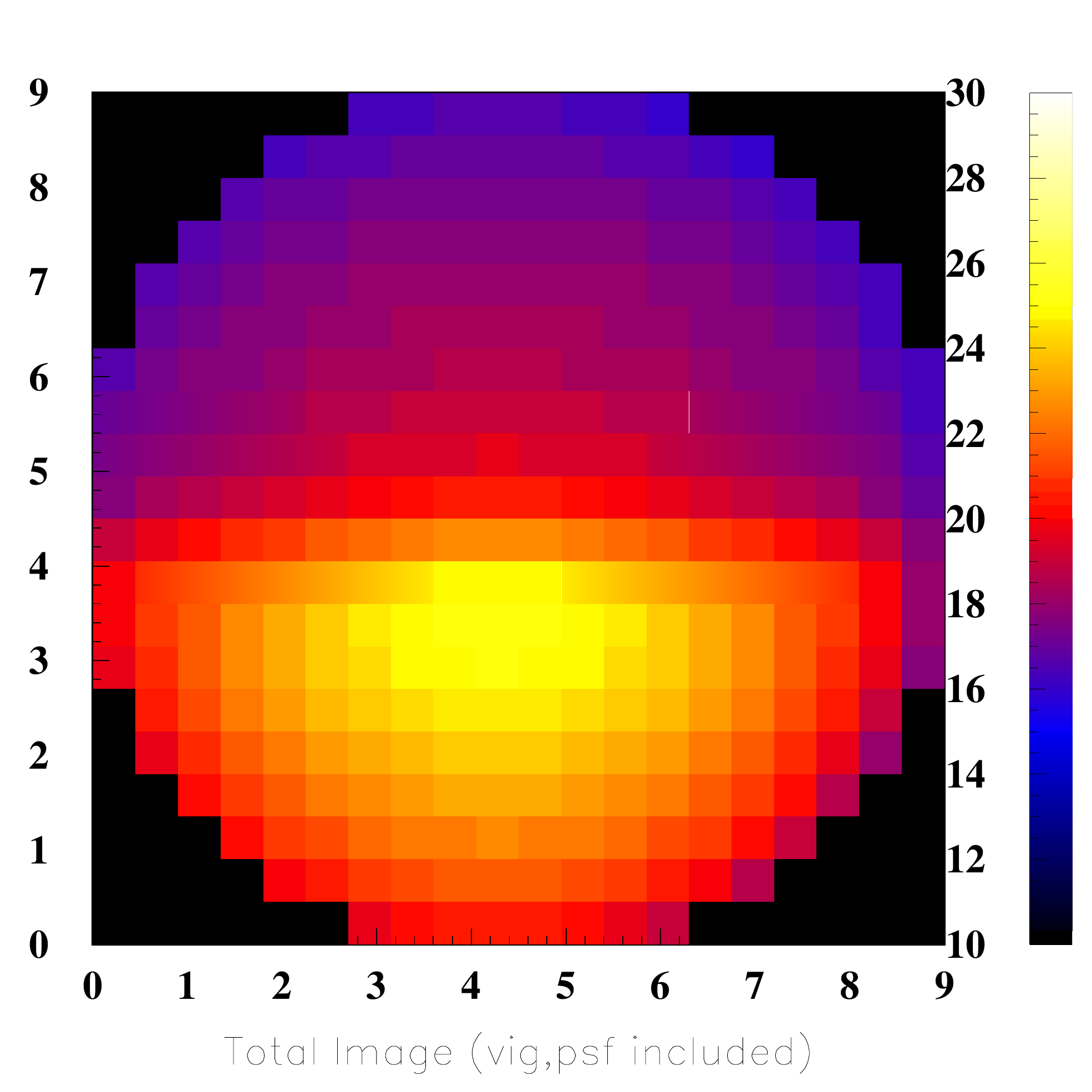}
  \includegraphics[width=0.45\hsize]{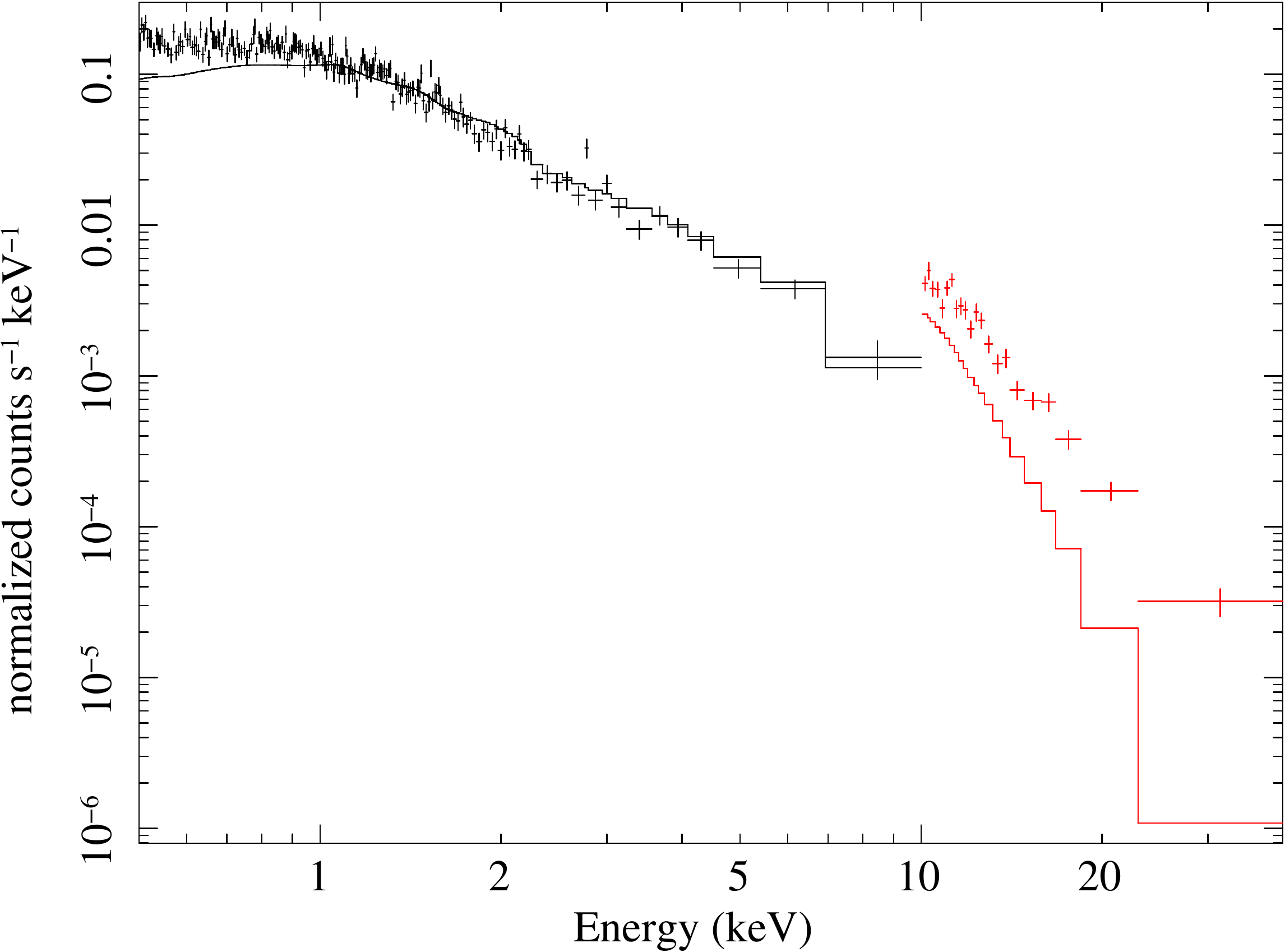}
  \caption{
Simulated observations of A3667 North West relic for a
200 ks exposure, including the non-thermal emission with $\Gamma=2.0$ as
described in the text. {\it Top left:} The HXI field of view ($9' \times
9'$) which covers the NW relic and its edge (red box), overlaid on the
20-cm ATCA image (R\"{o}ttgering et al. 1997). {\it Top right:}
Simulated HXI raw image (NXB and CXB are included) of the potential
sharp edge in hard X-ray band. The image is binned to 20 $\times$ 20
pixels. The color scale is in counts per pixel and indicated in the
legend on the right. The PSF and vignetting effects are taken into
account. {\it Bottom:} Background (NXB+CXB) subtracted SXI (black cross)
and HXI (red cross) spectra. For reference, solid lines show the thermal
spectra with $kT=5$ keV. } 
\label{fig-highe1}
\end{center}  
\end{figure}
\begin{figure}[t]
 \begin{center}
  \includegraphics[width=0.55\textwidth,clip,angle=0]{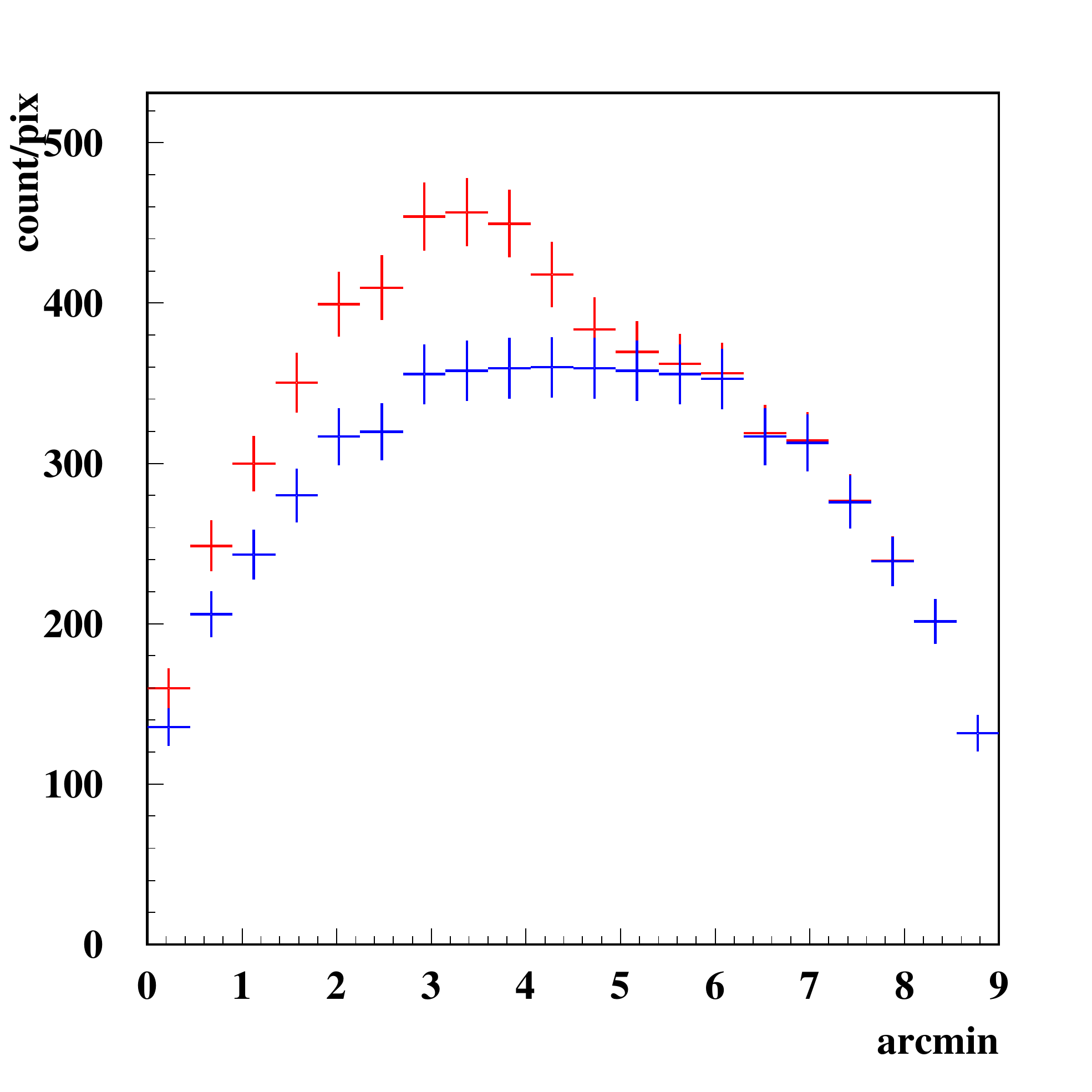}  
\caption{
Projection to Y-axis of Figure ~\ref{fig-highe1}. Profiles of the total
count and background count are shown in red and blue,
respectively. Error bars indicate statistical 1$\sigma$ errors.
  }
 \label{fig-highe2}
\end{center}
\end{figure}

\subsection{Targets and Feasibility}

\subsubsection{The NW relic of Abell 3667} 

The North West (NW) radio relic of A3667 (R\"ottgering
et al.\ 1997; see Figure 2 of Finoguenov et al. 2010) is the brightest
such object in the sky, which makes it one of the best targets for the
detection of inverse Compton radiation in hard X-rays.  Its angular size
also perfectly matches the HXI FOV and resolution.
From soft X-ray observations by \xmm (Finoguenov et al. 2010), it is
difficult to determine if the X-ray emission associated with this relic
is thermal or nonthermal. Using \suzaku data, Nakazawa et al. (2009)
reported an upper limit on the non-thermal flux from the {\it entire}
cluster region as $7.3 \times 10^{-13}$ erg s$^{-1}$ cm$^{-2}$ at 10--40
keV and the corresponding lower limit on the magnetic field of 1.6
$\mu$G. The upper limit on the surface brightness is about $1.8 \times
10^{-15}$ erg s$^{-1}$ cm$^{-2}$ arcmin$^{-2}$ (10--40 keV) in the NW
relic region. 

In our simulations for HXI, we extrapolate the soft X-ray emission
observed by \xmm to hard X-rays assuming a power-law spectrum with the
photon index $\Gamma=2.0$, yielding the relic surface brightness of $1.7
\times 10^{-15}$ erg s$^{-1}$ cm$^{-2}$ arcmin$^{-2}$ in the 10--40 keV
band. Note that this value lies just below the \suzaku upper limit
mentioned above.  We also assume that the X-ray brightness traces the
spatial distribution of the radio synchrotron
emission. Figure \ref{fig-highe1} shows that, with an
exposure time of 200 ks, the assumed brightness edge can be detected and
HXI plays a crucial role in identifying the nonthermal spectrum.  For
clarity, we also plot the projected emission profile onto the Y-axis in
Figure \ref{fig-highe2}, exhibiting that the non-thermal emission
exceeds the background inside the radio edge.

Accurate prediction of the sensitivity of HXI before launch is
challenging since it depends largely on various systematic effects
including the background levels in orbit. Current estimates are that the
sensitivity will be better than that of \suzaku by about a factor of
$\sim 6$ (see Sec. 6.3.3 for details). 
If we do not detect the inverse Compton signal from this
radio relic with such sensitivity, a combination of its radio brightness
and our X-ray upper limit will yield a lower limit on the strength of
the magnetic field $B\gax 4\,\mu$G. Such a strong magnetic filed would
be very unexpected so far away from the cluster center.

\subsubsection{Very hot gas at the center of Abell 3667}

\suzaku revealed the presence of very hot thermal gas in the
central region of A3667, with temperature higher than 13 keV (Nakazawa
et al. 2009). With the improved sensitivity and imaging capability of
{\it ASTRO-H}, we will be able to determine the temperature and 
map the location of this ICM component.

\begin{figure}[t]
\begin{center}
\includegraphics[width=0.42\textwidth]{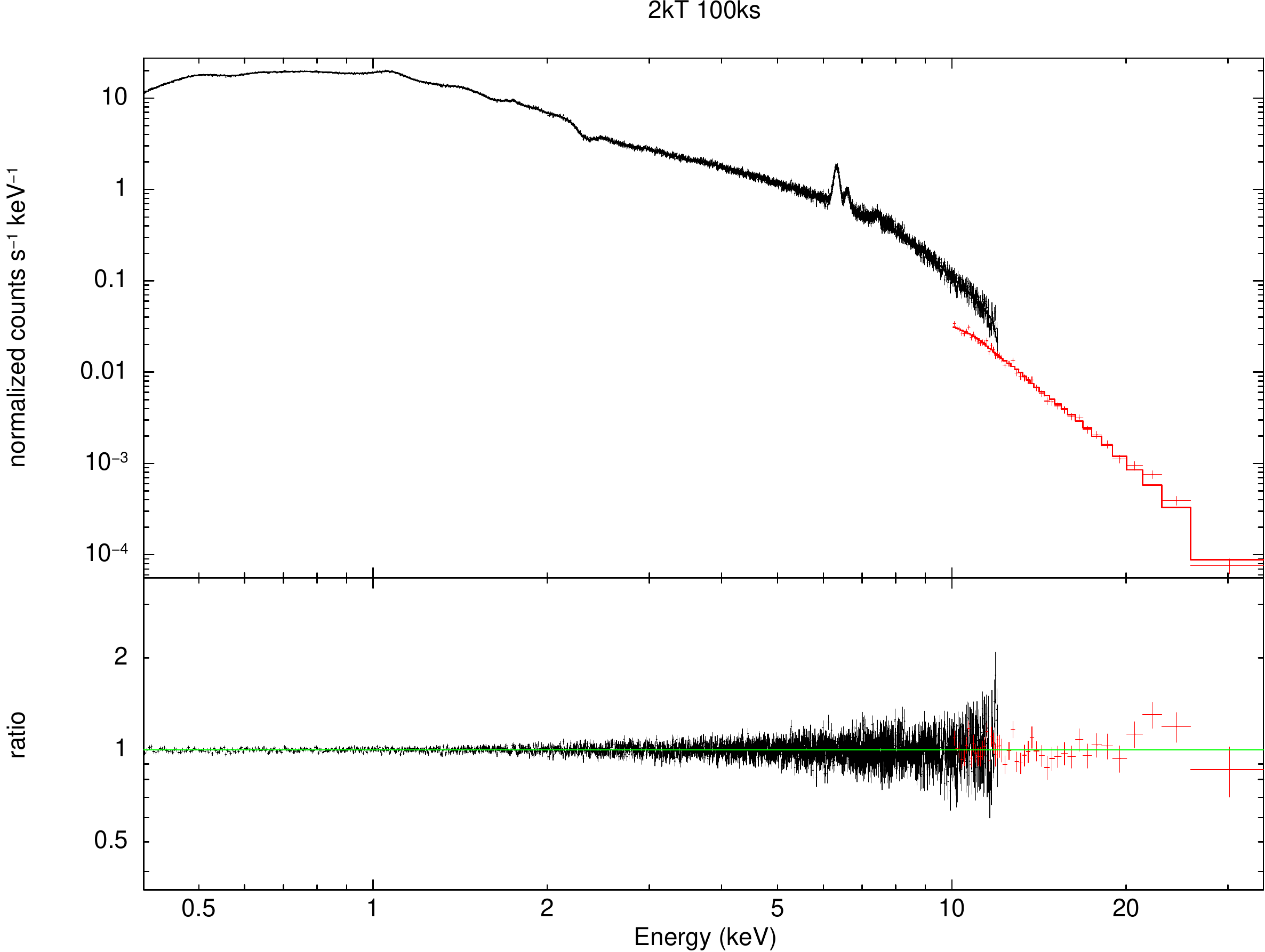}
\includegraphics[width=0.42\textwidth]{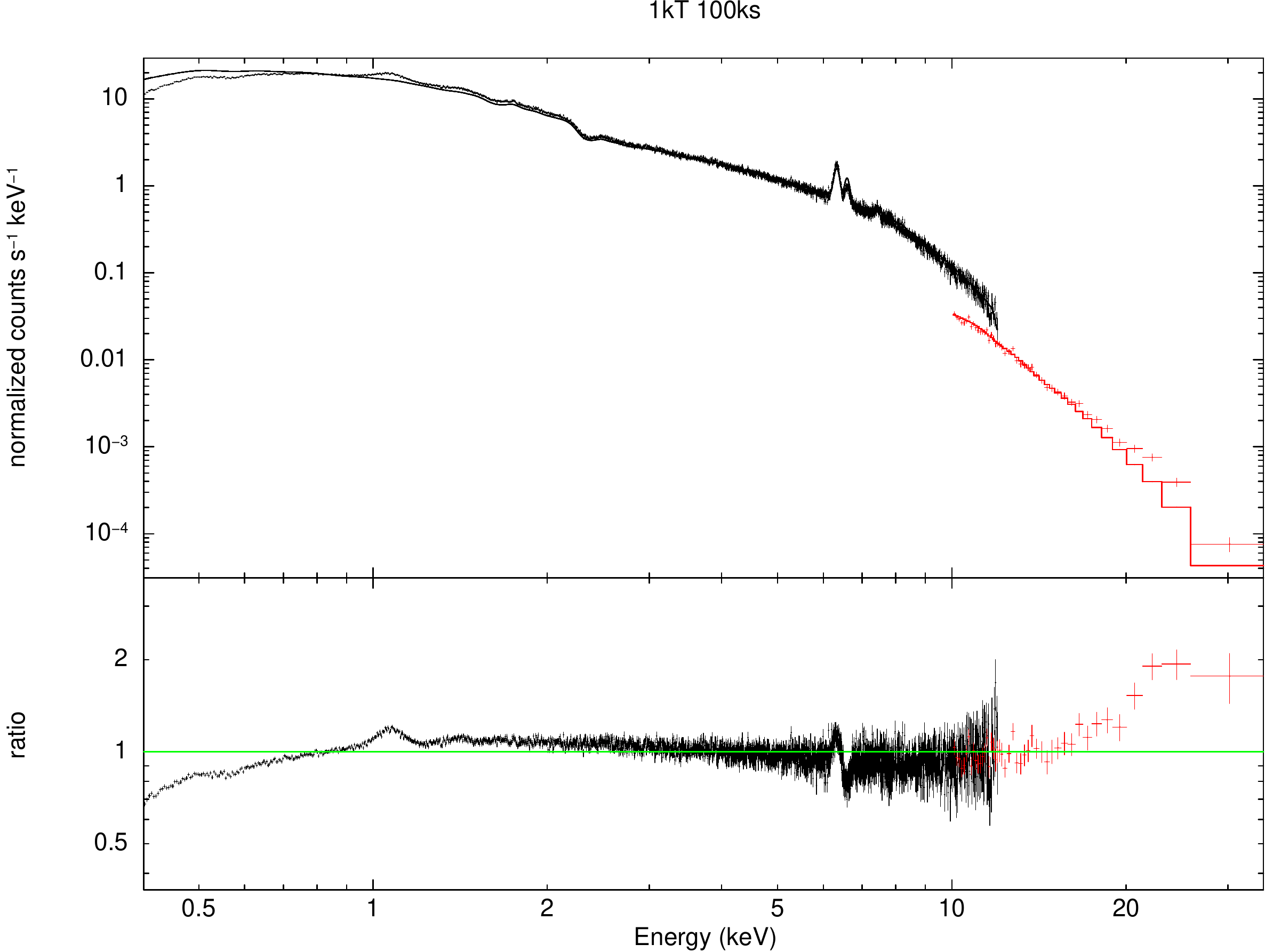}
\includegraphics[width=0.42\textwidth]{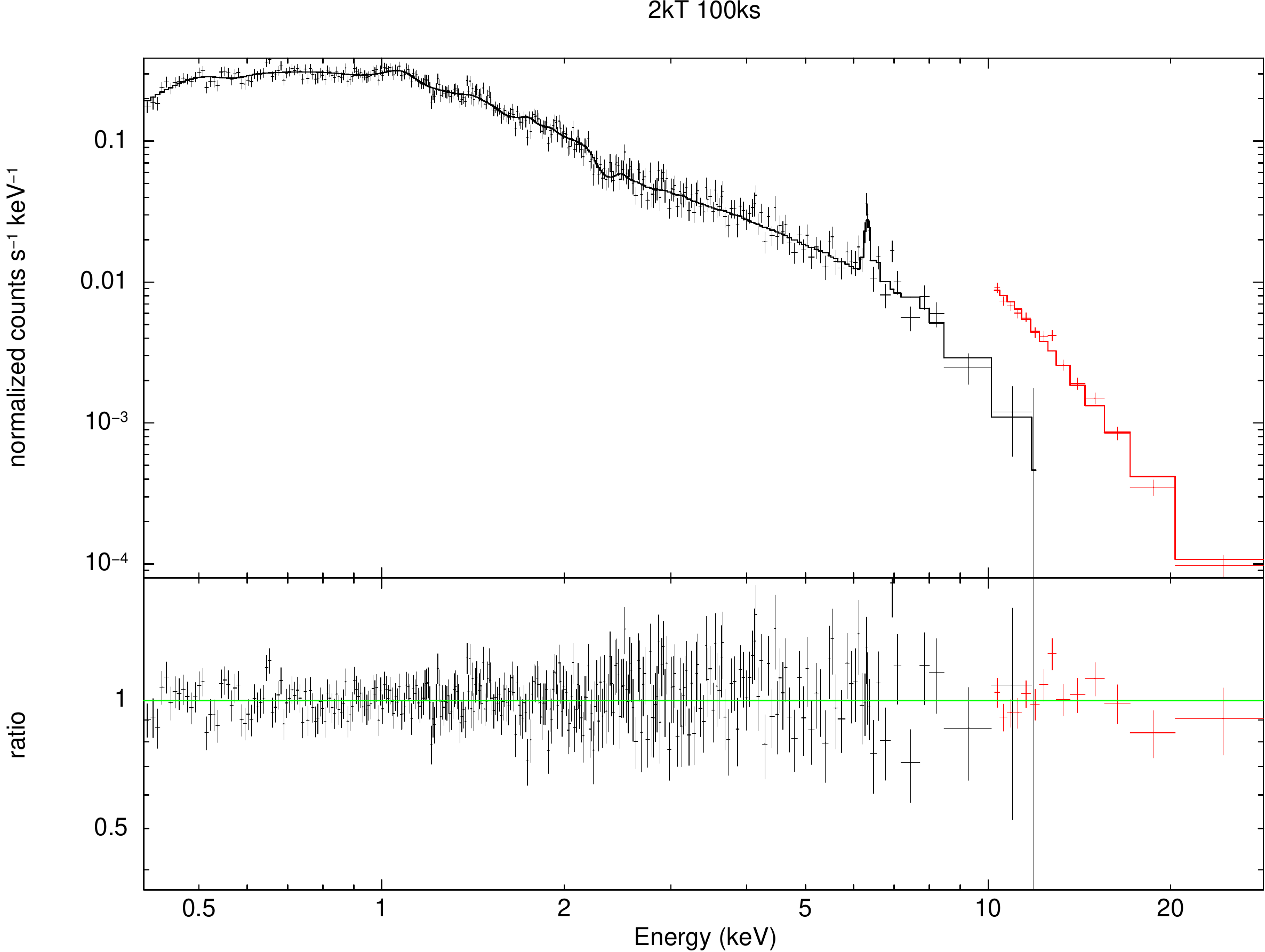}
\includegraphics[width=0.42\textwidth]{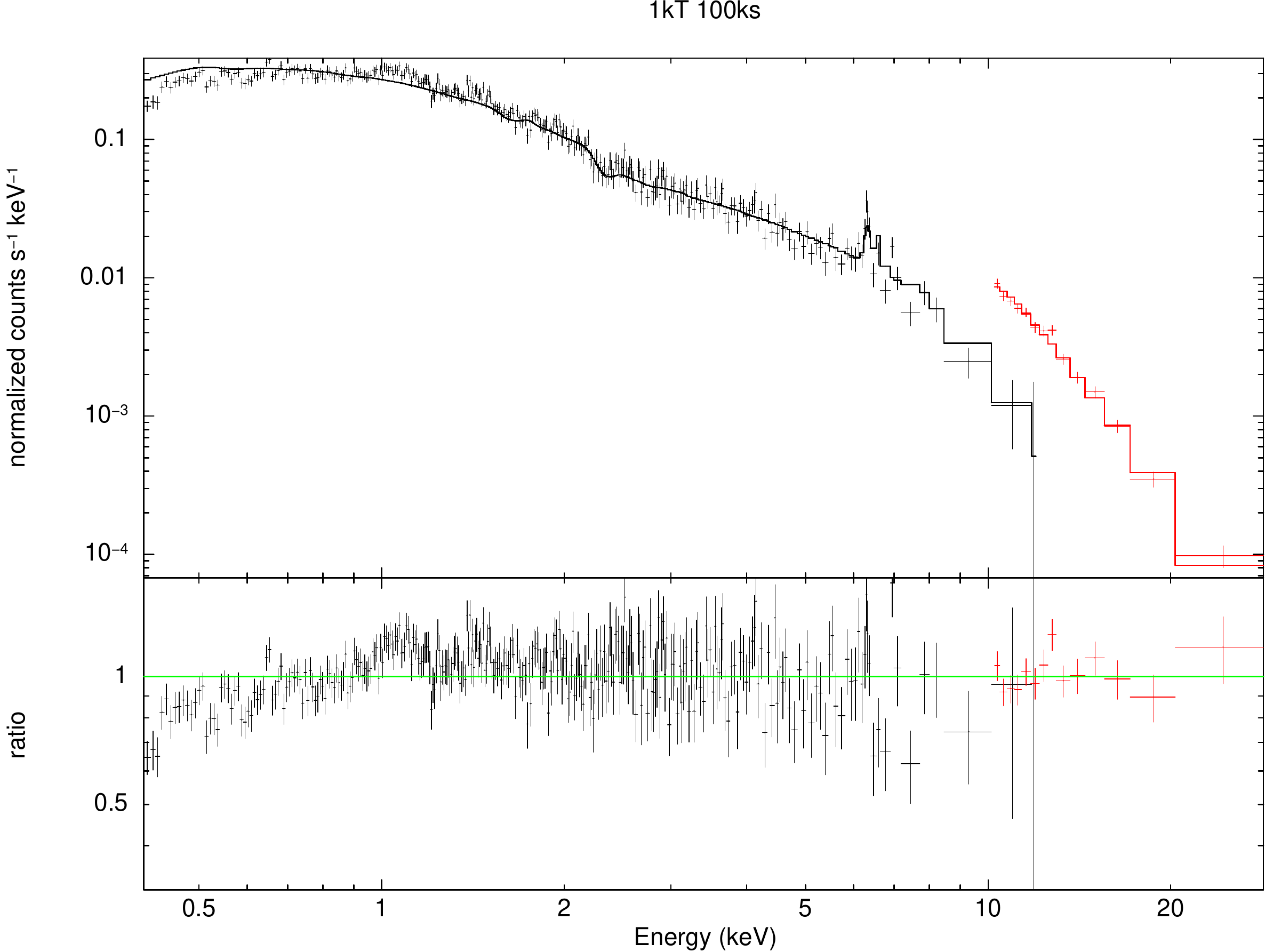}
\caption{
Simulated SXI and HXI spectra (100 ks) of the two-temperature ICM (4.7
keV and 19.2 keV) at the center of A3667, fitted with 
two-temperature model (top left, $\chi^2 / d.o.f = 2500.7 / 2519 =
0.993$; null hypothesis probability = 0.598) and one-temperature
model (top right, $\chi^2 / d.o.f = 37572.6 / 2522 = 14.9$; null
hypothesis probability = 0.000).  Another simulation at the adjacent HXI
region is also fitted with two-temperature model (bottom left,
$\chi^2 / d.o.f = 580.8 / 572 = 1.015$; null hypothesis probability =
0.391) and one-temperature model (bottom right, $\chi^2 / d.o.f =
960.9 / 575 = 1.671$; null hypothesis probability = 0.000).  }
\label{fig-highe4}
\end{center}
\end{figure}

To simulate HXI and SXI spectra from the center of A3667, we extract the
parameters from Nakazawa et. al.  (2009) assuming a two-temperature
thermal emission (2kT) model with temperatures of 4.7 keV and 19.2 keV
(the most probable temperature of the very hot gas inferred in this
paper). We also simulate the spectra toward an adjacent region $9'$ away
from the cluster center, assuming an azimuthally
averaged brightness profile, to examine if the fainter emission from
this region is still detectable. The simulated spectra are fitted either
with a 2kT model or a single-temperature (1kT) model; the temperature,
abundance and normalization are free parameters, whereas the neutral
hydrogen column density ($N_{\rm H} = 4.7 \times 10^{20}$ cm$^{-2}$) and
the redshift are fixed. Figure \ref{fig-highe4} shows that the
broad-band HXI and SXI data will enable us to clearly disentangle the
two temperature components not only from the cluster center but also
from the fainter adjacent region.

\subsubsection{Coma cluster} 

\begin{figure}[t]
\begin{center}
\includegraphics[width=0.5\textwidth,clip,angle=-90]{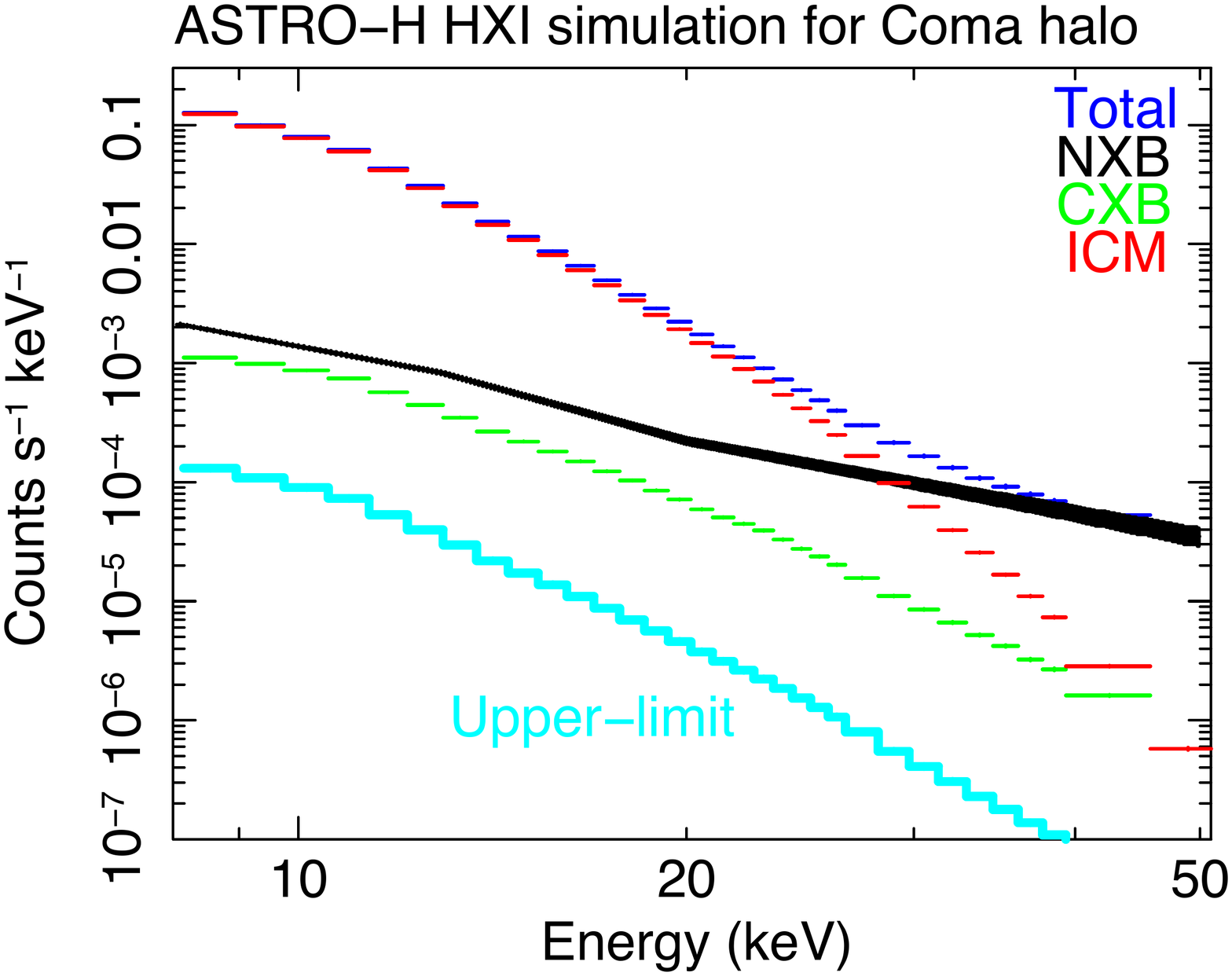}
\caption{Simulated HXI spectrum (blue) from the center of the Coma
cluster (500 Ms exposure) with 7\%, 5\%, and 7\% uncertainties in the
NXB, the CXB, and the cross normalization between the SXI and the HXI,
respectively. Also shown is each spectral component, the ICM (red), NXB
(black), CXB (green), and 90\% upper limit of non-thermal power-law
emission (cyan).}
\label{fig-highe3}
\end{center}
\end{figure}

We further explore the detectability of inverse-Compton radiation from
the radio halo of the Coma cluster. To quantify the impacts of
systematics that dominate the error budget, we first create simulated
spectra of SXI and HXI for an extremely long (500 Ms) exposure taking
into account the NXB, the CXB, and the thermal ICM component with
$kT=8.62$ keV. The CXB is modeled in the same manner as Nakazawa et
al. (2009).  Figure \ref{fig-highe3} illustrates the simulated HXI
spectra of each component together with the 90\% upper limit that can be
placed on the non-thermal component with $\Gamma=2.0$. Table
\ref{tab-upperlimit} further shows how the expected sensitivity from the
HXI and SXI data varies with uncertainties of the backgrounds
and the cross normalization between the SXI and HXI.
We also performed simulations for 
500 ks (which is close to the exposure discussed in
Section \ref{subsec-coma} for an SXS investigation of the ICM
turbulence) and checked that statistical errors will increase the upper
limits listed in this Table only by $\sim 0.2 \times 10^{-16}$ erg
s$^{-1}$ cm$^{-2}$ arcmin$^{-2}$.

In the most conservative case with 7\%,  5\%, and 7\% 
uncertainties in the NXB, the CXB, 
and the cross normalization between the SXI and HXI,
respectively, the expected upper-limit (90\%) on the
non-thermal intensity is 
$8.9 \times 10^{-16}$ erg s$^{-1}$ cm$^{-2}$ arcmin$^{-2}$
at 20--80 keV for a 500 ks exposure. This will be a factor of 
$\sim 6$
improvement over the previous limit by \suzaku for the same
energy band, $5.3 \times 10^{-15}$ erg s$^{-1}$ cm$^{-2}$ arcmin$^{-2}$
(Wik et al. 2009), and will correspond to the lower bound on the
magnetic field of $B \sim 0.4~\mu$G for $\Gamma=2.0$.  
Using Faraday rotation measurements, Bonafede et
al. (2010) estimate the magnetic field strength at the center of the
Coma cluster to be $B \sim 5~\mu$G, which would suggest that, given the
sensitivity limit of \astroh presented above, inverse-Compton radiation
is unlikely to be detected from the giant halo in Coma. Estimates for
other clusters with radio halos suggest that the expected inverse
Compton signal is similarly below the HXI sensitivity, if the current
Faraday rotation estimates of $B$\/ are representative of the
volume-averaged field. However, the X-ray signal may be considerably
higher if, for example, the magnetic field and the cosmic ray population
are distributed differently in the cluster volume, which is not an
improbable scenario. Since merging clusters with giant radio halos (such
as Coma, A754, A2319, Bullet) are likely to be the targets of long SXS
observations, with time we should accumulate enough HXI data to test
such scenarios.

  \begin{table}[t]
  \begin{center}
  \caption{Sensitivity of HXI+SXI on the non-thermal emission with
   $\Gamma=2.0$ from the center of Coma for a range of systematic
   errors, neglecting statistical errors (assuming an exposure time of
   500 Ms).}
  \begin{tabular}{cc}
  \hline
  \hline
Systematic errors & Upper limit (90\%, 20--80 keV) \\ 
 &  [erg s$^{-1}$ cm$^{-2}$ arcmin$^{-2}$] \\
  \hline
 NXB 3\% & $4.0 \times 10^{-16}$\\
 NXB 5\% & $5.3 \times 10^{-16}$\\
 NXB 7\% & $6.5 \times 10^{-16}$\\
 NXB 3\%, CXB 5\% & $4.6 \times 10^{-16}$\\
 NXB 5\%, CXB 5\% & $5.8 \times 10^{-16}$\\
 NXB 7\%, CXB 5\% & $7.1 \times 10^{-16}$\\
 NXB 7\%, CXB 5\%, Cross-norm 1\% & $7.4 \times 10^{-16}$\\
 NXB 7\%, CXB 5\%, Cross-norm 5\% & $8.3 \times 10^{-16}$\\
 NXB 7\%, CXB 5\%, Cross-norm 7\% & $8.7 \times 10^{-16}$\\
  \hline 
  \end{tabular}
\label{tab-upperlimit}
  \end{center}
  \end{table}

\ \\



\subsection{Lessons Learned from \nustar}

It is important to take advantage of the lessons learned from the
cluster observations with \nustar.  So far (as of April 1, 2014), there
is one publication reporting the observations of galaxy clusters with
\nustar (Wik et al. 2014).  That paper, reporting the observations of
the Bullet Cluster, 1E 0657--558 at $z=0.296$, contains a very
extensive discussion of background and systematic effects associated
with hard X-ray imaging observations of extended sources.  The \nustar
bandpass covers roughly 3 - 80 keV, so it is quite comparable to \astroh
HXI. The effective area is also comparable, peaking at $\sim 250$
cm$^{2}$ per module.  The Point-Spread Function of \nustar, $\sim 1'$,
is somewhat better than of the HXI, so the image quality is likely
somewhat better.  \nustar is on an equatorial orbit, so the particle and
activation backgrounds are lower than those for the orbit expected for
\astroh (here, the background rejection scheme adopted in the design of
the HXI detectors might mitigate the difference).  However, \nustar
lacks the light baffle in front of the instrument, and the stray Cosmic
X-ray Background partly contaminates the detector. In \astroh, the CXB
shield has been designed to fully block such stray cosmic X-rays.  All
those effects need to be taken into consideration in deriving the
expected sensitivity of the HXI for extended sources, and in the
comparison of the two instruments.  In any case, the \nustar
observations of the Bullet Cluster indicate a multi-temperature
structure, with the hot component at $kT \sim 15$ keV, and only a
marginal evidence of non-thermal emission.  All this means that the
very accurate determination of background from all sources will be key
in any conclusive hard X-ray measurements.

Again, two main ``direct'' goals resulting from the detection / measurement of non-thermal 
emission in clusters are:  the volume-integrated distribution of energetic 
particles in the cluster, and the average magnetic field.  However, those can be strongly enhanced 
by taking advantage of data in other bands.  

(1) Measurement of cluster magnetic fields:  here, the most important will be 
correlative studies with radio bands.  For relatively nearby clusters, the PSF 
of the Hard X-ray Imagers will be sufficient to perform detailed 
spatial cross-correlations with radio maps.  In addition, radio polarization observations 
will be very valuable.  Such observations provide a clue to the structure and 
physical scale of the coherence of magnetic fields.  

(2) Process of thermalization of the cluster plasma:  given the excellent complement of 
X-ray instruments on board of {\it ASTRO-H}, we should be able to address the question 
of thermalization of the intra-cluster gas.  One aspect that will weigh in on 
understanding this process is the strength and structure of magnetic field 
discussed above, but importantly, the distribution of temperature of the thermal 
plasma will play an important role.  Since the SXI and the SXS observations 
will be secured as all instruments are co-aligned, no additional data will be necessary.  
However, good measurements of the temperature structure will require that 
the instruments are accurately cross-calibrated, with an additional requirement 
that the off-axis effective area of {\it all} imagers is well-determined.

\bigskip

\section{Chemical Composition and Evolution}
\label{sec-chem}

\subsection*{Overview}
The chemical composition of the hot gas in galaxies and clusters of
galaxies contains valuable information about the origin of chemical
elements and their distribution during the evolution of the universe. It
is currently believed that most heavy elements were produced by type Ia
and core-collapse supernovae during the major epoch of star formation around
$z=2-3$. The chemical abundances 
provide clues about 
stellar populations, the history of star formation,  the Initial-Mass Function (IMF), and
the type Ia explosion mechanism.  We summarize the expected
 science results from spatially resolved abundance measurements in bright clusters
 of galaxies. Many of these clusters are also discussed in other
 scientific contexts in this paper.  Typically, 100 ksec exposures are
 required for detecting a range of elements in cluster cores.
In the data with the highest statistics we may detect rare elements such
as Na, Al, Cr, and Mn. Measuring the evolution of elements as a function
of redshift would be
challenging.\footnote{Coordinators of this section: J. de Plaa, K. Sato}





\subsection{Background and Previous Studies}

Chemical elements with atomic number 6 (C) and above have all 
been produced by stars and supernovae during the history of 
our Universe. The first stars that started the epoch of re-ionization 
around $z=10$, known as Population III stars, were also the first 
objects to create heavier metals. Their total contribution 
to the enrichment of the current-day Universe is, however, thought 
to be small \citep[e.g.][]{matteucci2005}. 
The bulk of the metals are produced by the stars formed around 
the peak of the universal star-formation rate at $z \sim 2-3$.
Core-collapse (SNcc), 
which would include the Population III stars,  and type Ia 
supernovae have very different abundance ratio patterns.
The initial metallicity of the progenitors, as well as the initial mass
function of the stars that explode as SNcc, also influence the relative
abundances in the present-day universe.  
The lighter metals, from O to Si and S, are mainly produced in massive
stars and ejected in SNcc at the end of their life time. Since the metal
yield of these supernovae depends on the mass of the progenitor star,
which ranges between roughly 8 M$_{\odot}$ and 100 M$_{\odot}$, the
total metal yield of the stellar population depends on the Initial-Mass
Function (IMF).  The metals from Si to Fe and Ni are
also produced during type Ia supernova explosions (SNIa).  The binary
configuration of SNIa progenitors and their explosion mechanism is,
however, still poorly known \citep[see e.g.][for a review]{howell2011}.
Current population synthesis models \citep{ruiter2009} suggest that most
of the type Ia supernovae have a white dwarf binary progenitor and
explode gigayears after the initial star burst.  Binaries with one white
dwarf and a different companion may produce a type Ia in the first few
gigayears after the star burst.  These models contain, however, several
assumptions and uncertainties.  In addition, uncertainties in the SNIa
explosion mechanism cause uncertainties in the yields of about a factor
of two.  Observational constraints, therefore, would help to understand
the SNIa origin and explosion mechanisms.

The ICM in clusters of galaxies serves as the repository for type
Ia and SNcc ejecta.  Since the formation of the ICM around $z \sim 2$,
star formation in the member galaxies has been suppressed. The hot gas
in clusters, which is heated through AGN feedback, also heats the cold
molecular clouds in the member galaxies, preventing them to collapse and
form stars \citep{gabor2010}.  Once the halo reaches a certain
temperature and mass (around $z \sim 2$ for most clusters), star
formation is effectively quenched.  Therefore, the cluster core mostly
contains elements produced by the stellar population formed during the
$z=2-3$ star burst, and has a different chemical history than our own
Galaxy.  Measuring the chemical composition of the ICM provides a
valuable insight in the production of metals by this early stellar
population. 
On the other hand, recent IR and radio observations have found that
some cD galaxies in the BCG have relatively higher 
star formation rate \citep[see e.g.][]{Odea08, McNamara13}, 
which are associated with the AGN activity and the feedback mechanism.


Measuring abundances in clusters has several advantages over
observations of individual supernovae or supernova remnants (SNRs).  In
our own Galaxy, supernova explosions are rare and there are only a
handful of bright SNRs available for abundance studies.  The
progenitor is usually difficult to determine, and the shocked ejecta
becomes mixed with material swept-up from the surrounding ISM, making it
difficult to estimate yields.  Furthermore, extragalactic supernovae are
faint in X-rays and are thus difficult to observe.  Optical abundance
measurements are often unreliable while the explosion is ongoing because
elements lying below the explosion's photosphere may be unobservable.
In contrast, the hot ICM of clusters is optically thin and contains the
accumulated ejecta of billions of supernovae, providing an average yield
from SN over time and space.  Most supernova products have not been
recycled into new stars since $z \sim 2$.  Therefore, the measured
abundances represent the average yield of the $z=$2--3 stellar
population.  In the end, the SN/SNR observations and the cluster
abundance measurements will play
complementary roles in 
providing information about supernova explosion physics.

The soft X-ray band is amenable to abundance studies. The ICM is
in or very close to collisional ionization equilibrium and is therefore
relatively easy to model. Moreover, all elements between C and Zn show
emission lines between 0.1 and 10 keV. With {\it ASCA}, it was possible
for the first time to resolve lines of highly abundant elements, like
Si and Fe \citep{mushotzky1996}.  \xmm and \suzaku enabled us
to study the abundance patterns of O, Mg, Si, S, Ar, Ca, Fe and Ni.  With
{\it Suzaku}, \citet{tamura2009} detected the He-like Cr line from the
cool core of the Perseus cluster.  Using {\it ASCA}, and later \xmm and 
{\it Suzaku}, several authors \citep[for
example][]{fukazawa1998,deplaa2007,sato2007} placed interesting constraints
on supernova models using abundance measurements. For example, \citet{deplaa2007} found that the
Ar/Ca ratio was sensitive to the
assumed SNIa model.  Because a number of common elements from O to
Ni were measured, these studies also provided estimates of the SNIa/SNcc
ratio. The SNIa contribution is estimated to be about 30\% of the total
contribution, which is consistent with the SNIa/SNcc ratio of 0.2-0.4 as
determined from studies at optical wavelengths.

The observed values of the Ni/Fe ratios in the ICM have a significant
scatter \citep[e.g.][]{degrandi2009, Matsushita13b}.  In contrast, in
our Galaxy, both SNcc and SNIa have synthesized Ni in the same way as
Fe, since the [Ni/Fe] of stars is $\sim$ 0, with no dependence on [Fe/H]
\citep{feltzing1998, Gratton2003}.  The observed scatter in the Ni/Fe
ratio observed in several cool core clusters indicates that Ni synthesis
in cD galaxies might be different from that in our Galaxy. However, with
CCD detectors, both Ni-L and K lines are blended into the Fe-L lines and
He-like Fe-K line at 7.9 keV, respectively, and the derived Ni abundance
might have some systematic uncertainties.

While most heavy elements are produced in supernovae, C and N are 
most likely produced by and ejected from
intermediate-mass and massive stars. Their origin is, however,
still subject to debate. Because C and N are relatively light elements, 
their strong K-shell transitions are found in the soft X-rays. 
They both ionize completely at high temperatures, therefore 
their line emission is best detected in elliptical galaxies and groups 
with temperatures below 2 keV. Using the RGS instrument on
\xmm, \citet{Werner06} and \citet{grange2011} detected N
in the spectra of M~87, NGC~5044, and NGC~5813. Its relatively high 
abundance would be difficult to produce by supernova explosions, which are
presumably rare in these objects, but
would be more easily produced by intermediate-mass or massive stars. 

Recent X-ray observations have yielded measurements of the metal abundance 
distributions in the ICM based on spatially resolved spectra.
The metals in the cool cores of clusters represent a mixture of those present in the ICM and later supplies from the cD galaxies. 
The latter contain Fe synthesized by SNIa and originating from stars through stellar mass loss.
However,  
in these regions, the distribution of metals is more extended compared
to that of the stars. Processes such as jets from a central AGN or the sloshing
of cD galaxies in the cluster's gravitational potential may eject metals from cD galaxies. 
Therefore, the metal distributions in the cool-cores
should be closely related with turbulence and bulk motions which will be
measured  with {\it ASTRO-H}

Measuring radial profiles of metals out to larger radii helps us understand the past
chemical evolution process. Outside the cool-cores, the metals are accumulated over much longer time
scales.  The gas in groups and clusters has a very similar value of the Si/Fe ratio, around 1--1.5
\citep[e.g.][]{sato2010, Matsushita13a, Matsushita13b}.  The radial
profiles of the Fe abundances in the ICM outside the core regions are
relatively flat \citep{Matsushita11, Matsushita13a, werner2013b} and 
both Si and Fe in the ICM are more extended
than stars \citep{Matsushita13a}. These results indicate that a
significant fraction of metals is synthesized in an early phase of
cluster evolution, certainly before the last merger epoch.  Since Si and
Fe are synthesized both by SNcc and SNIa, abundance measurements
of O, Ne and Mg are needed to obtain unambiguous information on their origin.
With some \suzaku observations, the O/Fe and Mg/Fe ratios
suggest an increase with radius.  However, with the CCD detectors, the
emission lines from our Galaxy and surrounding Fe-L lines cause
systematic uncertainties in the O and Mg abundance measurements.

Galactic winds, ram-pressure stripping, AGN activity,
and sloshing play a role \citep[e.g.][]{strickland2000, schindler2008}
in the enrichment, diffusion, and mixing of the metals, as do
contribution from intracluster stars \citep[e.g.][]{zhang2011,galyam2003}.
Although this stellar component is difficult to observe, it 
appears to contribute a substantial amount of metals to the ICM in the core.
Recent numerical simulations of the metal enrichment of 
the ICM indicate that it is difficult to know how the elements 
have been ejected from galaxies into the ICM;
numerical simulations are currently unable to reproduce
in detail the elemental distributions in the ICM  
\citep[e.g][]{kapferer2007,Planelles14}.

\subsection{Prospects and Strategy}

\astroh is well suited to abundance measurements.  {Although its
effective area is modest, deep observations of bright clusters
using the high-spectral resolution SXS
will provide accurate abundance measurements of the common elements from N to
Ni.}
Elements with strong lines, like O, Si, S,
and Fe, are relatively easy to measure. The challenge will be to measure
abundances of elements with weak and/or blended lines.
\astroh will be adept at the measurement of abundances of elements associated with
dense line complexes, such as Ne, which has lines in the Fe-L
complex.  

Abundance measurements of a wide range of heavy and
light elements will place tight constraints on the progenitor and current stellar populations in
clusters and the mechanisms that produced the chemical elements. 
O, Ne, and Mg are mainly ejected by SNcc. However, their relative abundances depend
not only on the details of the explosion, but on the initial
initial metallicity of the stars and the IMF.  Heavier elements
like Ar, Ca, Fe, and Ni, are produced in SNIa and their abundance ratios
depend on the SNIa explosion mechanism and perhaps also their initial
metallicity.  Using accurate \astroh abundances, we can learn more about
the dominant stellar population that was formed in the cluster around
$z=2-3$.  
Within cool cores, the contributions of ongoing enrichment via stellar
mass loss and SNe Ia from central galaxies may be significant 
 \citep[e.g.][]{Matsushita03,boehringer2004}, and can therefore
 be studied with \astroh .
 
SXS may detect weak lines from rare elements in the ICM
such as Cr, Mn, Al, and Na in high signal to noise observations. Nucleosynthesis models
 indicate that the abundances of odd-Z elements like Al and Na 
 depend on the stellar metallicity \citep{nomoto2006,Kobayashi06}. 
 These elements are enhanced by the surplus of neutrons
 in $^{22}$Ne, which is synthesized by the CNO cycle during He-burning.
 Therefore, the metallicity of Al and Na of the ICM may trace the metallicity
 of the underlying stellar population. Similarly, the Cr/Mn ratio is correlated
 with the initial metallicity of SNIa \citep{badenes2008}.  In
 addition, nucleosynthesis models and observations of Galactic stars
 indicate that Mn is mostly produced by SNIa rather than SNcc.  Therefore, 
 the Cr/Mn ratio can be used to probe the relative contribution of SNcc and SNIa 
\citep{nomoto2006}. Such abundance measurements  will be possible
in bright cD galaxies located in cluster cores, such as M~87, the Centaurus
cluster  and the Perseus cluster. Rare element abundances and the
 level of velocity broadening, which tends to wash-out the lines, are poorly
 understood, making feasibility estimates uncertain.  This problem
 is discussed in Section~\ref{sec:cluschem-rare}.

The SXS is a non-dispersive spectrometer that provides full spectral resolution 
for extended sources. In contrast, the spectral resolution of extended sources 
observed using dispersive spectrometers, such as the gratings on \xmm and {\it Chandra}, 
is eroded as the lines are blended with the spatial extent of the source. Therefore, 
the spectral resolution depends on the angular extent of the source. In addition, 
every spectral line emitted by the source may have its own spatial distribution, which 
complicates correct modeling of the line profiles. The \astroh SXS will not suffer 
this disadvantage, allowing much more accurate measurements of the strength of 
individual lines in extended sources.

A possible complication 
will be the currently unknown velocity profile of the spectral lines, which
will be detected for the first time in these high-resolution
spectra. Bulk motions of different pockets of gas in the core may cause
complicated line profiles, especially when these pockets have a
different enrichment history. Their velocity distributions
may be difficult to model using the observed the line profiles, resulting in
small systematic errors in the abundance measurements.

In order to investigate the role of elliptical and spiral galaxies, and
their (different) stellar populations, to the chemical enrichment of the ICM, it will
be necessary to observe individual elliptical and star-burst galaxies,
like NGC~5044 and NGC~253. Elliptical galaxies contain relatively old
stellar populations and would mainly produce SNe Ia and AGB star
products. Star-burst galaxies are dominated by SNcc and are likely to
have different chemical compositions. These differences will provide
better insight into the balance between the sources of metals.  As for
starburst galaxies, metals in the inter-stellar medium are thought to
escape into the inter-galactic space via outflows heated by SNcc.
Therefore, metal abundances and their ratios play key roles for
investigating how the metals are transported into the inter-galactic and ICM
space.

\subsection{Targets and Feasibility}
\label{sec:cluschem-targets}

Measurements of elemental abundances in bright clusters can be
done simultaneously with those of gas motions discussed in other
sections of this paper.  We intend to use this data supplemented by additional
observations described in detail below.  The proposed targets are 
nearby bright galaxies and galaxy clusters that span approximately a
decade in gas temperature ($kT\sim0.8$--8 keV). The sample is intended
to investigate metal enrichment processes over a wide range of
mass scales from individual galaxies to the largest galaxy clusters.
In this section, we use 68\% confidence limits, unless stated otherwise.

\subsubsection{The metals in the cool cores}

\begin{figure}[t]
\begin{center}
 \includegraphics[width=0.42\textwidth]{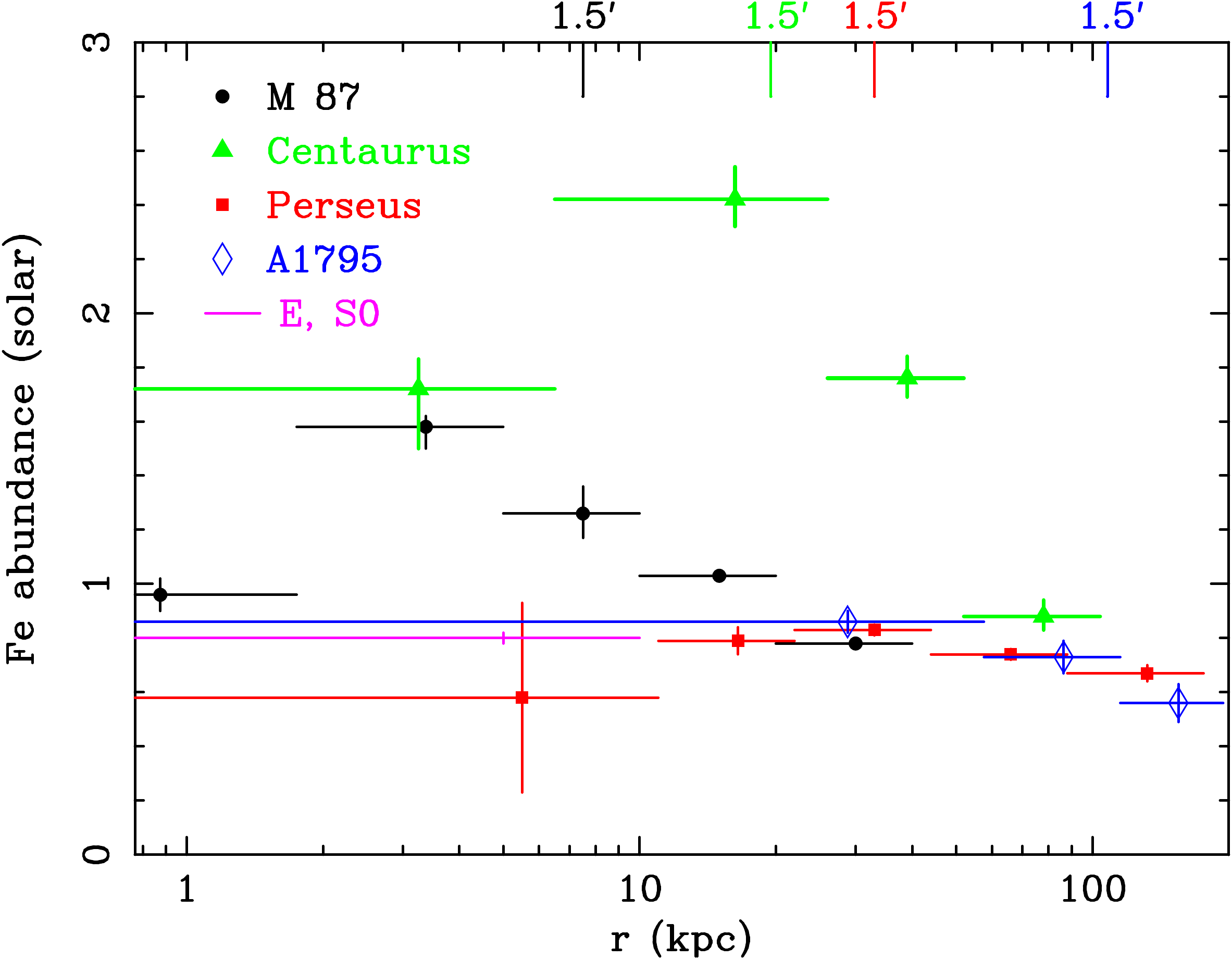}
  \includegraphics[width=0.42\textwidth]{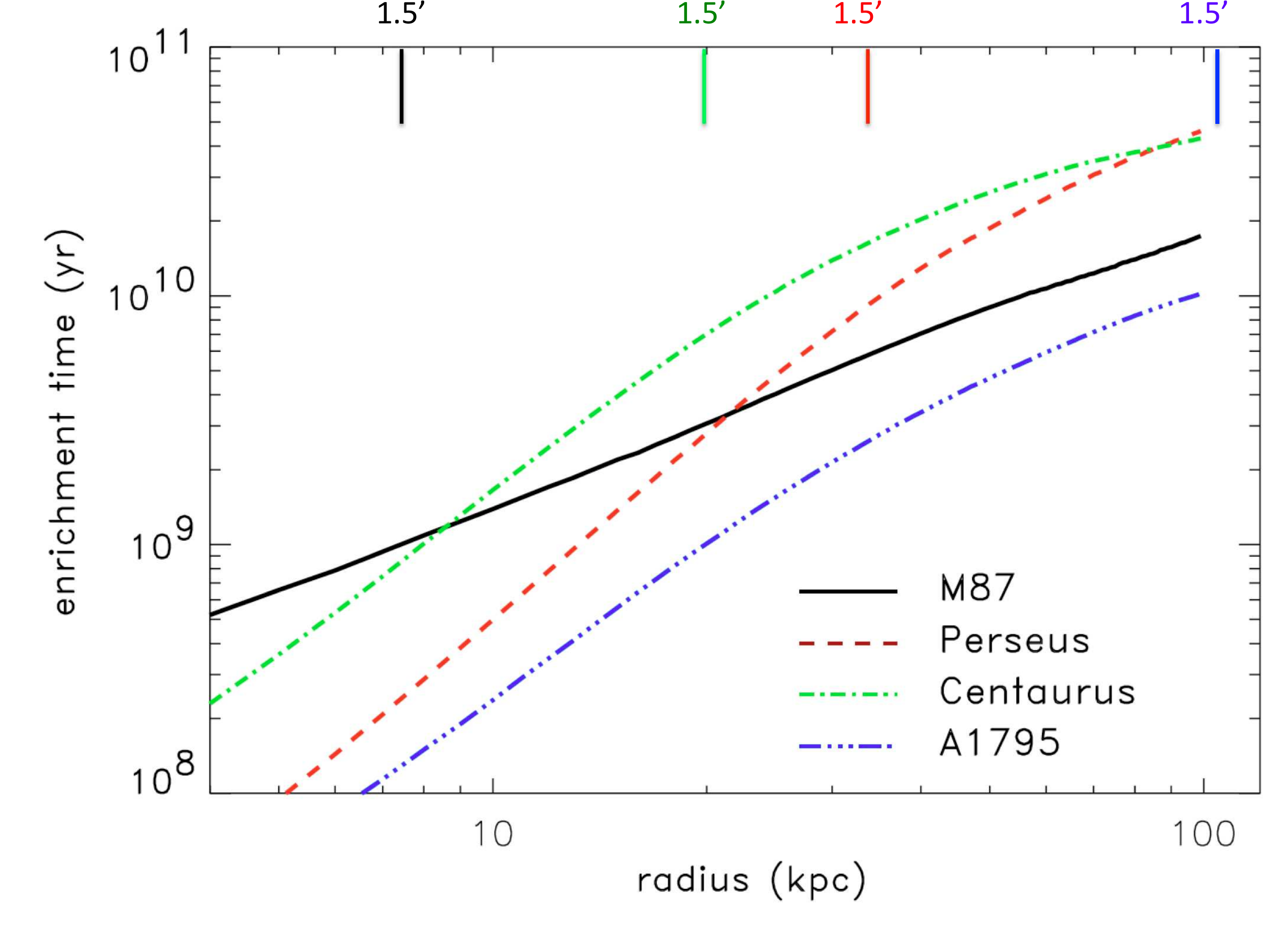}
  \caption{{\it Left}: The radial profiles of the Fe abundances in the
  center of the Virgo cluster (around M~87) \citep[closed black filled
  cirles,][]{Matsushita03}, the Centaurus cluster \citep[green filled
  triangles,]{Matsushita07}, Perseus cluster (red filled squares), and
  A1795 cluster \citep[blue open diamnods,][]{Matsushita11} observed
  with {\it XMM-Newton}.  The horizontal magenta lines show the weighted average of
  the Fe abundance in the hot ISM of 17 early-type galaxies observed
  with \suzaku \citep{Konami14}. The short vertical lines indicate
  1.5$'$ from the center of each cluster. ~{\it Right}: The ratio of the
  integrated Fe mass to the present Fe production rate by SN Ia for the
  central Fe abundance peak of four clusters by metals from the cD
  galaxies \citep{Boeringer04}.  The short vertical lines indicate
  1.5$'$ from the center of each cluster.  } \label{fig:Feprofile}
\end{center}
\end{figure}

The Virgo ($\sim$ 2 keV, $z$=0.004), Centaurus
($\sim$ 4 keV, $z$=0.010), and Perseus clusters ($\sim$ 6 keV,
$z$=0.018) are nearby, bright cool-core clusters.  Their metal abundance
patterns have been studied in detail using lower
resolution CCD spectra obtained with \xmm, \chandra, and \suzaku
\citep{Matsushita03, Sanders06, Matsushita07, tamura2009, Million11,
Sakuma11, Matsushita11}.  Figure \ref{fig:Feprofile} shows the radial
profiles of the Fe abundance of M~87, the Centaurus cluster, the Perseus
cluster, and a more distant cluster A~1795 at $z=0.062$, observed with
{\it XMM-Newton}.  Observations of relatively nearby clusters ($z<0.03$) 
are required to spatially resolve the cool cores and derive metal
distributions in these regions with SXS.
\astroh observations of the centers of these
clusters will provide the best statistics, and provide benchmark
measurements of the thermal and chemical structure of the intra-cluster
medium (ICM), and of the ICM dynamics as described in Sections
\ref{sec-perseus} and \ref{sec-virgo}.

\subsubsection*{Metal distribution in cool-cores and AGN feedback}

As shown in Figure \ref{fig:Feprofile}, within the central 1.5' of
M~87, the enrichment timescale of the Fe mass in the ICM,
or the ratio of integrated Fe mass to the present Fe production 
rate by SN Ia, is only
1 Gyr.  This short timescale indicates that the gaseous halo of M87
and its composition in this region may be explained by ongoing
stellar mass loss and SNIa.  \citet{simionescu2008a} suggested that AGN
activity in M~87 is able to lift enriched material from the center
further out into the ICM (see Sec.~\ref{sec-virgo} for details). 
The simulations  in Section \ref{sec-virgo} show that we will 
be able to accurately measure the
metallicity of the uplifted gas independently from that of the 
surrounding ICM, determining the Fe abundance of the cool 
gas to better than 10\% accuracy and thus robustly determining 
how gas motions induced by the 
AGN spread out metals produced in the central galaxy. 

The distributions of Fe in cool cores vary
from cluster to cluster (Figure \ref{fig:Feprofile}).  The enrichment
time scale within the central 1.5$'$ for the Centaurus cluster and the
Perseus cluster are an order of magnitude longer than that for M~87
(Figure \ref{fig:Feprofile}).  We will hence observe metals accumulated
over a much longer time scale in these clusters.  The Fe abundance within
1.5$'$ of the center of the Centaurus cluster is exceptionally
high, being roughly twice the solar value. 
 This is much higher than in other
clusters including the Perseus cluster. 
\citet{Panagoulia13} suggest that the Fe abundance has a sharp drop 
within the central 5 kpc region, and would be uplifted by the 
bubbling feedback process to 10--20 kpc, where the Fe abundance becomes high. 
 The difference in the Fe
distributions may be related to differences in mixing due to AGN feedback
from and sloshing motions.  Therefore, in combination with measurements
of turbulent velocities in these cool-cool cores, we will be able to
study the effect of the AGN feedback.

\subsubsection*{Abundance pattern of SNIa of the Centaurus cluster}

The Centaurus cluster is a compelling target 
for studying the abundance pattern synthesized by past SNIa.  
The  O/Fe and Mg/Fe ratios in the ICM of the core of the Centaurus cluster  
is the lowest among cool core clusters observed with {\it Suzaku} \citep{Sakuma11}.
The very high Fe abundance with the sharp Fe abundance peak and 
the low Mg/Fe ratios suggest that the metals in the center of the Centaurus 
cluster are more dominated from those synthesized by SNIa in the central galaxy
than those in other cool core clusters.

As shown in Figure \ref{fig:Feprofile}, the peak Fe abundance in the
Centaurus cluster is also significantly higher than that in the hot
interstellar medium, ISM, in early-type galaxies, which have much smaller gas
mass to light ratio than cool-core regions, and represent present metal
supply from these galaxies.  The O/Fe and Mg/Fe ratios in the ISM of these 
galaxies are close to the solar ratios and nearly a factor of two higher 
than those in the core of the Centaurus cluster \citep{Konami14}.
The difference in the Fe abundance and abundance ratios
indicates that the metals in the
center of the Centaurus cluster may not be a simple mixture of those in
the ICM and the current ISM in early-type galaxies.  A longer time scale
implies that the ratio of SNIa rate to stellar mass loss rate was higher
in the past.

\begin{figure}[t]
 \begin{center}
  \includegraphics[width=0.44\textwidth]{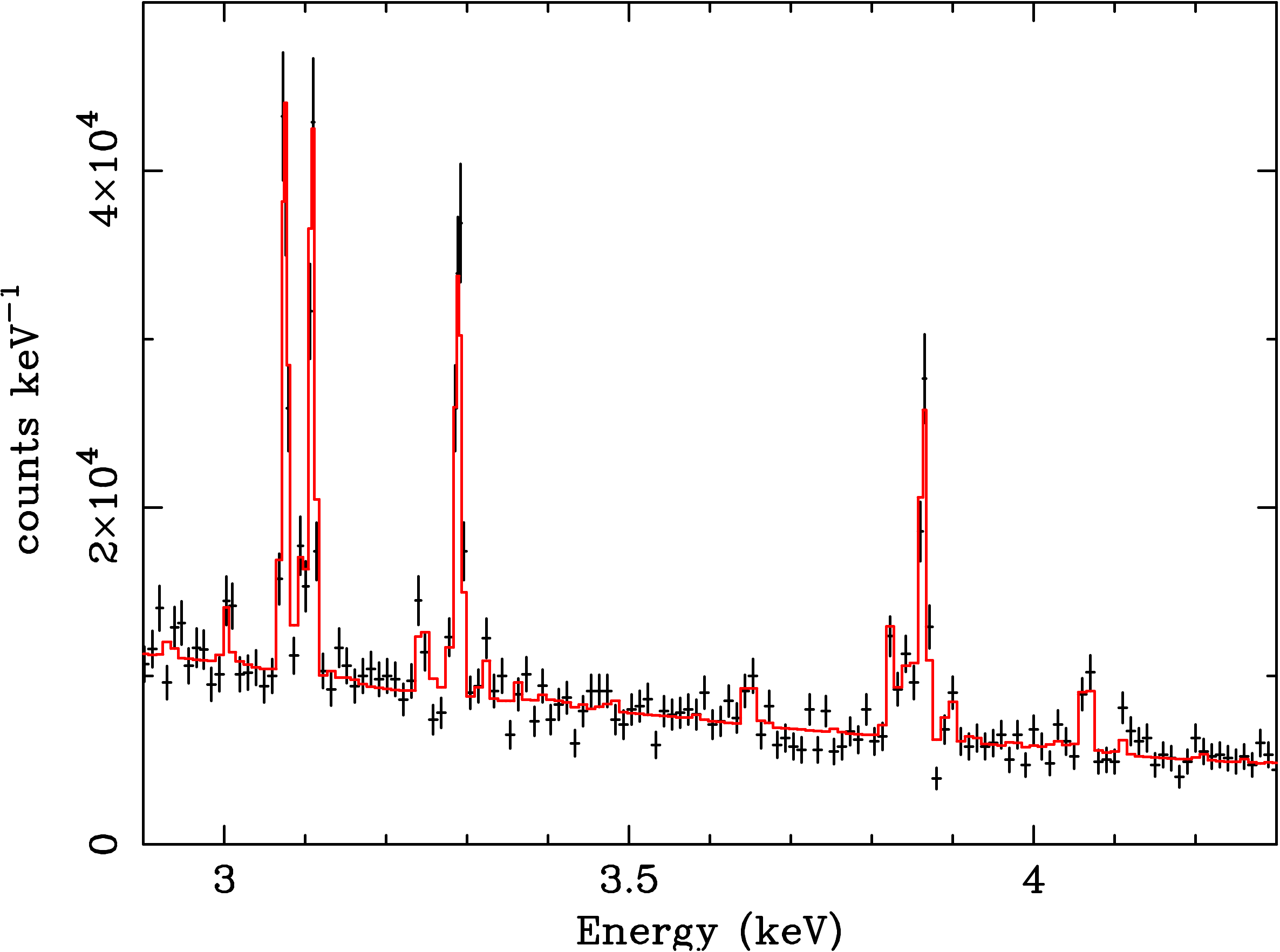}
\includegraphics[width=0.44\textwidth]{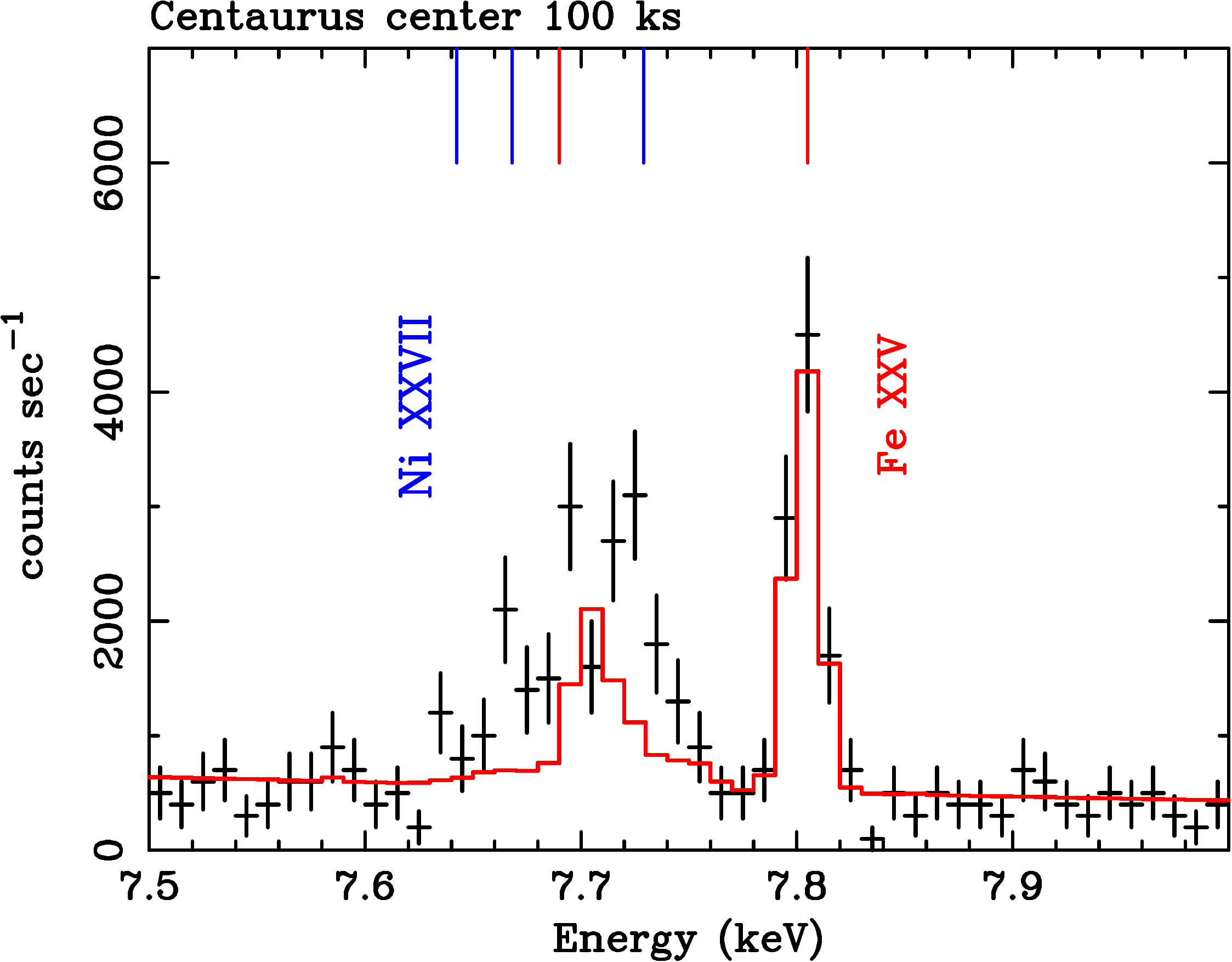}
 \caption{
The simulated SXS spectra at the center of the Centaurus cluster (100
ks) around the Ar and Ca lines (left panel) and the He-like Ni and Fe
lines (right panel).  The solid lines in the right panel correspond to
the models with zero Ni abundances.  The blue vertical lines indicate
the redshifted energies of strong lines in the He-like Ni triplets,
whereas the red vertical lines those of Fe lines.}
\label{crmnni}
\end{center}
\end{figure}

The higher SN Ia contribution to the enrichment in the Centaurus cluster
results in a higher abundance of Ar and Ca, whose abundance ratio is
sensitive to the SNIa model.  With a 100 ks exposure, both He-like and
H-like lines of Ar and Ca from the Centaurus cluster will be easily
detected with SXS as shown in Figure \ref{crmnni}. The total expected
number of photons from Ar and Ca lines are about 1000 and several
hundred, respectively.

With {\it XMM-Newton}, the derived Ni/Fe abundance in the Centaurus
cluster is significantly higher than the solar ratio, while those in the
Perseus cluster and the Coma cluster are more consistent with the solar
ratio \citep{Matsushita13b}.  With CCD detectors, the He-like Ni line at
7.8 keV (rest frame) and the He-like Fe line at 7.9 keV are blended into
a single bump, and the Ni abundance measurements couple with the effect
of resonant line scattering.  In contrast, as shown in Figure
\ref{crmnni}, with \astroh, these lines are separated from each
other. With a 100 ks exposure of the center of the Centaurus cluster, we
will expect a few hundred of photons for the He-like Ni line triplets
and $\sim$200 photons for the He-like Fe line at 7.9 keV.

\subsubsection*{Accurate measurements of the abundances of $\alpha$-elements}

\begin{figure}[t]
\begin{center}
 \includegraphics[width=0.65\textwidth]{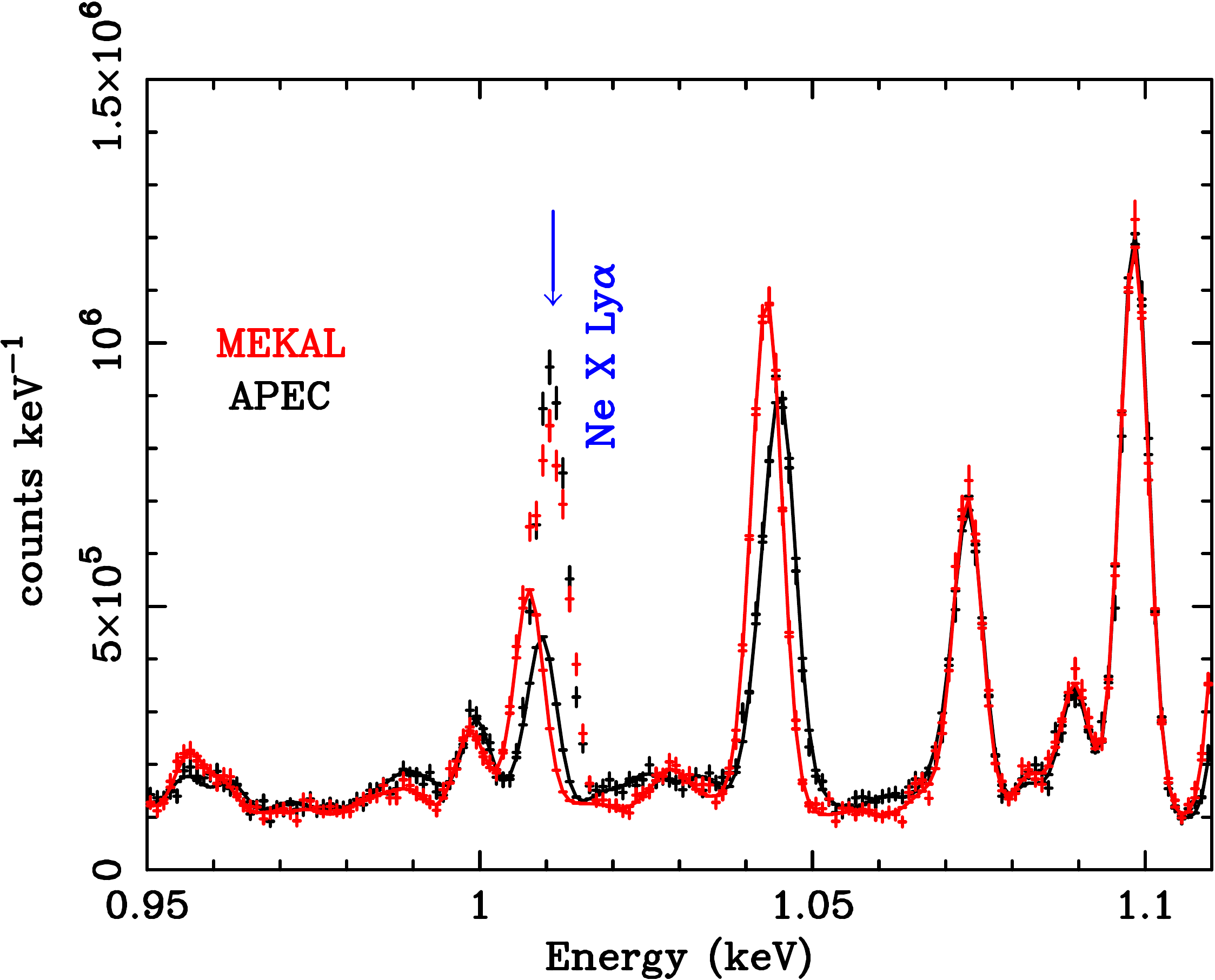}
 \smallskip\\
 \caption{The simulated SXS spectra of the Centaurus cluster
  (100 ks) around the Ly $\alpha$ line of
 H-like Ne (1.02 keV in the rest frame). We use the best-fit two temperature 
 vAPEC model (black, Fe=1.8 solar, Ne=2.4 solar) and vMEKAL model 
 (red, Fe=2.1 solar, Ne=2.1 solar) to the {\it XMM-Newton} MOS spectra 
 at 0.5'--1.5' from the X-ray peak of the Centaurus cluster.
 The solid lines show the best-fit models but the Ne abundance was changed to 0.
 }
\label{fig:nemgm87}
\end{center}
\end{figure}

\begin{figure}[t]
\begin{center}
\includegraphics[width=0.5\textwidth]{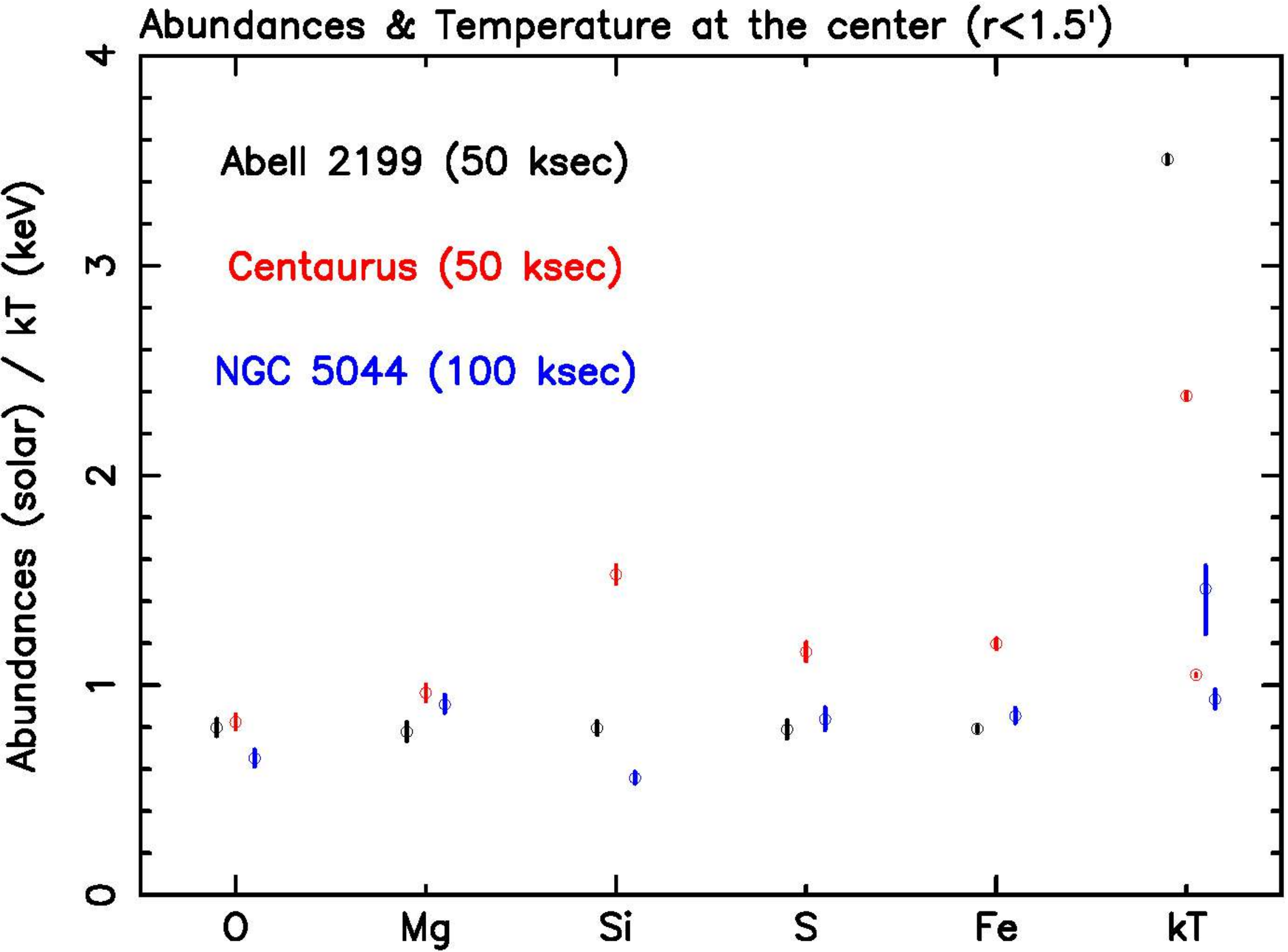}
\end{center}
 \caption{ Simulated abundances and temperature at the center
 within $r<1.5'$.  All the simulations are performed by the SIMX
 package, and the simulations include the current PSF status.}
\label{fig:abundances}
\end{figure}

The line diagnostics with SXS will constrain the temperature
structure in cool-core regions.  In addition, O, Ne
and Mg lines will be separated from the surrounding Fe-L lines.  With
the CCD detectors, different plasma codes sometimes yield significantly
different Ne and Mg abundances. As shown in Figure \ref{fig:nemgm87}, on the other
hand, the SXS will be able to derive line strengths of  
Ly$\alpha$ lines of H-like ions like Ne X and Mg XII. As a result, systematic
uncertainties in the abundance measurements caused by uncertainties in
the temperature structure and due to blending with other lines will be
significantly reduced.   
The abundances of O, Ne, Mg, Si, S, Ar, and Ca will be measured within a few
\% statistical accuracy in 100 ks exposures of 
cool-cores.  The expected statistical uncertainties of abundance
measurements in the Perseus cluster are listed in Tables \ref{simtable}
and \ref{simtable7ev}, and those for the Centaurus cluster and
Abell~2199 ($z=0.03$) are shown in Figure \ref{fig:abundances}.  

Using low-resolution \chandra data, \citet{Million11} found the
surprising trend that, in the Virgo cluster, the Si and S abundances are
{\it more} centrally peaked than Fe, challenging the standard picture of
chemical enrichment in galaxy clusters, wherein SNIa from an evolved
stellar population are thought to dominate the central enrichment.
\astroh will allow us to check these results.  As
shown in Table \ref{basesim}, we will measure the O, Ne, Mg, Si, and S
abundances in Virgo with an uncertainty of only $\pm0.1$ solar per 1x1
arcmin spatial resolution element in each of the central pointings with
the same 
data as simulated in Section \ref{sec-virgo},
providing a spatially resolved radial profile of the metal abundances
and relative SNIa/SNcc contributions out to a radius of 5 arcmin
(approximately 25~kpc).

\begin{figure}[t]
\begin{center}
 \includegraphics[width=0.44\textwidth]{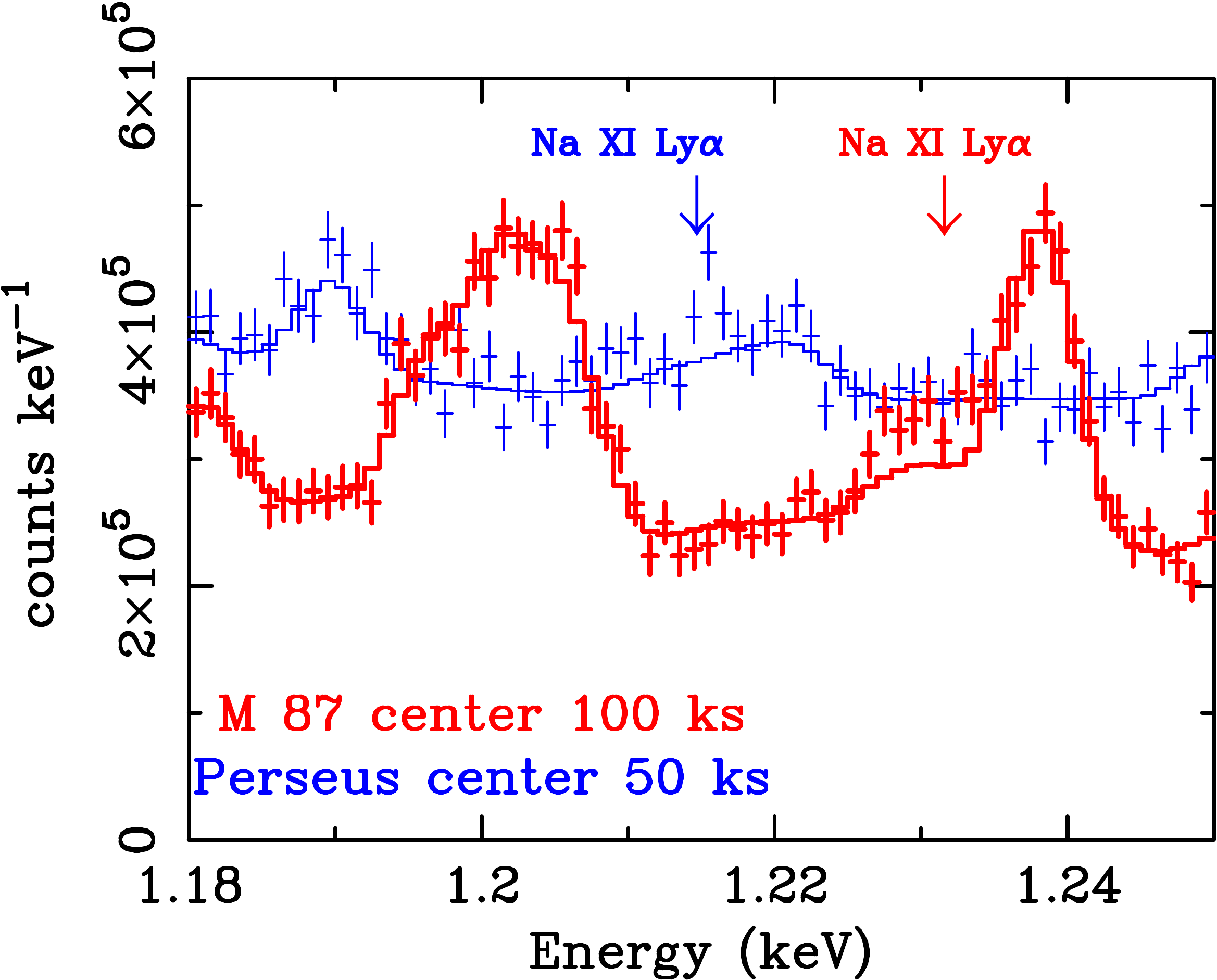}
 \hspace{0.5cm}
 \includegraphics[width=0.44\textwidth]{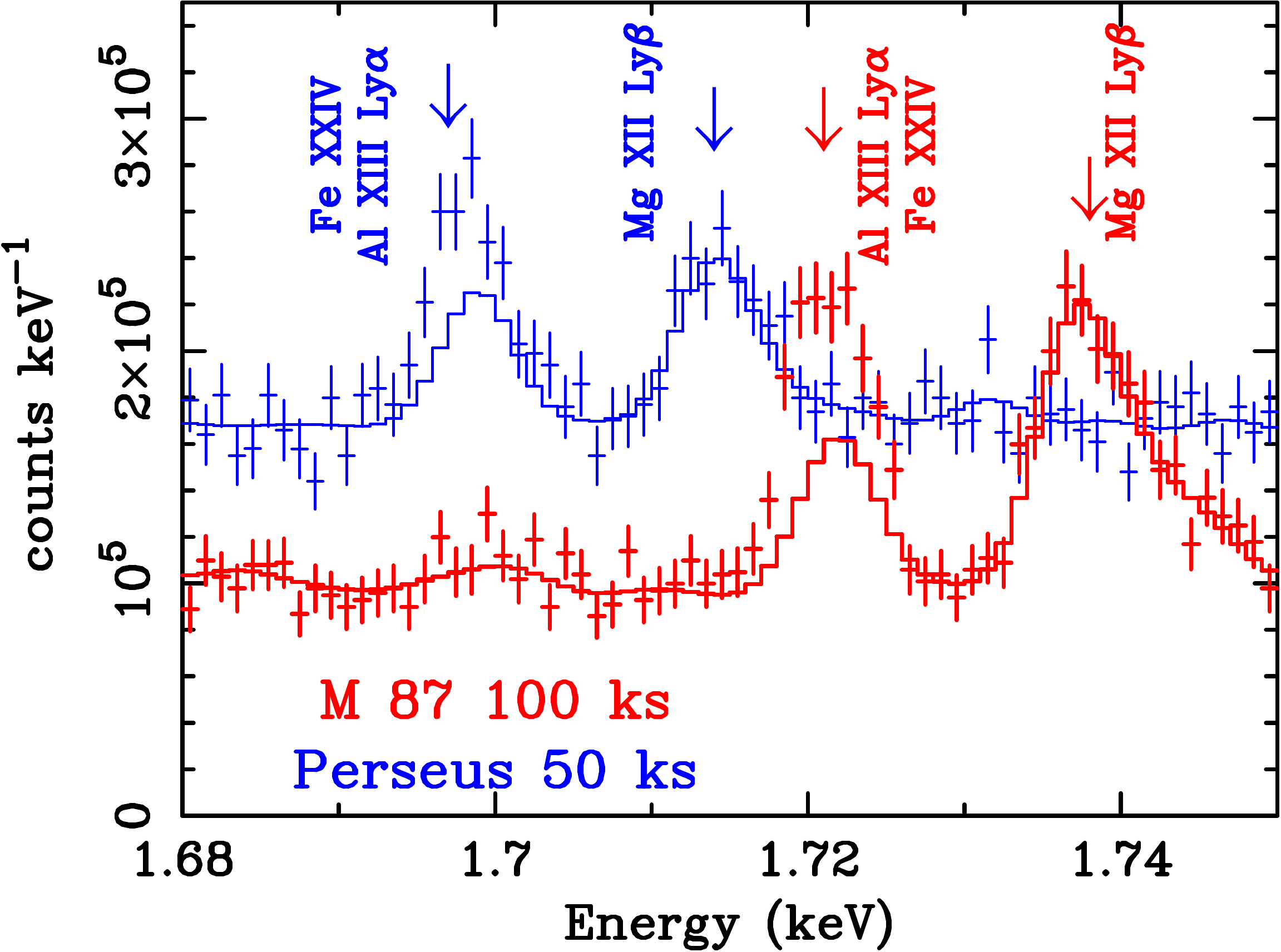}
 \caption{The simulated SXS spectra at the center of M 87 (red; 100 ks)
 and the Perseus (blue; 50 ks) cluster around Na (left), and Al (right)
 lines.  The solid lines correspond to the models with zero Na, and Al
 abundances.  The red (for M 87) and blue (for Perseus) down arrows
 indicate the redshifted line energies of Al XIII Ly$\alpha$ (1.729 keV
 in the rest frame), Fe XXIV (1.729 keV), Mg XII Ly $\beta$ (1.745 keV),
 and Na XI Ly $\alpha$ (1.237 keV).}
 \label{nnaal}
 \end{center}
 \end{figure}

\subsubsection*{Na and Al abundances of M~87}

\astroh will search for faint lines from rare elements.
Due to relatively low ICM temperatures, the low energy lines below 3 keV from
M~87 are much more prominent than those in hotter clusters.
M~87 is the best target to measure the metal abundances from N to Al.
Na and Al abundances in
the ISM of M~87 derived from the simulated spectrum (assuming that the
abundances of Na and Al are the same as that of Mg) are 0.92 $\pm$ 0.17
solar and 0.78$\pm$ 0.06 solar, respectively, with a 100 ks exposure,
considering only statistical errors.  Unfortunately, there are Fe and Ni
lines at almost same energies as the Ly$\alpha$ lines of Na and Al
(Figure \ref{nnaal}).  Here, we also show a simulated SXS spectrum of the
center of the Perseus cluster.  The contribution of Fe lines to the Al
and Fe line blend at 1.73 keV is higher in the hot clusters.  For
example, about a half of photons and a two-third of photons from the
1.73 keV line blend come from the Fe lines for M 87 and the Perseus
cluster, respectively.  The contribution of the Fe line should be much
smaller for groups of galaxies. The Ly $\alpha$ line of Na is seen at
the residual structure at 1.23 keV and 1.21 keV of the simulated spectra
of the M 87 and the Perseus cluster, respectively (Figure \ref{nnaal}).
Using other Fe-L lines, we may compare the Fe-L atomic data and then,
we will be able to constrain the systematic uncertainties due to
uncertainties in the Fe-L contributions to the Na and Al lines.

\subsubsection{Metal distributions outside cool-cores}

\begin{figure}[t]
\begin{center}
\includegraphics[width=0.5\textwidth,angle=-90]{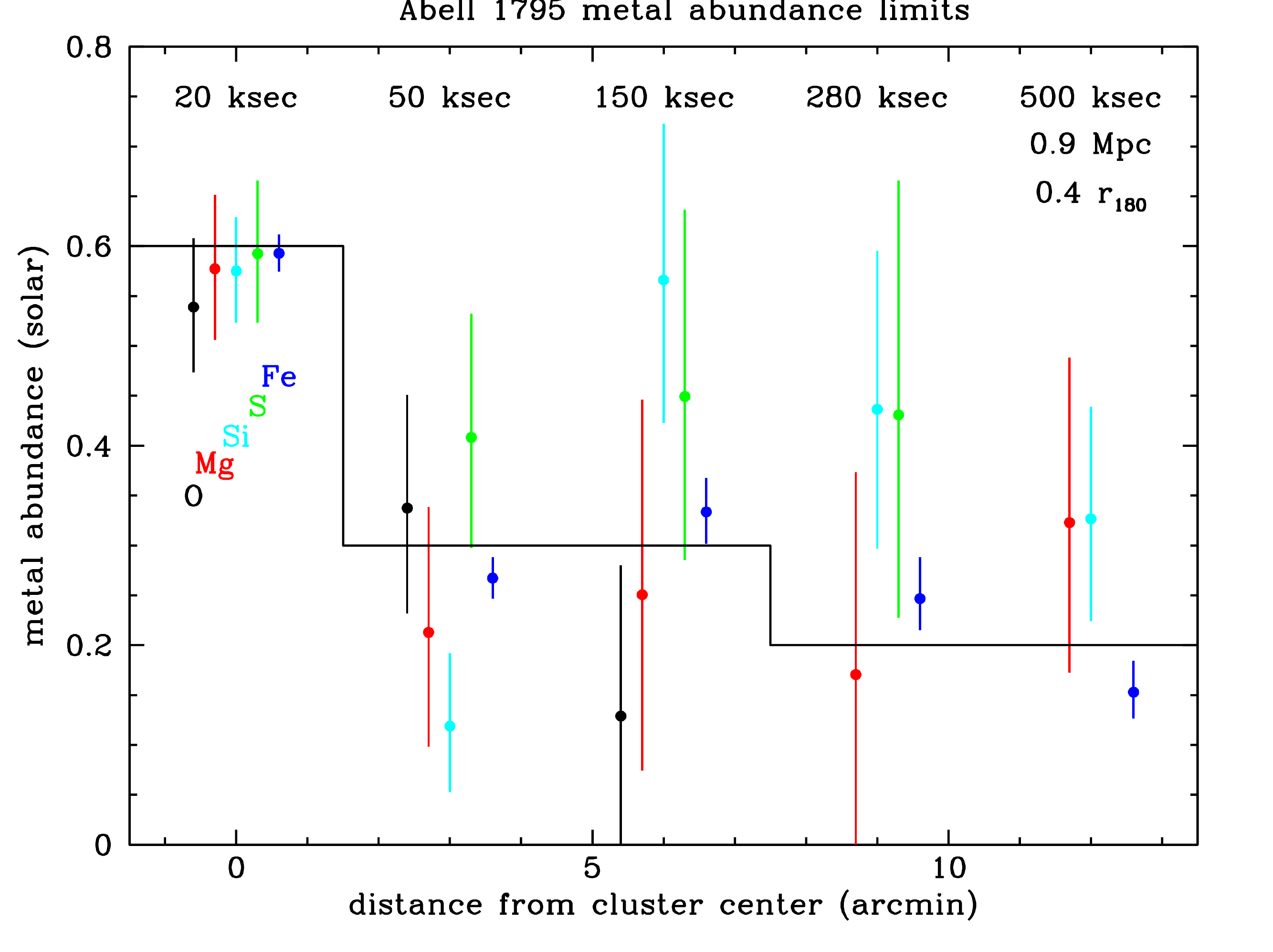}
\end{center}
\caption{Constraints on metal abundance profile expected for five pointings in Abell
1795. The exposure time of each region needed to reach the errors is
shown. The solid line shows the input metal abundance, assumed the same
for all elements in a given annulus.}
\label{fig:a1795_zprof}
\end{figure}

Primary targets for measuring the radial profile of elemental abundances
are those observed for measuring the gas motions out to large radii
discussed in Section \ref{sec-nonth}. They include Perseus with
$kT\simeq 5$ keV at $z = 0.0179$ (see also Sec. \ref{sec-perseus-chem}),
Abell 2199 with $kT\simeq 4$ keV at $z=0.030$, Abell 1795 with $kT\simeq
6$ keV at $z = 0.062$, and Abell~2029 with $kT\simeq 8$ keV at
$z=0.077$.  These clusters also span a range of gas temperatures, i.e.,
the mass scales of clusters. For Abell 2199 and Abell 1795, their
intermediate gas temperatures allow us to constrain abundances of
multiple elements both within and outside the cool cores (Figs
\ref{fig:abundances} and \ref{fig:a1795_zprof}).  For a higher
temperature system like Abell~2029, only the Fe lines are prominent, and its
abundance will be determined with $\pm 30$\% uncertainties out to
$0.5~r_{500}$ (Figs \ref{fig:a2029con}--\ref{fig:a2029neg}).  
Although the statistical uncertainties in O and Mg abundance
measurements are relatively large, we will be able to reduce systematic
uncertainties caused by the strong emission lines of O from our Galaxy
and systematic uncertainties in the Fe-L atomic data.

\subsubsection{NGC~5044 and galaxies}

NGC 5044 is a giant elliptical galaxy centered in a nearby group of galaxies at redshift 
$z=0.00902$. Its X-ray emission is spatially symmetric.
In contrast to clusters, observations of galaxy groups will
investigate differences in the chemical enrichment history between 
clusters and groups.
Figure \ref{fig:abundances} shows the resulting temperature  
and abundances with the errors from the simulations with a 2T model and 
a velocity of 200 km/s at the central region within $r<1.5'$ 
for a 100 ksec observation. 

The abundances measured in clusters of galaxies or elliptical galaxies 
can be compared to abundances in spiral galaxies, like NGC~253. 
Spiral galaxies tend to have a much younger stellar
populations and larger SNcc contributions. Due to the relatively 
complex structure of spiral galaxies, the presence of a halo, 
a disk component, and star-forming regions, the spectral model contains 
several components, which complicates abundance determinations. 
Using a 200 ks exposure, an estimate of the O/Fe abundance appears to be
feasible, which roughly gives an indication about the SNIa/SNcc
ratio. For the details about the simulations for elliptical and
starburst galaxies, see an accompanying \astroh white paper on ISM and
galaxies \citep{Paerels2014}.





\subsection{Beyond Feasibility}
\label{sec:cluschem-beyond}

\subsubsection{Detection of rare elements}
\label{sec:cluschem-rare}

\begin{figure}[t]
\begin{center}
\includegraphics[width=0.8\hsize]{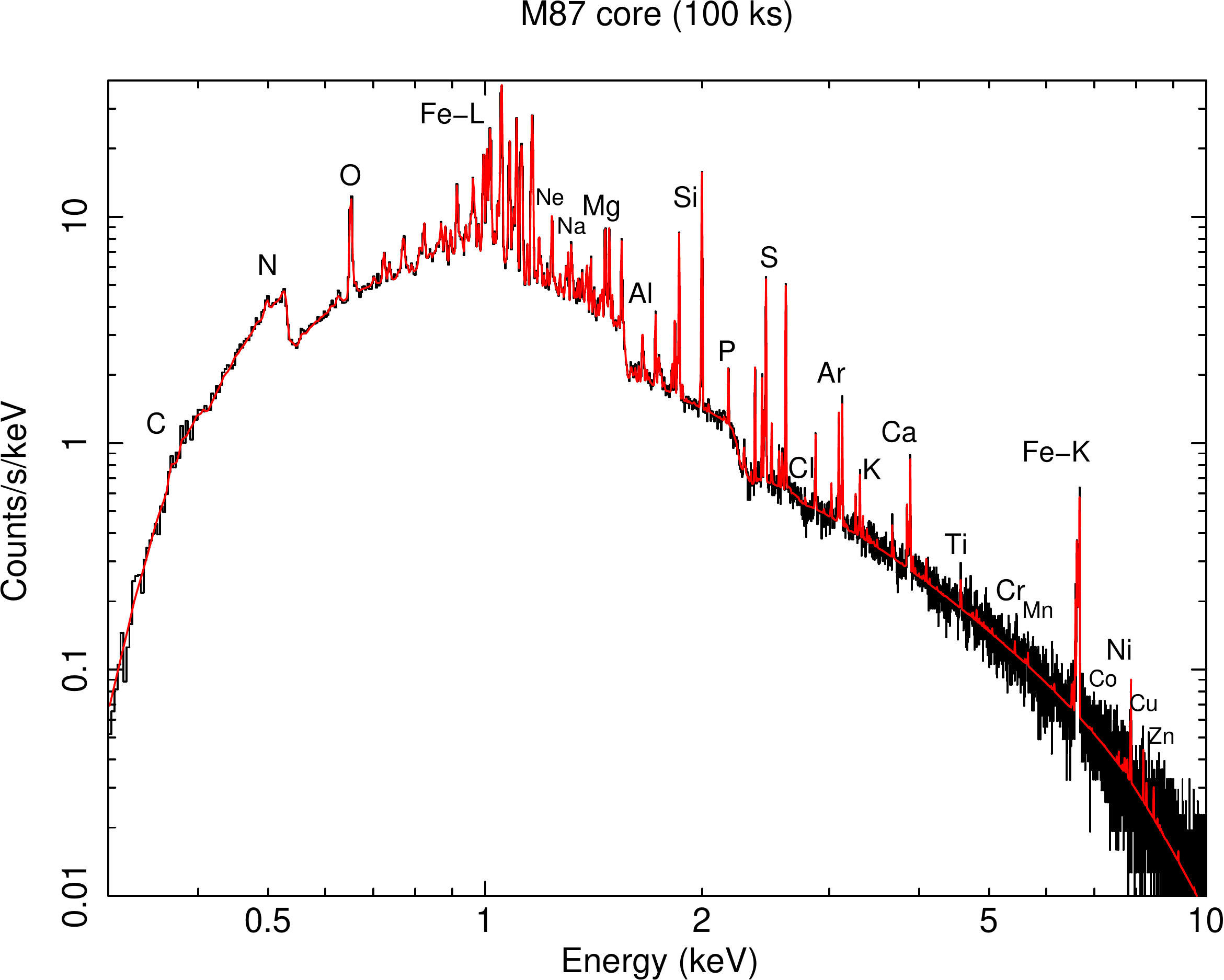}
\end{center}
\caption{Simulated 100 ks spectrum of the M87 core including lines of rare elements. The positions
of the strongest lines are labeled. It shows that a relatively cool objects like M87 produce
magnificent line-rich spectra.}
\label{fig:M87sim}
\end{figure}

The high spectral resolution of SXS in principle allows the detection of 
weak lines that place significant constraints on supernova models.
The Na/Al and Cr/Mn ratio are, for example, sensitive to the initial metallicity 
of the stellar population.  Weak lines are easier to detect
in bright clusters.  Nevertheless,  the cores of the brightest clusters
will require exposure times (well) above 100 ks  to accurately 
determine abundances of the weakest lined elements. Despite these caveats,
\astroh will in principle be capable to enlarge the number of detected elements 
from typically 10 with \suzaku and \xmm to about 20 (see Figure~\ref{fig:M87sim}), 
including elements that are key to understanding the SNIa explosion mechanism better. 
The feasibility of detecting weak lines depends on their abundance, the velocity 
structure of the core, and the availability of long SXS exposures of bright cluster cores.

\begin{table}[t]
\caption{Expected errors in solar units in the SXS Perseus core pointing
for several rare elements, assuming the abundance of 1 solar unit
for each element.
One would need an error smaller than 0.20 and 0.33 solar for a
detection better than $5\sigma$ and $3\sigma$, respectively. 
In the last column we provide an estimate of what exposure time is 
needed to detect an element at the 5$\sigma$ level.
}
\label{tab:rare_elements}
\begin{center}
\begin{tabular}{l|ccc}
		& \multicolumn{2}{c}{Exposure} & Exposure needed for\\
Element		& 50 ks   		& 100 ks   & detection at $>$5$\sigma$\\
\hline\hline
F		& 83			& 59       & - \\
Na		& 0.35 			& 0.25     & $\gtrsim$ 150 ks \\
Al		& 0.14 			& 0.10	   & $\gtrsim$ 25 ks  \\
P		& 0.7			& 0.5	   & $\gtrsim$ 625 ks \\
Cl		& 0.4			& 0.3	   & $\gtrsim$ 225 ks \\
K		& 0.7			& 0.5	   & $\gtrsim$ 625 ks \\
Sc		& 55			& 39	   & - \\
Ti		& 0.7			& 0.5	   & $\gtrsim$ 625 ks \\
V		& 6.0 			& 4.3	   & - \\
Cr		& 0.14			& 0.10	   & $\gtrsim$ 25 ks \\
Mn		& 0.27			& 0.19	   & $\gtrsim$ 90 ks \\
Co		& 0.7			& 0.5	   & $\gtrsim$ 625 ks \\
Cu		& 4.0			& 2.8	   & - \\
Zn		& 1.6			& 1.1	   & - \\	
\hline
\end{tabular}	
\end{center}
\end{table}

In Table~\ref{tab:rare_elements} we show the expected errors from a simulation
of the SXS spectrum of the Perseus core pointing of 50 ks, and what we would 
observe if the core pointing would be 100 ks. We need to stress that these errors
are based on a simulation assuming Gaussian line profiles, which will probably 
not hold in reality. On the other hand, unless the lines are blended with 
neighboring lines, the number of counts in the line could be estimated even 
for lines with a more complex line profile. There is a dependence of the error
on the velocity structure, but it is not very strong. In a 50 ks Perseus core 
exposure we will probably be able to detect Na, Al, Cr, and Mn at a 3$\sigma$ level.
With a 100 ks exposure of the core, we would also get Cl. Table~\ref{tab:rare_elements}
also shows that the Mn abundance, needed for the important Cr/Mn ratio, will 
probably need more than 50 ks exposure to reach a 5$\sigma$ significance level.
Combining the abundance measurement from the core of Perseus with the results 
from the off-set pointings will likely provide a Cr/Mn ratio with $>$5$\sigma$
significance. 

Table~\ref{tab:rare_elements} shows that many other elements 
are in principle detectable using SXS, but only by using exposure times 
of more than 0.5 Ms. Detections of these elements would therefore only be 
expected toward the end of the \astroh mission if sufficiently long observations 
of bright clusters can be stacked.

\subsubsection{Abundances as a function of redshift}

\begin{figure}[t]
\begin{center}
\includegraphics[angle=-90,width=0.6\textwidth]{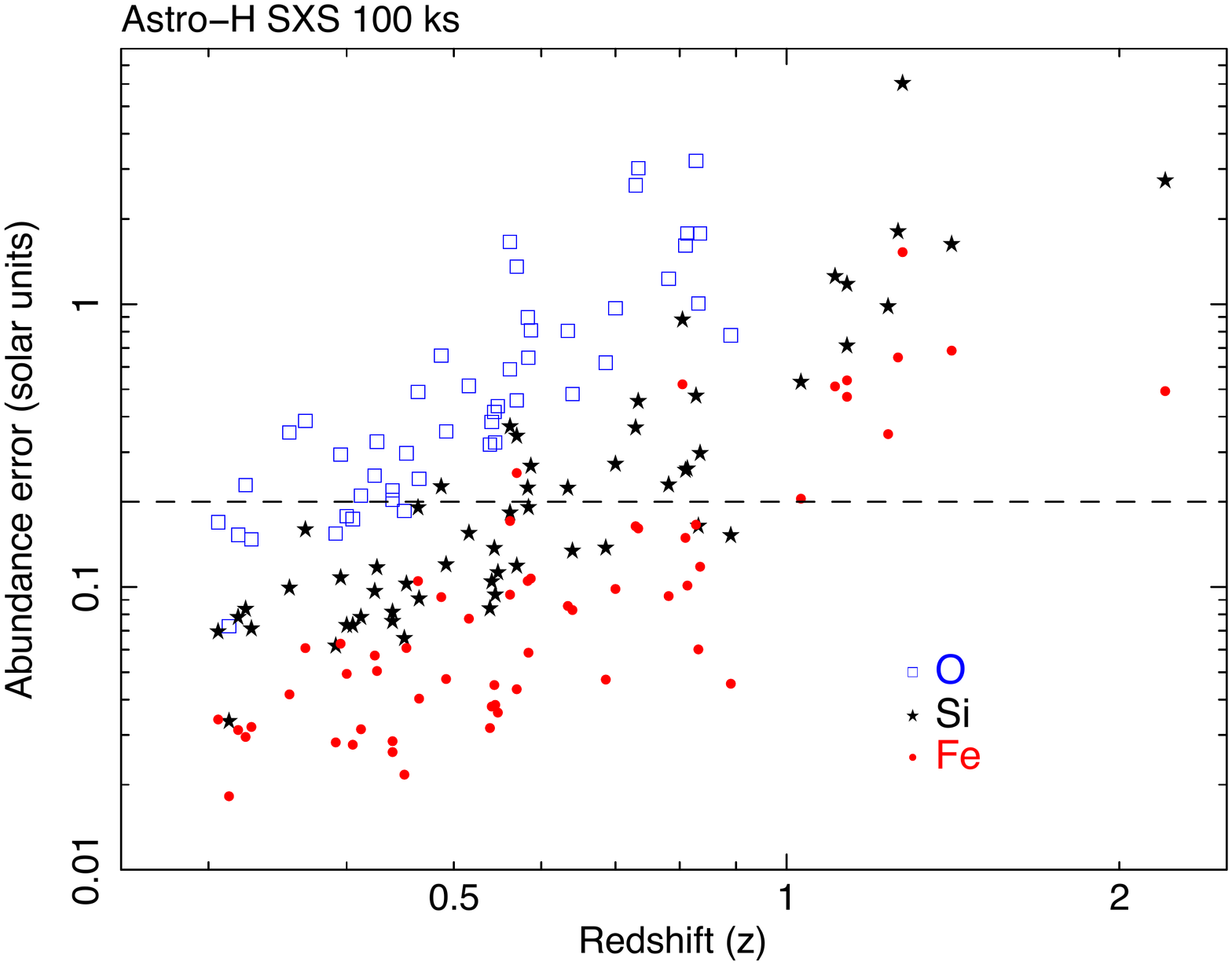}
\end{center}
\caption{Expected abundance errors of O, Si, and Fe as a function of
redshift.  Simulated clusters are based on the sample used by
\citet{balestra2007}. Each cluster has a simulated exposure time of 100
ks.}  \label{fig:abun_reds}
\end{figure}

It may be possible to observe high redshift clusters using long observations with \astroh .
The limited effective area and the PSF, however, were not optimized for faint cluster 
detection. 
In Figure \ref{fig:abun_reds},
we show the expected abundance errors (in solar units) for a sample of clusters 
lying between $z=0.25$ and $z=2.5$. Using the optimistic assumption that the abundances
of O, Si, and Fe are close to the solar value, a 5$\sigma$ detection would be feasible for points
below the dashed line. From the plot it is clear that iron abundances can 
be well determined only out to $z=1$ with a 100 ks exposure. Si abundances are feasible up to 
$z=0.6$. Oxygen abundances will be much more difficult and only feasible for 
a few systems between $z=0.2$ and $z=0.5$. 
In addition, the spatial resolution of \astroh will not permit the removal of
cool cores, which are found to have enhanced metal abundance compared to
the centers of non-cool core clusters and the outer ICM \citep[e.g.][]{degrandi2001}.  
This unresolved metal abundance profile, along with 
the unresolved core temperature structure, could complicate analysis
without accompanying high-spatial-resolution observations \citep[e.g.][]{maughan2008}.



\bigskip

\section{Detecting and Characterizing the Warm-Hot Intergalactic Medium}
\label{sec-whim}

\subsection*{Overview}

Understanding the discrepancy between the amounts of baryons measured at 
high and low redshifts has been the subject of extensive investigation 
in recent years both from the theoretical and observational point of view. 
In particular, the focus of the scientific community has been on the nature 
of the Warm-Hot Intergalactic Medium (WHIM), a warm, filamentary structure 
with density 20 to 1000 times the critical density of the universe and 
temperature greater than $5\times 10^5$~K filling the intergalactic space.
While the grasp (area times solid angle) of \astroh is too small for a 
thorough investigation and characterization of the WHIM, a small number 
of focused pointings with the SXS high resolution would significantly 
improve our understanding of the WHIM properties and evolution. 
In particular, we considered 3 pointings in directions
where WHIM filaments have either been detected or are expected.
The expected outcome of such investigation is a clear detection of at
a WHIM filament and its characterization.\footnote{Coordinators of this section: M. Galeazzi, T. Kitayama}

\subsection{Background and Previous Studies}

High redshift measurements point to about 4\% of the matter-energy
density of the Universe to be in the form of baryons, while the rest
consists of dark matter and dark energy
\citep{rauch,weinberg,burles,kirkman,bennett}.  In contrast, the amount
of baryons measured in the local Universe is less than 2\%
(e.g. \citealp{fukugita}).  Hydrodynamic simulations suggest that much
of the ``missing'' material lies in a hot ($10^5 - 10^7$~K) filamentary
gas at densities 20 to 1000 times the average baryon density of the
Universe, filling the intergalactic medium, the Warm-Hot Intergalactic
Medium (WHIM - \citealp{cen06, borgani}).

The hydrodynamic simulations all agree on the existence 
of the WHIM, and predict, on average, that about half of the 
baryons in the current universe are ``hidden'' in WHIM
filaments \citep{cen06,borgani}. However, there is still great variability 
in the predicted quantities, and detecting and characterizing the WHIM has 
been difficult. At the WHIM temperatures and densities the baryons 
are in the form of highly ionized plasma, making them 
essentially invisible to all but low energy X-ray and UV 
observations, mostly through excitation lines of highly ionized 
heavy elements.  However, the WHIM contribution to the total 
emission in these bands is weak and is expected to be on 
the order of 10-15\% in the soft X-ray band 
\citep{takey11, ursino2006, ursino10}. 
Highly ionized metals in the WHIM are also responsible for 
absorption features whose strength is rather weak \citep{branchini09}.

The current clearest detection of the WHIM comes from absorption 
lines in the FUV spectra of bright, distant sources 
\citep{danforth,Danforth08,tripp,thom,tilton}. 
However, FUV absorbers trace primarily the warm gas ($T<10^6$~K) 
and the bulk of the WHIM is expected to be at higher temperatures, 
characterized by soft X-ray features whose detection is more 
difficult. A small number of investigations have
successfully detected the WHIM signal through soft 
X-ray absorption and emission measurements. In particular, 
absorption features have been identified in the soft X-ray spectra 
of distant quasars \citep{nicastro05,buote09,fang10}, and emission from one 
filament between the clusters A222 and A223 has been detected \citep{werner}. 
The WHIM signal in emission using a statistical approach based on the angular 
Autocorrelation Function (AcF) of data from \xmm has also been
identified \citep{galeazzi09}.

Note that WHIM and cluster investigations conventionally
use different definitions for the overdensity, the baryon overdensity
$\delta\equiv \rho_{\rm b}/\bar{\rho}_{\rm b}$ and the mass overdensity
$\Delta\equiv \rho_{\rm m}/\rho_{\rm crit}$, where $\bar{\rho}_{\rm b}$
is the mean baryon density and $\rho_{\rm crit}$ is the critical density
of the Universe. Throughout this section, we use the former, which gives
a systematically larger value than the latter by $\delta \simeq 3.5
\Delta$ if baryon traces mass.

\subsection{Prospects and Strategy}

The \astroh grasp (area times solid angle) is too small for any
systematic study of the WHIM.  However, the SXS instrument will have
unprecedented combination of resolution and grasp to resolve emission
lines from individual filaments and it is possible to focus on a few
specific targets where WHIM filament are expected, to study their
density and temperature distribution.
In this paper we focus on how to take advantage of the energy
resolution of the SXS instrument to detect WHIM filaments in emission
and study the filament physical properties.  We also investigated the
potential of the SXS instrument to detect and characterize the WHIM in
absorption and we concluded that the performance of SXS to detect
absorption lines, while competitive with that of current instruments
(such as {\it Chandra}), does not improve what is currently
available. However, \astroh could have an advantage in observing
transient sources such as GRB and brightening blazars in timely
manner. For a discussion on the possibility of detecting WHIM filaments
in absorption with {\it ASTRO-H}, please refer to the accompanying
\astroh white paper on the high redshift chemical evolution
\citep{Tashiro2014}.

To have a high probability of detecting a WHIM filament with the
relatively small FOV of {\it ASTRO-H}, we focused on regions separating
virialized structures at the same redshift. Assuming that the virialized
structures represent the ``nodes'' of the cosmic web, if two structures
are sufficiently close in angle and redshift, there is a high likelihood
of filament connecting the two. In our search, we also required such
structures to be sufficiently apart to have a non-virialized region in
between, corresponding to a true WHIM filament.

We identified several targets matching our criteria, and focused our
attention on two targets, one where, as mentioned before, a high
density, high temperature filament has been identified with 
{\it XMM-Newton}, and the other where there is indirect evidence of a WHIM
filament (although a \suzaku investigation has not detected it, but set
upper limits on the filament's emission).  Based on our simulations
(described below), to have sufficient statistics on emission lines, we
need $\sim 200$~ks observation for each target. We point out that, due
to the SXS energy resolution, off-filament pointings to remove
background are not necessary.

\subsection{Targets and Feasibility}
 
We discuss here the general characteristics of the two identified targets.
Later, we will review each target in details, including the potential science outcome of 
such investigation.
\begin{enumerate}
\item Between A222 and A223 (observed with \xmm - \citealt{werner})
\begin{itemize}
\item $z=0.21$;
\item Excess emission detected at $kT=(0.91\pm 0.25)$~keV;
\item Baryon overdensity $\delta \sim 330 (L/3\mbox{ Mpc})^{-1/2}$,
      where $L$ is the line-of-sight extent of the filament. 
\end{itemize}
\item Between A3556 and A3558 in the Shapley supercluster 
(observed with \suzaku - \citealt{Mitsuishi12})
\begin{itemize}
\item $z=0.048$;
\item 
O VIII intensity $I< 1.5 \times 10^{-7}
$~ph~cm$^{-2}$~s$^{-1}$~arcmin$^{-2}$ ~ ($2\sigma$);
\item O VII intensity $I< 9.0 \times 10^{-7}
$~ph~cm$^{-2}$~s$^{-1}$~arcmin$^{-2}$ ~ ($2\sigma$);
\item Baryon overdensity $\delta \lesssim 220 (Z/0.3Z_\odot)^{-1/2} 
(L/3\mbox{ Mpc})^{-1/2}$ ~ ($2\sigma$);
\item An excess of point sources has been detected with \chandra; indications are that
the nature of the sources is a much higher than average density of galaxies that could be
associated to a WHIM filament; 
\item A possible overdensity of Neon is suggested.
\end{itemize}
\item Between A2804 and A2811 in the Sculptor supercluster (observed with \suzaku - \citealt{Sato10})
\begin{itemize}
\item $z=0.108$;
\item O VIII intensity $I<3.9\times 10^{-8}$ ~ph~cm$^{-2}$~s$^{-1}$~arcmin$^{-2}$ ~ ($2\sigma$);
\item Baryon overdensity $\delta \lesssim  100 (Z/0.3 Z_\odot)^{-1/2} 
(L/3\mbox{ Mpc})^{-1/2}$ ~ ($2\sigma$).
\end{itemize}
\end{enumerate}

\begin{figure}[t]
\begin{center}
\includegraphics[width=10cm]{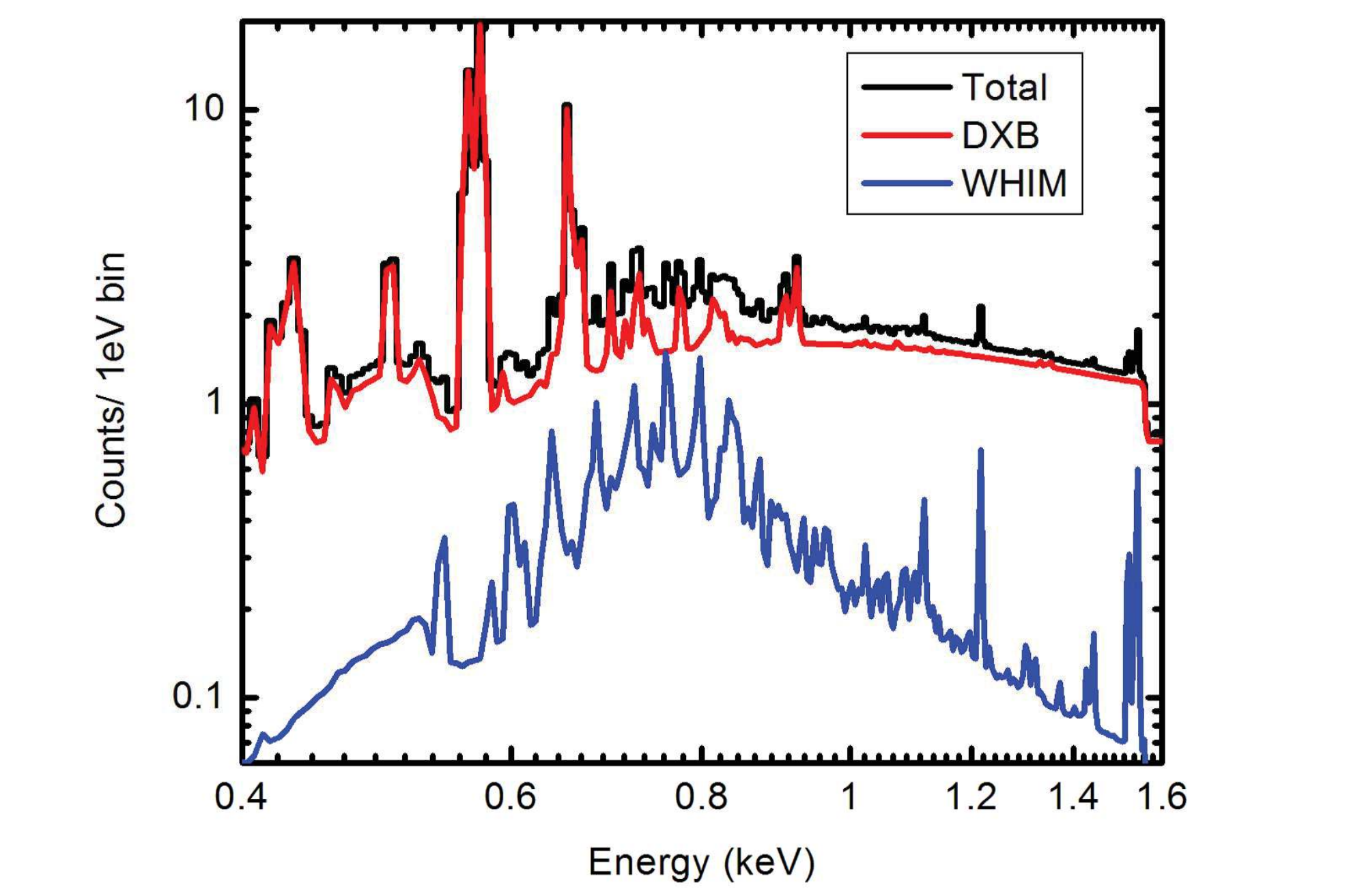} 
\caption{Simulated 
spectra for a 200~ks exposure of the filament between A222 and A223,
broken down into total X-ray emission (black), the WHIM component
(blue), and the Diffuse X-ray Background (DXB, red).}  
\label{f_a222}
\end{center}
\end{figure}

\subsubsection{Between A222 and A223}

An excess X-ray emission attributed to a WHIM filament has already been 
detected in this direction by \citet{werner}. The use of \astroh SXS 
would allow the determination with high accuracy of the redshift and temperature 
of the emitting plasma, to (a) confirm that it comes from a filament 
connecting the two clusters, and (b) determine the kinematic of the filament. 

We simulated \astroh SXS spectra for the filament connecting the clusters 
Abell 222/223, assuming a pointing between the clusters at coordinates 
01:37:45.00, 12:54:19.6. Furthermore, we assumed analyzing a spectrum 
extracted from the full field of view of \astroh SXS. 

The assumed properties of the X-ray foreground/background components and
of the filament were determined based on \xmm data of Abell 222/223 (see
\citealt{werner}).  We assumed that the filament is at the redshift of
the clusters ($z=0.21$), the hydrogen column density is $N_{\rm
H}=1.6\times 10^{20}$~cm$^{-2}$, the filaments emission measure is
$(n_e\times n_p\times V) = 1.72\times 10^{65}$~cm$^{-3}$, the filament
temperature is $kT=0.91$~keV, and the filament metallicity is 0.2 solar
(assuming proto-solar abundances of \citealt{lodders2003}).

Finally, we used the temperature $kT=0.08$~keV and the flux
$F(0.3-10 ~{\rm keV})=3.4\times 10^{-12}$~erg~s$^{-1}$~cm$^{-2}$ for the
soft X-ray foreground component, $kT=0.17$~keV and $F(0.3-10 ~{\rm keV})
= 2.9\times 10^{-12}$~erg~s$^{-1}$~cm$^{-2}$ for the galactic halo
component, and the photon index $\Gamma=1.41$ and the normalization
$2.2\times 10^{-11}$~erg~s$^{-1}$~cm$^{-2}$ for the power-law
component. The assumed metallicity of the Galactic foreground emitting
plasma is $Z=1$ solar.
 
The spectral simulations were done with the SPEX spectral fitting
package.  In the spectral fitting C-statistics was employed.  According
to the simulations a 200~ks \astroh observation will allow us to detect
the filament gas with a $\sim 9$~$\sigma$ significance and measure its
temperature with an expected $1$~$\sigma$ uncertainty of
0.06~keV. Furthermore, we will be able to measure the redshift of the
filament emission (and thus conclusively prove that it is not a local
foreground component) with an expected uncertainty of
$z=0.21^{+0.0023}_{-0.0011}$.  The simulated spectrum, including
background, is reported in Figure \ref{f_a222}.
\begin{figure}[t]
\includegraphics[width=8cm]{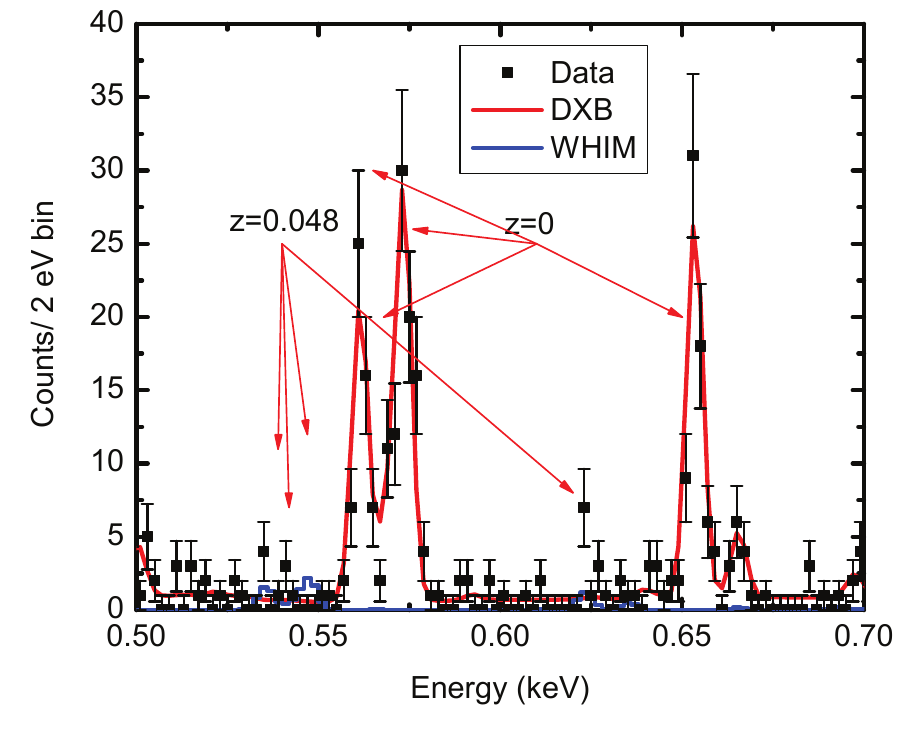}
\includegraphics[width=8cm]{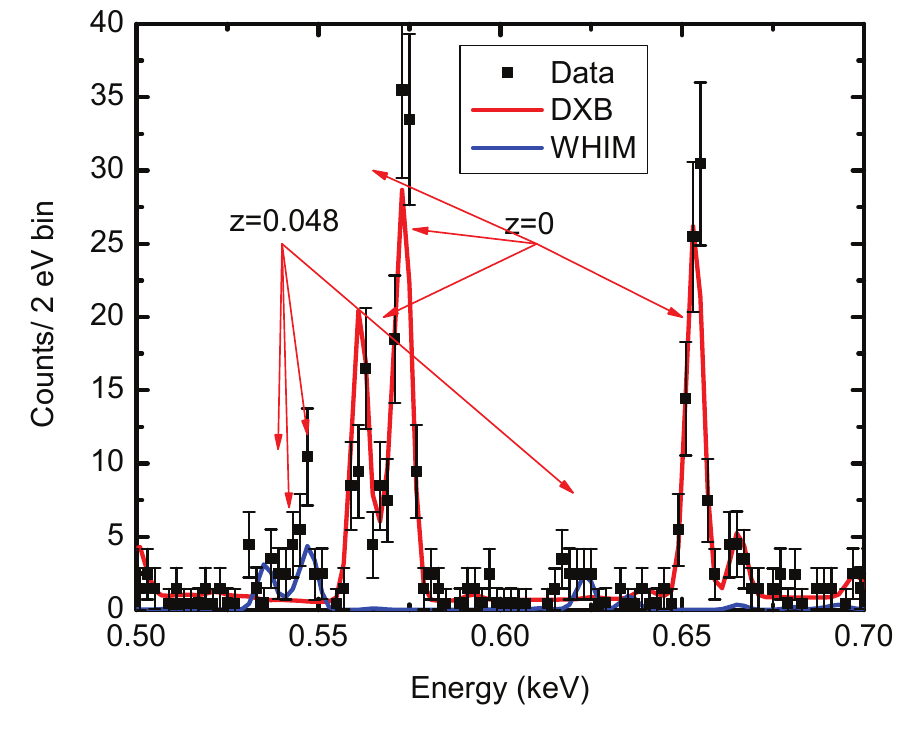} 
\caption{Simulated
spectra for a 200~ks exposure of the filament between A3556 and
A3558 in the Shapley supercluster. The figures included a potential
filament at the 3~$\sigma$ detection threshold (left) and at twice that
intensity (right).  Each plot includes total emission (black
squares), the WHIM component (blue line), and the DXB (red line).}
\label{f_shapley}
\end{figure}

\subsubsection{Between A3556 and A3558 in the Shapley supercluster}

A \suzaku investigation in this direction in search of a WHIM filament 
emission did not find it, and set upper limits on the oxygen emission 
and ``filament'' density. However, the investigation also found possible 
evidence of Ne IX emission and a stronger than usual ``power-law'' 
emission, associated with unresolved point sources. 
The same strong component is not present in two control observations 
one less than 2~deg away.
The excess point source emission was confirmed and investigated by a subsequent 
short \chandra pointing. The \chandra pointing indicates that the excess
is not simply due to cosmic fluctuations and that the source population
responsible for the excess seems different from the typical one, dominated by AGNs. 
In fact, the characteristics of the population is in better agreement with
a significantly higher than usual number of galaxies, which could be 
associated with a WHIM filament. 

As no filament has been detected so far in this direction, we focused on the 
detectability of one and its accuracy. 
The redshift of the clusters is sufficient to separate the oxygen 
emission from any local component, making any detection and plasma study 
much simpler. The excess point sources mentioned before is 
represented by a larger than usual number of faint sources, which will
appear in the SXS spectrum as a power-law distribution, roughly
doubling the continuum term of the CXB, but marginally
affecting the line detection capabilities.

Using the \suzaku and \chandra data toward the ``filament'' for the
evaluation of the CXB, the expected SXS detector background, and
assuming a 200~ks pointing, the $3$~$\sigma$ detection limit for the
combined oxygen lines is $0.4$~ph~s$^{-1}$~cm$^{-2}$~sr$^{-1}$, 
which improves the current limits \citep{Mitsuishi12} by nearly an order
of magnitue for OVIII and by a factor of 50 for OVII. Note that the
above limit is the minimum brightness required for an uncontroversial
detection of the filament.  If the filament is only twice as bright as
the threshold or brighter (i.e., 5--25 times dimmer than the
current limits), any positive detection will be supported by the SXS
high energy resolution, which, in addition to making the detection
uncontroversial, will also provide detailed information about plasma
conditions (e.g., temperature and pressure).  Two simulated spectra,
zoomed on the Oxygen lines, assuming a oxygen emission equal to the
detection threshold (left) and twice the detection threshold (right),
are shown in Figure \ref{f_shapley}.

To estimate the likelihood of detecting X-ray emission from a filament,
we used multi-wavelength observations of the region to estimate the
density of galaxies and, from there, the total gas density.  Data from
NASA/IPAC Extra-galactic Database (NED) indicate a significantly higher
than average density of galaxies in the region between A3556 and A3558.
The stellar mass of the NED galaxies in a field of radius $15'$ centered
between the two clusters goes from 3 to 7 times the average baryon
density $\bar{\rho}_{\rm b}$.  Including estimates of the galaxy halo mass (and
possibly the Circumgalactic Medium - CGM) the estimated gas density in
galaxies and their surroundings is between 25 and 100 times
$\bar{\rho}_{\rm b}$. Adding gas not associated with the galaxies brings the
expected gas density in the region between 50 and 200 $\bar{\rho}_{\rm b}$.

\begin{figure}[t]
\begin{center}
\includegraphics[width=10cm]{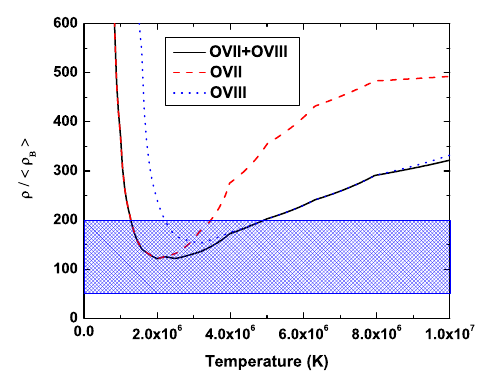} 
\caption{Expected gas overdensity in the region between A3556 and A3558
(blue shaded area), compared with the expected sensitivity $(3\sigma)$
of SXS for a 200 ks pointing, as a function of the filament temperature.
The curves represent the limit assuming that only OVII or OVIII is
detected and for a combination of the two. The derived overdensity
depends on metallicity and the line-of-sight extent of WHIM as $(Z/0.3
Z_\odot)^{-1/2}(L/3{\rm ~ Mpc})^{-1/2}$.}  \label{f_shapley2}
\end{center}

\end{figure}

We compared such estimates of the density with the density necessary to
produce OVII and OVIII lines above the calculated SXS detection limit
for a 200~ks pointing, assuming the gas with metallicity $Z=0.3
Z_{\odot}$ and a line-of-sight extent $L=3$ Mpc.  The result is
shown in Figure \ref{f_shapley2} in a plot of baryon overdensity
$\delta$ versus gas temperature. In the plot, the area in blue is our
estimated filament density.  The curves represent the SXS 3$\sigma$
detection limit for OVII and OVIII alone and for the combination of the
two. For reasonable values of the filament temperature (1-3 million
degrees), we expect a 200~ks pointing with SXS to be sensitive to baryon
overdensities of about 100 or higher, which is well within the range of
expected densities for the filament.

\subsection{Between A2804 and A2811 in the Sculptor supercluster}

Most of the considerations for this target are similar to those discussed in 
the previous subsection and won't be repeated here. 
A \suzaku investigation has not found any evidence of 
excess emission attributable to the filament and upper limits have been set. 
Using \suzaku and data toward the ``filament'' for the
evaluation of the CXB, the expected SXS detector background, and
assuming a 200~ks pointing, the $3$~$\sigma$ detection limit for the
combined oxygen lines is $0.6$~ph~s$^{-1}$~cm$^{-2}$~sr$^{-1}$, 
which improves the current limits \citep{Sato10} by about a factor of 1.5.

It is important to notice that, while a factor of 1.5 may not seem like
a significant improvement compared to previous investigations, there are
compelling factors making this a quite interesting target and SXS a
unique instrument for such investigation.  First of all, the \suzaku
limits were obtained by looking at the shape of the oxygen lines,
searching for redshifted lines ``hiding'' behind the local
emission. However, SXS will be able to clearly resolve and identify, if
present, the redshifted oxygen lines due to the WHIM filament, with a
significant reduction in systematic errors, and the possibility of
combining multiple lines, beyond just oxygen, to improve the
sensitivity.  Moreover, the \suzaku investigation used the full
$18'\times 18'$ XIS field of view, which is optimized for the filament
detection. With the smaller $3'\times 3'$ SXS field of view it is
possible to effectively go much deeper for a region right at the core of
the filament.

\subsubsection{Science outcome}

Studying the filament
between A222 and A223 would guarantee a positive result and significantly 
improve our understanding of the filament. However, its temperature and
density are somewhat outside what is believed to be the bulk of the WHIM,
limiting the broader impact of such investigation.

The pointing toward Shapley provides the most improvement in sensitivity 
compared to current limits. This is a high appealing target 
and could provide the first evidence of a filament between the two clusters.
However, the proximity between clusters places the pointing at the limit
of the virial radius of the two clusters. While such gas has not been investigated 
before, any positive detection would bear the question on whether it is due
to true WHIM or the clusters.

On the other hand, the chance of detecting and studying a true WHIM
filament in Sculptor is high risk (there is a chance that the filament
is below the SXS detection threshold), but also high reward. The
pointing in this case will be clearly outside the virial region of
either cluster, guaranteeing that any detection will be associated with
the WHIM filament.  Moreover, the SXS energy resolution would guarantee
a clean, uncontroversial detection of the filament and provides the
possibility of further studying its characteristics.

\bigskip

\section{A Spectroscopic Search for Dark Matter}
\label{sec-dm}

\subsection*{Overview}

X-ray spectroscopic observations provide a unique probe of direct
signatures of dark matter, such as a decay line of a hypothetical
sterile neutrino in the $\sim$ keV mass range. In the event that any
candidate emission line is detected in the $1-10$ keV energy band,
\astroh SXS will offer the first opportunity to resolve its shape and
distinguish it from plasma lines and instrumental effects.  The
significance of dark matter identification will be improved crucially if
the line is detected from multiple sources with distinguishable
differences in redshifts and velocity dispersions. Plausible targets
include nearby galaxy clusters, the Milky Way Galaxy, and dwarf
spheroidal galaxies, many of which will be observed by SXS for other
purposes.  
\footnote{Coordinators of this section: T. Kitayama, T. Tamura, S. W. Allen}

\subsection{Background and Previous Studies}

The nature of dark matter is one of the fundamental unsolved problems in
physics.  Despite intensive search both in laboratories and from
celestial objects, a direct identification is yet to be made.  X-ray
observations provide a direct probe of dark matter particles in the
$\sim$ keV mass range, which often comprise warm dark matter. They are
of particular astrophysical interest because they alleviate several
known problems of the conventional cold dark matter model
\citep[e.g.,][]{Weinberg2013}.  In addition, the mass of any fermionic
dark matter should lie above the robust Tremaine-Gunn bound of $\sim
0.3$ keV \citep{Tremaine79,Boyarsky09}.

A representative candidate of $\sim$ keV mass dark matter is a sterile
neutrino which arises in some extensions of the standard model of
particle physics \citep{Dodelson94,Shi99}.  The dark matter sterile
neutrino is predicted to decay and emit a line in the X-ray energy band
\citep{Abazajian2001}; even if its lifetime is much longer than the age
of the Universe, the decay signal may reach a level accessible by
existing and future instruments, owing to the high concentration of
dark matter over the large volume of galaxies and galaxy clusters.
Other candidates that can be searched for in X-rays include moduli
\citep{Kusenko12} and axions \citep[e.g.,][]{Higaki14}.

While most of previous dark matter searches in X-rays resulted only in
upper limits \citep[see][for reviews]{Abazajian2012,Boyarsky12}, there
have also been reports on unidentified emission lines
at $\sim
3.5$ keV from a sample of galaxy clusters and M31
\citep{Bulbul14,Boyarsky14}.
For any candidate line that has been (or will be) inferred, the spectral
resolution of current X-ray CCD detectors is still insufficient for
resolving its shape and distinguishing it from plasma lines and
instrumental effects.  Key quantities in this regard are accurate
centroid energy and width of the line.  The former is crucial for
identifying the line in a variety of astronomical objects at various
redshifts, whereas the latter for extracting the velocity dispersion of
dark matter which in general is different from thermal and turbulent
velocities of metals in the plasma.  The \astroh SXS will provide the
first opportunity to explicitly probe these quantities simultaneously.

\subsection{Prospects and Strategy}

For an arbitrary dark matter particle of mass $m_{\rm dm}$ that decays
into a photon of energy $E_\gamma$ at the rate per unit time $\Gamma$,
the number of photons observed at the energy $E_{\rm obs} =
E_\gamma/(1+z)$ per unit area, unit solid angle, and unit time is given
by
\begin{eqnarray}
{\cal N}_\gamma  &=& \frac{\Sigma_{\rm dm}}{4\pi (1+z)^3}
\frac{\Gamma}{m_{\rm dm}}
\label{eq-nphdm1} \\
&\simeq & 9.3 \times 10^{-5} \mbox{cm$^{-2}$~sr$^{-1}$~s$^{-1}$}
\frac{1}{(1+z)^3}
\left(\frac{\Sigma_{\rm dm}}{10^{3}M_\odot {\rm pc}^{-2}}\right)
\left(\frac{\Gamma}{10^{-32} {\rm ~s}^{-1}}\right) 
\left(\frac{m_{\rm dm}}{\rm keV}\right)^{-1},
\label{eq-nphdm2}
\end{eqnarray}
where $z$ is the source redshift, 
and $\Sigma_{\rm dm}$ is the dark matter mass column density. 
For
example, a sterile neutrino is expected to decay into an active neutrino
and a photon of energy $E_\gamma=m_{\rm dm}c^2/2$ at the rate
\begin{eqnarray}
  \Gamma \simeq  1.4 \times 10^{-32} \mbox{s}^{-1} \left(\frac{\sin^2
				2\theta}{10^{-10}}\right) 
\left(\frac{m_{\rm dm}}{\rm keV}\right)^5, 
\label{eq-sterilerate}
\end{eqnarray}
where $\theta$ is the mixing angle with the active neutrino
\citep{Pal82,Loewenstein2009}.  Given the knowledge of $\Sigma_{\rm dm}$ toward the source, the intensity and the energy of the decay line hence give a measure of
$m_{\rm dm}$ and $\Gamma$ (or $\theta$ for the sterile neutrino). 

The FWHM of the observed line is given by a convolution of the
instrumental width $W_{\rm inst}$ (eq. [\ref{eq-winst}]) and the width
determined by the line-of-sight velocity dispersion $\sigma_{\rm dm}$ of
the dark matter:
\begin{eqnarray}
W_{\rm dm} 
\simeq  7.9  ~ {\rm eV} 
\left(\frac{\sigma_{\rm dm}}{1000 ~{\rm km/s}} \right)
\left(\frac{E_{\rm obs}}{\rm keV}  \right).
\label{eq-wdm}
\end{eqnarray}
It follows that $W_{\rm inst} <10$ eV is essential for resolving the
decay line and distinguishing it from plasma lines that are broadened
separately by thermal and turbulent motions of the gas (eqs
[\ref{eq-wtion}] and [\ref{eq-wturb}]). The high spectral resolution is
also crucial for eliminating the instrumental effects intrinsic to
particular energies by comparing the positions of the line centroid from
multiple sources with redshift differences down to $\Delta z < 10^{-2}$.

Equation (\ref{eq-nphdm1}) further indicates that, for given
$\Gamma/m_{\rm dm}$, stronger signals are expected from the targets with
higher mass column within the field-of-view ($3'\times 3'$) of SXS. As
shown in the next section, they include nearby galaxy clusters, the
Milky Way Galaxy, and dwarf spheroidal galaxies. Long exposures are
still required to detect (or place meaningful limits on) yet
unidentified lines owing to the limited grasp (effective area times the
field-of-view) of \astroh SXS. A practical strategy will be to
make use of the data obtained for other purposes and, if necessary,
perform additional observations covering a range of redshifts and
velocity (or mass) scales. In what follows, we examine the feasibility
in more detail by taking into account relevant plasma emission,
background components, and instrumental capabilities.

\subsection{Targets and Feasibility}

\begin{figure}[t]
\begin{center}
\centerline{\includegraphics[width=0.6\textwidth]{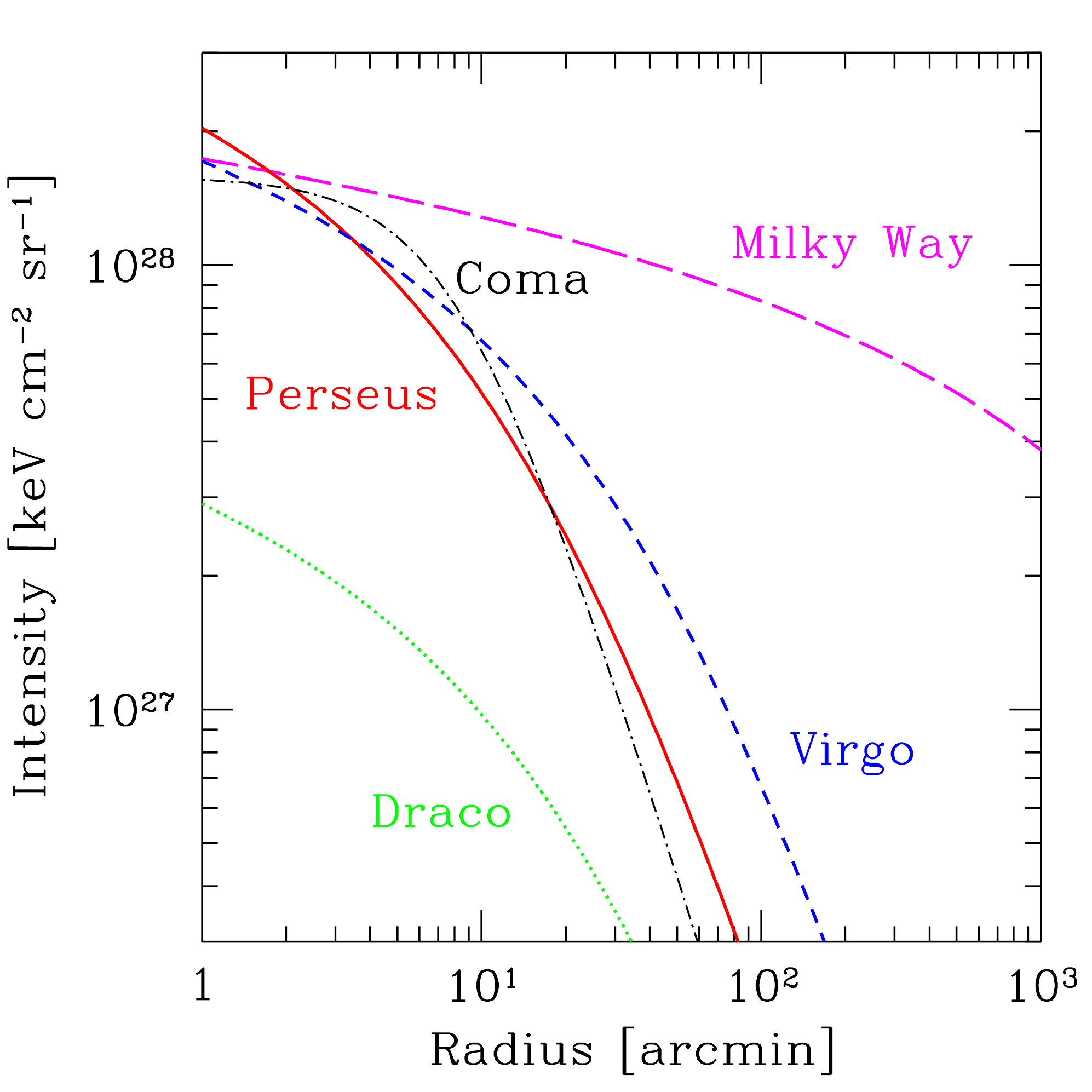}}
\caption{Normalized intensity (eq.~[\ref{eq-nphdm1}] divided by
$\Gamma/m_{\rm dm}$) of a dark matter decay line from various targets as
a function of the angle from the center (L. Strigari, private
communication).  The NFW dark matter density profile $\rho(r)=\rho_{\rm
s} (r/r_{\rm s})^{-1}(1+r/r_{\rm s})^{-1}$ is assumed with $(r_{\rm
s},~\rho_{\rm s}) = (22~\mbox{kpc}, ~4.9\times
10^{15}M_\odot\mbox{Mpc}^{-3})$ for the Milky Way \citep[magenta long
dashed line,][]{Boyarsky07}, $(0.8~\mbox{kpc}, ~6.0\times
10^{15}M_\odot\mbox{Mpc}^{-3})$ for the Draco dwarf galaxy \citep[green
dotted line,][]{Strigari07}, $(430~\mbox{kpc}, ~1.0 \times
10^{15}M_\odot\mbox{Mpc}^{-3})$ for Perseus (red solid line), and
$(250~\mbox{kpc}, ~1.0 \times 10^{15}M_\odot\mbox{Mpc}^{-3})$ for Virgo
(blue short dashed line). For Coma (black dot-dashed line), the density
profile $\rho(r)=\rho_0 [1+(r/r_{\rm c})^{2}]^{-3/2}$ is assumed with
$(r_{\rm c},~\rho_0)=(230~\mbox{kpc}, 3.9 \times
10^{15}M_\odot\mbox{Mpc}^{-3})$. The density profiles of Perseus, Virgo,
and Coma are consitent with the observed mass-temperature relation of
local clusters \citep{Vikhlinin09a}.} \label{fig-dmdecay}
\end{center}
\end{figure}

Figure \ref{fig-dmdecay} compares the normalized intensity, the quantity
given by equation (\ref{eq-nphdm1}) divided by $\Gamma/m_{\rm dm}$, from
various targets (L. Strigari, private communication).  Under the assumed
density profile, the Milky Way appears to be a plausible target for a
detection of the decay signal.  The caveats are the high absorption
column density and the bright X-ray emission in the vicinity of the
Galactic center. One should hence choose carefully the regions of low
Galactic absorption like the ``1.5 degree Window'' at
$(l,b)=(0.08^{\circ},-1.42^{\circ})$ with $N_{\rm H} \sim 7.5\times
10^{21}$ cm$^{-2}$ or those in the lower brightness Galactic bulge (see
also an accompanying \astroh white paper \citep{Koyama14} on the
feasibility toward the bulge regions).  It should also be noted that the
uncertainty in the dark matter mass profile toward the Galactic center
will be a major source of systematic errors in the measured (or
constrained) values of $\Gamma$ (or $\theta$ for the sterile neutrino)
from the data.

Clusters of galaxies provide complementary probes of the decay signal
with better knowledge of the mass profile, lower absorption columns, and
larger velocity dispersions.  Observations of galaxy clusters at
different redshifts and/or comparisons with the Milky Way will also be
crucial for removing any instrumental effects. The pointings toward
nearby clusters such as Perseus ($z=0.0179$; Sec. \ref{sec-perseus}),
Virgo ($z=0.0043$; Sec. \ref{sec-virgo}), and Coma ($z=0.0231$;
Sec. \ref{sec-coma}) discussed in other sections of this paper will
automatically provide useful data sets for this purpose.

\begin{figure}[t]
\begin{center}
\includegraphics[width=0.6\hsize]{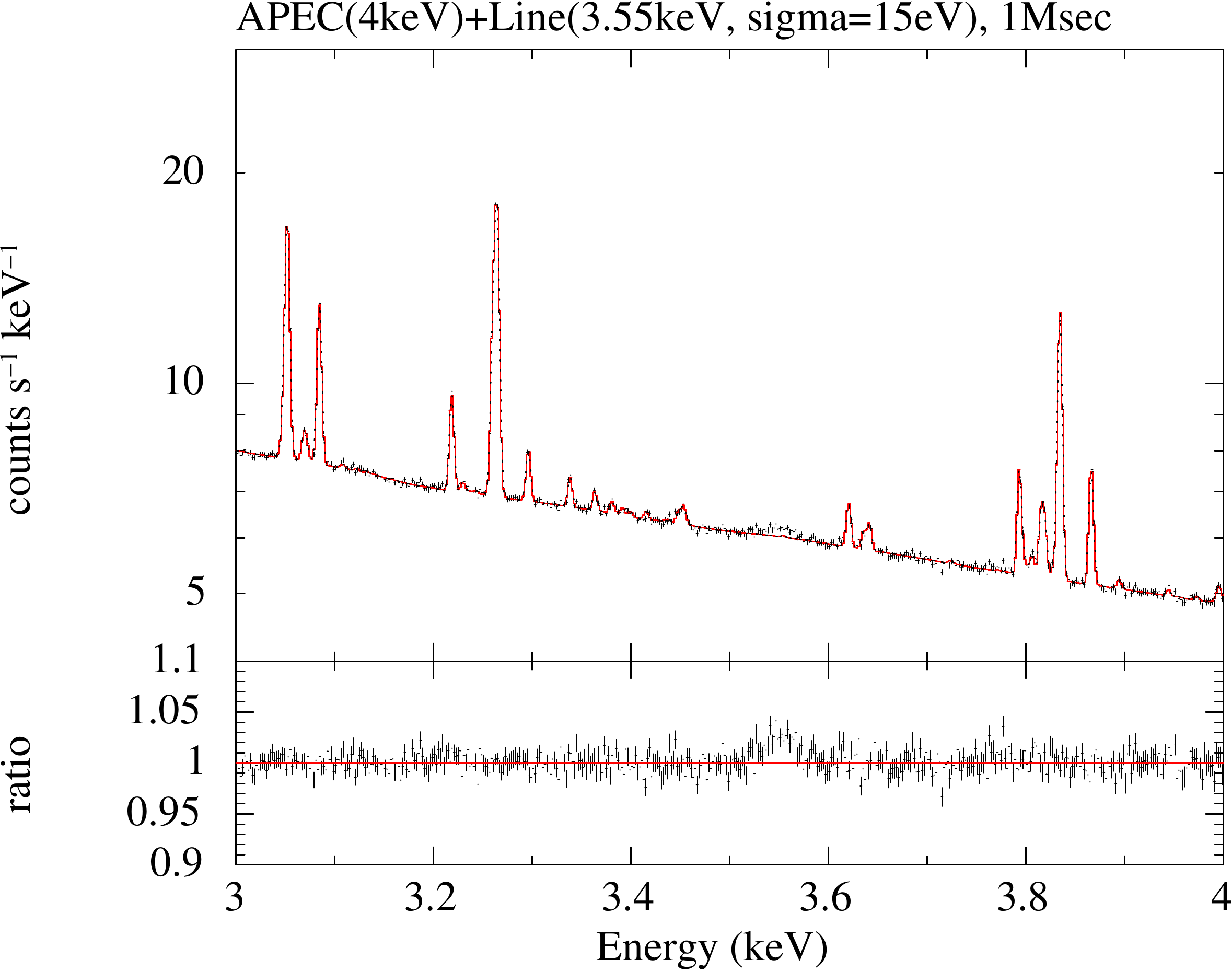}
\caption{Simulated spectra of the Perseus core at $z=0.0178$ with
(black) and without (red) a dark matter line at $3.55$~keV after an
exposure of 1~Msec by SXS. For the dark matter line, we adopt the flux
$3\times 10^{-5}$ ph s$^{-1}$ cm$^{-2}$ within the field-of-view of SXS
from Table 5 of \cite{Bulbul14} and $W_{\rm dm} =35$~eV corresponding
to the velocity dispersion $\sigma_{\rm dm}=1300$~km s$^{-1}$.  For the ICM
thermal emission, we assume $kT=4$~keV and $Z=0.7$~solar with no
turbulent broadening.  } \label{fig-dmspectra}
\end{center}
\end{figure}

Figure \ref{fig-dmspectra} shows simulated spectra toward the core of
Perseus cluster. We confirm the results of \cite{Bulbul14} that a 1 Ms
exposure by SXS through this cluster will allow us to resolve the shape
of the dark matter line inferred from the \xmm data and distinguish it
from plasma lines and instrumental effects. Note that the bright thermal
emission from the intracluster plasma dominates the background toward
the cluster cores. The relative importance of the dark matter line over
the thermal emission is likely to increase with the distance from the
core because the emissivity of the plasma is proportional to the gas
density squared whereas that of the decay line depends linearly on
the dark matter density. The spatial variation of the ratio between the
two can in turn be used to test the origin of the line.

\begin{figure}[t]
\begin{center}
\centerline{\hbox{
\includegraphics[width=0.44\hsize]{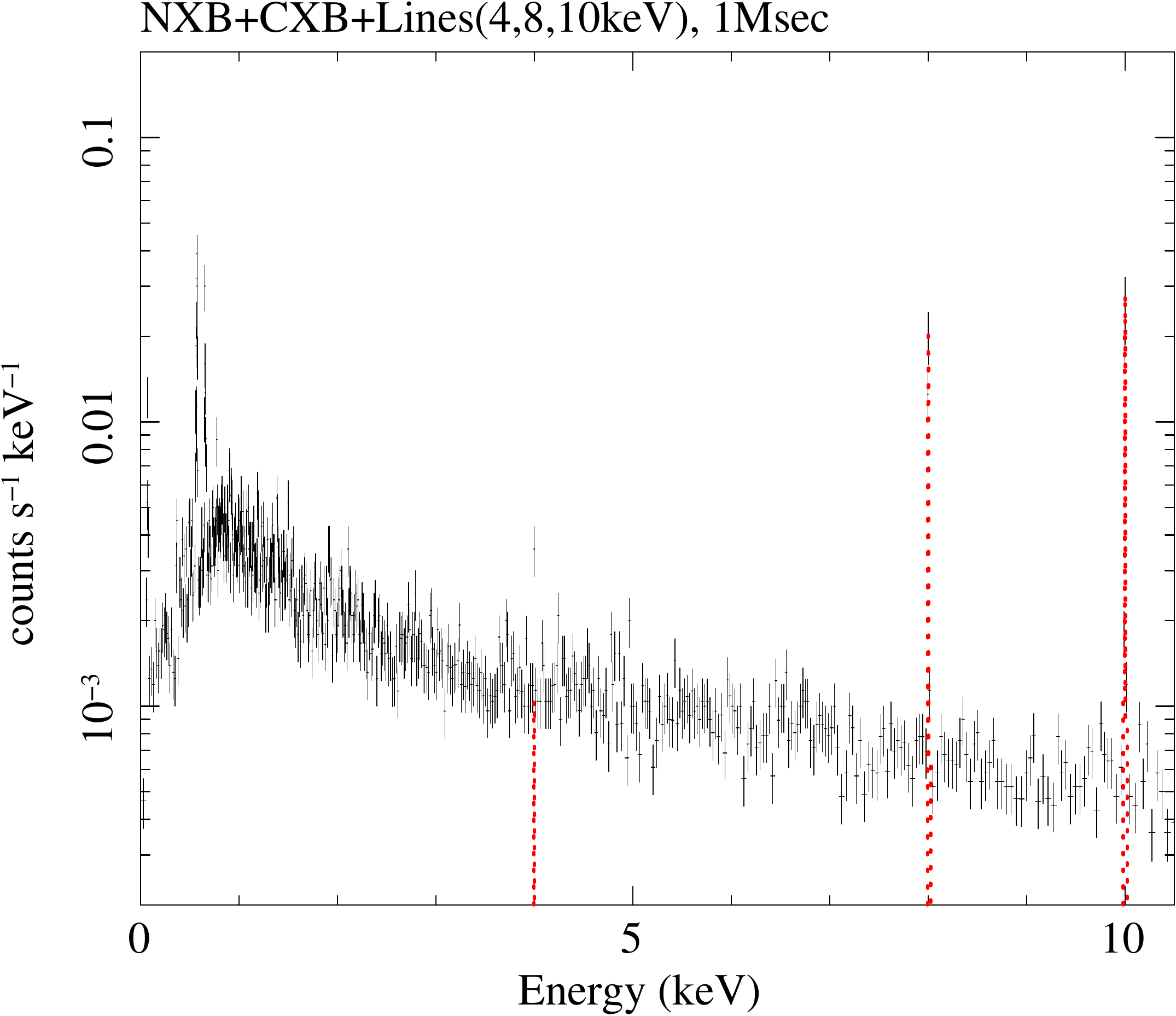}
\includegraphics[width=0.44\hsize]{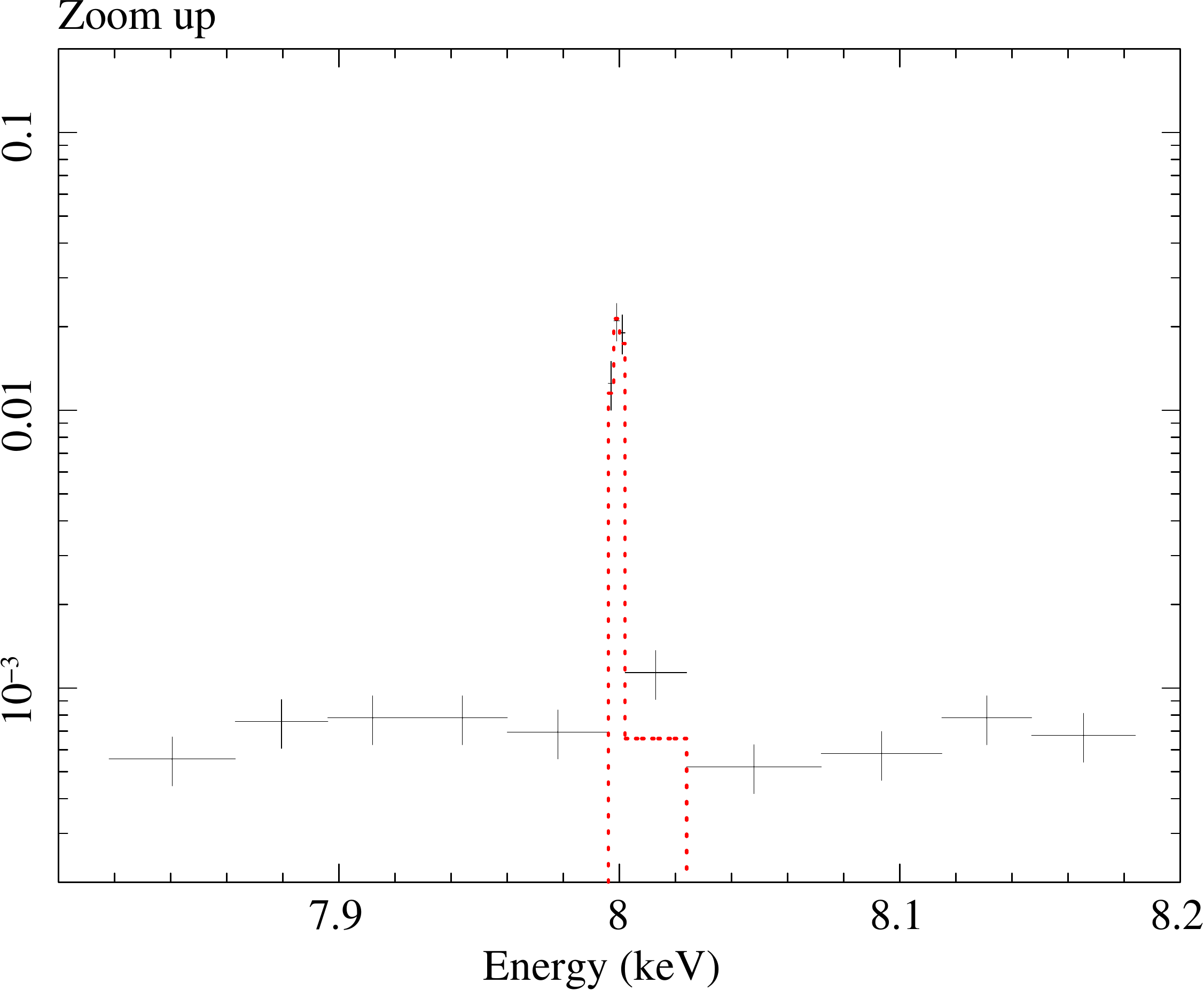}
}} \caption{{\it Left:} A 1~Msec SXS simulated spectrum of a dwarf
galaxy. We incorporate a range of hypothetical sterile neutrino lines
(red dotted) at 4.0, 8.0, and 10.0~keV for $\Sigma_{\rm dm}= 100
\,M_\odot$ pc$^{-2}$ and $\sin^2 2\theta = 10^{-10}$ corresponding to
the flux 0.04, 0.65, and 1.6 $\times 10^{-6}$ ph s$^{-1}$ cm$^{-2}$,
respectively, within the field-of-view of SXS. The adopted value of
$\Sigma_{\rm dm}$ is typical of local dwarf spheroidals and that of
$\sin^2 2\theta$ lies close to the current observational limits. No
diffuse galactic emission nor observable line broadening is assumed. The
sum (black solid) includes instrumental and cosmic background models
taken from the \astroh internal release ({\tt
sxs\_cxb$+$nxb\_7ev\_20110211\_1Gs.pha}).  {\it Right:} Same as the left
panel, except that the line at 8 keV is zoomed in.}
\label{fig-dmspectra-gal}
\end{center}
\end{figure}

Dwarf spheroidal galaxies such as Draco  
will offer yet further tests of a dark matter line against diffuse
plasma emission and instrumental effects. Their observed stellar
velocity dispersions of $10-30$ km s$^{-1}$ \citep{Mateo1998} imply that
1) broadening of the line should be well below the instrumental width
(eq.~[\ref{eq-wdm}]) and 2) contamination by plasma emission is
very low since the diffuse X-ray gas cannot be sustained by their
shallow gravitational potential.
The caveats are the weaker decay signal and the larger uncertainty of
the mass profile than the other targets mentioned above.  Simulated
spectra of a typical dwarf galaxy with a range of hypothetical sterile
neutrino lines are shown in Figure \ref{fig-dmspectra-gal}.  The
continuum is dominated by instrumental and cosmic X-ray backgrounds with
the expected brightness lower by about three orders of magnitude than
the Perseus core (Figure~\ref{fig-dmspectra}). For a given value of
$\theta$, one expects a stronger line from the sterile neutrino decay at
higher energies up to $\sim 10$ keV. This is a result of two
competing effects; $\Gamma$ increases rapidly with $m_{\rm dm}$
(eq.~[\ref{eq-sterilerate}]) whereas the effective area of SXS decreases
with energy.  It follows that the stronger limit on $\theta$ will be
derived at higher energies in the case of no detection.

\begin{figure}
\begin{center}
\centerline{\hbox{
\includegraphics[width=0.44\hsize]{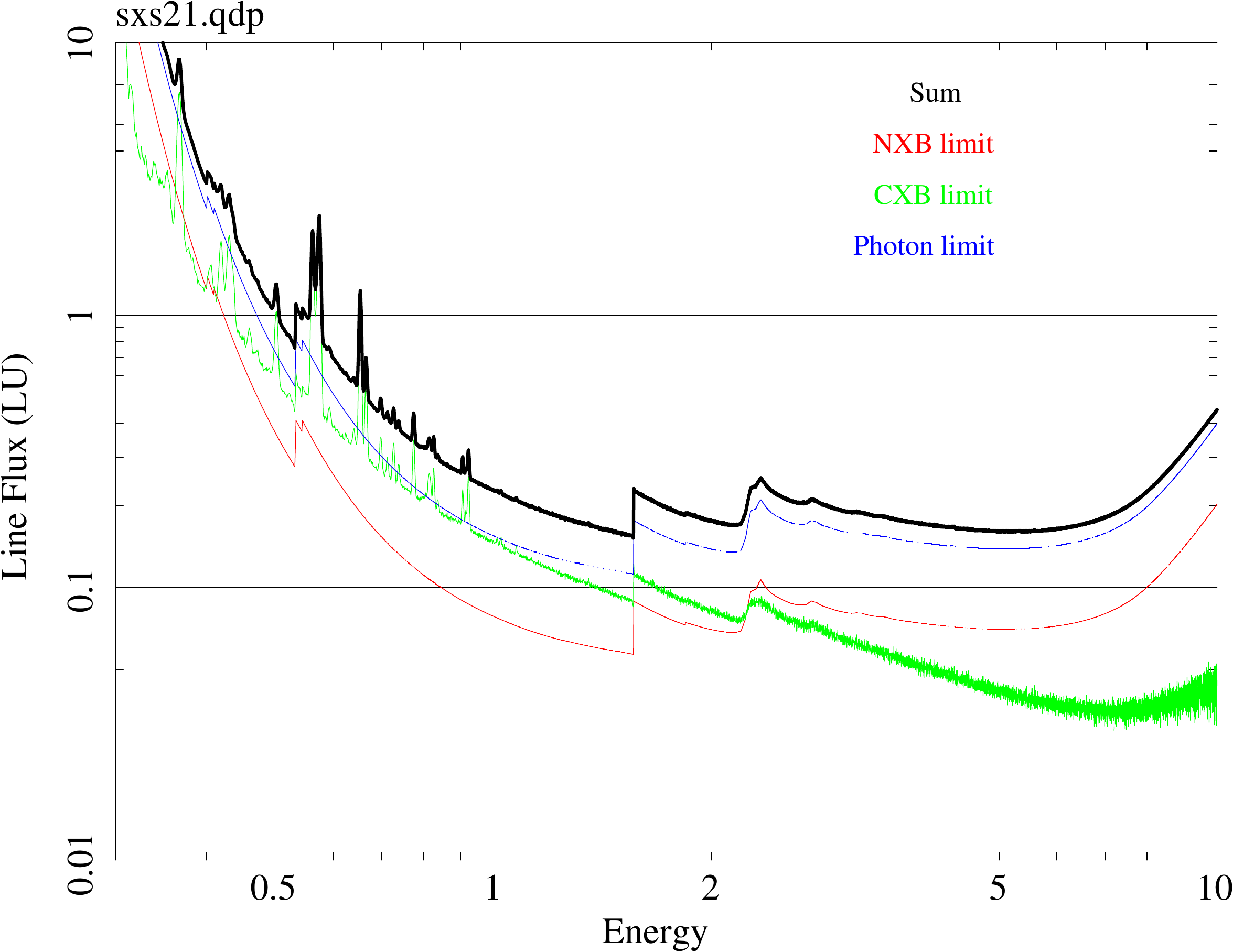}
\includegraphics[width=0.44\hsize]{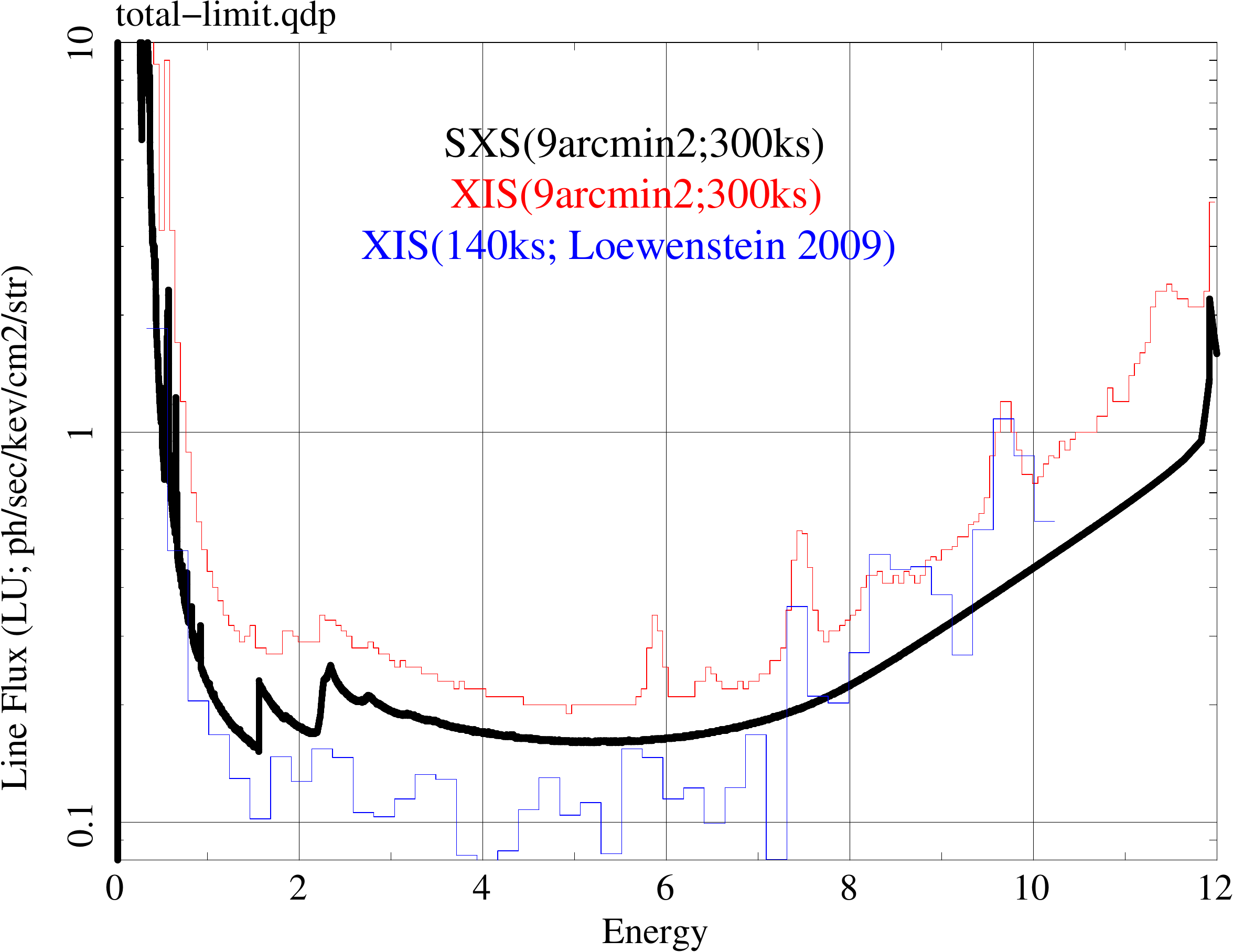}}}
\caption{{\it Left:} The $3\sigma$ line sensitivities over a sky
coverage $\Omega=9$ arcmin$^2$ after a 300 ks exposure in unit of LU
(line unit; photons cm$^{-2}$ s$^{-1}$ str$^{-1}$) for \astroh  SXS.
We assume that the signal-to-noise ratio (S/N) in an energy bin is given
by $\sqrt{C_{\rm Line}^2/(C_{\rm Line}+2C_{\rm B})}$, where $C_{\rm
Line}$ and $C_{\rm B}$ denote the counts of a line and background,
respectively \citep[e.g.][]{Bradt2004}.  The 'photon limit' (blue)
assumes $C_{\rm Line} \gg C_{\rm B}$, whereas 'Background limits' (red
and green) $C_{\rm Line} \ll C_{\rm B}$.  The 'Sum' (black) is
calculated by adding these components in quadrature.  The instrumental 
background (NXB) and the cosmic X-ray background (CXB) models are taken
from the \astroh SWG internal release.  {\it Right:} The same quantity
as the left panel for \suzaku XIS (2FI; red) and \astroh  SXS
(black). Also shown for reference is the limits for \suzaku XIS with
$\Omega=240$arcmin$^2$ (blue) from \cite{Loewenstein2009} obtained using
a Monte Carlo simulation. } \label{fig:sxs-limit}
\end{center}
\end{figure}

Figure \ref{fig:sxs-limit} further illustrates the predicted line
sensitivity of \astroh SXS as compared to an existing X-ray CCD on
board \suzaku (XIS).  Note that this figure ignores any diffuse emission
other than the instrumental background and the cosmic X-ray background;
it corresponds to the limits that can be derived for X-ray faint sources
such as dwarf spheroidal galaxies. In such cases, the sensitivity above
$\sim 1$ keV is essentially photon limited after $\sim 100$ ks
exposures, whereas the sensitivity in the softer energies remains
background limited mainly due to the Galactic line emission.  A major
improvement in the sensitivity is expected in the hard band for the flux
within the field-of-view of SXS, whereas the sensitivity is largely
limited by the small grasp of SXS for the flux from the larger sky
area. We stress that a highly improved spectral resolution will
still be indispensable for identifying or rejecting any candidate lines
once they are suggested.

\bigskip
\bigskip 
\section*{Acknowledgments}

We thank Louis Strigari, Ayuki Kamada, and Naoki Yoshida for many useful
discussions on the dark matter search and their considerable input to
Section \ref{sec-dm}.

\bigskip

\appendix 
\section*{Appendix}

\section{Systematic Errors in Gas Velocities}
\label{sec-sys}

For bright X-ray sources such as cores of nearby galaxy clusters, the
accuracy of gas velocity measurements by \astroh SXS can be limited by
systematic errors rather than statistical errors. This section
summarizes potential sources of the systematic errors and how they
affect the measurements of bulk and turbulent velocities.

\subsection{Bulk Velocity}

Calibration errors in the energy gain $\Delta E_{\rm gain}$ directly
lead to the uncertainty in the line-of-sight bulk velocity measured by a
line shift as
\begin{eqnarray}
\Delta v_{\rm bulk} = c \frac{\Delta E_{\rm gain}}{E_{\rm obs}} 
= 45 ~\mbox{km/s} ~ \left(\frac{\Delta E_{\rm gain}}{\mbox{eV}}\right) 
\left(\frac{E_{\rm obs}}{6.7~\mbox{keV}}\right) ^{-1},    
\end{eqnarray}
where $E_{\rm obs}$ is the observed energy of the line.


\subsection{Turbulent Velocity}

The line-of-sight component of the turbulent velocity measured by line
broadening can be affected by calibration errors in instrumental
broadening due to the line spread function and uncertainties in thermal
broadening characterized by the ion temperature.

To quantify their impacts, let us decompose the observed FWHM of a
spectral line, assuming a Gaussian profile for each component, as
\begin{eqnarray}
\label{eq-wobs}
 W_{\rm obs}^2 &=& W_{\rm inst}^2 + W_{\rm therm}^2 + W_{\rm turb}^2 
+ \cdots, 
\label{eq-decomp}
\end{eqnarray}
where $W_{\rm inst}$, $W_{\rm therm}$, and $W_{\rm turb}$ are the FWHMs
of instrumental broadening, thermal broadening, and turbulent
broadening, respectively. Their nominal values are
\begin{eqnarray}
\label{eq-winst}
W_{\rm inst} &\simeq& 5 ~\mbox{eV}, \\
W_{\rm therm} &=& 
\sqrt{8 \ln 2} 
\left(\frac{kT_{\rm ion}}{m_{\rm ion}c^2}\right)^{1/2}
E_{\rm obs}, \nonumber \\
&=& 4.9 ~ {\rm eV} 
\left(\frac{k T_{\rm ion}}{\rm 5~keV} \right)^{1/2}
\left(\frac{m_{\rm ion}}{56 ~m_{\rm p}} \right)^{-1/2}
\left(\frac{E_{\rm obs}}{6.7 {\rm keV}} \right), 
\label{eq-wtion} \\
W_{\rm turb} &=& \sqrt{8 \ln 2}\left(
\frac{v_{\rm turb}}{c}\right)E_{\rm obs}, \nonumber \\  
&=& 5.3  ~ {\rm eV} 
\left(\frac{v_{\rm turb}}{100 ~{\rm km/s}} \right)
\left(\frac{E_{\rm obs}}{6.7 ~{\rm keV}}  \right), 
\label{eq-wturb}
\end{eqnarray}
where $T_{\rm ion}$ and $m_{\rm ion}$ are the temperature and the mass
of the ion producing the line observed at $E_{\rm obs}$, $m_{\rm p}$ is
the proton mass, and $v_{\rm turb}$ is the RMS turbulent velocity along
the line of sight. Note that natural line broadening takes a Lorentzian
profile with the FWHM of
\begin{eqnarray}
W_{\rm nat} = \hbar A = 0.31 ~{\rm eV} \left(\frac{A}{4.67 \times
					  10^{14} ~{\rm s}^{-1}}\right),
\end{eqnarray}
where $A$ is the Einstein coefficient and the quoted value is for the Fe
XXV resonant line at 6.7 keV. The presence of natural broadening has
little effect on the decomposition given by equation (\ref{eq-decomp})
and on the results presented in this section.

In the following, we quantify the errors in $W_{\rm turb}$ arising from
those in $W_{\rm inst}$ and $W_{\rm therm}$. Note that a conventional
error propagation law of the form
\begin{equation}
\Delta W_{\rm turb} = \left|
\frac{\partial W_{\rm turb}}{\partial W_{\rm inst}}
\right|\Delta W_{\rm inst} 
= \frac{W_{\rm inst}}{W_{\rm turb}} 
\Delta W_{\rm inst}
\label{eq-prop}
\end{equation}
is NOT valid for $\Delta W_{\rm turb} > W_{\rm turb}$ because equation
(\ref{eq-prop}) is based on a linear approximation; e.g., $\Delta W_{\rm
turb}$ diverges as $W_{\rm turb}\rightarrow 0$, even if $W_{\rm inst}$
and $\Delta W_{\rm inst}$ are both finite. We may readily face such a
situation if the observed turbulence is small and only an upper limit is
to be placed on $W_{\rm turb}$.  We hence present more general
expressions of $\Delta W_{\rm turb}$. The conversion to $\Delta v_{\rm
turb}$ is done by
\begin{eqnarray}
\Delta v_{\rm turb} &=& 
\frac{c}{\sqrt{8 \ln 2}}
\frac{\Delta W_{\rm turb}}{E_{\rm obs}}
= 19  ~{\rm km/s}
\left(\frac{\Delta W_{\rm turb}}{{\rm eV}}\right)
\left(\frac{E_{\rm obs}}{6.7 {\rm keV}}  \right)^{-1}.
\label{eq-w2v}
\end{eqnarray}

\subsubsection{Instrumental broadening}

Equation (\ref{eq-wobs}) implies that $\Delta W_{\rm inst}$ alone gives
rise to $\Delta W_{\rm turb}$ via
\begin{eqnarray}
\Delta (W_{\rm turb}^2)&=& \Delta (W_{\rm inst}^2),
\label{eq-quadinst}
\end{eqnarray}
where, for an arbitrary component $j$,
\begin{equation}
\Delta (W_j^2) \equiv (W_j+\Delta W_j)^2 - W_j^2 =  2 W_j \Delta W_j + (\Delta W_j)^2.   
\label{eq-quad}
\end{equation}
Equation (\ref{eq-quadinst}) is quadratic in $\Delta W_{\rm turb}$ and 
yields
\begin{eqnarray}
\Delta W_{\rm turb} &=& 
- W_{\rm turb} + \sqrt{W_{\rm turb}^2 + 
\Delta (W_{\rm inst}^2)} , 
\label{eq-deltawt} 
\end{eqnarray}
whereas the other solution, $- W_{\rm turb} - \sqrt{W_{\rm turb}^2 +
\Delta (W_{\rm inst}^2)}$, is irrelevant here (it is nonzero for $\Delta
(W_{\rm inst}^2) =0$). Equation (\ref{eq-deltawt}) depends on both
$W_{\rm inst}$ and $\Delta W_{\rm inst}$. Its asymptotes are:
\begin{enumerate}
\item For 
$\Delta W_{\rm turb} \ll W_{\rm turb}$ 
and  $\Delta W_{\rm inst} \ll W_{\rm inst}$, 
\begin{eqnarray}
\Delta W_{\rm turb} &\simeq& 
\frac{W_{\rm inst}}{W_{\rm turb}} 
\Delta W_{\rm inst}, 
~~~~~ \mbox{(in agreement with eq.~[\ref{eq-prop}])}
\nonumber \\  
&=& 0.95  ~ {\rm eV} 
\left(\frac{v_{\rm turb}}{100 {\rm km/s}} \right)^{-1}
\left(\frac{W_{\rm inst}}{5 {\rm eV}} \right)
\left(\frac{\Delta W_{\rm inst}}{{\rm eV}}  \right)
\left(\frac{E_{\rm obs}}{6.7 {\rm keV}}  
\right)^{-1} , \\ 
\Delta v_{\rm turb} 
&\simeq& 18 ~{\rm km/s}   
\left(\frac{v_{\rm turb}}{100 {\rm km/s}} \right)^{-1}
\left(\frac{W_{\rm inst}}{5 {\rm eV}} \right)
\left(\frac{\Delta W_{\rm inst}}{{\rm eV}}\right)
\left(\frac{E_{\rm obs}}{6.7 {\rm keV}}  \right)^{-2}.
\label{eq-deltavt2}
\end{eqnarray}

\item For $\Delta W_{\rm turb} \gg W_{\rm turb}$ 
and $\Delta W_{\rm inst} \ll W_{\rm inst}$,  
\begin{eqnarray}
\Delta W_{\rm turb} &\simeq&  \sqrt{2  W_{\rm inst}\Delta W_{\rm inst}}, 
~~~~~ \mbox{(independent of $v_{\rm turb}$)} \nonumber \\
&=& 3.2 ~ {\rm eV} \left(
\frac{W_{\rm inst}}{5 {\rm eV}} \right)^{1/2} 
\left(
\frac{\Delta W_{\rm inst}}{{\rm eV}}  
\right)^{1/2}, 
\label{eq-wcriti}\\
\Delta v_{\rm turb} 
&\simeq& 60 ~{\rm km/s}   
\left(
\frac{W_{\rm inst}}{5 {\rm eV}} \right)^{1/2} 
\left(
\frac{\Delta W_{\rm inst}}{{\rm eV}}  
\right)^{1/2}
\left(\frac{E_{\rm obs}}{6.7 {\rm keV}}  
\right)^{-1}.
\label{eq-deltavt1}
\end{eqnarray}
\end{enumerate}

\begin{figure}[t]
\includegraphics[width=7.8cm]{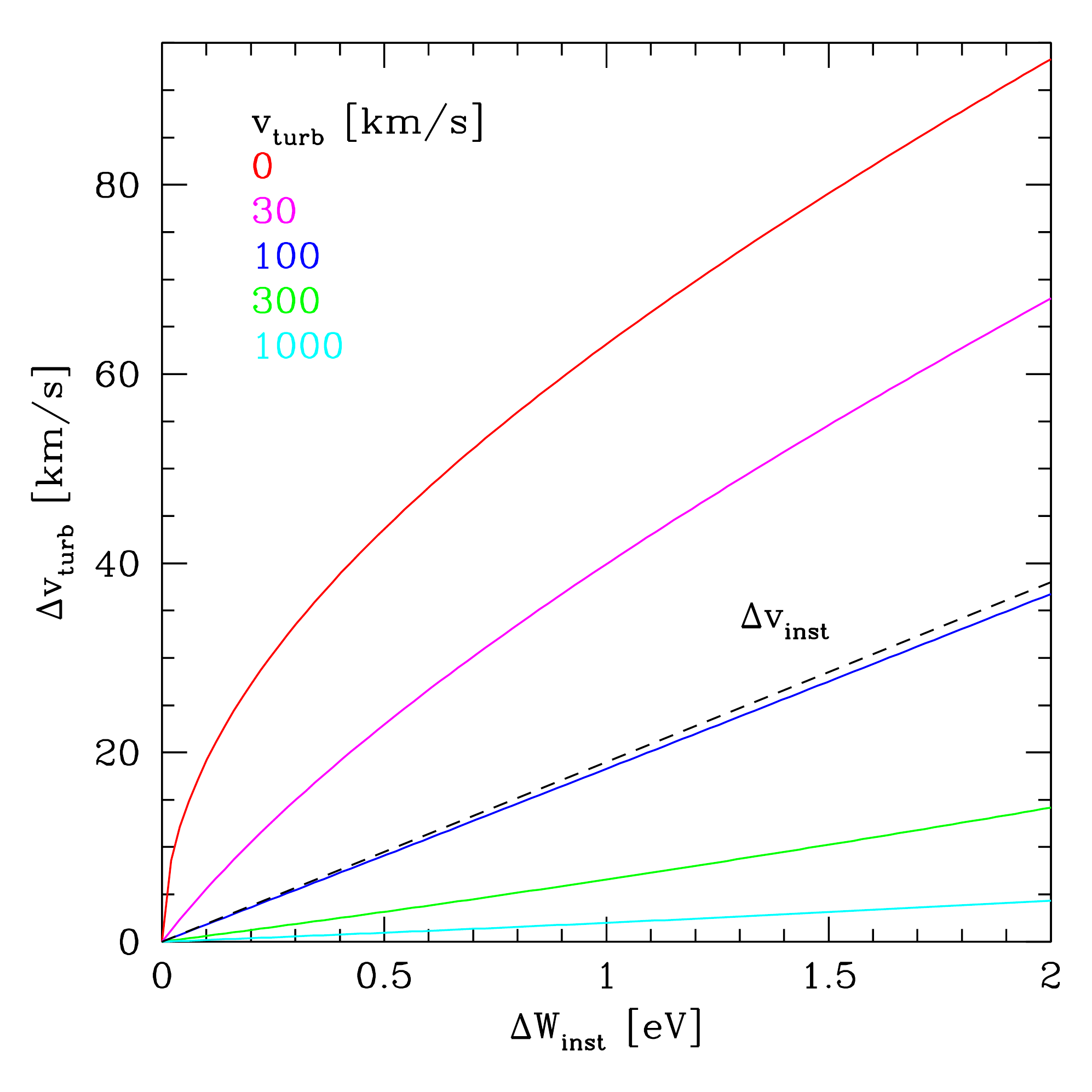}
\includegraphics[width=7.8cm]{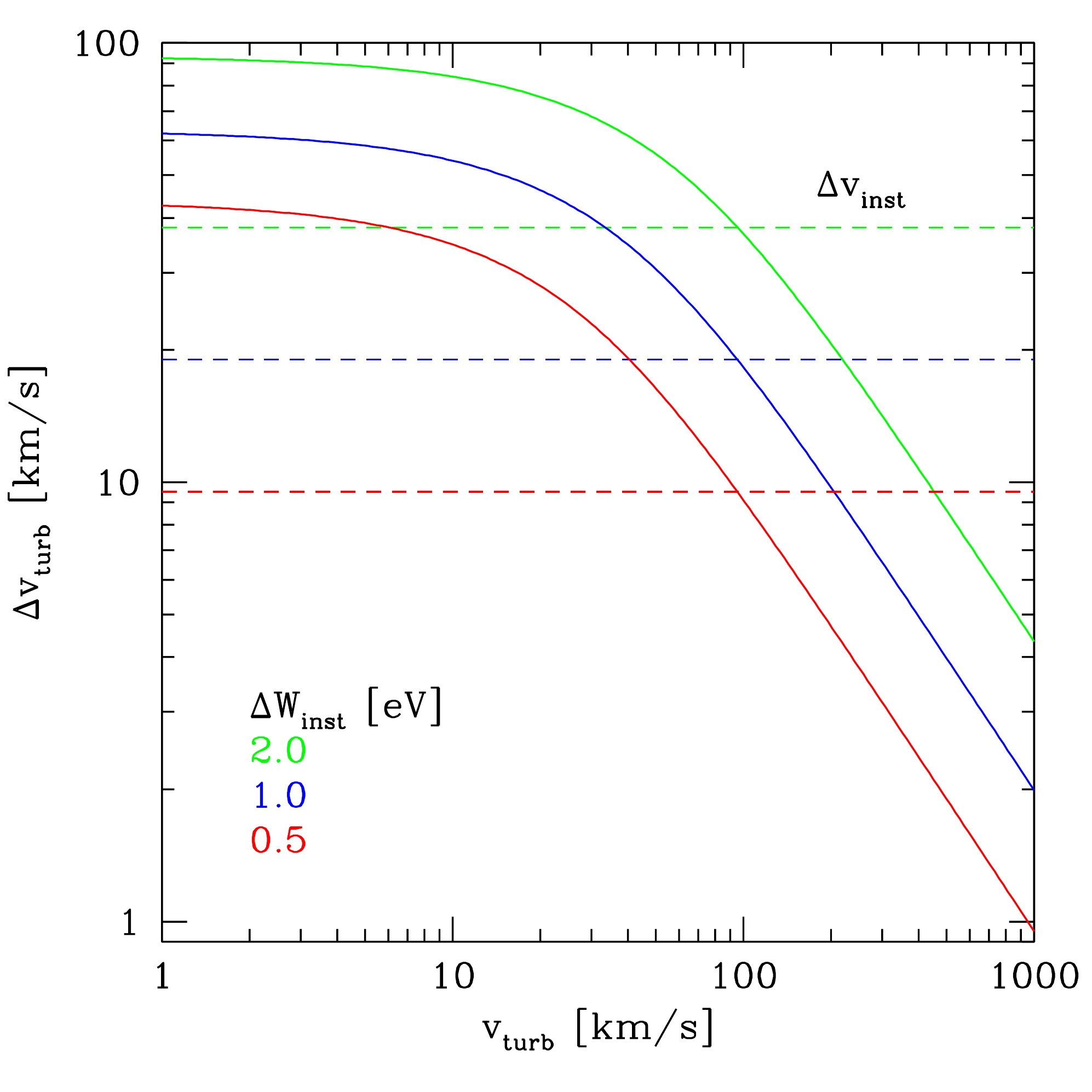} 
\caption{ $\Delta v_{\rm turb}$ originating from instrumental broadening
(eq.~[\ref{eq-deltawt}]) as a function of $\Delta W_{\rm inst}$ (left
panel) and $v_{\rm turb}$ (right panel) in the case of $W_{\rm inst}=5$
eV and $E_{\rm obs}=6.7$ keV.  For reference, dashed lines show $\Delta
v_{\rm inst}$ defined by equation (\ref{eq-dvi}). The color assignment
is indicated in the panels.}  \label{fig-vlimit}
\end{figure}

Figure \ref{fig-vlimit} illustrates $\Delta v_{\rm turb}$ given by
equations (\ref{eq-deltawt}) and (\ref{eq-w2v}). At large $v_{\rm
turb}$, $\Delta v_{\rm turb}$ is nearly proportional to $\Delta W_{\rm
inst}/v_{\rm turb}$ in agreement with the error propagation law of
equation (\ref{eq-prop}). At the smallest $v_{\rm turb}$, on the other
hand, $\Delta v_{\rm turb}$ does not diverge and is bounded by equation
(\ref{eq-deltavt1}). Equation (\ref{eq-deltavt1}) corresponds to the
``tightest upper limit on $v_{\rm turb}$'' that can be inferred in the
absence of other errors. In any case, it should be distinguished from
\begin{eqnarray}
\Delta v_{\rm inst} &\equiv & 
\frac{c}{\sqrt{8 \ln 2}}
\frac{\Delta W_{\rm inst}}{E_{\rm obs}}
= 19  ~{\rm km/s}
\left(\frac{\Delta W_{\rm inst}}{{\rm eV}}\right)
\left(\frac{E_{\rm obs}}{6.7 {\rm keV}}  \right)^{-1}.
\label{eq-dvi}
\end{eqnarray}

\subsubsection{Thermal broadening}

Similarly, $\Delta W_{\rm therm}$ contributes to $\Delta W_{\rm turb}$
as
\begin{eqnarray}
\Delta W_{\rm turb} &=& 
- W_{\rm turb} + \sqrt{W_{\rm turb}^2 
+ \Delta (W_{\rm therm}^2) }, 
\label{eq-deltawtb} 
\end{eqnarray}
where, from equation (\ref{eq-wtion}),  
\begin{eqnarray}
\Delta (W_{\rm therm}^2) &=& 4.7 ~ {\rm eV}^2 
\left(\frac{k\Delta T_{\rm ion}}{\rm keV} \right)
\left(\frac{m_{\rm ion}}{56 m_{\rm p}} \right)^{-1}
\left(\frac{E_{\rm obs}}{6.7 {\rm keV}} \right)^2.
\label{eq-dtherm}
\end{eqnarray}
Equation (\ref{eq-deltawtb}) depends on $\Delta T_{\rm ion}$ but not on
$T_{\rm ion}$. Its asymptotes are: 
\begin{enumerate}
\item For $\Delta W_{\rm turb} \ll W_{\rm turb}$, 
\begin{eqnarray}
\Delta W_{\rm turb} &\simeq & 
\frac{1}{2} \frac{\Delta (W_{\rm therm}^2)}{W_{\rm turb}}
, ~~~~~ \mbox{(in agreement with eq.~[\ref{eq-prop}])} \nonumber \\  &= & 
0.45  ~ {\rm eV} 
\left(\frac{v_{\rm turb}}{100 {\rm km/s}} \right)^{-1}
\left(\frac{k\Delta T_{\rm ion}}{\rm keV} \right)
\left(\frac{m_{\rm ion}}{56 m_{\rm p}} \right)^{-1}
\left(\frac{E_{\rm obs}}{6.7 {\rm keV}} \right), \\ 
\Delta v_{\rm turb} 
&\simeq & 
8.6 ~{\rm km/s}   
\left(\frac{v_{\rm turb}}{100 {\rm km/s}} \right)^{-1}
\left(\frac{k\Delta T_{\rm ion}}{\rm keV} \right)
\left(\frac{m_{\rm ion}}{56 m_{\rm p}} \right)^{-1},
\label{eq-deltavt2b}
\end{eqnarray}
\item  For $\Delta W_{\rm turb} \gg W_{\rm turb}$, 
\begin{eqnarray}
\Delta W_{\rm turb}&\simeq&  \sqrt{\Delta (W_{\rm therm}^2)}, 
~~~~~ \mbox{(independent of $v_{\rm turb}$)} \nonumber \\&=& 
2.2 ~ {\rm eV} 
\left(\frac{k\Delta T_{\rm ion}}{\rm keV} \right)^{1/2}
\left(\frac{m_{\rm ion}}{56 m_{\rm p}} \right)^{-1/2}
\left(\frac{E_{\rm obs}}{6.7 {\rm keV}} \right),
\label{eq-wcritt} \\
\Delta v_{\rm turb} 
&\simeq& 
41 ~{\rm km/s}   
\left(\frac{k\Delta T_{\rm ion}}{\rm keV} \right)^{1/2}
\left(\frac{m_{\rm ion}}{56 m_{\rm p}} \right)^{-1/2}.
\label{eq-deltavt1b}
\end{eqnarray}
\end{enumerate}

\subsubsection{Combined error}

Since uncertainties in the instrumental and thermal widths are in
principle independent of each other, their contribution to $\Delta
(W_{\rm turb}^2)$ can be added in quadrature:
\begin{eqnarray}
\left\{
\Delta (W_{\rm turb}^2) \right\}^2 &=& 
\left\{\Delta(W_{\rm inst}^2) \right\}^2 
+ \left\{
\Delta(W_{\rm therm}^2) \right\}^2+ \cdots, \\
\label{eq-deltawtc}
\rightarrow ~ \Delta W_{\rm turb} &=& -W_{\rm turb}
+\sqrt{W_{\rm turb}^2 + 
\left[\left\{\Delta(W_{\rm inst}^2) \right\}^2 
+ \left\{
\Delta(W_{\rm therm}^2) \right\}^2+ \cdots \right]^{1/2}}, \\
\label{eq-deltawtc0}
&\simeq & \left\{\begin{array}{ll}
\frac{\left[\left\{\Delta(W_{\rm inst}^2) \right\}^2 
+ \left\{\Delta(W_{\rm therm}^2) \right\}^2+ \cdots \right]^{1/2}
}{2W_{\rm turb}}, & ~ (\Delta W_{\rm turb} \ll W_{\rm turb})\\
 \left[\left\{\Delta(W_{\rm inst}^2) \right\}^2 
+ \left\{\Delta(W_{\rm therm}^2) \right\}^2+ \cdots \right]^{1/4},& 
~ (\Delta W_{\rm turb} \gg  W_{\rm turb})
\end{array}
\right. 
\end{eqnarray}
where $\Delta (W_{\rm turb}^2)$ and $\Delta(W^2_{\rm inst})$ are given
by equation (\ref{eq-quad}), $\Delta(W^2_{\rm therm})$ by equation
(\ref{eq-dtherm}), and the dots represent the other random uncertainties
including statistical errors.

\begin{figure}[t]
\centering \includegraphics[width=7.8cm]{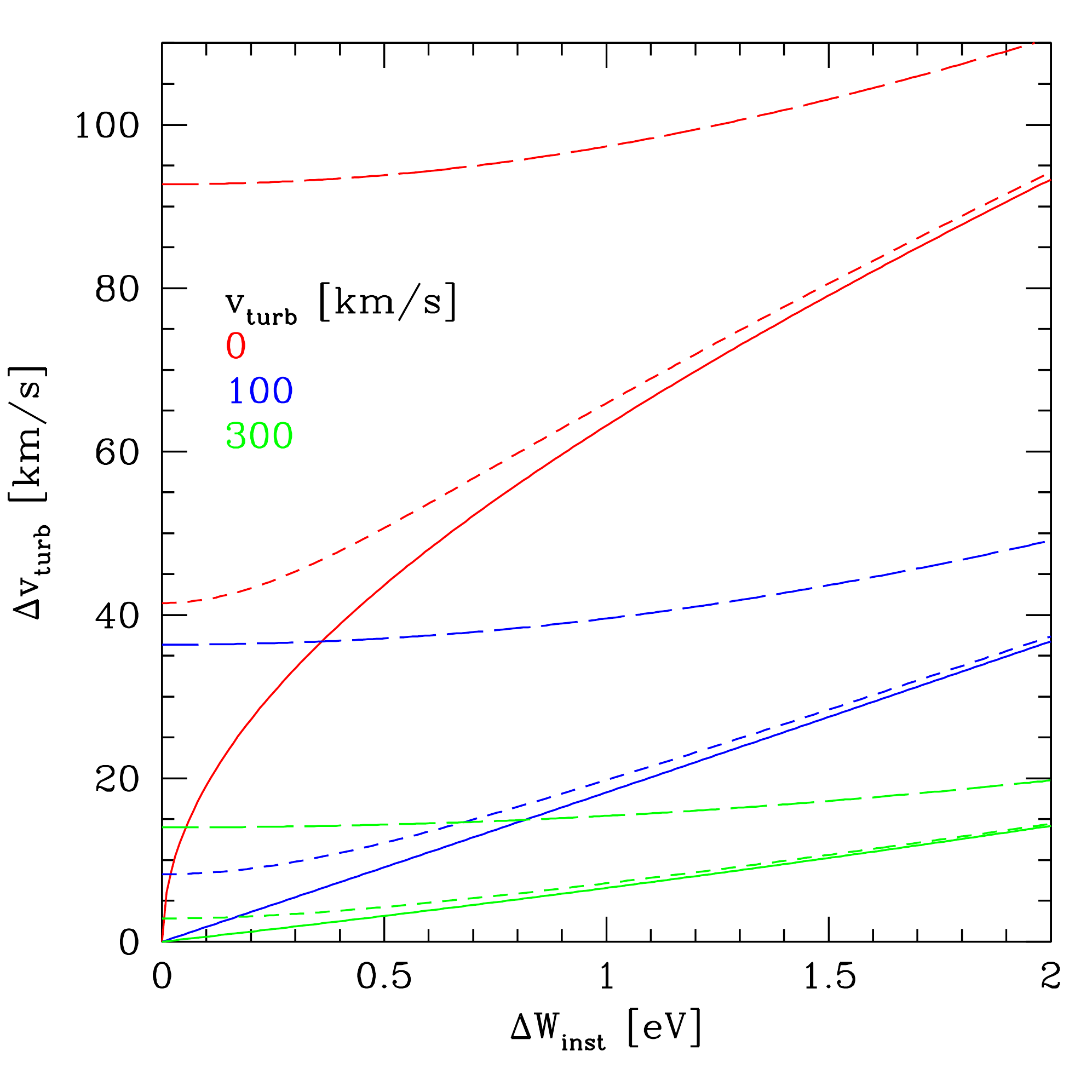}
\includegraphics[width=7.8cm]{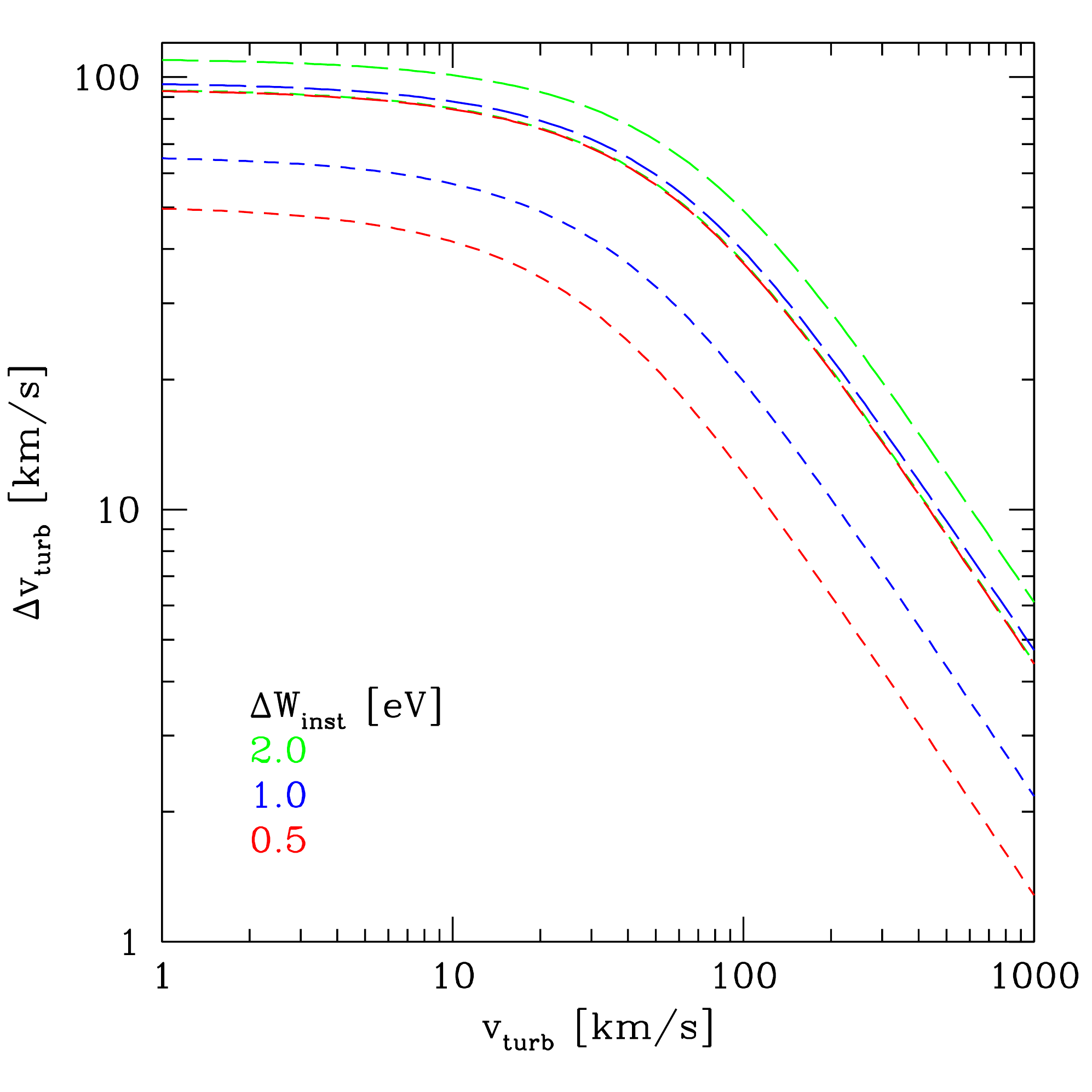}
 \caption{Same
as Figure \ref{fig-vlimit} except for adding uncertainties in thermal
broadening (eq. [\ref{eq-deltawtc}]) with $k\Delta T_{\rm ion}=5$ keV
(long dashed), 1 keV (short dashed), and 0 (solid; omitted for clarity
in the right panel since the results are identical to
Figure~\ref{fig-vlimit}).}  \label{fig-vlimit2}
\end{figure}

Figure \ref{fig-vlimit2} shows $\Delta v_{\rm turb}$ given by equations
(\ref{eq-deltawtc}) and (\ref{eq-w2v}).  It follows that $\Delta v_{\rm
turb}$ increases with decreasing $v_{\rm turb}$ and hampers the
measurement of $v_{\rm turb}$ less than $\sim$70 km/s, if $W_{\rm inst}
\simeq 5$ eV, $\Delta W_{\rm inst} \simeq 1$ eV, and $k\Delta T_{\rm
ion} \simeq 1$ keV.

\bigskip

\clearpage
\begin{multicols}{2}
\end{multicols}

\end{document}